\documentclass[useAMS]{mn2e}
\usepackage[tight]{subfigure}
\usepackage{longtable}
\usepackage{lscape}
\usepackage{graphicx}
\usepackage{amssymb}

\title[Stellar populations in BCGs]{Stellar populations in the centres of brightest cluster galaxies}
\author[Loubser et al.]{S. I. Loubser$^{1,2}$\thanks{E-mail:
siloubser@uclan.ac.uk (SIL)}, P. S\'{a}nchez-Bl\'{a}zquez$^{1,3}$, A. E. Sansom$^{1}$, I. K.
Soechting$^{4}$\\
$^{1}$Jeremiah Horrocks Institute for Astrophysics and Supercomputing, University of Central Lancashire, Preston, PR1 2HE, UK\\
$^{2}$Physics Department, University of the Western Cape, Cape Town 7535, South Africa\\
$^{3}$Instituto de Astrofisica de Canarias, Via Lactea S/N, 38205, La Laguna, Tenerife, Spain\\
$^{4}$Oxford Astrophysics, Department of Physics, University of Oxford, Oxford, OX1 3RH, UK}
\begin{document}

\date{Accepted 2009 June 01. Received 2009 May 30; in original form 2009 January 02}

\pagerange{\pageref{firstpage}--\pageref{lastpage}} \pubyear{2009}

\maketitle

\label{firstpage}

\begin{abstract}
This paper is part of a series devoted to the study of the stellar populations in brightest cluster galaxies (BCGs), aimed at setting constraints on the formation and evolution of these objects. We have obtained high signal-to-noise ratio, long-slit spectra of 49 BCGs in the nearby Universe. Here, we derive Single Stellar Population (SSP)-equivalent ages, metallicities and $\alpha$-abundance ratios in the centres of the galaxies using the Lick/IDS system of absorption line indices. We systematically compare the indices and derived parameters for the BCGs with those of large samples of ordinary elliptical galaxies in the same mass range. We find no significant differences between the index--velocity dispersion relations of the BCG data and those of normal ellipticals, but we do find subtle differences between the derived SSP-parameters. The BCGs show, on average, higher metallicity ([Z/H]) and $\alpha$-abundance ([E/Fe]) values. We analyse possible correlations between the derived parameters and the internal properties of the galaxies (velocity dispersion, rotation, luminosity) and those of the host clusters (density, mass, distance from BCG to X-ray peak, presence of cooling flows), with the aim of dissentangling if the BCG properties are more influenced by their internal or host cluster properties. The SSP-parameters show very little dependence on the mass or luminosity of the galaxies, or the mass or density of the host clusters. Of this sample, 26 per cent show luminosity-weighted ages younger than 6 Gyr, probably a consequence of recent -- if small -- episodes of star formation. In agreement with previous studies, the BCGs with intermediate ages tend to be found in cooling-flow clusters with large X-ray excess. 
\end{abstract}


\section{Introduction}

The assembly history of the most massive galaxies in the Universe and the influence of the cluster environment are very important, but poorly understood, aspects of galaxy formation. The most direct route to investigate the evolution of early-type galaxies is to observe them at different redshifts. Unfortunately, it is very difficult to find the progenitors of early-type galaxies by direct observations, and this method also demands large amounts of observing time even on the current generation of telescopes. An alternative approach is to infer the star formation histories (SFHs) of large samples of nearby galaxies by studying their stellar population properties and the relationships between these and the structural and kinematic properties of the galaxies.

In the context of the, now widely accepted, $\Lambda$CDM model of structure formation, Dubinski (1998) showed that a central cluster galaxy forms naturally when a cluster collapses along the filaments. Gao et al.\ (2004), using numerical simulations, predicted that central galaxies in clusters experienced a significant number of mergers since $z \sim 1$. De Lucia $\&$ Blaizot (2007) provided a complete quantitative estimate of the formation of BCGs using the Millennium simulation. They predicted that the stars that will end up in a BCG formed at high redshift (with 50 per cent of the stars already formed $\sim$ 12.5 Gyr ago). However, the BCG continues to assemble at much lower redshifts (with 50 per cent of the mass assembling after  $z\sim0.5$). The nature of these mergers (or accretion) is dissipationless and, therefore, no new stars are formed in the process. Thus, according to these simulations, we expect to see evidence of dissipationless mergers, little dependence of metallicity on mass, and old stellar populations in these galaxies. 

Because of their position in the cluster, the mergers forming BCGs are expected to be with preferentially radial orbits. Boylan-Kolchin, Ma $\&$ Quataert (2006) showed that these types of mergers, in absence of dissipation, create systems that depart from the Faber--Jackson relationship in the same way as BCGs do (Tonry 1984; Oegerle $\&$ Hoessel 1991; Bernardi et al.\ 2007; Desroches et al.\ 2007; Lauer et al.\ 2007; Von der Linder et al.\ 2007). Observations that support BCG formation by predominantly radial mergers include those of Bernardi et al.\ (2008), who showed that the shapes of the most massive ($\sigma \geq 350$ km s$^{-1}$), high luminosity objects (with properties similar to BCG properties) are consistent with those expected if the objects formed through radial mergers. Observations of the luminosity functions of clusters put additional constraints on BCG evolution. Loh $\&$ Strauss (2006) found the luminosity gap between the first and second brightness-ranked galaxies to be large ($\sim$ 0.8 mag), larger than could be explained by an exponentially decaying luminosity function of galaxies. Loh $\&$ Strauss (2006) found that the large luminosity gap showed little evolution with redshift since $z = 0.4$, and suggest that the BCGs must have become the dominant cluster members by $z > 0.4$.

Various photometric studies have found correlations between the luminosity of the BCG and the mass or density of the host cluster, measured through the cluster velocity dispersion and X-ray luminosity (e.g.\ Edge $\&$ Stewart 1991; Burke, Collins $\&$ Mann 2000; Brough et al.\ 2002; Whiley et al.\ 2008; Stott et al.\ 2008). Taken together, these observations point to an evolutionary history of BCGs which are closely connected to the evolution of the cluster, as predicted by the hierarchical models of galaxy formation, although it is not clear if the amount of galaxy growth due to accretion agrees with these models (e.g.\ Whiley et al.\ 2008).

The mass growth of BCGs with redshift have been well-studied. Arag\'on-Salamanca, Baugh $\&$ Kauffmann (1998) investigated the evolution of $K$-magnitudes since z $\sim$ 1 and found the lack of evolution in the magnitudes compatible with mass growth of a factor of two to four since z $\sim$ 1. However, other studies have found a more modest stellar mass growth of BCGs since those redshifts (e.g.\ Collins $\&$ Mann 1998; Whiley et al.\ 2008), although it is known that the amount of mass growth in BCGs depends on the luminosity of the host clusters (Burke et al.\ 2000; Brough et al.\ 2002). No significant BCG stellar mass growth is observed in the most X-ray luminous clusters (L$_{\rm X}>1.9\times 10^{44}$ erg s$^{-1}$) since $z\sim$ 1, whereas BCGs in less X-ray luminous clusters experience an increase in their mass by a factor of four in that redshift range (Brough et al.\ 2002). However, more recent studies using the velocity dispersion of the cluster instead of X-ray luminosity, have not confirmed this trend (Whiley et al.\ 2008).

The contradictory results might be partially due to the difficulty in the sample selection and in the comparison between aperture magnitudes (observations) and total magnitudes (models; e.g.\ Whiley et al.\ 2008). Aperture magnitudes used so far in the literature include less than 50 per cent of the total mass of the BCGs. Furthermore, it is not clear if  intracluster light should be considered when comparing with the models (Gonz\'{a}lez, Zaritsky $\&$ Zabludoff 2007).

An alternative to measuring the BCG mass growth with redshift to study their evolution, is to analyse their dynamical, structural and stellar population properties, and investigate if these are compatible with that expected from remnants of multiple dry mergers over a large redshift range. 

Cooling flows are very common at low redshifts (Chen et al.\ 2007; Edwards et al.\ 2007), but their role in shaping the stellar populations of BCGs is not fully understood. The lack of widespread detection of iron lines, expected from cluster gas cooling below 1 -- 2 keV in \textit{XMM-Newton} observations of cool-core clusters, contradicted the model that BCG formation is a consequence of cooling flows (Jord\'an et al.\ 2004; see also discussion in Loubser et al.\ 2008, hereafter Paper 1). However, it is possible that star formation is ongoing in cool-core clusters at a much reduced rate (Bildfell et al.\ 2008). Several studies reported examples of recent or ongoing star formation in BCGs hosted by cooling-flow clusters (Cardiel, Gorgas $\&$ Arag\'{o}n-Salamanca 1998a; Crawford et al.\ 1999; McNamara et al.\ 2006; Edwards et al.\ 2007; O'Dea et al.\ 2008; Bildfell et al.\ 2008; Pipino et al.\ 2009). However, the origin of the gas fuelling the recent star formation in some BCGs is not yet known. The competing explanations include cooling flows, or cold gas deposited during a merging event (Bildfell et al.\ 2008). 

This paper is the second in a series of papers investigating a new, large sample of BCGs, their kinematic and stellar population properties and the relationships between these and the properties of the host clusters. The first paper was devoted to the spatially resolved kinematics of 41 BCGs (Paper 1). Here, we measure and interpret BCG spectral line strengths to gain insights into their stellar populations. The stellar populations in early-type galaxies have been studied by numerous authors (Gonz\'alez 1993; Fisher, Franx $\&$ Illingworth 1995; J\o{}rgensen 1999; Mehlert et al.\ 2000; Trager et al.\ 2000a,b; Kuntschner et al.\ 2001; Moore et al.\ 2002; Caldwell, Rose $\&$ Colcannon 2003; Nelan et al.\ 2005; Thomas et al.\ 2005; S\'{a}nchez-Bl\'{a}zquez et al.\ 2006a, 2006b, 2006c; Trager, Faber $\&$ Dressler 2008; Ogando et al.\ 2008 and many others). However, as discussed in detail in Paper 1, very little is known about the stellar population properties of BCGs. Recently, Von der Linden et al.\ (2007) carried out a study of 625 brightest group and cluster galaxies, taken from the SDSS, to contrast their stellar population properties with those of elliptical galaxies with the same mass. They found that stellar populations of BCGs are not different from the stellar populations of ordinary elliptical galaxies, except for the $\alpha$-enhancement, which is higher in BCGs. Brough et al.\ (2007), with a much smaller sample, did not find this difference.

The study of Von der Linden et al.\ (2007) constitutes a benchmark in the study of stellar populations in BCGs. However, they did not have spatial information. The merger history of a galaxy determines the kinematical and stellar population properties and these can, therefore, be used as a probe for the assembly history of those galaxies. Brough et al.\ (2007) showed that BCGs present a large spread in their metallicity gradients, probably reflecting differences in their assembly history and the dissipation during the interactions. In Paper 1 we showed that our sample of BCGs shows great variety in the galaxies' dynamical and kinematic properties (see also Brough et al.\ 2007).

In this paper, we concentrate on the central properties of these galaxies. The stellar population gradients and reconstructed SFHs will be investigated in future papers in the series. This paper is structured as follows: Section 2 contains the details of the sample selection, observations and data reduction. Section 3 contains the central index measurements and their relations with the velocity dispersions of the BCGs. The single stellar populations (SSPs) are derived, and compared with those of ordinary elliptical galaxies in Section 4. Section 5 shows the relations of the derived properties with the galaxy kinematics, and Section 6 details the context of the cluster environment. The conclusions are summarised in Section 7.

\section{Sample, observations and data reduction}
This study was initially intended to investigate a subsample of BCGs with extended haloes (cD galaxies). However, due to the difficulties in the classification of cD galaxies and the very inhomogeneous definitions in the literature, we cannot be confident that all the galaxies in our sample are cD galaxies. Instead, we can say that our sample comprises of the dominant galaxies closest to the X-ray peaks in the centres of clusters. For consistency, we call these galaxies BCGs to comply with recent literature (for example De Lucia $\&$ Blaizot 2007; Von der Linden et al.\ 2007). For a small fraction of clusters, the BCG might not strictly be the brightest galaxy in the cluster, but they are always the galaxy closer to the X-ray peak. The sample selection, observations and data reduction procedures were detailed in Paper 1. In summary: these 49 galaxies were classified as cD either in NED (in the morphological classification or in the notes of previous observations) and/or have profiles breaking the $r^{\frac{1}{4}}$ law in the external parts. In addition, NGC4946 (an ordinary elliptical) and NGC6047 (an E/S0) were also observed with the same observational set-up, and are included in this project as control galaxies.

The galaxies were observed on the WHT and Gemini North and South telescopes. In addition to the 41 BCGs described in Paper 1, long-slit spectra were obtained for eight more BCGs with Gemini South in the 2007B (July 2007 to January 2008) observing semester. Thus, 49 BCGs were observed in total (for details of these galaxies see table 1, Paper 1). The instrumental set-up and data reduction procedure for the Gemini South 2007B observations were the same as used for the data taken in the previous semesters. The data reduction procedures are described in Paper 1 and will not be repeated here.

To analyse the central parts of the BCGs, we extracted spectra inside apertures of $a_{\rm e}/8$ along the slit. The effective half-light radius was calculated as
$a_{\rm e} = \frac{r_{\rm e}(1-\epsilon)}{1-\epsilon \ \mid cos(\mid PA - MA \mid)\mid}$, with $\epsilon$ the ellipticity (data from NED), $r_{\rm e}$ the radius containing half the light of the galaxy (computed from the 2MASS $K$-band 20th magnitude arcsec$^{-2}$ isophotal radius as described in Paper 1), PA the slit position axis, and MA the major axis. For old stellar populations, these half-light radii do not differ much from those derived using the optical bands (Jarrett et al.\ 2003). The central values are for an aperture of $1 \times a_{\rm e}/8$ arcsec$^{2}$ for the WHT data and $0.5 \times a_{\rm e}/8$ arcsec$^{2}$ for the Gemini data, as the slit widths were 1 and 0.5 arcsec, respectively. The signal-to-noise ratios per \AA{} around the H$\beta$ region of the central spectra ranged between 16 (ESO541-013) and 502 (NGC1399), with an average of 87. The central kinematics used in the present paper are taken from our BCG kinematic study in Paper 1 (table 4). 

\subsection{Transformation to the Lick/IDS system}

A widely used set of spectral absorption indices is the Lick system based on a large survey of individual stars in the solar neighbourhood carried out with the image dissecting scanner (IDS) at the Lick Observatory (Burstein et al.\ 1984; Faber et al.\ 1985; Gorgas et al.\ 1993; Worthey et al.\ 1994). The original Lick system consisted of 21 indices from CN$_{1}$, at $\sim$ 4150 \AA{}, to TiO$_{2}$, at $\sim$ 6230 \AA{} (Faber et al.\ 1985; Worthey 1994; Trager et al.\ 1998). Worthey $\&$ Ottaviani (1997) later contributed four more indices centred on the Balmer lines H$\delta$ and H$\gamma$. This collection of 25 indices will be referred to as the Lick indices in this study. The advantages of using this set of indices are: they are well calibrated against globular clusters; the indices are not affected by dust (MacArthur 2005); and their sensitivity to different chemical elements have been calculated using model atmospheres (Tripicco $\&$ Bell 1995; Korn, Maraston $\&$ Thomas 2005).

Line strength indices depend on the broadening of lines caused by the velocity dispersion of the galaxies and the instrumental resolution. In order to use model predictions based on the Lick/IDS system, spectra need to be degraded to the wavelength dependent resolution of the Lick/IDS spectrograph and indices need to be corrected for the broadening caused by the velocity dispersion of the galaxies. This has been done using the prescriptions of Worthey $\&$ Ottaviani (1997). The detailed procedure follows in Appendix \ref{Appendix2}. 

\subsection{Emission correction}

The presence of emission lines leads to problems in analysing the absorption lines in stellar populations. Key absorption indices like H$\beta$, H$\gamma$ and H$\delta$ suffer from emission line in-filling. Fe5015 is also affected by [OIII]$\lambda$5007\footnote{The standard notation is used, where the spectral identification is written between two square brackets for forbidden lines.} emission, while Mg$_{\rm b}$ is affected by [NI]$\lambda$5199 emission. In the case of the Balmer lines, emission fill-in can weaken the line strength and lead to older derived ages. 

\begin{figure}
   \centering
   \includegraphics[scale=0.38]{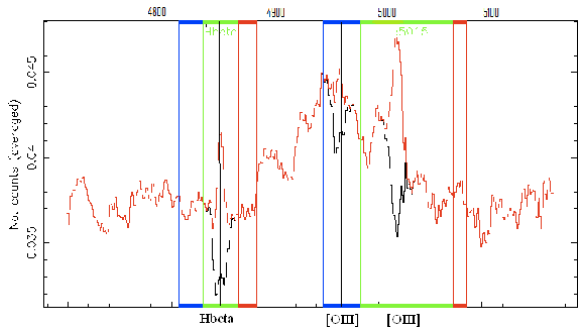}
   \caption[Illustration of emission line correction (NGC6166).]{Illustration of emission line correction of NGC6166. The original spectrum is plotted over the emission-corrected spectrum with the affected indices (H$\beta$ and Fe5015) and their sidebands indicated. The H$\beta$ and [OIII]$\lambda\lambda 4958,5007$ emission lines are also indicated by vertical lines.}
   \label{fig:Emission_NGC6166}
\end{figure}

To measure the emission-line fluxes of the BCGs in this study, the \textsc{gandalf} routine (Sarzi et al.\ 2006) was used. This software treats the emission lines as additional Gaussian templates, and solves linearly at each step for their amplitudes and the optimal combination of stellar templates, which are convolved by the best stellar line-of-sight velocity distribution. The stellar continuum and emission lines are fitted simultaneously. The stellar templates used were based on the MILES stellar library (S\'{a}nchez-Bl\'{a}zquez et al.\ 2006d). The [OIII]$\lambda$5007 line was fitted first. Where H$\beta$ emission was relatively weak, the kinematics of all the other lines were tied to [OIII]$\lambda$5007, following the procedure described in Sarzi et al.\ (2006). This was done to avoid any spurious detections of H$\beta$ lines that might have been caused by the presence of a number of metal features around 4870 \AA{}. However, in cases were H$\beta$ was strong enough to measure its kinematics, this was calculated independently as there is no a priori reason to expect the kinematics measured from the [OIII]$\lambda$5007 and H$\beta$ lines to be the same (as they can originate in different regions). This procedure was used in order to derive the best-fitting emission-line spectrum in the galaxies where emission was detected, and enables us to derive a purely-stellar spectrum for these galaxies. The spectra of ESO349-010, MCG-02-12-039, NGC0541, NGC1713, NGC3311, NGC4874, NGC4946, NGC6166, NGC6173, NGC7012, NGC7649, NGC7720 and PGC044257 have detectable emission lines. For these galaxies, a purely-stellar spectrum was derived by subtracting the best-fitting emission-line spectrum from the observed one. Figure \ref{fig:Emission_NGC6166} shows the H$\beta$ region of the spectrum of NGC6166 before and after the emission line correction. 

\begin{figure}
   \centering
   \includegraphics[scale=0.35]{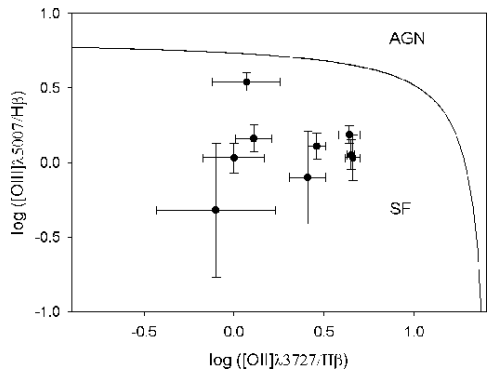}
   \caption[Diagnostic diagram separating star forming galaxies from AGN.]{Diagnostic diagram using emission lines to separate star forming galaxies from AGN. Star forming galaxies are below the line (given by Lamareille et al.\ 2004).}
   \label{fig:Emission}
\end{figure}

Emission lines originate from the hot ionised gas in the galaxies which can indicate an active galaxy (such as AGN, LINER) or active star formation in the galaxy. We could not measure [NII]$\lambda$6584 or H$\alpha$, as they were outside our wavelength range. Therefore, we use the diagnostic diagram [OIII]$\lambda$5007/H$\beta$ against [OII]$\lambda$3727/H$\beta$ to separate the two major origins of emission: star formation and AGN (Baldwin, Phillips $\&$ Terlevich 1981; Lamareille et al.\ 2004). Figure \ref{fig:Emission} shows the galaxies for which the H$\beta$, [OII]$\lambda$3727 and [OIII]$\lambda$5007 lines could be measured (within at least 2$\sigma$ detections) on the diagnostic plot. 

All nine emission-line galaxies in the present work for which these three lines could be measured should be star forming galaxies according to this test. BCGs are known to be more likely to host radio-loud AGN than any other galaxy with the same stellar mass (Burns 1990; Best et al.\ 2007). This is especially true if they are hosted by cooling-flow clusters (Burns et al.\ 1997). However, Best et al.\ (2007) argue that radio-loud AGN and emission-line AGN (detectable through optical emission lines) are independent, unrelated phenomena. Edwards et al.\ (2007) found, in their sample of BCGs, that the emission was mostly LINER-like or a combination of LINER and star formation, and concluded that this emission was directly related to the cooling of X-ray gas at the cluster centre. Von der Linden et al.\ (2007) found emission lines in more than 50 per cent of their sample, although the four lines that they used as a diagnostic could only be measured with a S/N $>$ 3 in 30 per cent of their sample. From this subsample, only six per cent were star forming galaxies, 70 per cent LINERs and 24 per cent composite objects (star forming and LINER). 

Because the diagnostic diagram used here in Figure \ref{fig:Emission} is much less effective at separating different sources of ionisation (Stasi\'{n}ska et al.\ 2006), and because the fraction of the current sample containing emission lines is relatively small, and with weak emission lines, it is too early to draw detailed conclusions about the nature of emission lines in BCGs.

\section{Central BCG indices} 

We measured line-strength indices from the flux-calibrated spectra and calculated the index errors according to the error equations presented in Cardiel et al.\ (1998b), Vazdekis $\&$ Arimoto (1999) and Cenarro et al.\ (2001). 

The wavelength coverage allowed us to measure all the Lick indices, with the exception of TiO$_{2}$ for most galaxies (and in a few cases also TiO$_{1}$) in the Gemini data and five indices of NGC6047 (Mg$_{1}$ to Fe5335). TiO$_{1}$ is very close to the edge of the spectrum and therefore not reliable. On very close inspection, five galaxies (Leda094683, NGC7649, NGC6166, PGC025714 and PGC030223) showed either residuals of sky line removal or residuals where the gaps in between the GMOS CCDs were inside the band definitions of the indices used to derive the SSP parameters, hence they were also excluded from any further analysis using those indices.

\begin{figure*}
   \centering
   \mbox{\subfigure{\includegraphics[width=4.5cm,height=4.5cm]{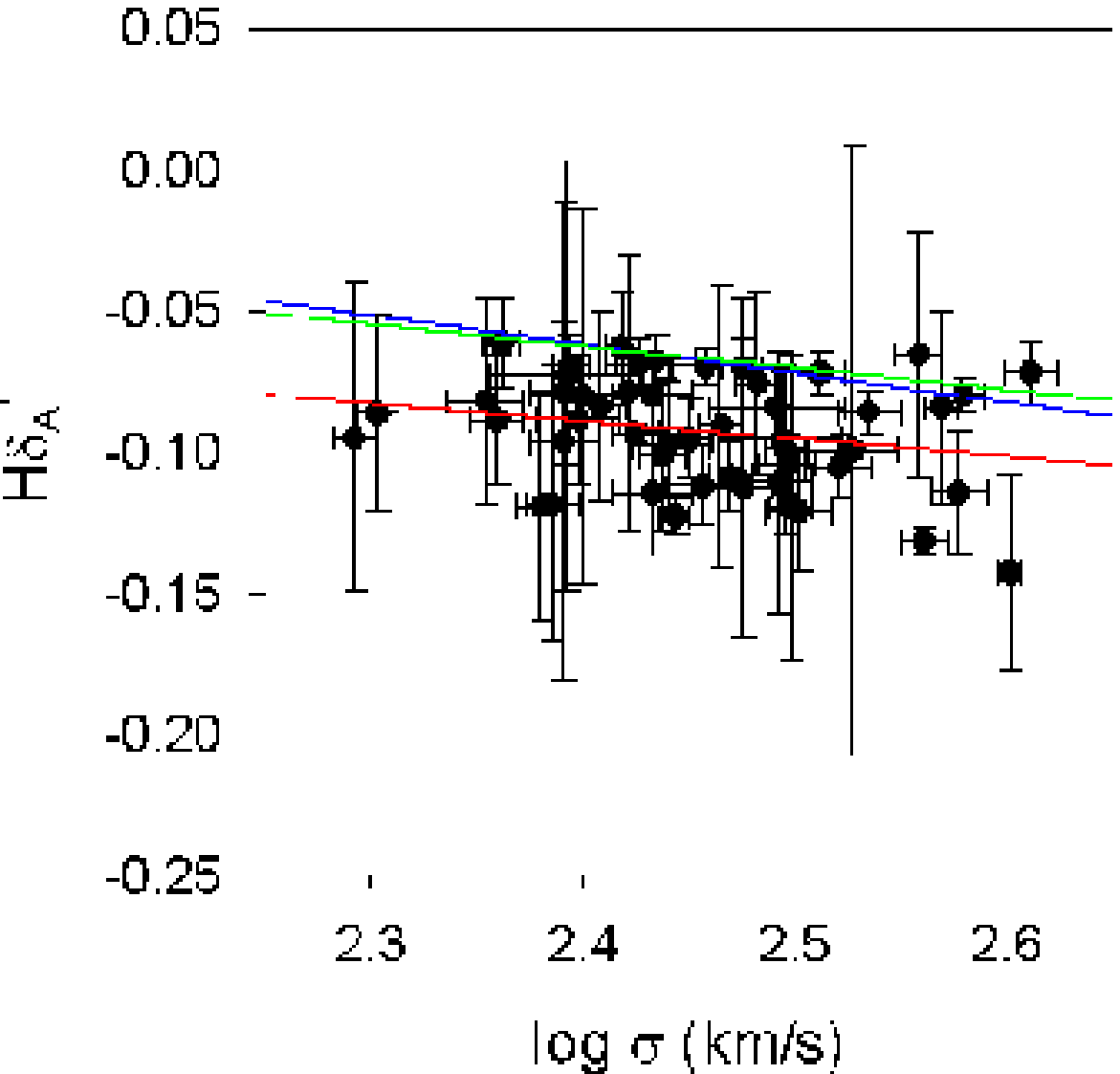}}\quad
         \subfigure{\includegraphics[width=4.5cm,height=4.5cm]{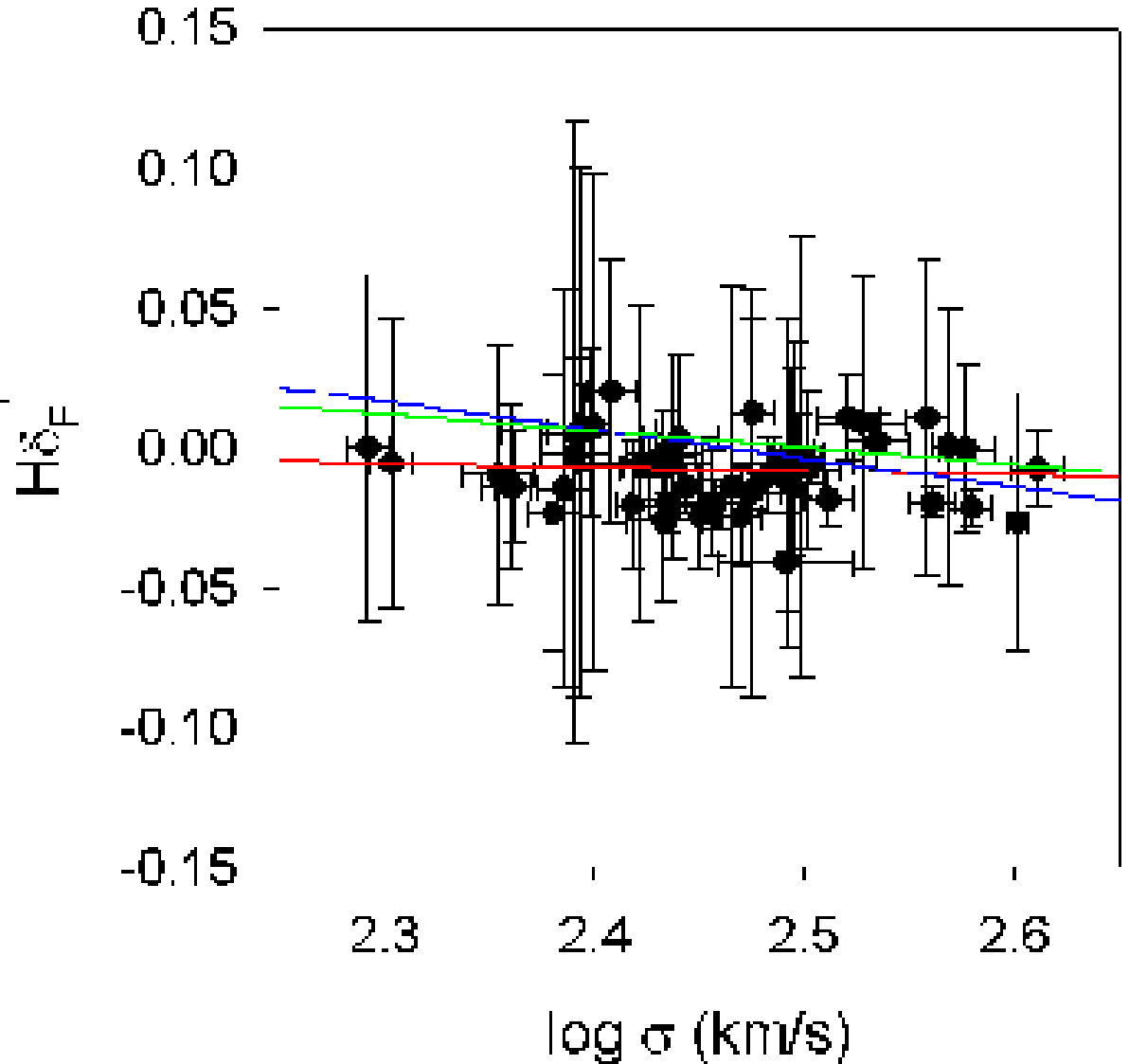}}\quad
         \subfigure{\includegraphics[width=4.5cm,height=4.5cm]{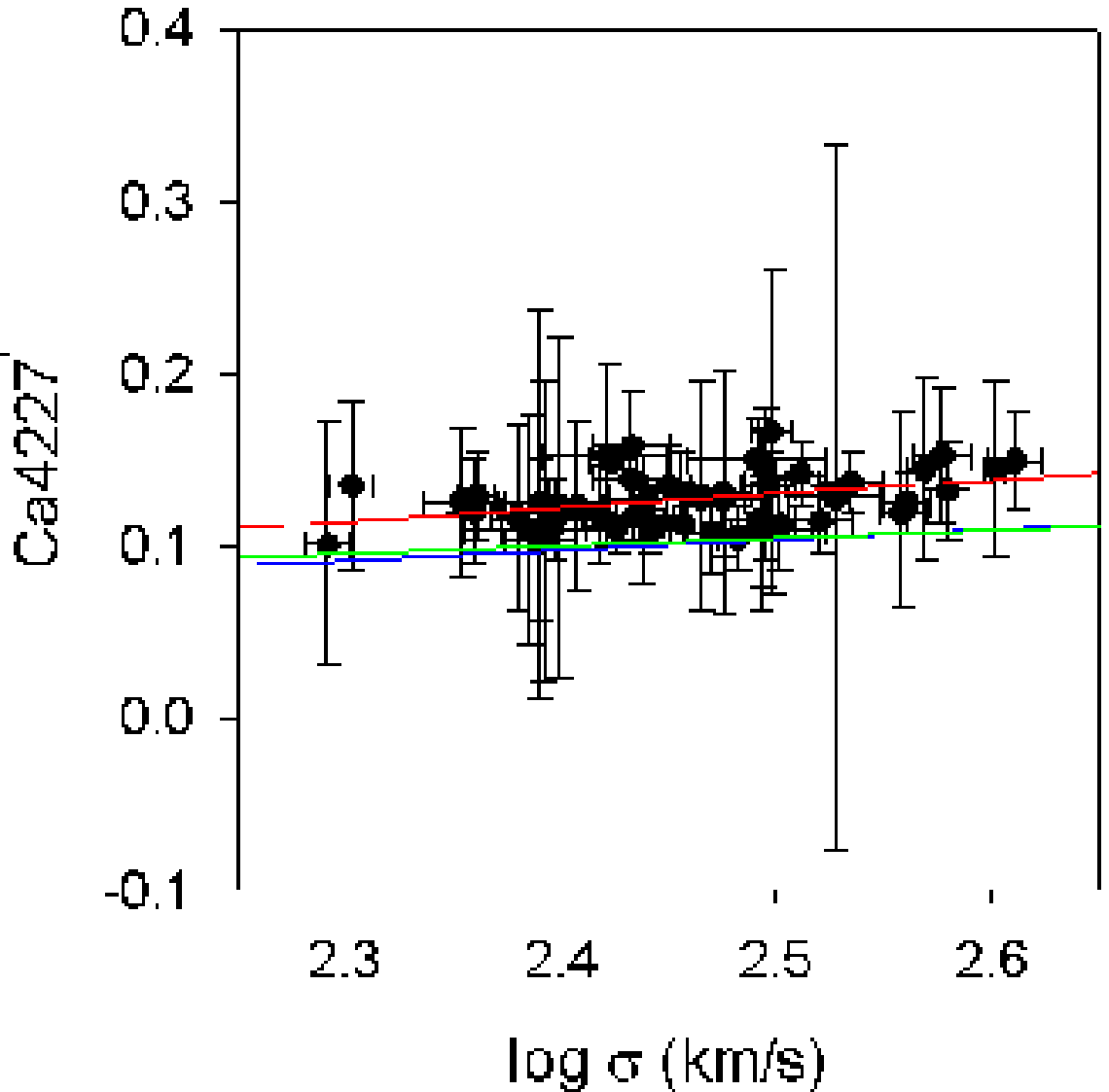}}}
\mbox{\subfigure{\includegraphics[width=4.5cm,height=4.5cm]{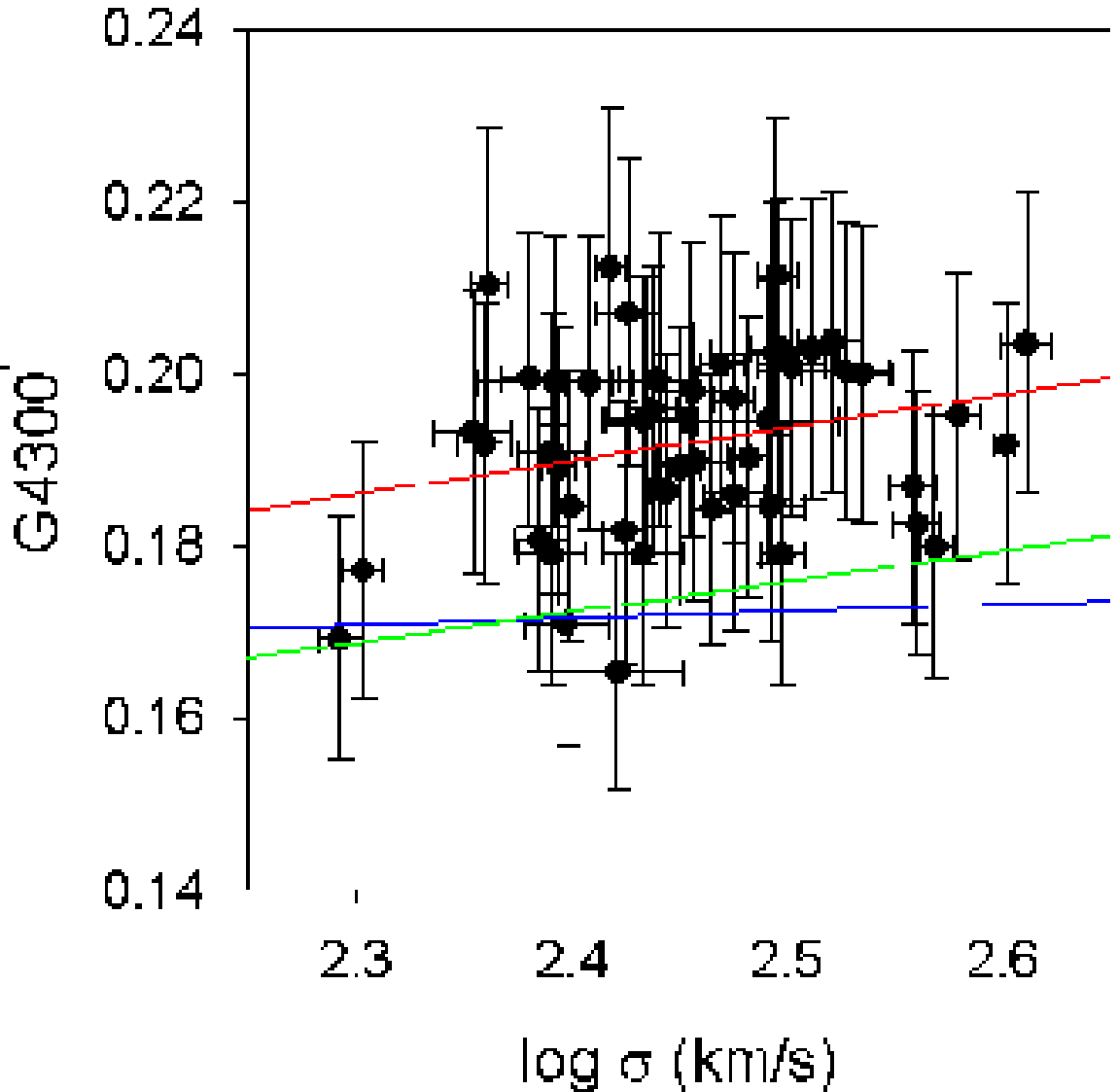}}\quad
         \subfigure{\includegraphics[width=4.5cm,height=4.5cm]{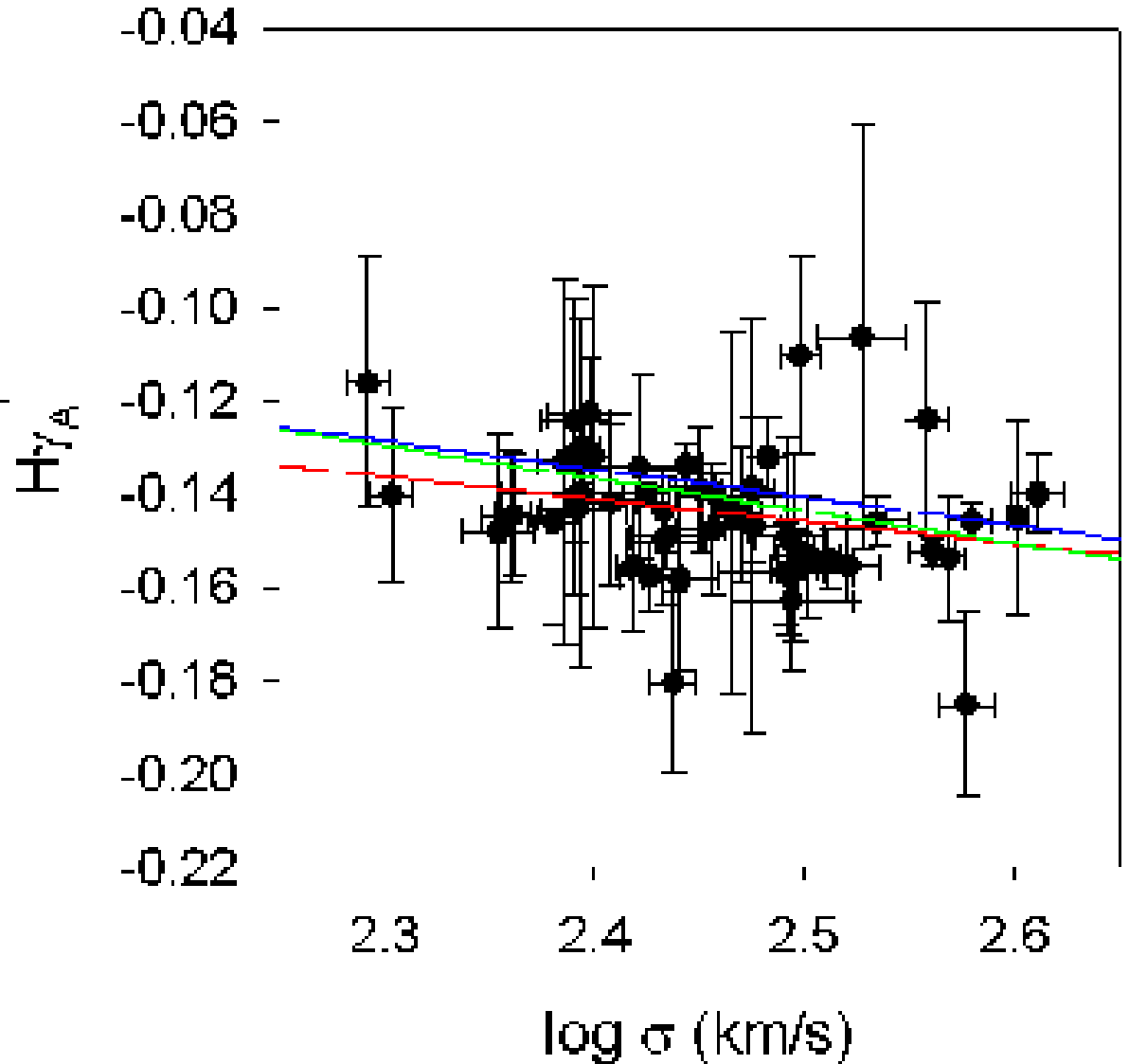}}\quad
         \subfigure{\includegraphics[width=4.5cm,height=4.5cm]{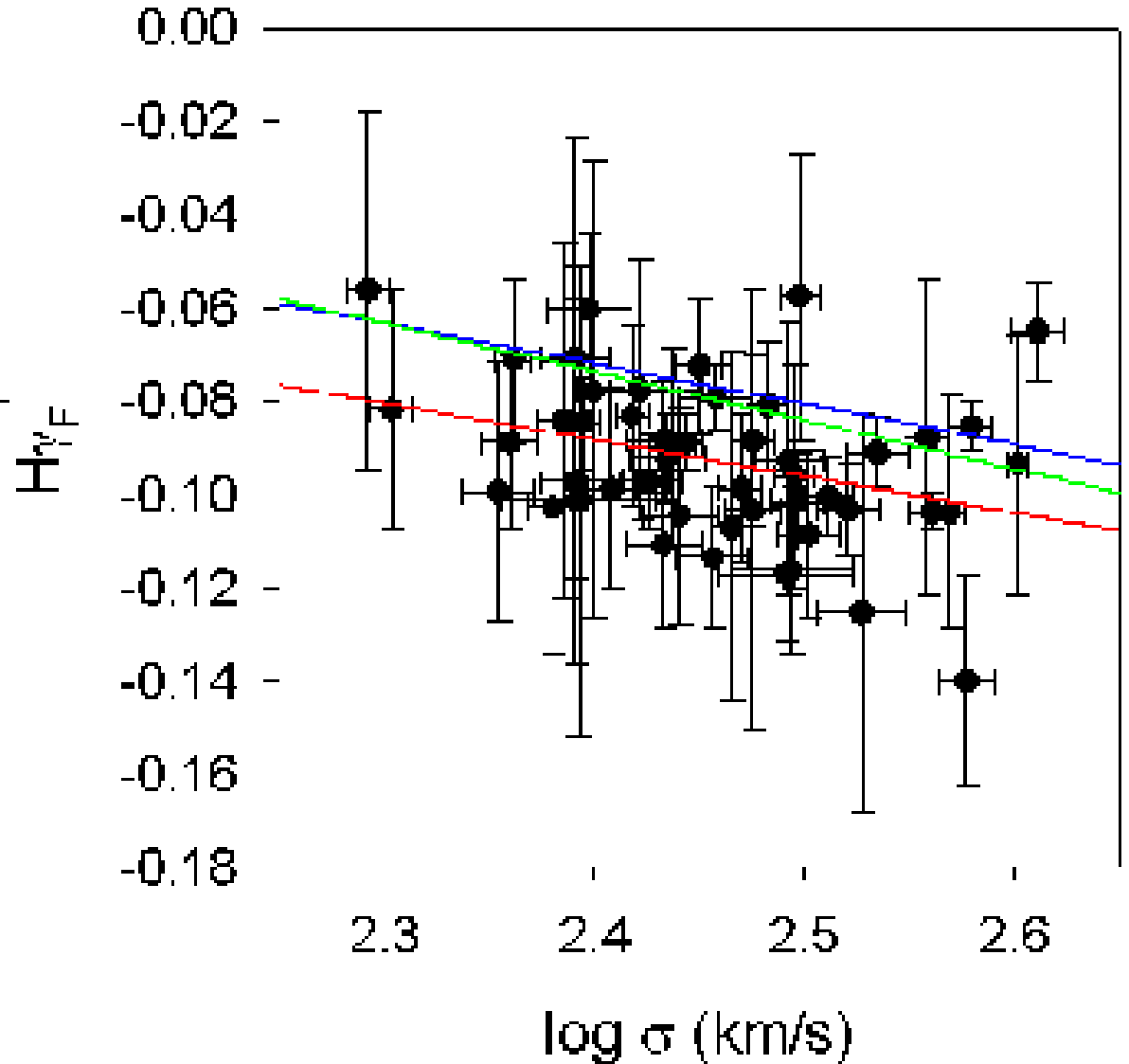}}}
   \mbox{\subfigure{\includegraphics[width=4.5cm,height=4.5cm]{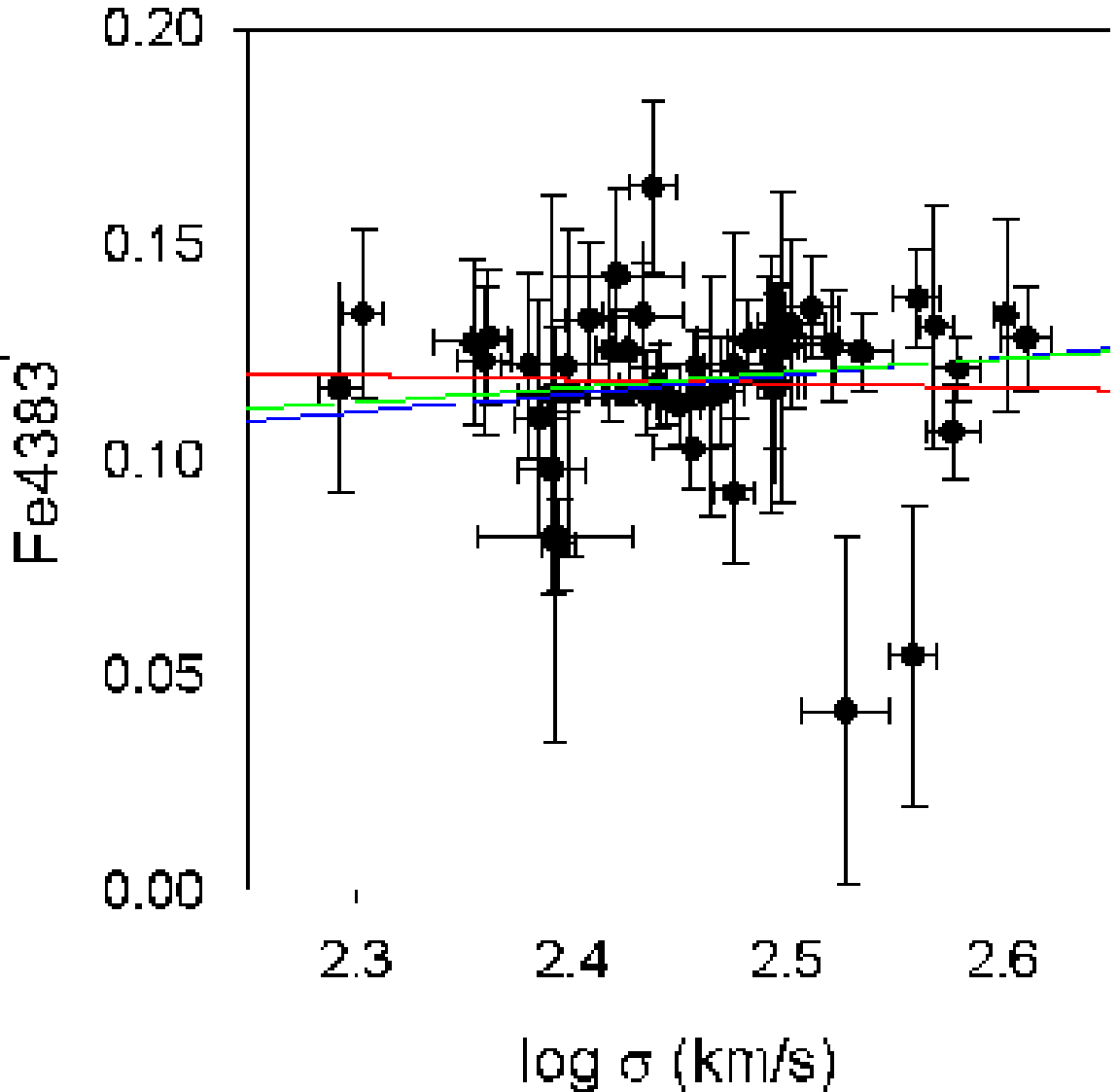}}\quad
         \subfigure{\includegraphics[width=4.5cm,height=4.5cm]{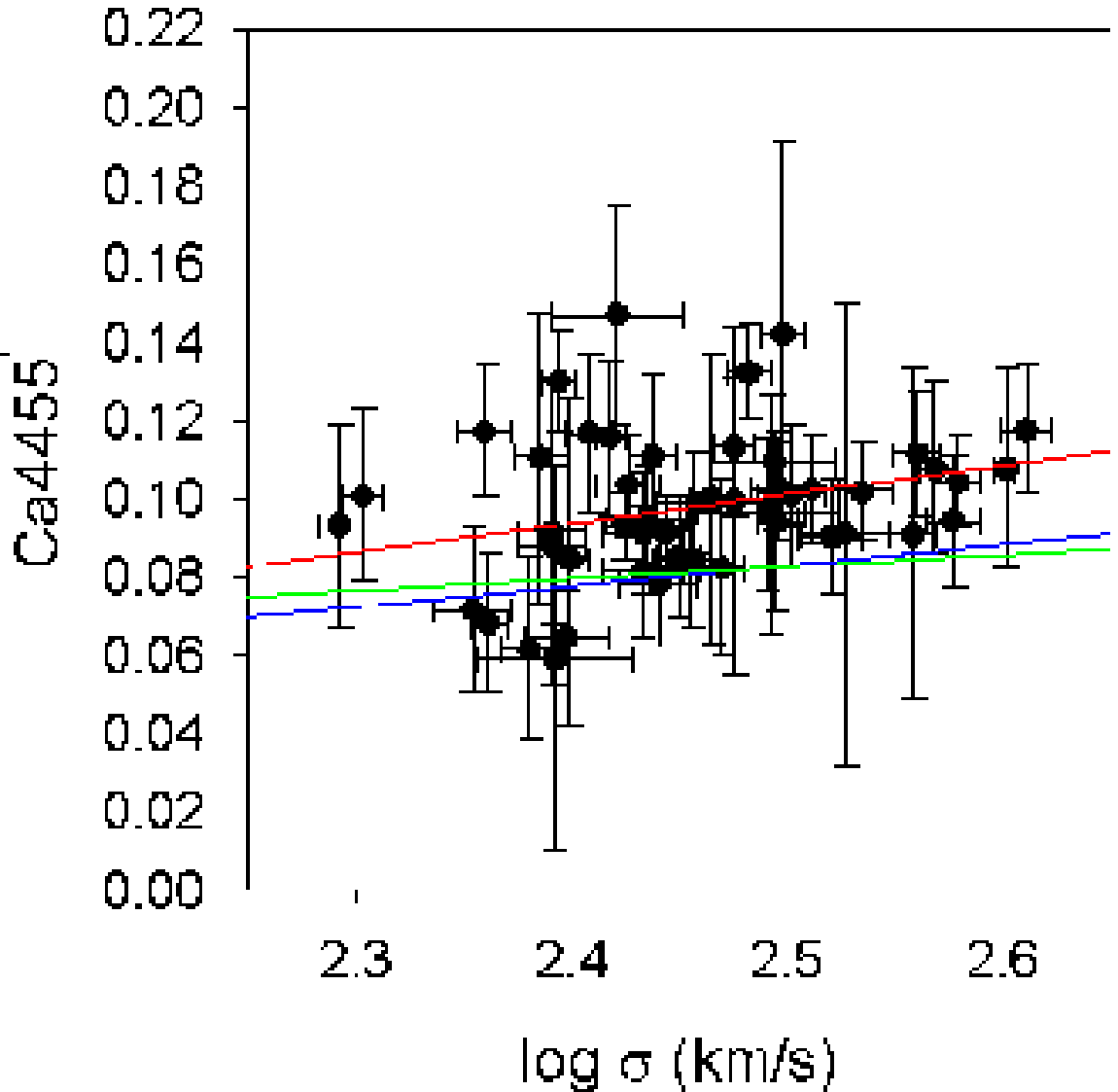}}\quad
         \subfigure{\includegraphics[width=4.5cm,height=4.5cm]{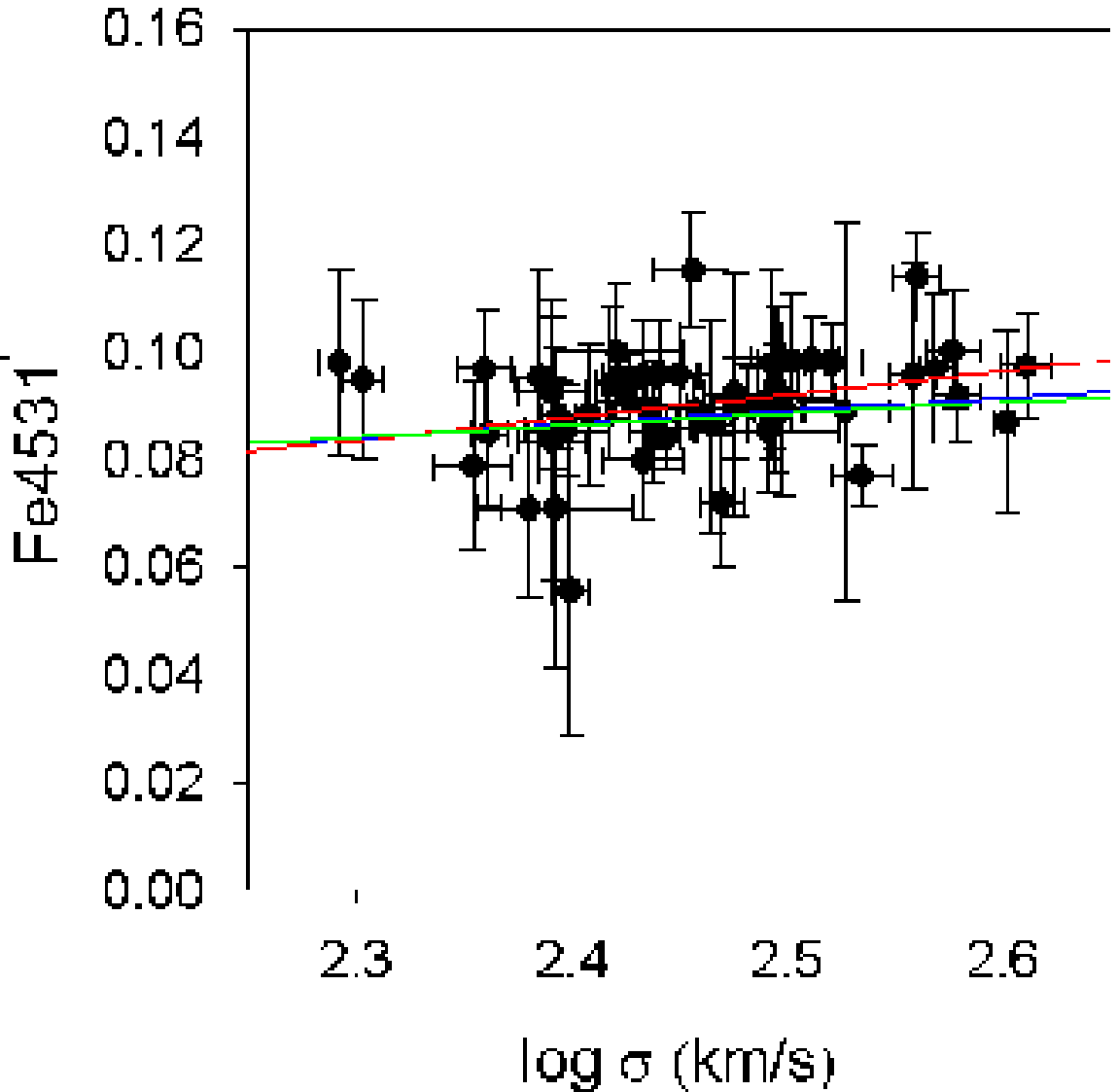}}}
\mbox{\subfigure{\includegraphics[width=4.5cm,height=4.5cm]{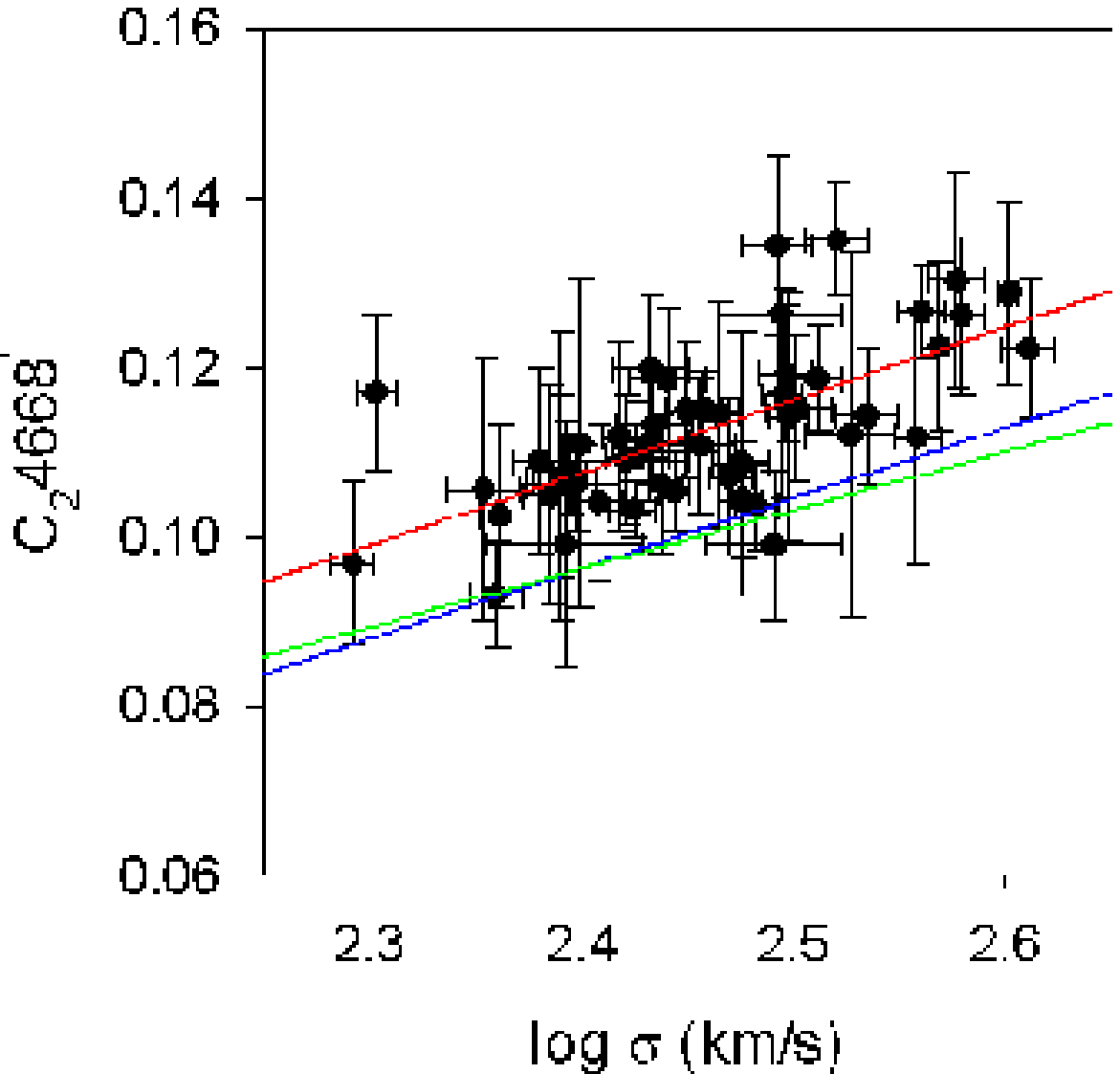}}\quad
         \subfigure{\includegraphics[width=4.5cm,height=4.5cm]{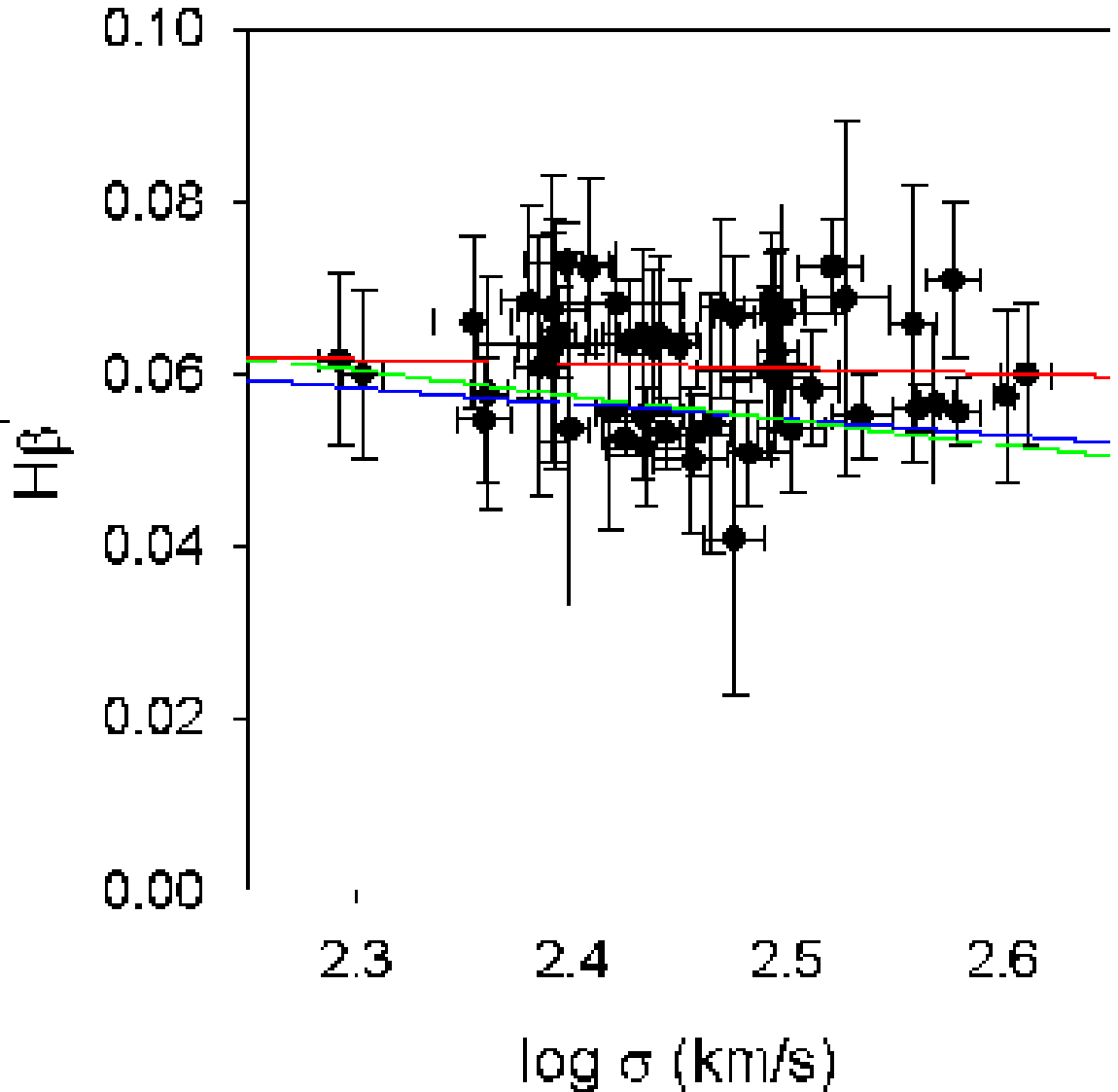}}\quad
         \subfigure{\includegraphics[width=4.5cm,height=4.5cm]{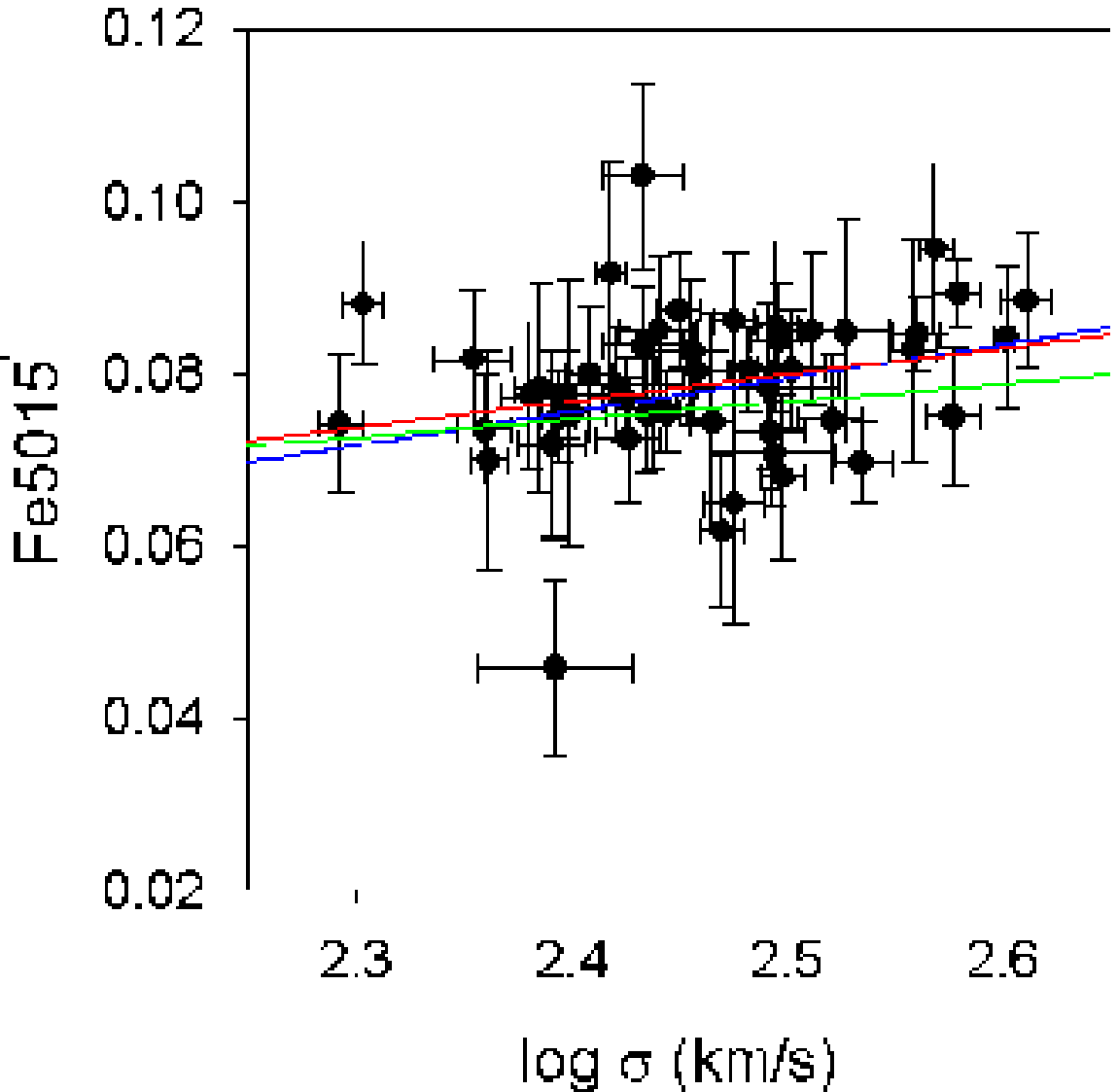}}}
\caption[Central Index Measurement Against Velocity Dispersion]{Central index measurements against velocity dispersion. The straight lines fitted to the BCG data are in red. The blue lines denote the relations found for the SB06 elliptical sample in the same mass range, and the green line the relations for the complete SB06 elliptical sample (high and low density samples combined).}
   \label{Ind_Sigma}
\end{figure*} 

\begin{figure*}
   \centering
  \mbox{\subfigure{\includegraphics[width=4.5cm,height=4.5cm]{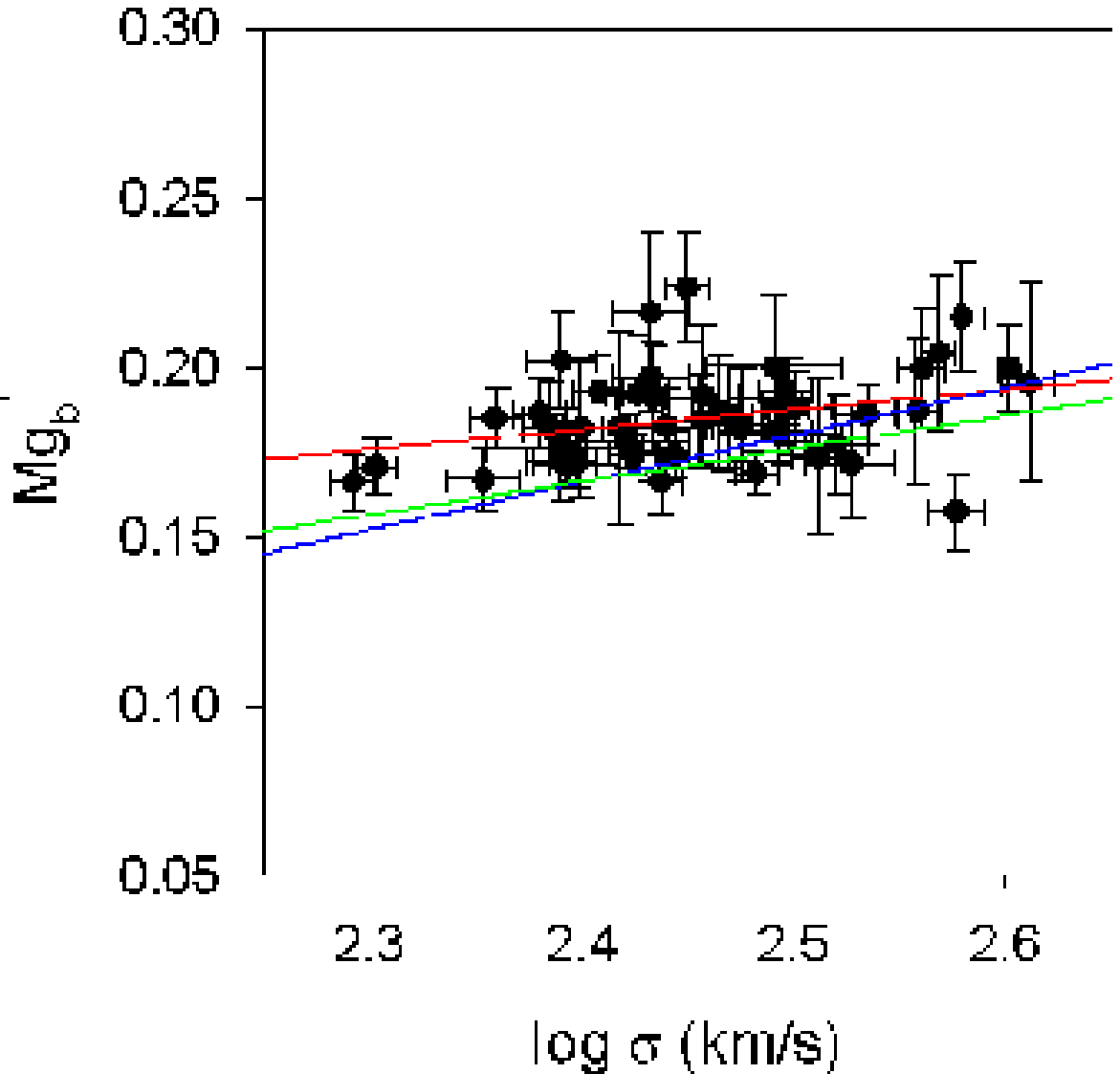}}\quad
        \subfigure{\includegraphics[width=4.5cm,height=4.5cm]{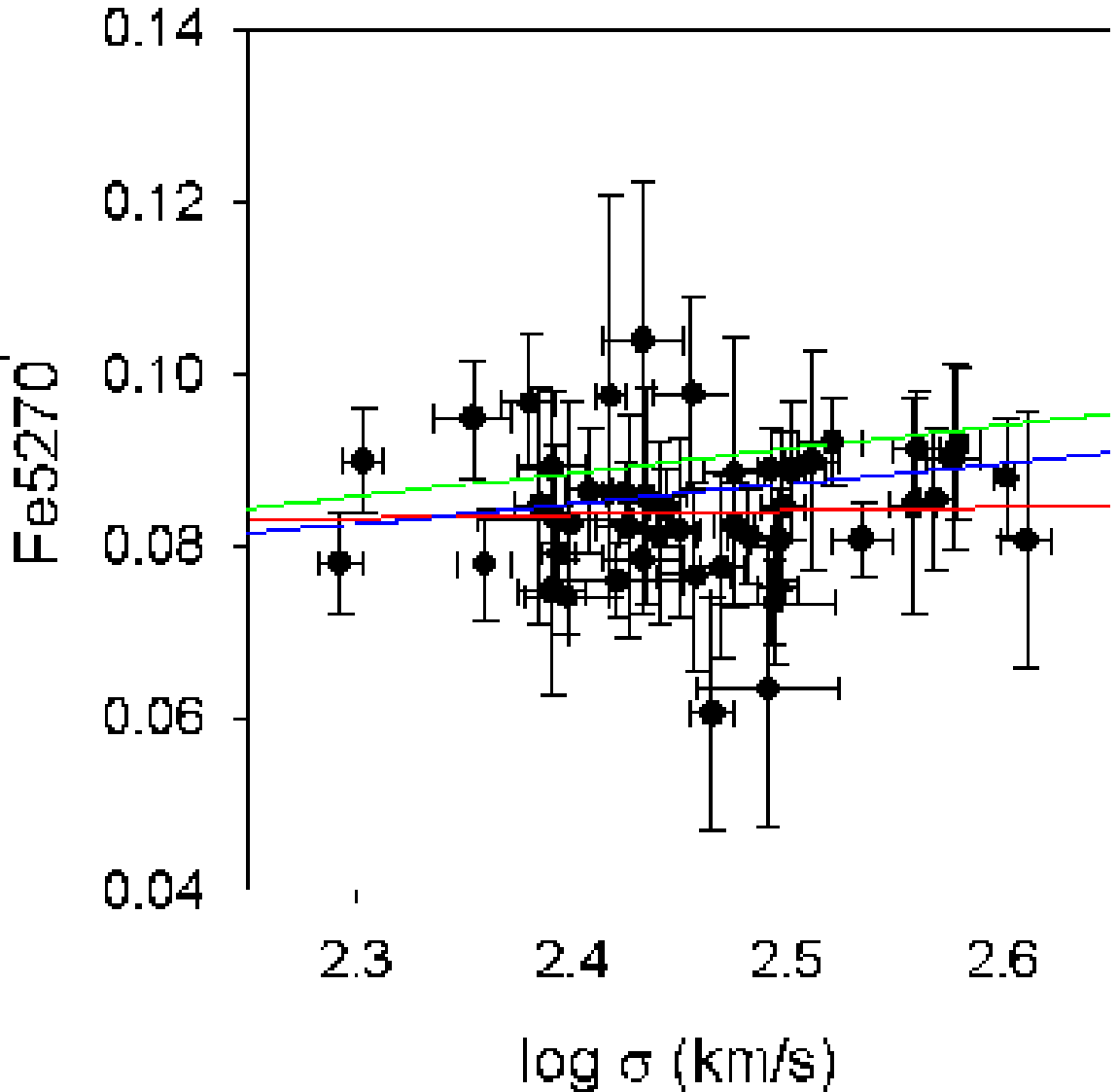}}\quad
        \subfigure{\includegraphics[width=4.5cm,height=4.5cm]{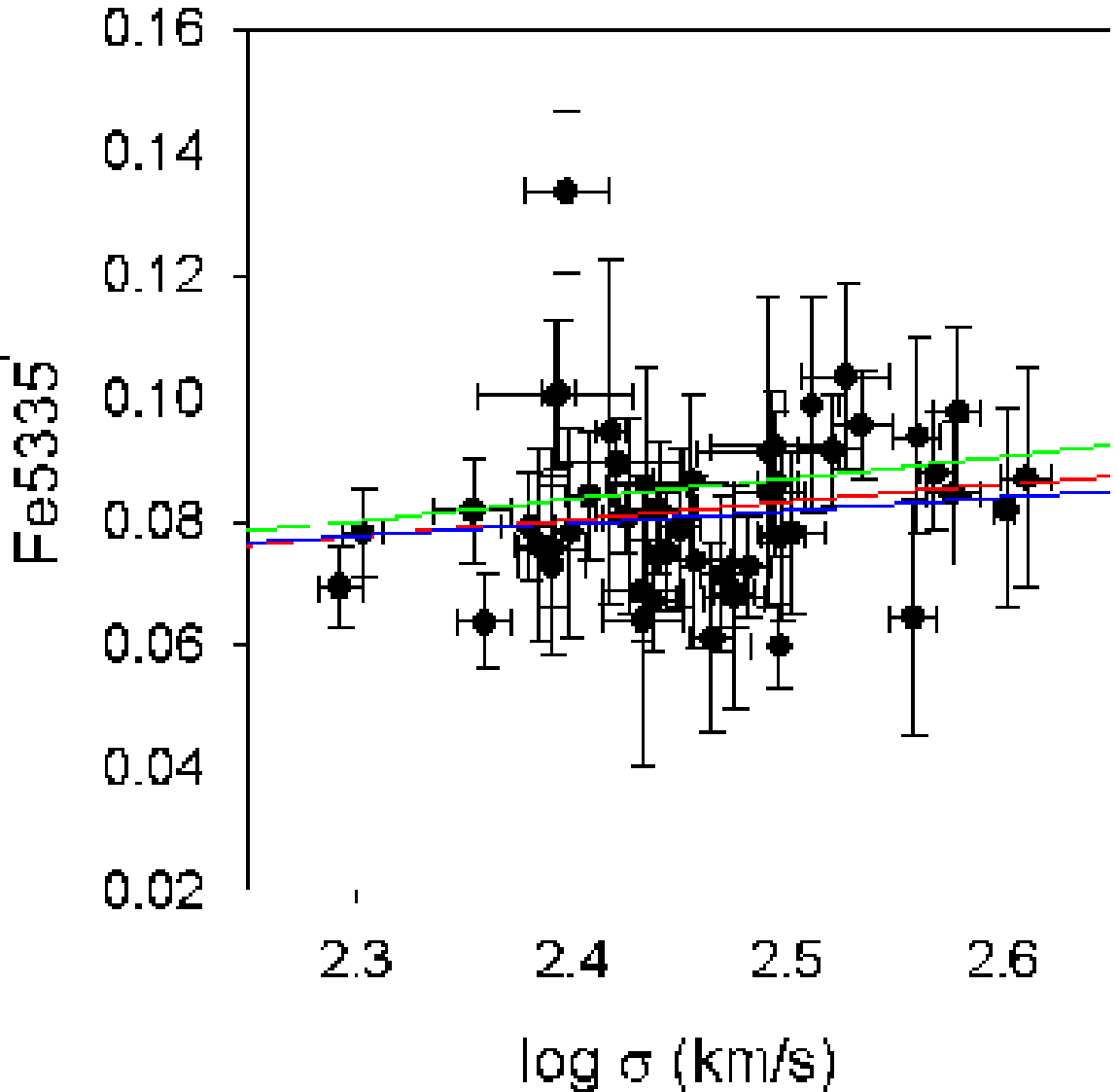}}}
\mbox{\subfigure{\includegraphics[width=4.5cm,height=4.5cm]{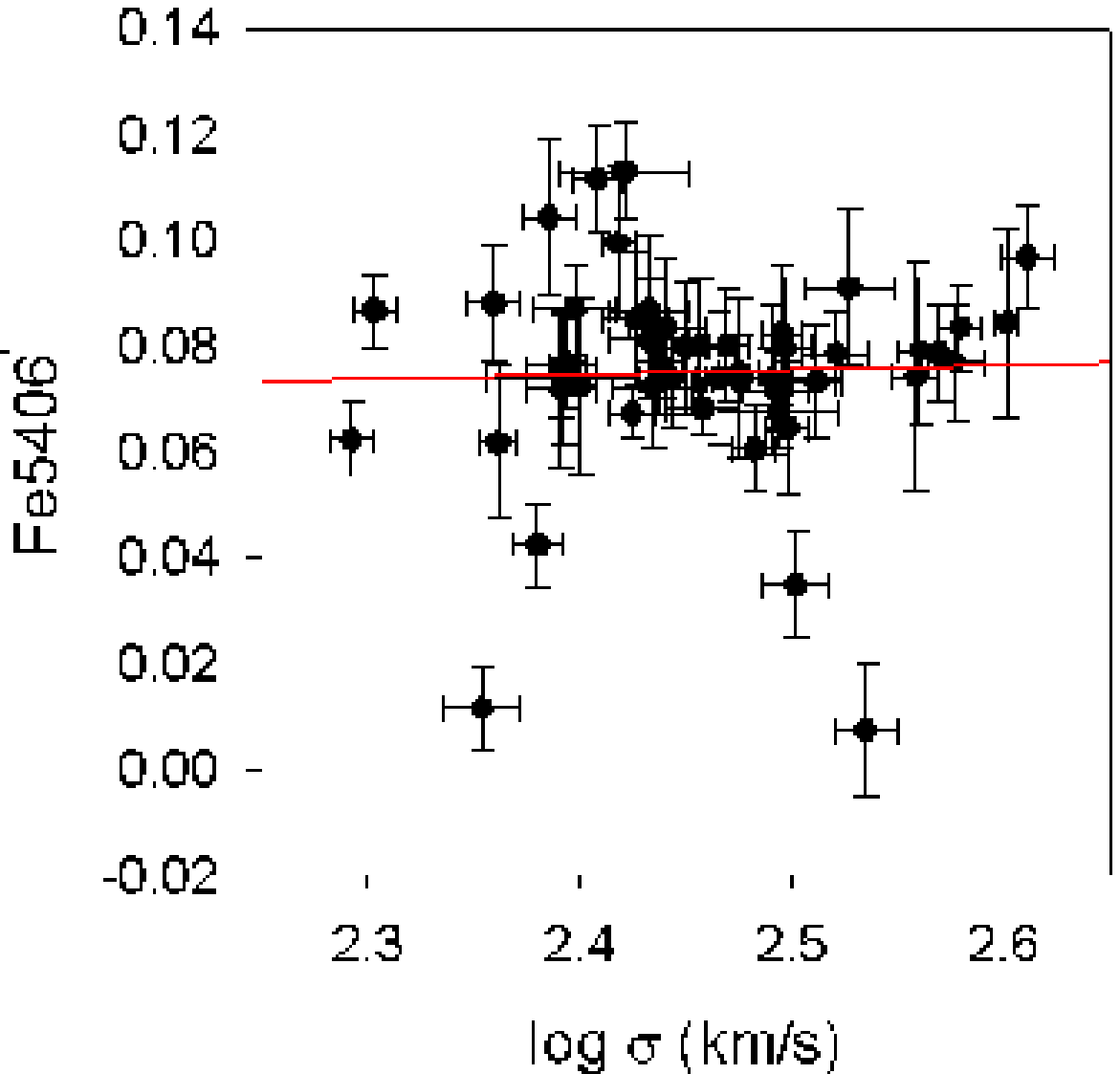}}\quad
        \subfigure{\includegraphics[width=4.5cm,height=4.5cm]{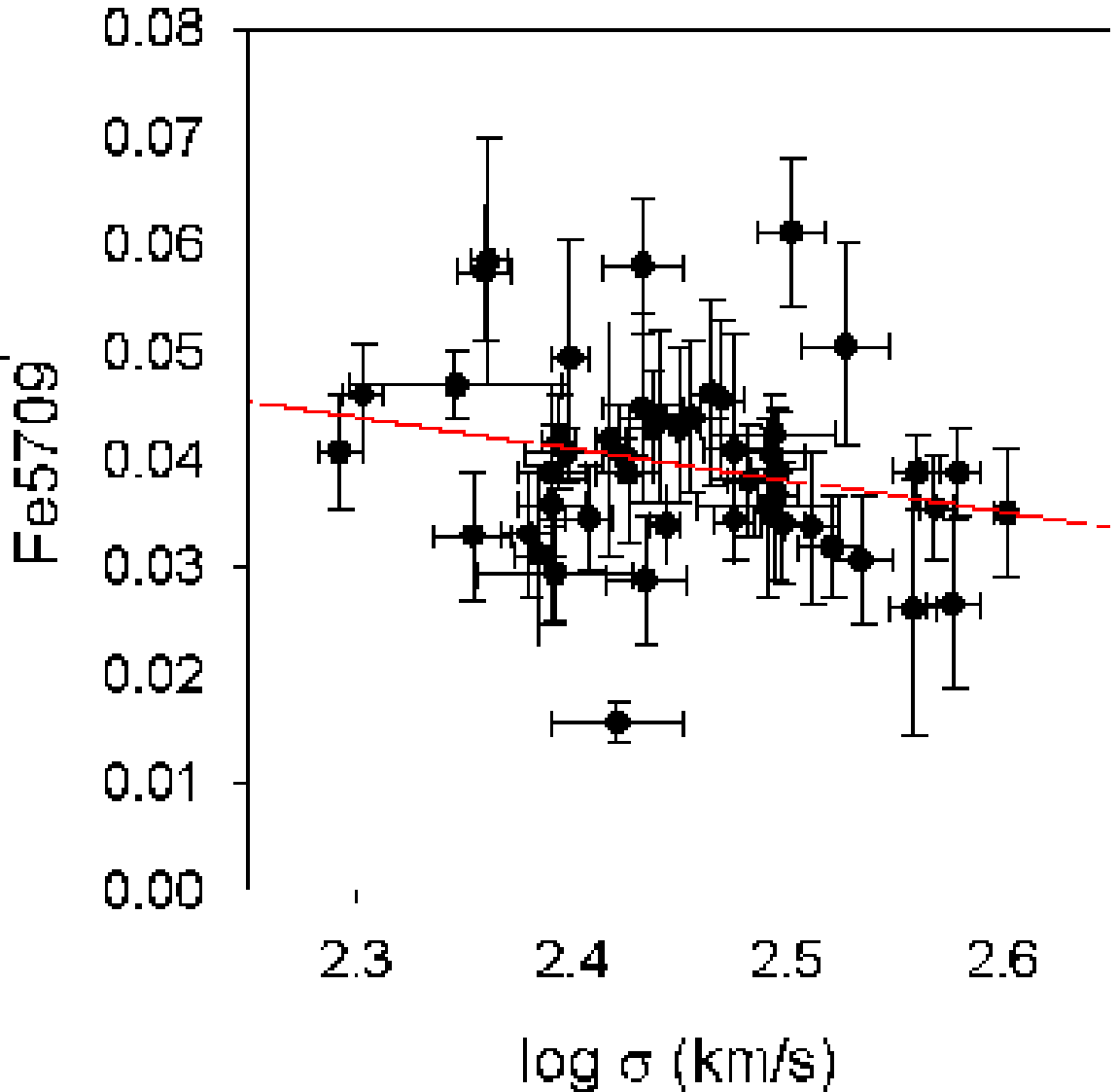}}\quad
        \subfigure{\includegraphics[width=4.5cm,height=4.5cm]{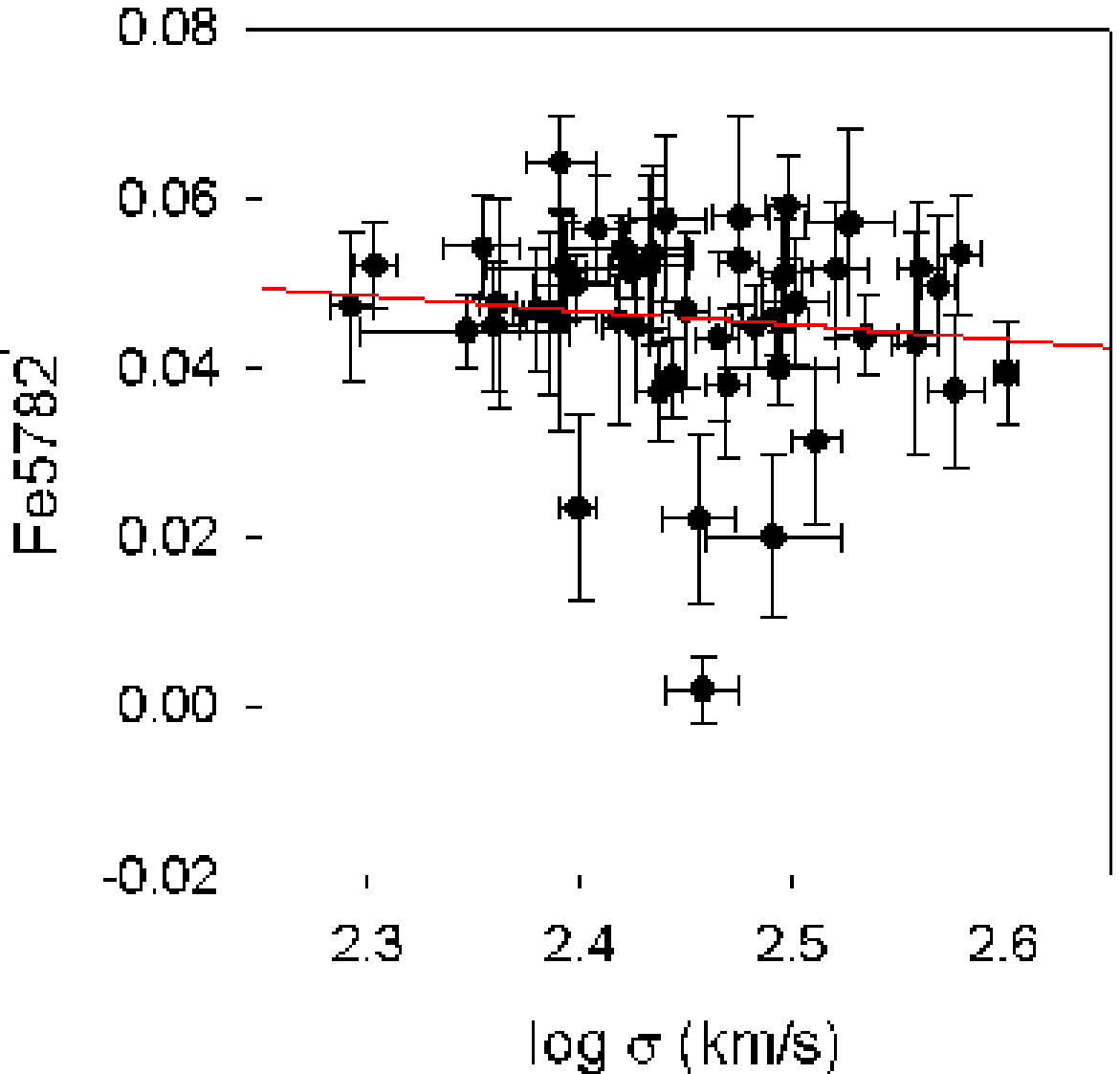}}}
\caption{Central index measurements against velocity dispersion. The lines are as in Figure \ref{Ind_Sigma}.}
   \label{Ind_Sigma3}
\end{figure*} 

The derived errors on the indices take the Poisson and systematic errors (flux calibration effects; velocity dispersion corrections; sky subtraction uncertainties; wavelength calibration and radial velocity errors) into account. We do not take into account the errors derived on the offsets to transform the indices to the Lick system (shown in Table \ref{table:GeminitoLick}). The central index measurements compared very well with previous measurements from the literature, as shown in Appendix \ref{appendix3}.

Lick offsets are applied to correct measurements that were flux-calibrated using the spectrophotometric system to the Lick/IDS system and, therefore, should be identical for all flux-calibrated studies. Thus, we compare our derived offsets with other sources to test the robustness of the Lick transformation. The Lick offsets derived here for the Gemini data were compared to independently derived offsets from the same Lick star dataset (M. Norris, private communication), as described in Appendix \ref{Appendix2}. Most of the offsets are in agreement, with the exception of C$_{2}$4668 and Ca4455 which are marginally higher in this study. Since the BCG data are being compared with SB06 (Section \ref{Next}), the Lick offsets were also compared. All the Lick offsets agree within the errors, and no real outliers were found. As a final test, we also compare our offsets with the offsets derived by comparing all the stars in common between the flux-calibrated MILES library (S\'{a}nchez-Bl\'{a}zquez et al.\ 2006d; S\'{a}nchez-Bl\'{a}zquez et al.\ 2009) and the Lick/IDS library. Any remaining systematics would not have been corrected by the usual Lick calibration offsets, as the systematic would have had a different effect on stars and on galaxies (at different redshifts).

\subsection{Results: Index -- velocity dispersion relations}
\label{Next}

Following various other authors, and to compare with S\'{a}nchez-Bl\'{a}zquez et al.\ (2006a, hereafter SB06), the atomic indices will be expressed in magnitudes when correlated with the velocity dispersion. These will be denoted by the name followed by a prime sign, and were obtained using $I_{\rm Mag}=2.5\log\left[\frac{I_{\rm Ang}}{(\lambda_{1}-\lambda_{2})}+1 \right]$, where the wavelength range is $\lambda_{1}-\lambda_{2}$ in the central bandpass, and $I_{\rm Ang}$ and $I_{\rm Mag}$ are the index measurement in \AA{} and magnitudes respectively.

Straight line fits ($I' = a + b \times \log \sigma$) were made to the BCG data with a least-squares fitting routine, and are shown in Figures \ref{Ind_Sigma} and \ref{Ind_Sigma3} and given in Table \ref{table:Slopes}. Statistical t-tests were run to explore the presence of correlations. Only six out of 18 indices possess a statistically significant slope different from zero (as can be seen from Table \ref{table:Slopes}). These are Ca4227, H$\gamma_{\rm F}$, Ca4455, Fe4531, C$_{2}$4668 and Mg$_{\rm b}$. 

\begin{figure}
   \centering
   \includegraphics[scale=0.35]{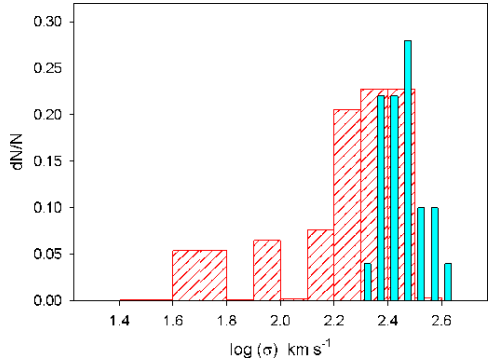}
   \caption[The velocity dispersion distributions of the complete SB06 and the BCG sample.]{The velocity dispersion distributions of the complete SB06 (red) and the BCG sample (cyan). The vertical line indicates the lower limit in velocity dispersion (log $\sigma$ = 2.3 km s$^{-1}$) of the SB06 subsample used here for comparison.}
   \label{fig:Sig_distr}
\end{figure}

The scatter in the CN$_{1}$, CN$_{2}$, Mg$_{1}$, Mg$_{2}$, NaD and TiO$_{1}$ indices were found to be large. The molecular indices are known to show only a small amounts of scatter, but are frequently affected by flux calibration uncertainties, since the index definitions span a broad wavelength range. For 15 of the indices plotted, the index--velocity dispersion relations found by SB06 for a large sample of elliptical galaxies are also indicated. The complete SB06 sample consists of 98 galaxies, of which 35 belong to the Coma cluster, and the rest are galaxies in the field, in groups or in the Virgo cluster. Two relations per index were derived by SB06: one for ellipticals in a higher density environment, and one for ellipticals in a lower density environment. Here, both the higher and lower density samples are combined for comparison with the BCG data, since some of the BCGs are in Virgo-equivalent environments. As mentioned in Appendix \ref{appendix3}, five galaxies were in common with SB06. Four of those are in the higher density sample and one in the lower density sample. The BCG data are spread over a narrower range in velocity dispersion ($\log \sigma$ = 2.3 to 2.6 km s$^{-1}$) than the SB06 sample ($\log \sigma$ = 1.4 to 2.6 km s$^{-1}$), which meant that their slopes could possibly be heavily influenced by the lower velocity dispersion galaxies. The velocity dispersion distributions of the two samples are shown in Figure \ref{fig:Sig_distr}. We performed a Kolmogorov--Smirnov test on the velocity dispersion distributions of the two samples within the $\log \sigma$ = 2.3 to 2.6 km s$^{-1}$ range, where the null hypothesis is that the distributions were drawn from an identical parent population. Within this limited range, the two velocity dispersion distributions are consistent (the test value is 0.260, where a test value larger than $D=0.290$ indicates that the two samples compared are significantly different from each other at the 95 per cent confidence level). New relationships between the indices and velocity dispersion were derived for the SB06 sample to compare with the BCG data, only including the elliptical galaxies in the same mass range and excluding the five known BCGs. These new relations and their errors, derived using a subsample of 45 galaxies from SB06, are given in Table \ref{table:Slopes}. Two relations derived for the SB06 sample are shown in Figures \ref{Ind_Sigma} and \ref{Ind_Sigma3}: the relation found for the SB06 elliptical sample in the same mass range (blue line), as well as the relation for the complete SB06 elliptical sample (green line). Both lines are for the high and low density samples combined.

\begin{table*}
\begin{scriptsize}
\begin{tabular}{l r@{$\pm$}l r@{$\pm$}l c c r@{$\pm$}l  r@{$\pm$}l r@{$\pm$}l} 
\hline Index & \multicolumn{8}{c}{ BCG Galaxies } & \multicolumn{4}{c}{Ellipticals} \\  
  & \multicolumn{2}{c}{$a\ \pm$ std err($a$)} & \multicolumn{2}{c}{$b\ \pm$ std err($b$)} & $t$ & $P$ & \multicolumn{2}{c}{Mean $\pm$ std err} & \multicolumn{2}{c}{$a\ \pm $ std err($a$)} & \multicolumn{2}{c}{$b\ \pm$ std err($b$)}  \\ 
\hline H$\delta_{\rm A}^{'}$ & 0.0598 & 0.0903 & --0.0620 & 0.0368 & 1.687 & 0.098 & --0.093 & 0.020 & 0.184 & 0.016 & --0.102 & 0.007 \\
H$\delta_{\rm F}^{'}$ & 0.0280 & 0.0592 & --0.0145 & 0.0241 & 0.603 & 0.550 & --0.008 & 0.016 & 0.251 & 0.018 & --0.102 & 0.008 \\
Ca4227$^{'}$ & --0.0700 & 0.0681 & 0.0800 & 0.0277 & 2.887 & 0.006 & 0.127 & 0.016 & --0.044 & 0.015 & 0.059 & 0.007 \\ 
G4300$^{'}$ & 0.0973 & 0.0500 & 0.0385 & 0.0204 & 1.893 & 0.065 & 0.192 & 0.011 & 0.152 & 0.016 & 0.008 & 0.007 \\
H$\gamma_{\rm A}^{'}$ & --0.0261 & 0.0669 & --0.0479 & 0.0272 & 1.758 & 0.085 & --0.144 & 0.015 & 0.009 & 0.017 & --0.060 & 0.007 \\
H$\gamma_{\rm F}^{'}$ & 0.0991 & 0.0746 & --0.0781 & 0.0304 & 2.570 & 0.013 & --0.093 & 0.017 & 0.135 & 0.023 & --0.086 & 0.010 \\
Fe4383$^{'}$ & 0.1403 & 0.0950 & --0.0091 & 0.0387 & 0.235 & 0.815 & 0.118 & 0.020 & 0.014 & 0.008 & 0.042 & 0.003 \\
Ca4455$^{'}$ & --0.0826 & 0.0843 & 0.0735 & 0.0343 & 2.142 & 0.037 & 0.098 & 0.019 & --0.049 & 0.009 & 0.053 & 0.004 \\
Fe4531$^{'}$ & --0.0140 & 0.0460 & 0.0424 & 0.0187 & 2.266 & 0.028 & 0.090 & 0.010 & 0.027 & 0.005 & 0.025 & 0.002 \\
C$_{2}$4668$^{'}$ & --0.0985 & 0.0334 & 0.0859 & 0.0136 & 6.317 & $<$0.001 & 0.113 & 0.009 & --0.103 & 0.011 & 0.083 & 0.005 \\
H$\beta^{'}$ & 0.0754 & 0.0327 & --0.0060 & 0.0133 & 0.448 & 0.656 & 0.061 & 0.001 & 0.100 & 0.012 & --0.018 & 0.005 \\
Fe5015$^{'}$ & 0.0035 & 0.0413 & 0.0306 & 0.0168 & 1.819 & 0.075 & 0.079 & 0.010 & --0.018 & 0.013 & 0.039 & 0.006\\
Mg$_{\rm b}^{'}$ & 0.0405 & 0.0606 & 0.0588 & 0.0247 & 2.383 & 0.021 & 0.185 & 0.014 & --0.169 & 0.015 & 0.140 & 0.007 \\
Fe5270$^{'}$ & 0.0738 & 0.0396 & 0.0041 & 0.0161 & 0.253 & 0.801 & 0.084 & 0.008 & 0.029 & 0.007 & 0.023 & 0.003 \\
Fe5335$^{'}$ & 0.0128 & 0.0632 & 0.0281 & 0.0257 & 1.093 & 0.280 & 0.082 & 0.014 & 0.030 & 0.009 & 0.021 & 0.004 \\
Fe5406$^{'}$ & 0.0545 & 0.0913 & 0.0084 & 0.0372 & 0.225 & 0.823 & 0.075 & 0.019\\
Fe5709$^{'}$ & 0.1109 & 0.0401 & --0.0291 & 0.0163 & 1.783 & 0.081 & 0.040 & 0.009\\
Fe5782$^{'}$ & 0.0877 & 0.0517 & --0.0171 & 0.0210 & 0.812 & 0.421 & 0.046 & 0.011\\
\hline
\end{tabular}
\end{scriptsize}
\caption[Parameters of the Indices vs $\log \sigma$ Comparison]{Parameters of the indices against velocity dispersion comparison between the BCGs and elliptical galaxies. T-tests were run on all the slopes to assess if a real slope are present or if $b=0$ (as a null hypothesis). A $t$ value larger than 1.96 means that there is a true correlation between the variables ($b \neq$ 0), at a 95 per cent confidence level. $P$ is the probability of being wrong in concluding that there is a true correlation (i.e.\ the probability of falsely rejecting the null hypothesis). The average index measurements (mean $\pm$ std err) for the BCGs are also given.}
\label{table:Slopes}
\end{table*}				  

\begin{table*}
\centering
\begin{tabular}{l c c c c c}
\hline \multicolumn{4}{c}{Scatter in BCG data points.} & Scatter around the slope (BCGs) & Scatter around the slope (Ellipticals) \\
Index &  $\sigma_{\rm std}$ & $\sigma_{\rm exp}$ & $\sigma_{\rm res}=\sqrt{\sigma_{\rm std}^{2}-\sigma_{\rm exp}^{2}}$ & $\sigma_{\rm res}$ & $\sigma_{\rm res}$ \\
\hline H$\delta_{\rm A}^{'}$ & 0.020 & 0.029 & $\star$ & 0.023 & 0.005\\
H$\delta_{\rm F}^{'}$ & 0.016 & 0.036 & $\star$ & $\star$ & 0.006\\
Ca4227$^{'}$ & 0.016 & 0.042 & $\star$ & $\star$ &0.006\\
G4300$^{'}$ & 0.011 & 0.016 & $\star$ & 0.013 &0.006\\
H$\gamma_{\rm A}^{'}$ & 0.015 & 0.017 & $\star$& 0.021 &0.006\\
H$\gamma_{\rm F}^{'}$ & 0.017 & 0.022 & $\star$ & 0.021 &0.008\\
Fe4383$^{'}$ & 0.020 & 0.019 & 0.006 & 0.034 &$\star$\\
Ca4455$^{'}$ & 0.019 & 0.023 & $\star$ & 0.025 &$\star$\\
Fe4531$^{'}$ & 0.010 & 0.016 & $\star$ & 0.010 & $\star$\\
C$_{2}$4668$^{'}$ & 0.009 & 0.010 & $\star$ & 0.009 & $\star$\\
H$\beta^{'}$ & 0.001 & 0.010 & $\star$ & 0.009 &0.004\\
Fe5015$^{'}$ & 0.010 & 0.009 & 0.004 & 0.014 &0.005\\
Mg$_{\rm b}^{'}$ & 0.014 & 0.014 & $\star$ & 0.020 &$\star$\\
Fe5270$^{'}$ & 0.008 & 0.009 & $\star$ & 0.013 &$\star$\\
Fe5335$^{'}$ & 0.014 & 0.013 & 0.005 & 0.022 &$\star$\\
Fe5406$^{'}$ & 0.019 & 0.011 & 0.015 & 0.036 &\\
Fe5709$^{'}$ & 0.009 & 0.006 & 0.007 & 0.015 &\\
Fe5782$^{'}$ & 0.011 & 0.008 & 0.008 & 0.019 &\\
\hline
\end{tabular} 
\caption[Scatter of the index measurements.]{Scatter of the index measurements (in magnitudes) compared to that expected from the errors. $\sigma_{\rm std}$ is the standard deviation on the mean index value, $\sigma_{\rm exp}$ is the standard deviation expected from the mean errors on the index values, and $\sigma_{\rm res}=\sqrt{\sigma_{\rm std}^{2}-\sigma_{\rm exp}^{2}}$ is the residual scatter not explained by the errors on the index measurements. The indices marked with $\star$ have $\sigma_{\rm std} \leq \sigma_{\rm exp}$. The first three columns are for the scatter in the BCG data points. The last two columns are for the intrinsic scatter around the slope for the BCG and elliptical data, respectively, over the same mass range.}
\label{Scatter}
\end{table*}

Most of the fitted relations agree with those of elliptical galaxies. The only indices where a notable difference is detected are G4300, Ca4455 and C$_{2}$4668 where the zero points of the BCG data are higher than for the SB06 elliptical data. For Ca4455 and C$_{2}$4668, this is explained by the offsets applied to the data to transform it onto the Lick system. But, this does not explain the discrepancy in the G4300 index. Thus, there are no significant discrepancies between the index--velocity dispersion relations of the BCG data and that of normal ellipticals in the same mass range, with the exception of G4300. 

We calculate the intrinsic scatter (not explained by the errors) for both the BCGs and the SB06 elliptical sample in the mass range considered here. All the indices, for which the intrinsic scatter around the slope could be calculated for both samples, showed that the BCGs are intrinsically more scattered than the ellipticals (column 5 compared to column 6 in Table \ref{Scatter}). This might indicate that the BCGs are more affected by properties other than mass, than elliptical galaxies over this mass range.

\section{Derivation of luminosity-weighted properties}

\begin{table}
\centering
\begin{scriptsize}
\begin{tabular}{l r@{$\pm$}l r@{$\pm$}l r@{$\pm$}l r@{$\pm$}l}
\hline Galaxy  & \multicolumn{2}{c}{$\log$ ($\sigma$)} & \multicolumn{2}{c}{$\log$ (age)} & \multicolumn{2}{c}{[Z/H]} & \multicolumn{2}{c}{[E/Fe]}\\
 & \multicolumn{2}{c}{km s$^{-1}$} & \multicolumn{2}{c}{Gyr} & \multicolumn{2}{c}{} & \multicolumn{2}{c}{}\\ 
\hline ESO146-028&	299 & 3 &0.97& 0.18&  --0.12&	0.06&	0.60&  0.03 \\
ESO202-043& 256 & 3&	         0.58&	0.18&	0.58&	0.06&	0.45&	0.03\\
ESO303-005& 276 & 5&	         0.81&	0.23&	0.34&	0.05&	0.41&	0.03\\
ESO346-003& 226 & 4&	         0.63&	0.43&	0.43&	0.09&	0.30&	0.03\\
ESO349-010& 282 & 3&	         0.89&	0.22&	0.21&	0.08&	0.30&	0.05\\
ESO444-046& 292 & 3&	         1.25&	0.20&  --0.18&	0.12&	0.59&	0.05\\
ESO488-027& 248 & 2&	         0.85&	0.13&	0.42&	0.08&	0.32&	0.03\\
ESO541-013& 295 & 3&	         0.75&	0.38&	0.30&	0.09&	0.49&	0.04\\
ESO552-020& 229 & 3&	         1.12&	0.19&	0.02&	0.07&	0.48&	0.03\\
GSC555700266& 312 & 9 &	         0.66&	0.37&	0.40&	0.05&	0.41&	0.03\\
IC1101& 378 & 5&	                 0.70&	0.15&	0.41&	0.08&	0.27&	0.04\\
IC1633& 400 & 2&	                 0.94&	0.19&	0.38&	0.06&	0.44&	0.04\\
IC4765& 286 & 5&	                 1.01&	0.23&	0.36&	0.07&	0.30&	0.04\\
IC5358& 243 & 3&	                 0.92&	0.32&	0.28&	0.10&	0.41&	0.05\\
MCG-02-12-039 & 271 & 5&      	 1.05&	0.21&	0.14&	0.07&	0.51&	0.03\\
NGC0533& 299 & 4&	         1.25&	0.10&	0.14&	0.10&	0.38&	0.05\\
NGC0541& 246 & 4&	         0.66&	0.55&	0.37&	0.08&	0.38&	0.04\\
NGC1399& 371 & 3&	         0.93&	0.10&	0.42&	0.09&	0.45&	0.05\\
NGC1713 & 251 & 2 &	         1.03&	0.34&	0.19&	0.11&	0.39&	0.07\\
NGC2832 & 364 & 4 &	         0.93&	0.02&	0.48&	0.06&	0.38&	0.04\\
NGC3311 & 196 & 2&	         0.94&	0.27&	0.12&	0.07&	0.40&	0.03\\
NGC3842& 287 & 5&	         1.10&	0.17&	0.11&	0.09&	0.47&	0.05\\
NGC4839& 278 & 2 &	         1.07&	0.12&	0.13&	0.05&	0.35&	0.03\\
NGC4874 & 267 & 4 &	         0.89&	0.12&	0.35&	0.05&	0.46&	0.05\\
NGC4889& 380 & 4&	         0.92&	0.04&	0.57&	0.05&	0.42&	0.04\\
NGC6034& 325 & 4 &	         0.92&	0.07&	0.42&	0.10&	0.26&	0.06\\
NGC6086& 318 & 5&	         1.00&	0.14&	0.28&	0.07&	0.39&	0.04\\
NGC6160& 266 & 3&	         1.05&	0.06&	0.19&	0.04&	0.32&	0.02\\
NGC6173& 304 & 3&	         0.90&	0.14&	0.20&	0.05&	0.39&	0.02\\
NGC6269& 343 & 5&	         0.95&	0.08&	0.36&	0.04&	0.36&	0.03\\
NGC7012 & 240 & 3&	         0.69&	0.19&	0.51&	0.08&	0.39&	0.03\\
NGC7597& 264 & 8&	         0.67&	0.29&	0.40&	0.05&	0.40&	0.02\\
NGC7647 & 271 & 5&	         0.86&	0.11&	0.48&	0.12&	0.54&	0.07\\
NGC7720& 409 & 5&	         0.92&	0.13&	0.36&	0.07&	0.44&	0.07\\
NGC7768& 272 & 5&	         1.02&	0.24&	0.29&	0.10&	0.38&	0.04\\
PGC004072& 313 & 3 &	         0.90&	0.20&	0.31&	0.07&	0.45&	0.04\\
PGC044257 & 247 & 9 &	         0.63&	0.19&	0.50&	0.10&	0.31&	0.04\\
PGC071807& 315 & 3 &	         0.63&	0.40&	0.42&	0.09&	0.46&	0.05\\
PGC072804& 311 & 5&	         0.69&	0.37&	0.48&	0.05&	0.38&	0.03\\
UGC00579& 246 & 4 &	         0.91&	0.27&	0.26&	0.10&	0.56&	0.04\\
UGC02232& 314 & 4&	         0.99&	0.17&	0.13&	0.06&	0.54&	0.03\\
UGC05515& 362 & 4&	         0.88&	0.34&	0.25&	0.11&	0.50&	0.07\\
UGC10143& 262 & 2&	         0.93&	0.21&	0.45&	0.09&	0.28&	0.06\\
\hline
\end{tabular}
\end{scriptsize}
\caption[Central Values for the SSP-equivalent Parameters.]{\begin{footnotesize} Central values for the SSP-equivalent parameters derived using H$\beta$, Mg$_{\rm b}$, Fe5270 and Fe5335.\end{footnotesize}}
\label{SSPS}
\end{table} 

To calculate the ages, metallicities ([Z/H]), and $\alpha$-enhancement ratios ([E/Fe]), we compare our derived line-strength indices with the predictions of Thomas, Maraston $\&$ Bender (2003) and Thomas, Maraston $\&$ Korn (2004). These models are based on the evolutionary population synthesis models of Maraston (1998; 2005).
Variations of the indices with chemical partitions departing from solar are computed with the Tripicco \& Bell (1995)
and  Korn et al.\ (2005) model atmospheres, with a slightly modified method from the one presented
by Trager et al.\ (2000a). The models are presented for six different metallicities  [Z/H] = $-$2.25, $-$1.35, $-$0.33, 0.0, 0.35, 0.67, ages between 
1 and 15 Gyr, evenly spaced in logarithmic steps of 0.025, and [E/Fe] = 0.0, 0.3 and 0.5. The ``E'' group contains O, Ne, Mg, Si, S, Ar, Ca, Ti, Na and N. The models
are computed at constant metallicity in such a way that an increase in the ``E'' group
is compensated by a decrease of the abundances of the elements Fe and Cr (see Thomas et al.\ 2003
and Trager et al.\ 2000a for a more detailed discussion), because the total metallicity is dominated by oxygen (included in the E group).

Ages, [Z/H] and [E/Fe] were derived using the indices $<$Fe$>$\footnote{Where $<$Fe$>$ is defined as (Fe5270 + Fe5335)/2 (Gonz\'alez 1993).}, H$\beta$ and Mg$_{\rm b}$. We started by interpolating the model grids in increments of 0.05 in [Z/H] and [E/Fe] and 0.1 dex in age. Then we 
applied a $\chi^{2}$ technique to find the combination of SSP that best reproduced the four indices simultaneously. Errors in the parametres were calculated performing 50 Monte--Carlo simulations
in which, each time, the indices were displaced by an amount given by a Gaussian probability 
distribution with a width equal to the errors on these indices.

\begin{figure*}
   \centering
  \mbox{\subfigure{\includegraphics[scale=0.25]{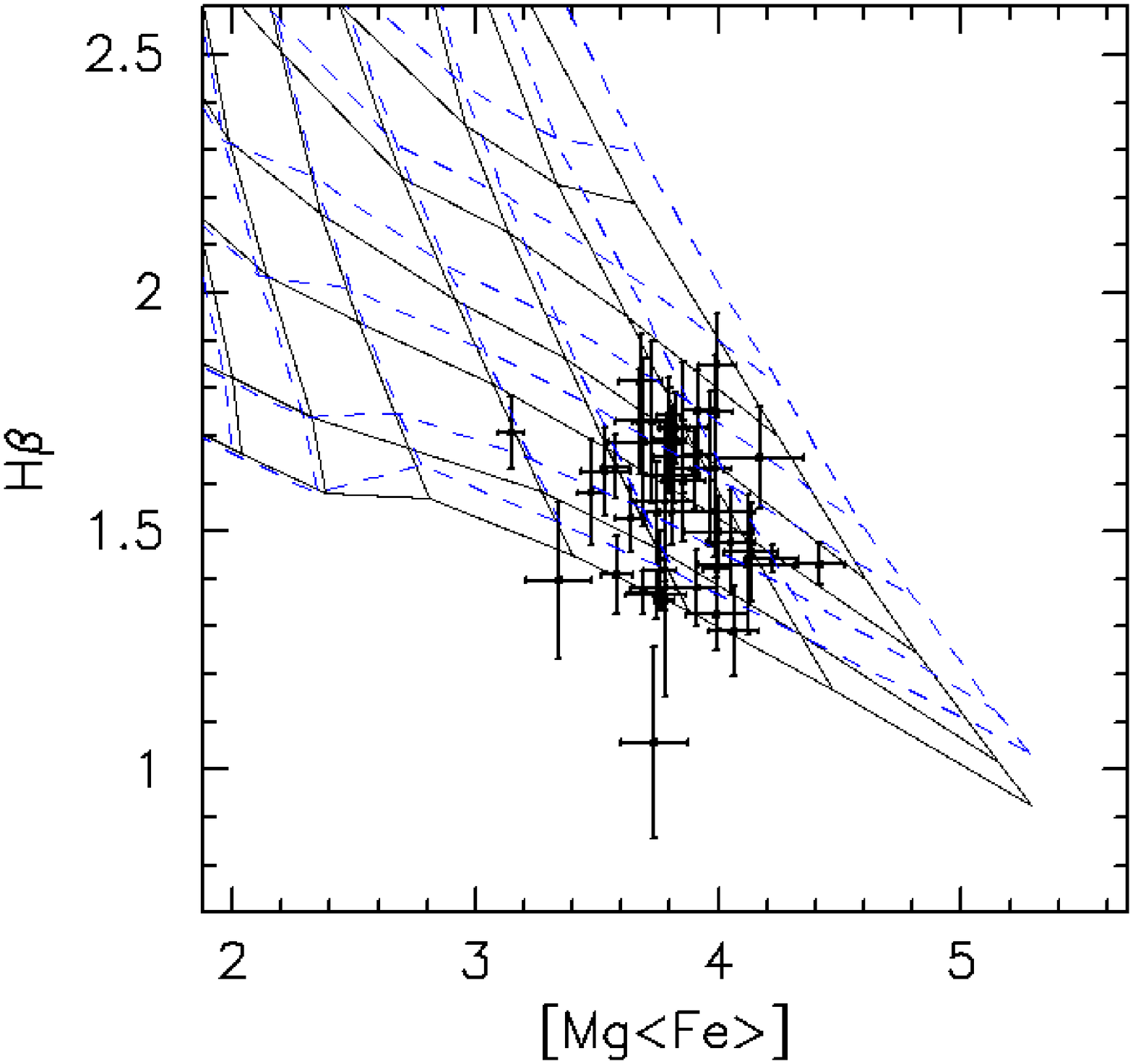}}\quad
        \subfigure{\includegraphics[scale=0.25]{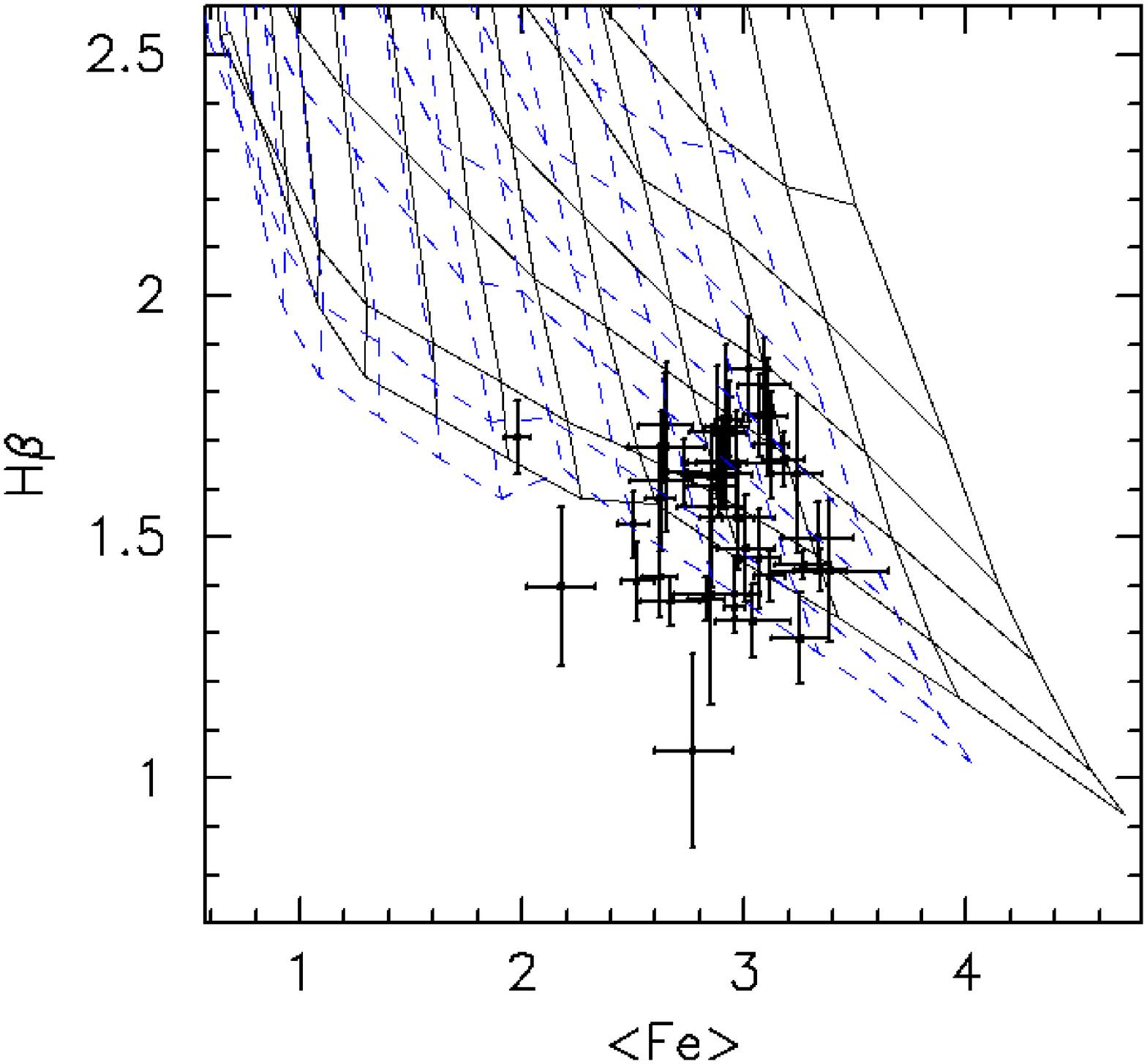}}\quad
        \subfigure{\includegraphics[scale=0.25]{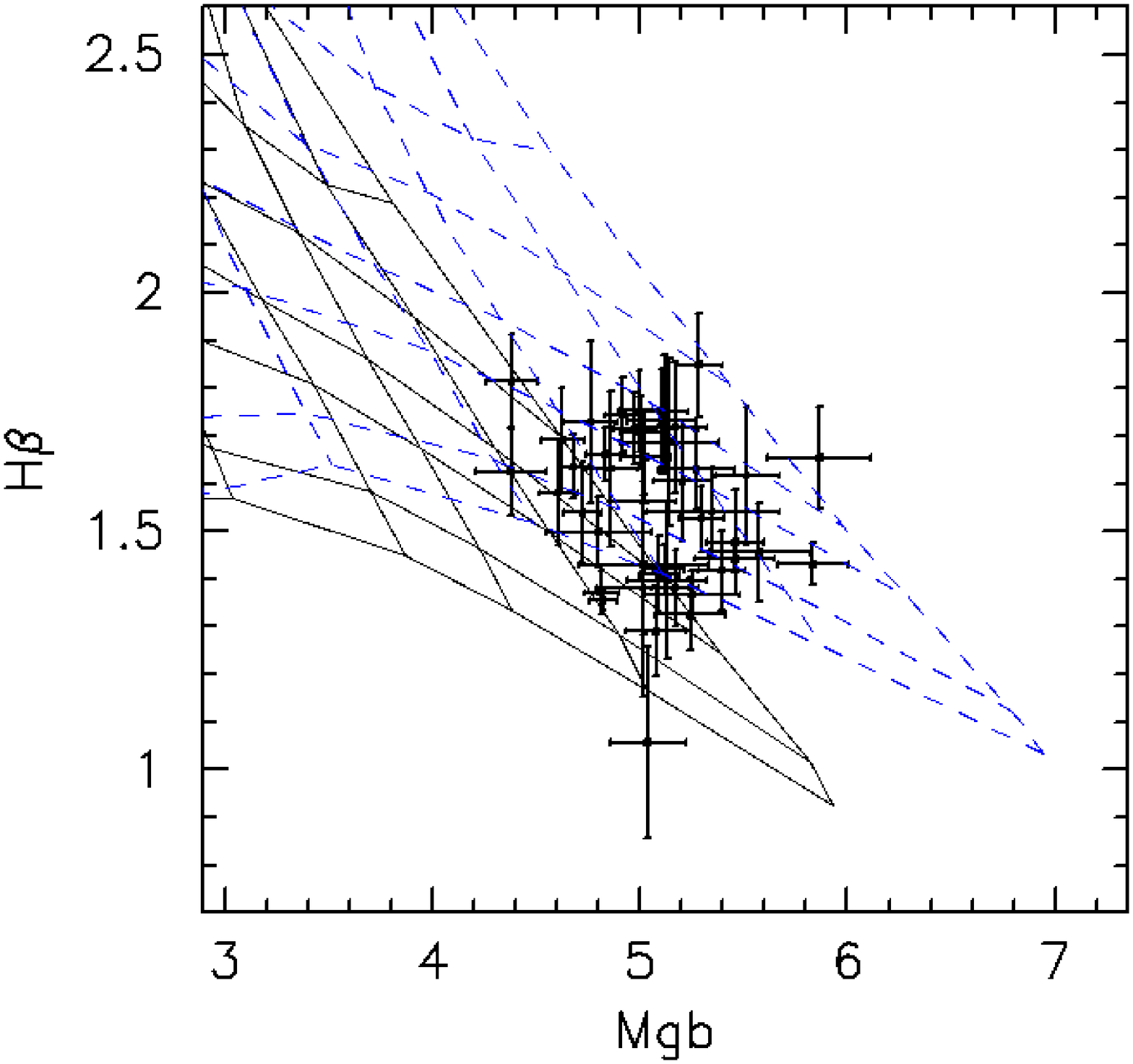}}}
\caption{Index-index plots. The grids correspond to the Thomas et al.\ (2003) models, with $\alpha$-enhancement [E/Fe] = 0 (solid) and 0.3 (dashed). Age lines are 15, 12, 8, 5 and 3 Gyr (from the bottom) and metallicities from [Z/H] = 0.80 decreasing in steps of 0.25 dex towards the left. All index measurements are in \AA{}. The galaxy point which lies below the H$\beta$ grid is that of NGC0533, which reached the upper age limit of the models but does not contain emission.}
   \label{Grids}
\end{figure*}

All SSP parameter results are shown in Table \ref{SSPS}. The errors for the galaxy PGC026269 are unusually high and, therefore, this galaxy is excluded from further analysis. As mentioned before, five galaxies (Leda094683, NGC7649, NGC6166, PGC025714 and PGC030223) showed sky line or CCD gap residuals in Mg$_{\rm b}$, Fe5270 or Fe5335. Hence the derived parameters using these indices were deemed unreliable for these galaxies, and they were also excluded from the SSP analysis. Thus, with NGC6047 and NGC4946 (the two ordinary elliptical galaxies) already excluded, our final sample for which SSP-analysis are carried out, contains 43 BCGs. Index-index plots are shown in Figure \ref{Grids}\footnote{[Mg$<$Fe$>$]=$\sqrt{\rm Mg_{\rm b} \times <Fe>}$}.

Two galaxies (ESO444-046 and NGC0533 -- neither of which show nebular emission) reached the upper age limit of the KMT models, and 11 galaxies have $\log$ (age) $<$ 0.8 Gyr: ESO202-043, ESO346-003, ESO541-013, GSC555700266, IC1101, NGC0541, NGC7012, NGC7597, PGC044257, PGC071807 and PGC072804. Emphasis is placed on the fact that these are SSP-equivalent ages. If a galaxy has experienced a more complicated SFH than a single burst of star formation, then the age derived here will be biased towards the age of the younger stars and not the dominating stellar populations by mass (Li $\&$ Han 2007). On the other hand, the SSP-metallicity will be more biased to the metallicity of the old population, depending on the mass fraction of the burst (Serra $\&$ Trager 2007; S\'{a}nchez-Bl\'{a}zquez et al.\ 2007). Thus, the derived age should not be interpreted as the time that passed since formation of most stars in that galaxy. However, Balmer-line based ages allow us to detect minor amounts of recent star formation in generally old galaxies. Several authors (Maraston et al.\ 2003; Trager et al.\ 2005) pointed out that another complication to the age and metallicity analysis of non-star forming galaxies is the possible presence of hot populations of stars not included in the models (such as blue stragglers and horizontal branch stars). However, Trager et al.\ (2005) showed that these stars affect the inferred metallicities more than ages. 

Figure \ref{Histogram_age} shows the age, [Z/H] and [E/Fe] distributions for our sample of BCGs. We compare the observed scatter with that expected from the errors (see Table \ref{Scatter_SSP}) to investigate if the distributions are compatible with a single value of age, [Z/H] and [E/Fe] for all the BCGs. We found that, while the errors in the age explain the observed scatter, the scatter in [Z/H] and [E/Fe] is much larger, possibly indicating a real variation in the mean stellar abundances of these galaxies.

\begin{table}
\centering
\begin{tabular}{l c c c}
\hline Parameter &  $\sigma_{\rm std}$ & $\sigma_{\rm exp}$ & $\sigma_{\rm res}=\sqrt{\sigma_{\rm std}^{2}-\sigma_{\rm exp}^{2}}$ \\
\hline $\log$ (age) & 0.1616 & 0.2016 & $\star$ \\
$[$Z/H$]$ & 0.1669 & 0.0740 & 0.1496 \\
$[$E/Fe$]$ & 0.0849 & 0.0393 & 0.0752 \\
\hline
\end{tabular} 
\caption[Scatter of the SSP Parameters.]{Scatter of the SSP parameters compared to that expected from the errors. $\sigma_{\rm std}$ is the standard deviation on the mean SSP-parameter value, $\sigma_{\rm exp}$ is the standard deviation expected from the mean errors on the parameter values, and $\sigma_{\rm res}=\sqrt{\sigma_{\rm std}^{2}-\sigma_{\rm exp}^{2}}$ is the residual scatter not explained by the errors on the parameters.}
\label{Scatter_SSP}
\end{table}

\subsection{SSP-equivalent parameters: comparison with ellipticals}

Several authors have found a decreasing amount of scatter in the age and metallicity parameters with increasing velocity dispersion for early-type galaxies (Caldwell et al.\ 2003; Nelan et al.\ 2005; S\'{a}nchez-Bl\'{a}zquez et al.\ 2006b). Thus, more massive elliptical galaxies seem to be a much more homogeneous family of galaxies than less massive ones. Is this also true for the galaxies in the centre of the clusters?

\begin{figure}
   \centering
   \mbox{\subfigure{\includegraphics[scale=0.33]{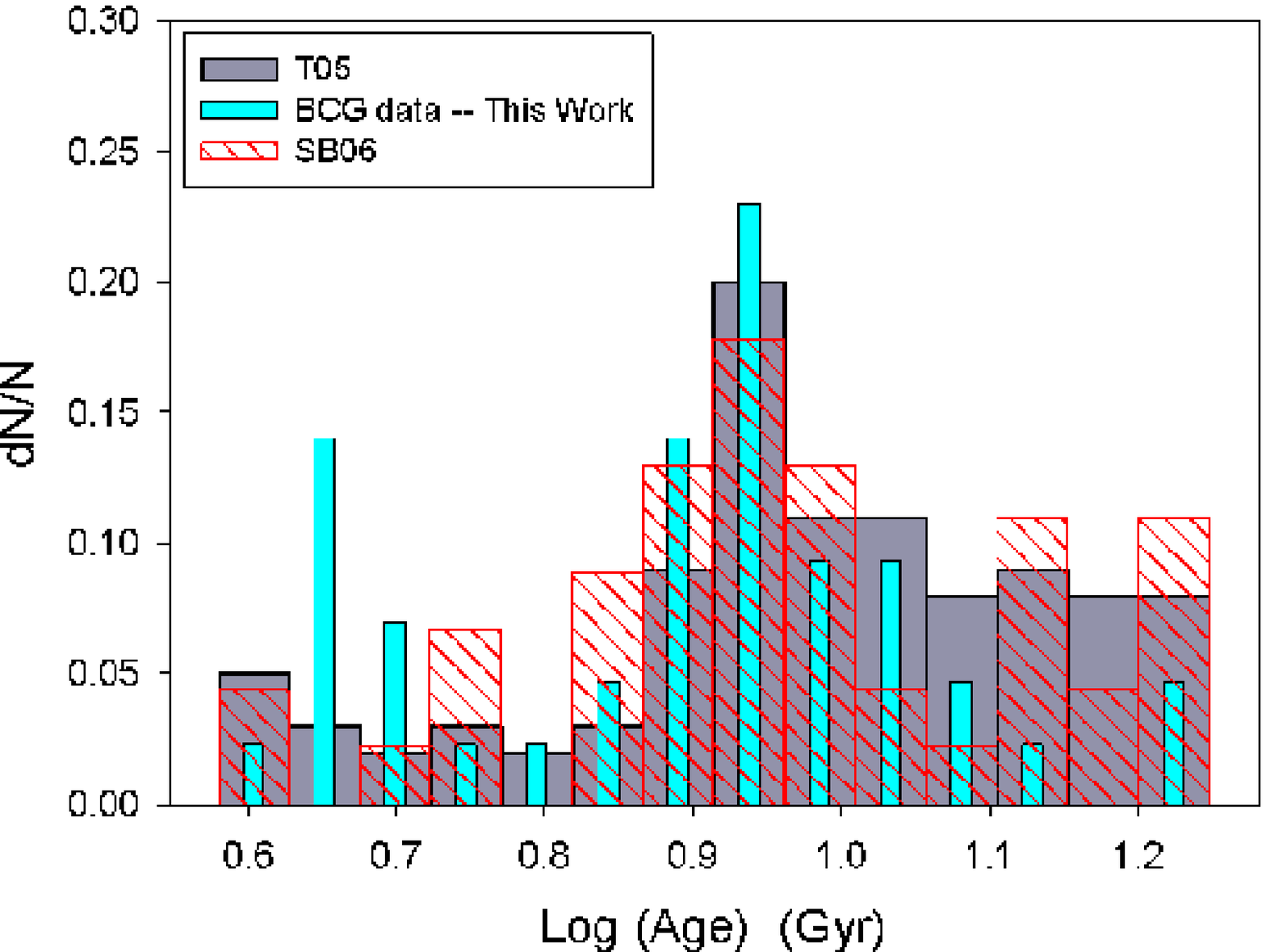}}}
   \mbox{\subfigure{\includegraphics[scale=0.33]{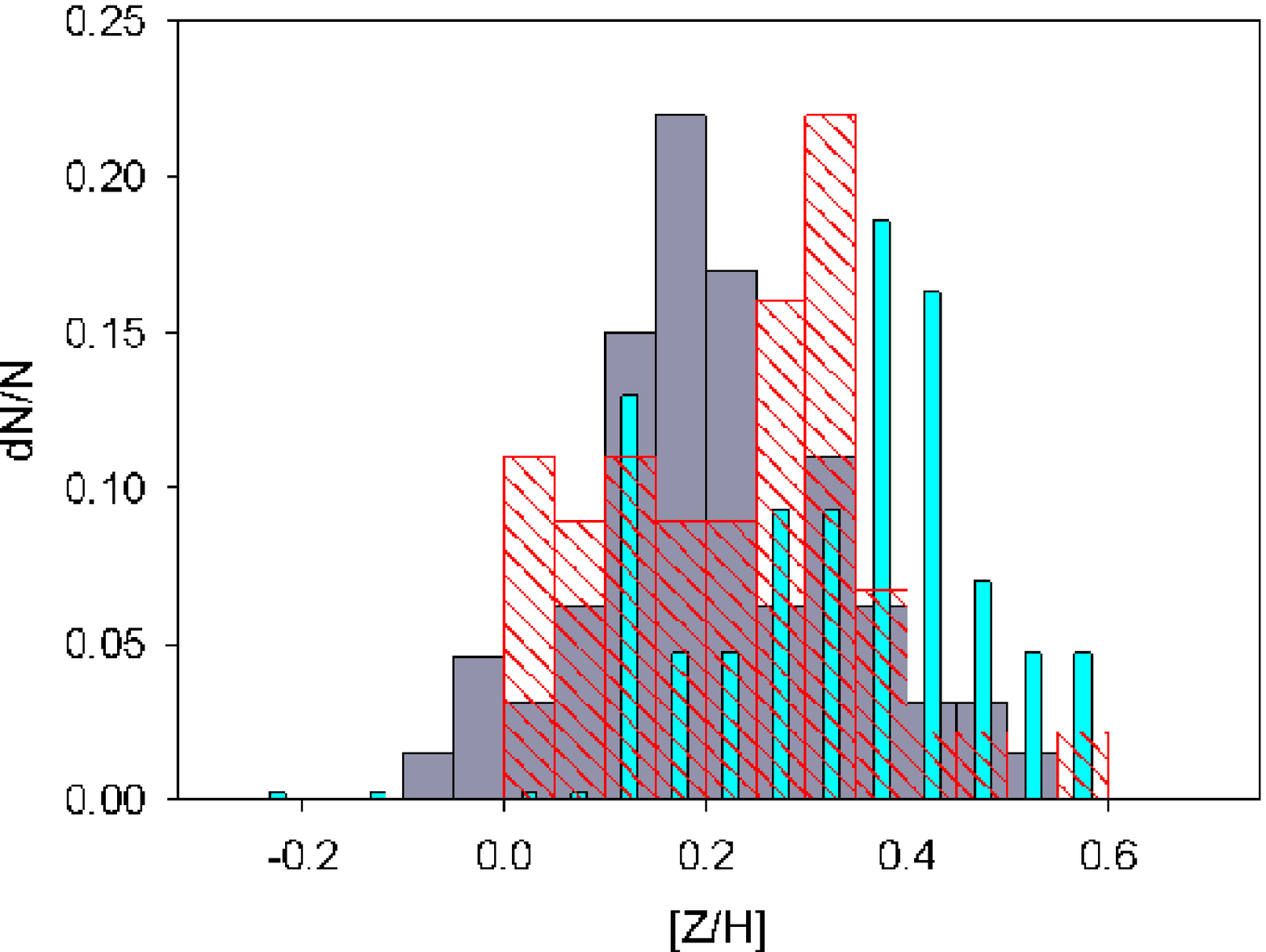}}}
   \mbox{\subfigure{\includegraphics[scale=0.33]{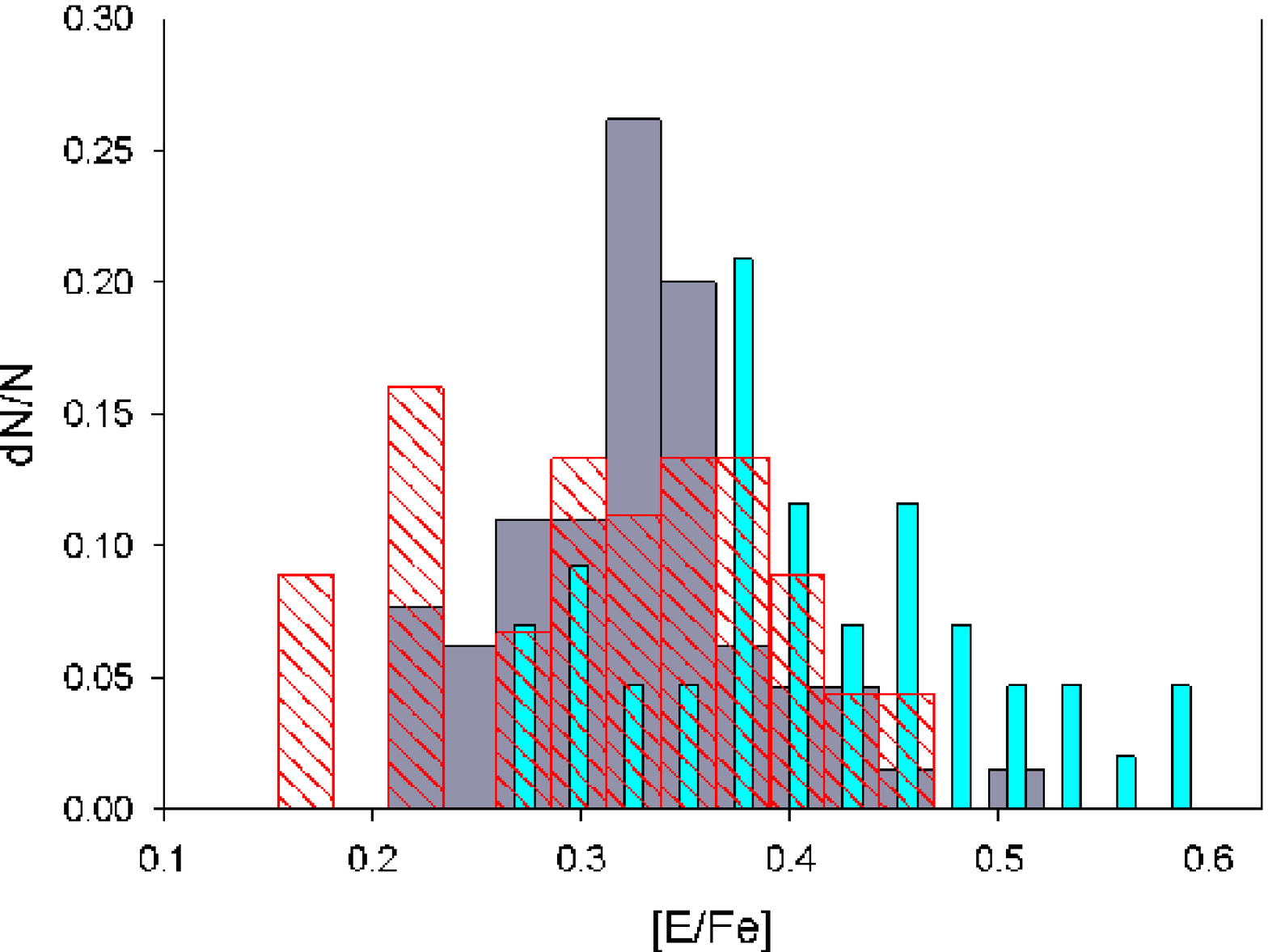}}}
   \caption{Distributions of the SSP-equivalent parameters of the BCGs (cyan), compared to that of ordinary ellipticals (T05 -- grey; SB06 -- red), over the same mass range.}
   \label{Histogram_age}
\end{figure}

Figure \ref{Histogram_age} compares the distribution of SSP-equivalent parameters with that of ordinary ellipticals from Thomas et al.\ (2005 -- T05 hereafter) and SB06. To avoid artificial offsets between samples due the use of different techniques to calculate the SSP-parameters, we recalculated these parameters using exactly the same indices and method as used for our sample of BCGs. Nevertheless, small offsets can remain due the use of different apertures. The T05 central indices were measured within 1/10 of the effective radius, and the SB06 indices within an equivalent aperture of 4'' at a redshift of $z=0.016$. The uncertainties in assuming gradients to perform these aperture corrections are large because of the great variety of gradients found for elliptical galaxies. These gradients seem to be uncorrelated with other galaxy properties, such as mass, and mean gradients are usually calculated (S\'{a}nchez-Bl\'{a}zquez et al.\ 2009). For the samples used here, which all contain nearby galaxies, these aperture differences produce a negligible effect in the indices, and hence, given the uncertainties in the aperture corrections, these corrections were not applied here. From the T05 and SB06 studies, we select those galaxies with central velocity dispersions between $\log \sigma$ = 2.3 and 2.6 km s$^{-1}$ to match the same range as in our BCGs. We also excluded their known BCGs (four in T05 and five in SB06). This left subsamples of 65 and 45 elliptical galaxies from T05 and SB06, respectively.

Systematic differences were detected between the two samples of ordinary elliptical galaxies. The T05 and SB06 samples used here have 19 galaxies in common. In addition, the three Coma BCGs (NGC4839, NGC4874 and NGC4889) form part of the BCG sample and the original T05 and SB06 samples. Thus, offsets in H$\beta$, Mg$_{\rm b}$ and $<$Fe$>$ could be derived between the two samples of elliptical galaxies used. The indices of the three Coma BCGs measured here were in better agreement with the measurements from SB06 than from T05. For example, the average difference between the H$\beta$ measurements was 0.064 \AA{} compared to the SB06 sample and 0.35 \AA{} compared to the T05 sample. Thus, the following offsets in the indices (derived using the galaxies in common) were applied to the data from T05: H$\beta$ --0.158 \AA{}; Mg$_{\rm b}$ --0.139; $<$Fe$>$ +0.013 before the derivation of the SSP parameters. This normalisation placed the two elliptical samples in very good agreement, as confirmed by a Kolmogorov--Smirnov test (see the last column of Table \ref{KStest}, described below).

We explore possible differences in the distributions of the SSP-parameters derived for BCGs and elliptical galaxies by performing a Kolmogorov--Smirnov test on the three samples (see Table \ref{KStest}), where the null hypothesis is that the distributions were drawn from an identical parent population. We comment on the results in the following paragraphs. 

\begin{table}
\centering
\begin{tabular}{l c c c}
\hline & BCGs   & BCGs  & SB06  \\
                         & compared to          & compared to & compared to \\
                         & SB06                     & T05 & T05 \\
                         & $D=0.290$             & $D=0.267$                & $D=0.264$\\
\hline $\log$ (age) & 0.197 & 0.236 & 0.097 \\
$[$Z/H$]$ & 0.378 & 0.390 & 0.203 \\
$[$E/Fe$]$ & 0.440 & 0.560 & 0.122 \\
\hline
\end{tabular} 
\caption[Kolmogorov-Smirnov tests on the SSP parameter distributions.]{Kolmogorov--Smirnov tests on the SSP parameter distributions. The critical value of the statistical test, at a 95 per cent confidence level, is $D=1.36\sqrt{\frac{m+n}{m \times n}}$, where $m$ and $n$ are the number of galaxies in the sample. Thus, if the test value is larger than $D$ (given in the heading of the table for the three different comparisons), then the samples compared are significantly different from each other at a 95 per cent confidence level.}
\label{KStest}
\end{table} 

\paragraph*{Age:}

As can be seen in Figure \ref{Histogram_age}, and confirmed in Table \ref{KStest}, the peaks and dispersions of the age distributions of all three samples coincide. Nevertheless, the BCG age distribution show a second, smaller peak at $\log$ (age) $\sim$ 0.65, which although not statistically significant, is absent in the elliptical galaxy distributions.

\paragraph*{Metallicity:}
The models provide an estimate of [Z/H] which includes all the elements heavier than H and He. The distribution of the BCG metallicities peaks at a higher value (average [Z/H] = 0.31 $\pm$ 0.17) than the ordinary ellipticals (average SB06 [Z/H] = 0.24 $\pm$ 0.13; T05 [Z/H] = 0.21 $\pm$ 0.13). Table \ref{KStest} confirms that the BCG metallicity distribution is significantly different from both elliptical samples. 

\paragraph*{$\alpha$-enhancement:}
The [E/Fe] ratio is often used as an indicator for the time-scales of star formation (Worthey, Faber $\&$ Gonz\'{a}lez 1992; Trager 2006), as it serves as a crude estimation of the ratio of SNII to SNIa. The high values of [E/Fe] detected in massive early-type galaxies has been commonly interpreted in terms of star formation time-scales, i.e.\ the star formation stop before SNIa have time to contribute significantly with their products (Tinsley 1980). However, a high [E/Fe] can also be the consequence of differences in the initial mass function (IMF) where it is skewed towards massive stars, differences in the binary fractions, or to selective winds that drive most of the Fe-group elements to the intracluster medium.

Figure \ref{Histogram_age} shows that the BCG sample has slightly higher $\alpha$-enhancement values (average BCG [E/Fe] = 0.41 $\pm$ 0.09; SB06 [E/Fe] = 0.30 $\pm$ 0.10; T05 [E/Fe] = 0.33 $\pm$ 0.06). Table \ref{KStest} confirms that the BCG $\alpha$-enhancement distribution is significantly different from both elliptical samples.

This result agrees with that of Von der Linden et al.\ (2007), who studied brightest group and cluster galaxies in the SDSS. They found that at the same stellar mass, the stellar populations of BCGs and non-BCGs are similar with the exception of their $\alpha$-element enhancement ratios, which were found to be higher in BCGs. These authors interpret their results in terms of star formation time-scales (star formation occurring in shorter time-scales in the BCGs than in ellipticals), but it is possible that other mechanisms, related to the cluster environment and the privileged position of BCGs are acting to influence their chemical abundances ratios. Other scenarios causing higher [E/Fe] values such as differences in the IMF, or in the binary fractions, or selective winds cannot be conclusively eliminated.

\section{Correlation between kinematics and derived properties}

Recent studies have shown that the lack of rotation found in massive ellipticals is compatible with the idea that these objects formed through dissipationless mergers (e.g.\ Naab $\&$ Burkert 2003; Boylan-Kolchin et al.\ 2006). The remnants left by mergers with or without dissipation are expected to differ in their kinematical structure. For example, in a merger where dissipationless processes dominate, the remnant will show little or no rotation, whereas rotation is expected in remnants left by mergers involving gas (Paper 1, and references therein). In Paper 1, we showed that our sample of BCGs have great variety in their kinematical and dynamical properties, and that a number of BCGs show clear rotation contrary to what is expected if all BCGs formed by radial accretion of satellites without gas. If rotating BCGs are the consequence of dissipational mergers, and these happened relatively recently, we would expect younger ages for these systems. As numerical simulations have shown, when the gas is present in merging systems, it is very effectively funnelled toward the centre of the remant where star formation occurs (e.g.\ Mihos $\&$ Hernquist 1994). 

\begin{figure}
   \centering
   \includegraphics[scale=0.35]{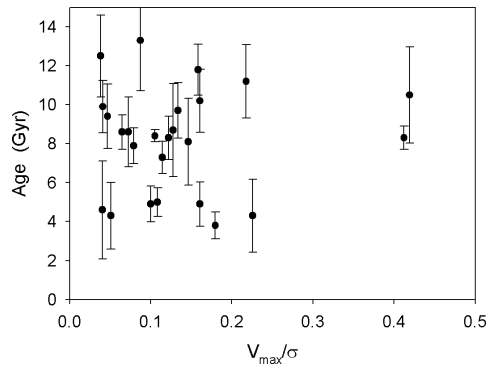}
   \caption[Age vs Rotation.]{The derived ages versus anisotropy parameters for the BCGs (major axis data).}
   \label{SSP_rotation}
\end{figure}

Figure \ref{SSP_rotation} shows the anisotropy parameters (where $V_{\rm max}$ is half the difference between the peaks of the rotation curve) of the BCGs for which major axis spectra were observed against the ages. No real difference is visible in the ages of rotating and non-rotating galaxies, and a whole range of ages were found for galaxies showing a lack of rotation. In fact, the two galaxies rotating the fastest are amongst the oldest in our sample.

\begin{figure}
   \centering
   \mbox{\subfigure{\includegraphics[scale=0.42]{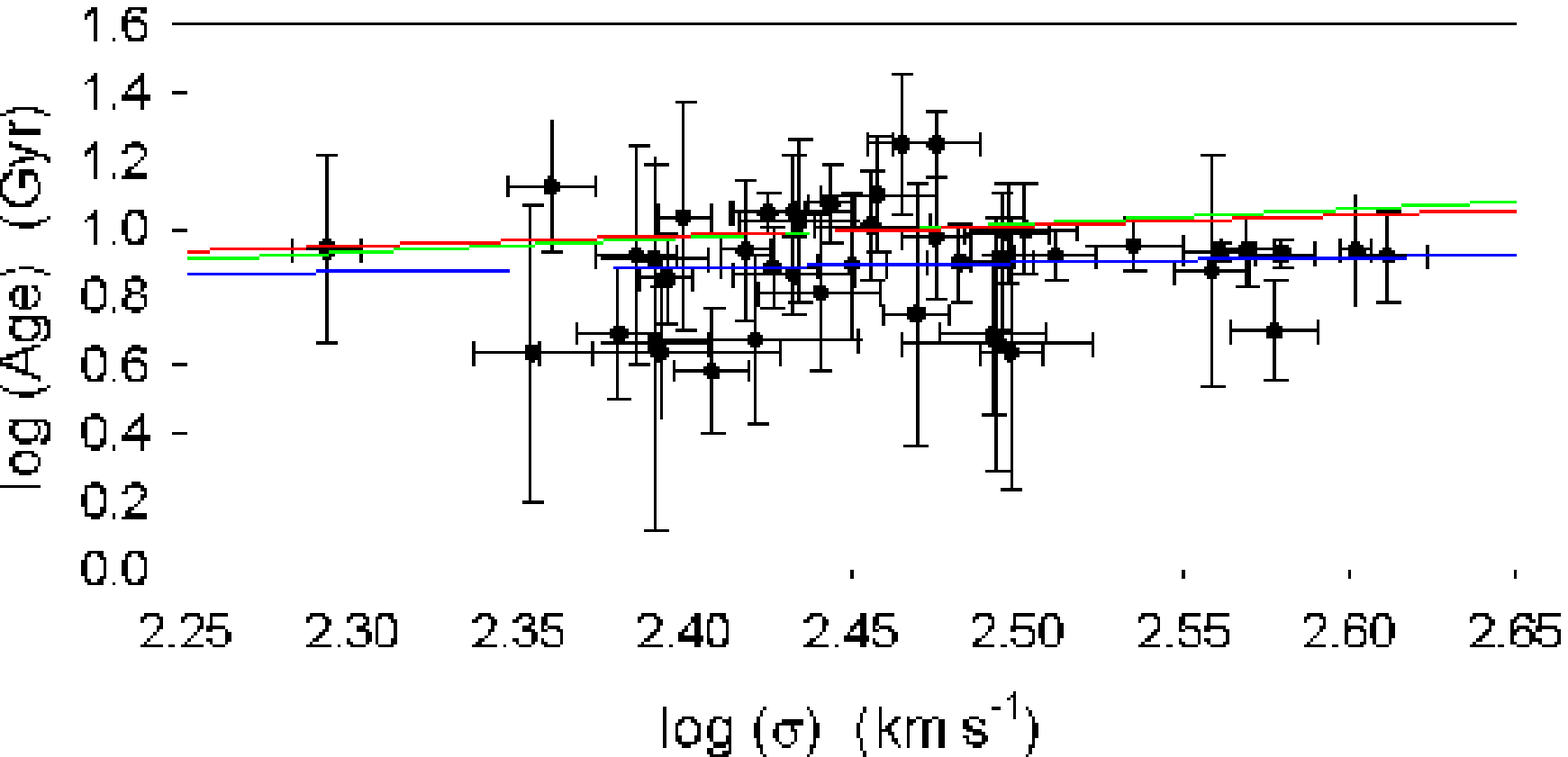}}}
   \mbox{\subfigure{\includegraphics[scale=0.42]{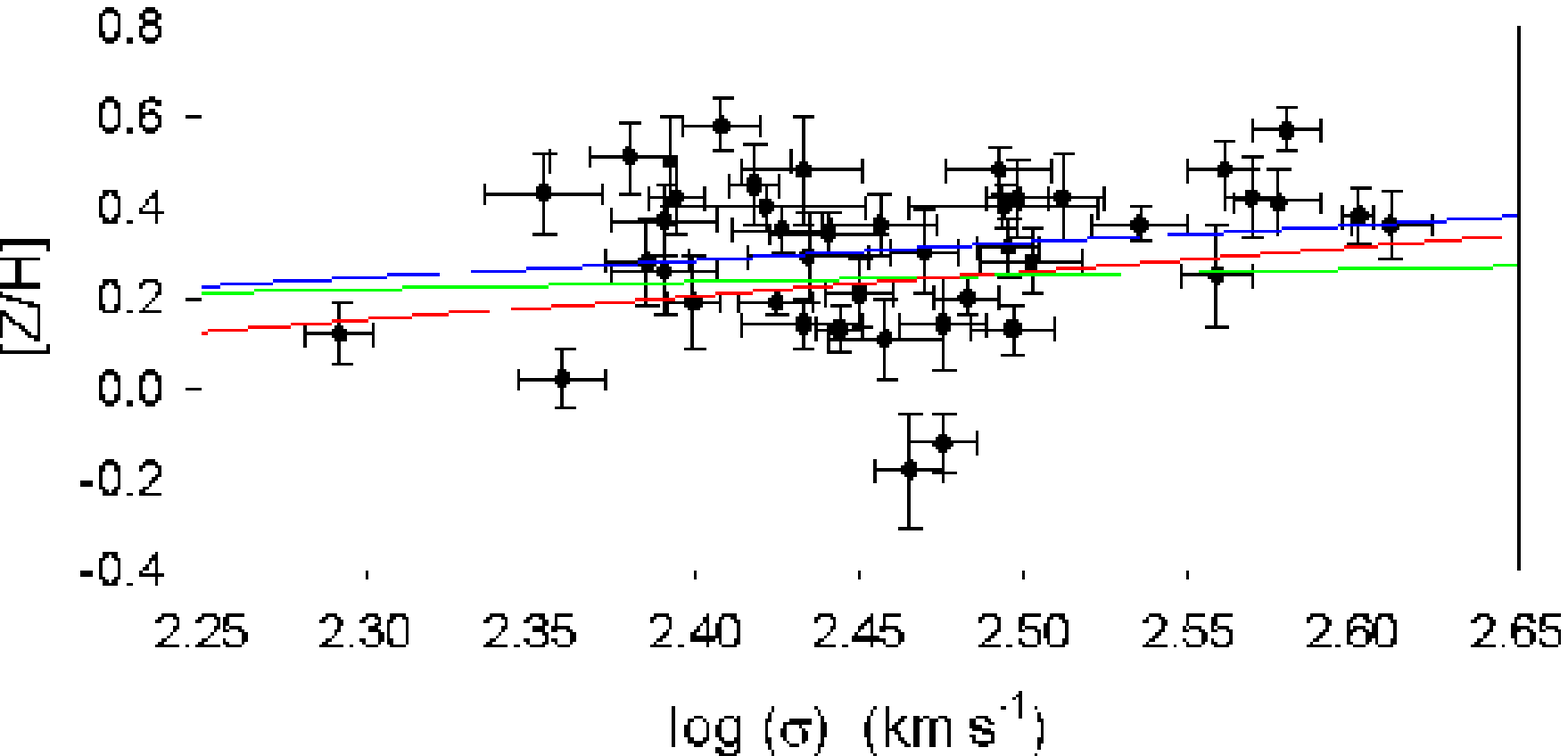}}}
   \mbox{\subfigure{\includegraphics[scale=0.42]{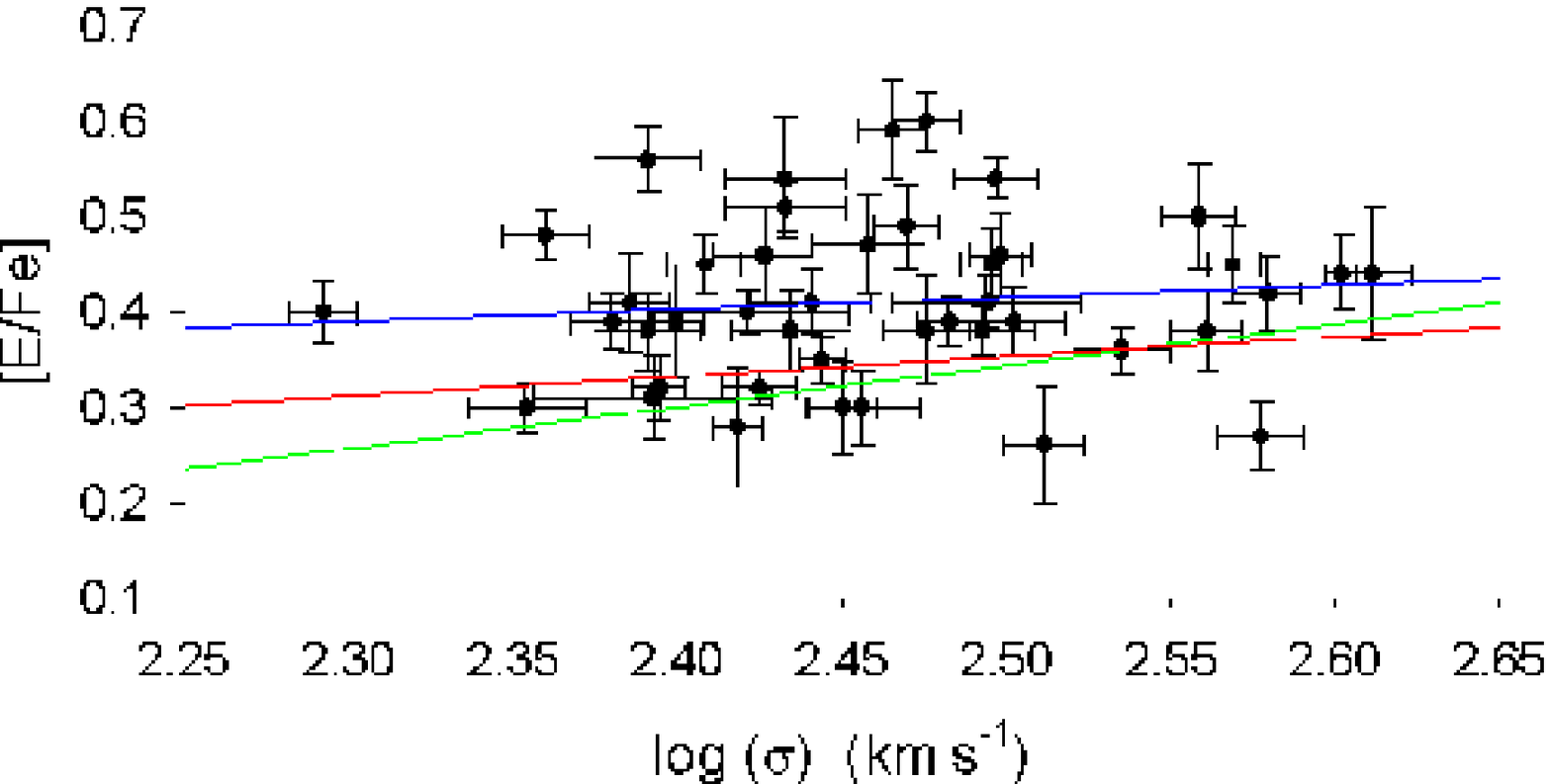}}}
   \caption{Correlations between derived SSP-equivalent parameters and velocity dispersions. The blue line denotes the correlation found for the BCGs whereas the red and green line denotes the correlations found for the samples of T05 and SB06, respectively, over the same mass range and using the same SSP models and procedure.}
   \label{Correlation}
\end{figure} 

\begin{table}
\centering
\begin{tabular}{l c c}
\hline Relationship & $t$ & $P$ \\
\hline $\log$ (age) = 0.530 + 0.148 $\log \sigma$ & 0.461 & 0.648 \\
$[$Z/H$] =$ --0.643 + 0.386 $\log \sigma$ & 1.176 & 0.246 \\
$[$E/Fe$] =$ 0.098 + 0.127 $\log \sigma$ & 0.755 & 0.455 \\
\hline
\end{tabular} 
\caption[SSP Parameters: Correlations with Velocity Dispersion.]{SSP parameters: best-fitting straight-line relations with velocity dispersions derived for BCGs.}
\label{Against_sigma}
\end{table}

Figure \ref{Correlation} shows the derived SSP-parameters $\log$ (age), [Z/H] and [E/Fe] against the central velocity dispersion. To test for the presence of correlations, linear relations of the form $P = a + b\times \log \sigma$ were fitted to the BCG data and t-tests were performed to test the null hypothesis $b=0$ (see Table \ref{Against_sigma}). The $t$ and $P$ values have the same meaning as previously (in Table \ref{table:Slopes}). The relations derived for the BCG data are compared with those derived for the ordinary elliptical galaxy samples of T05 (red line) and SB06 (green line) in Figure \ref{Correlation}. The SSP-parameters of the elliptical samples were recalculated using H$\beta$, Mg$_{\rm b}$, Fe5270 and Fe5335 with the KMT models in exactly the same way as for the BCG data.

It can be seen in Figure \ref{Correlation} and Table \ref{Against_sigma} that we do not find significant relations between the derived SSP-parameters and the velocity dispersions in our sample of galaxies, although the slopes are similar to that of normal elliptical galaxies over the same mass range. This result, in principle, contrasts with the relationships obtained for normal elliptical galaxies, for which several authors have found an increase in the mean age with the velocity dispersion (Thomas et al.\ 2005; Nelan et al.\ 2005; S\'anchez-Bl\'azquez et al.\ 2006b). However, the mass range spanned by our sample is too narrow to be able to see the differences. Even if the BCG were simply the extension towards the massive end of normal elliptical galaxies, we would not be able to detect significant correlations in this mass range. Indeed, there are virtually no difference between the relationship obtained for the BCGs and those for the two normal elliptical samples shown in Figure \ref{Correlation}. Several authors have also found the existence of a positive correlation between the degree of $\alpha$-enhancement in the central parts of early-type galaxies and velocity dispersion (Worthey et al.\ 1992; Kuntschner 2000; Trager et al.\ 2000b; S\'{a}nchez-Bl\'{a}zquez et al.\ 2007), which is not visible here because of the narrow mass range. Figure \ref{Correlation} again shows that the BCGs have, on average, higher [E/Fe] values than the ellipticals.

We also investigated the possibility of correlations with the SSP-parameters when the total magnitude is used instead of velocity dispersion. None of the derived parameters show a potential correlation with $K$-band magnitudes from 2MASS, and this was not investigated further. 

\subsubsection*{Other correlations: age--metallicity}

A strong age--metallicity anti-correlation was found for the BCG sample, but is likely to be an artifact of the degeneracy between age and metallicity since the errors on these stellar population parameters are not independent (Kuntschner et al.\ 2001). To check this for the present study, 50 Monte--Carlo simulations were performed. The mean values of the indices were taken and moved randomly using a Gaussian distribution with a width equal to the typical error on the indices. The ages and metallicities were then derived with the same procedure used for the BCG data. The differences between the BCG data and the simulated data are marginal in both the slope of the best-fitting correlation as well as the standard deviation from the relation. Thus, the age--metallicity anti-correlation can almost entirely be explained by the correlation of the errors on the parameters. Table \ref{Scatter_SSP} also showed that there is no intrinsic scatter (other than expected from the errors) in the ages derived for the BCGs, whereas the metallicities do show inherent scatter.

\section{Context of the environment}

Hierarchical models of galaxy formation predict that the formation of the central galaxy is closely connected with the evolution of the host cluster. In principle, this would not necessarily be reflected by differences in the SFH, as stars might have formed before the formation -- or assembly -- of the real galaxy. However, it is interesting to investigate whether, and to what extent, the characteristics of the host cluster influence the stellar populations of the BCG.

The X-ray properties of the host clusters are given in Table \ref{XRAY}. All the X-ray luminosity and temperature (L$_{\rm X}$ and T$_{\rm X}$) values are from spectra observed in the 0.1 -- 2.4 keV band, and using the same cosmology namely H$_{0}$ = 50 km s$^{-1}$ Mpc$^{-1}$, $\Omega_{\rm m}$ = 1 and $\Omega_{\Lambda}$ = 0. 

\begin{table*}
\centering
\begin{scriptsize}
\begin{tabular}{l r c r r r r r r r r}
\hline Galaxy  & Cluster & \multicolumn{2}{c}{L$_{\rm X} \times 10^{44}$} & T$_{\rm X}$ & \multicolumn{2}{c}{Cooling Flow} & \multicolumn{2}{c}{$\sigma_{\rm cluster}$} & \multicolumn{2}{c}{R$_{\rm off}$} \\ 
               &         & (erg s$^{-1}$) & ref & (keV) &  & ref & (km s$^{-1}$) & ref& (Mpc) & ref \\
\hline ESO146-028 & RXCJ2228.8-6053 & 0.17 & b & -- & -- & -- & -- & -- & 0.051 & cb\\
 ESO202-043 & A S0479 & -- & -- & -- & -- & -- & -- & -- & -- & -- \\
 ESO303-005 & RBS521 & 0.79 & b & -- & -- & -- & -- & -- & 0.010 & cb \\
 ESO346-003 & A S1065 & 0.096 & r & -- & -- & -- & -- & -- & 0.032 & cr\\ 
 ESO349-010 & A4059 & 2.80$\pm$0.06 & a & 3.5 & $\checkmark$ & e & 845 & w & 0.019 & e\\
 ESO444-046 & A3558 & 6.56$\pm$0.04 & a & 3.8 & X & e & 986 & w & 0.019 & e\\
 ESO488-027 & A0548 & 0.21 & b & 2.4 & $\checkmark$ & w & 853 & w & $\star$ & cb\\
 ESO541-013 & A0133 & 2.85$\pm$0.04 & a & 3.8 & $\checkmark$ & w & 767 & w & 0.017 & e\\
 ESO552-020 & CID 28 & 0.16 & b & -- & -- & -- & -- & -- & 0.013 & cb \\
 GSC555700266 & A1837 & 1.28 & b & 2.4 & $\checkmark$ & w & 596 & w & 0.020 & cb\\
 IC1101 & A2029 & 17.07$\pm$0.18 & a & 7.8 & $\checkmark$ & w & 786 & w & 0.131 & p\\
 IC1633 & A2877 & 0.20 & b & 3.5 & X & w & 738 & w & 0.015 & cb\\
 IC4765 & A S0805 & 0.03 & b & -- & -- & -- & -- & -- & 0.007 & cb\\
 IC5358 & A4038 & 1.92$\pm$0.04 & a & -- & $\checkmark$ & c & 891 & m & 0.002 & cb\\
 Leda094683 & A1809 & -- & -- & 3.7 & $\checkmark$ & w & 249 & w & 0.044 & p\\
 MCG-02-12-039 & A0496 & 3.77$\pm$0.05 & a & 4.7 & $\checkmark$ & w,e & 705 & w & 0.031 & e \\
 NGC0533 & A0189B & 0.04 & b & -- & -- & -- & -- & -- & 0.004 & cb \\
 NGC0541 & A0194 & 0.14 & b & 1.9 & X & w & 480 & w & 0.037 & cb \\
 NGC1399 & RBS454 & 0.08$\pm$0.01 & a & -- & X & c & 240 & w & $<$ 0.001 & cb\\
 NGC1713 & CID 27 & -- & -- & -- & -- & -- & -- & -- & -- & -- \\
 NGC2832 & A0779 & 0.07 & b & 1.5 & $\checkmark$ & w & 503 & w & 0.038 & cl\\
 NGC3311 & A1060 & 0.56$\pm$0.03 & a & 3.3 & $\checkmark$ & w & 608 & w & 0.015 & pe\\
 NGC3842 & A1367 & 1.20$\pm$0.02 & a & 3.5 & X & w,e,g & 822 & w & 0.252 & e\\
 NGC4839 & A1656 & -- & -- & -- & -- & -- & -- & -- & $\star$ & --\\
 NGC4874 & A1656 & 8.09$\pm$0.19 & a & 8.0 & X & e,g & 1010 & w & 0.038 & cb\\
 NGC4889 & A1656 & 8.09$\pm$0.19 & a & 8.0 & X & e,g & 1010 & w & 0.169 & e\\
 NGC4946 & A3526 & 1.19$\pm$0.04 & a & -- & -- & -- & -- & -- & \multicolumn{2}{c}{not BCG} \\
 NGC6034 & A2151 & 0.98 & c & 3.5 & X & g & 827 & w & $\star$ & -- \\
 NGC6047 & A2151 & 0.98 & c & -- & X & g & 827 & w & \multicolumn{2}{c}{not BCG} \\
 NGC6086 & A2162 & -- & -- & -- & X & g & 323 & s & 0.053 & cl\\
 NGC6160 & A2197 & 0.13 & c & 1.6 & $\checkmark$ & w,g & 564 & w & 0.017 & cc\\
 NGC6166 & A2199 & 4.20$\pm$0.12 & a & 4.7 & $\checkmark$ & w,e,g & 794 & w & 0.007 & e\\
 NGC6173 & A2197 & -- & -- & -- & -- & -- & -- & -- & $\star$ & --\\
 NGC6269 & AWM5 & 0.36 & c & -- & -- & -- & -- & -- & 0.002 & cc \\
 NGC7012 & A S0921 & -- & -- & -- & -- & -- & -- & -- & -- & -- \\
 NGC7597 & A2572 & 0.58 & c & -- & -- & -- & 676 & st & 0.048 & cc\\
 NGC7647 & A2589 & 1.87$\pm$0.04 & a & 3.7 & X & e & 500 & w & 0.073 & e\\
 NGC7649 & A2593 & -- & -- & 3.1 & X & w & 690 & w & 0.020 & cl\\
 NGC7720 & A2634 & 0.99$\pm$0.03 & a & 3.4 & X & e,g & 744 & w & 0.018 & e\\
 NGC7768 & A2666 & -- & -- & 1.6 & X & g & 476 & w & 0.006 & cl \\
 PGC004072 & A0151 & 0.99 & b & -- & -- & -- & 715 & s & 0.006 & cb\\
 PGC025714 & A0754 & 3.97$\pm$0.11 & a & 8.7 & X & e & 747 & w & 0.328 & e\\
 PGC026269 & A0780 & 5.61 & b & -- & $\checkmark$ & e & 641 & e & 0.015 & e\\
 PGC030223 & A0978 & 0.50 & b & -- & -- & -- & 498 & st & 0.027 & cb \\
 PGC044257 & A1644 & 3.92$\pm$0.34 & a & 4.7 & $\checkmark$ & w & 933 & w & 0.009 & pe\\
 PGC071807 & A2622 & -- & -- & -- & -- & -- & 942 & s & 0.249 & cc \\
 PGC072804 & A2670 & 2.70 & b & 3.9 & $\checkmark$ & w & 1038 & w & 0.035 & cb\\
 UGC00579 & A0119 & 3.34$\pm$0.05 & a & 5.1 & X & w,e & 863 & w & 0.054 & e\\ 
 UGC02232 & A0376 & 1.36 & c & 5.1 & X & e & 903 & w & 0.136 & cc\\
 UGC05515 & A0957 & 0.81 & b & 2.9 & X & w & 669 & w & 0.037 & cb \\
 UGC10143 & A2147 & 2.87$\pm$0.15 & a & 4.4 & X & e,g & 1148 & w & 0.082 & e\\
\hline
\end{tabular}
\end{scriptsize}
\caption[X-ray Properties of the Host Clusters.]{\begin{footnotesize} X-ray properties and velocity dispersions of the host clusters for all 49 BCGs and two ellipticals. The $\sigma_{\rm cluster}$ values are in km s$^{-1}$ and the projected distance between the galaxy and the cluster X-ray peak (R$_{\rm off}$) is in Mpc. The $\star$ marks at R$_{\rm off}$ indicate the galaxy is not in the centre of the cluster but closer to a local maximum X-ray density, different from the X-ray coordinates given in the literature. The references are: a = Chen et al.\ (2007); b = Bohringer et al.\ (2004); c = Bohringer et al.\ (2000); r = Cruddace et al.\ (2002); w = White, Jones $\&$ Forman (1997); e = Edwards et al.\ (2007); g = Giovannini, Liuzzo $\&$ Giroletti (2008); cc = Calculated from Bohringer et al.\ (2000); cl = Calculated from Ledlow et al.\ (2003); cb = Calculated from Bohringer et al.\ (2004); cr = Calculated from Cruddace et al.\ (2002); m = Mahdavi $\&$ Geller (2001); st = Struble $\&$ Rood (1999); s = Struble $\&$ Rood (1991); p = Patel et al.\ (2006); pe = Perez et al.\ (1998). All the values for T$_{\rm X}$ are from White et al.\ (1997). \end{footnotesize}}
\label{XRAY}
\end{table*} 

\subsection{Velocity dispersion -- log L$_{\rm X}$}

\begin{figure}
   \centering
   \includegraphics[scale=0.4]{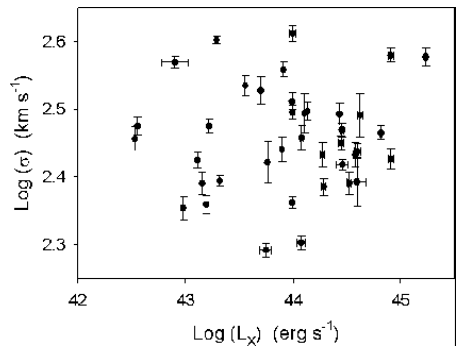}
   \caption[$\log$ ($\sigma$) vs $\log$ L$_{\rm X}$.]{Velocity dispersion for the BCGs plotted against $\log$ L$_{\rm X}$ for the host clusters.}
   \label{logsigma_loglx}
\end{figure}

Cluster X-ray luminosity is directly proportional to the square of the density of the intracluster medium and thus provides a measure of environmental density (Reiprich $\&$ Bohringer 2002). We do not find a correlation between the X-ray luminosities of the clusters (i.e.\ density of the cluster) and the velocity dispersions of the BCGs (i.e.\ the mass of the BCG), as shown in Figure \ref{logsigma_loglx}. Brough et al.\ (2007) noted a weak trend (only 2$\sigma$) between these two properties, in the sense that galaxies in higher density clusters are more massive. However, their sample consisted of only six galaxies. It can be seen from Figure \ref{logsigma_loglx} that this correlation is not found when this much larger sample of BCGs is used. Nevertheless, it is well-known that BCG luminosity does correlate with host cluster mass (Lin $\&$ Mohr 2004; Popesso et al.\ 2006; Hansen et al.\ 2007; Whiley et al.\ 2008).

\subsection{Indices -- log L$_{\rm X}$} 

\begin{figure}
   \centering
   \mbox{\subfigure{\includegraphics[scale=0.4]{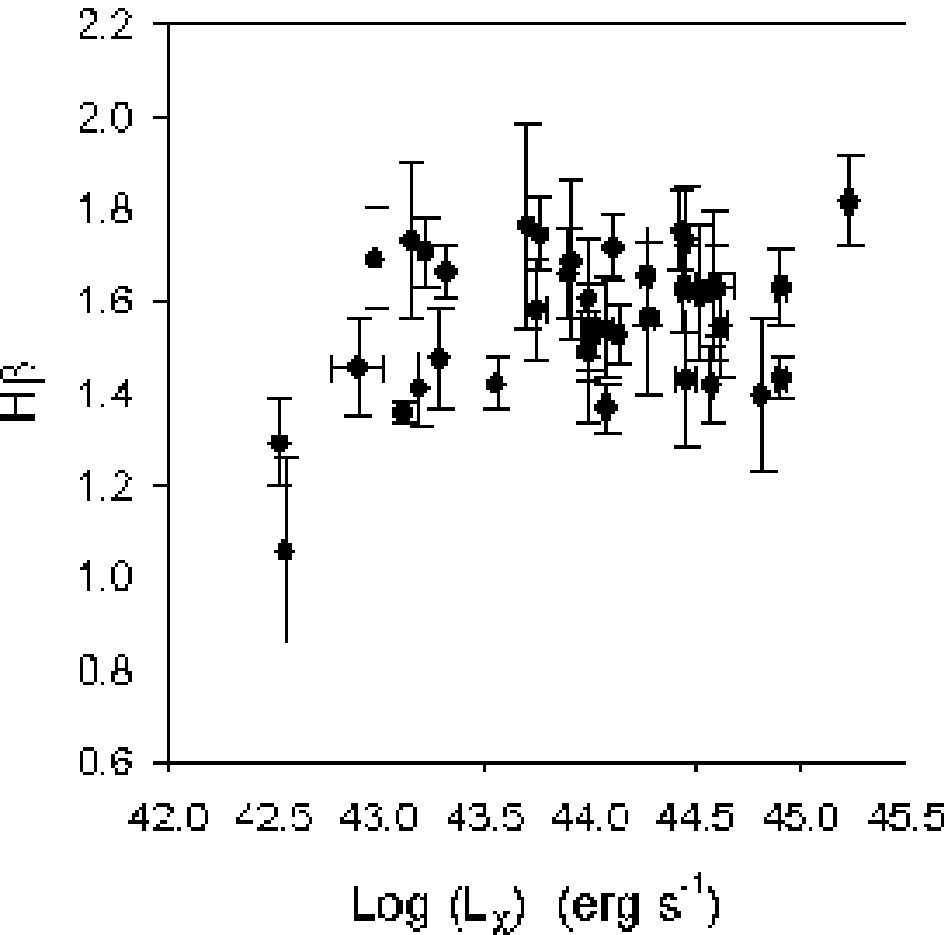}}}
   \mbox{\subfigure{\includegraphics[scale=0.4]{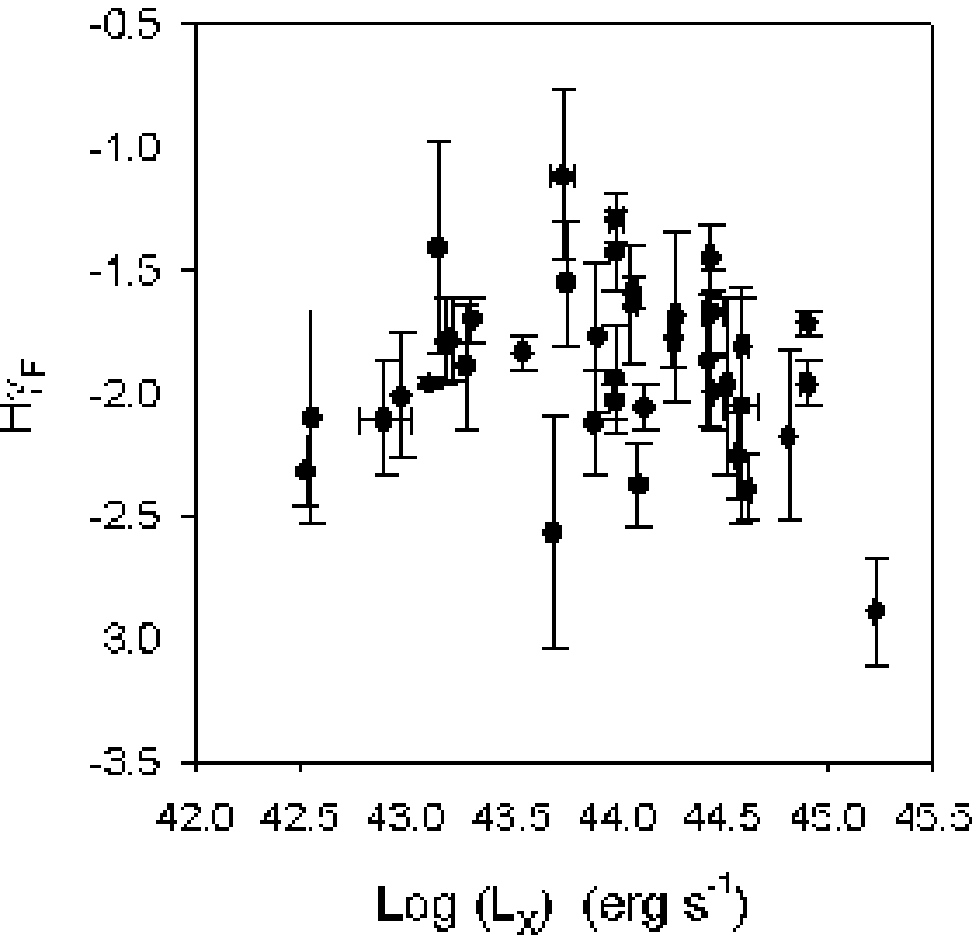}}}
    \mbox{\subfigure{\includegraphics[scale=0.4]{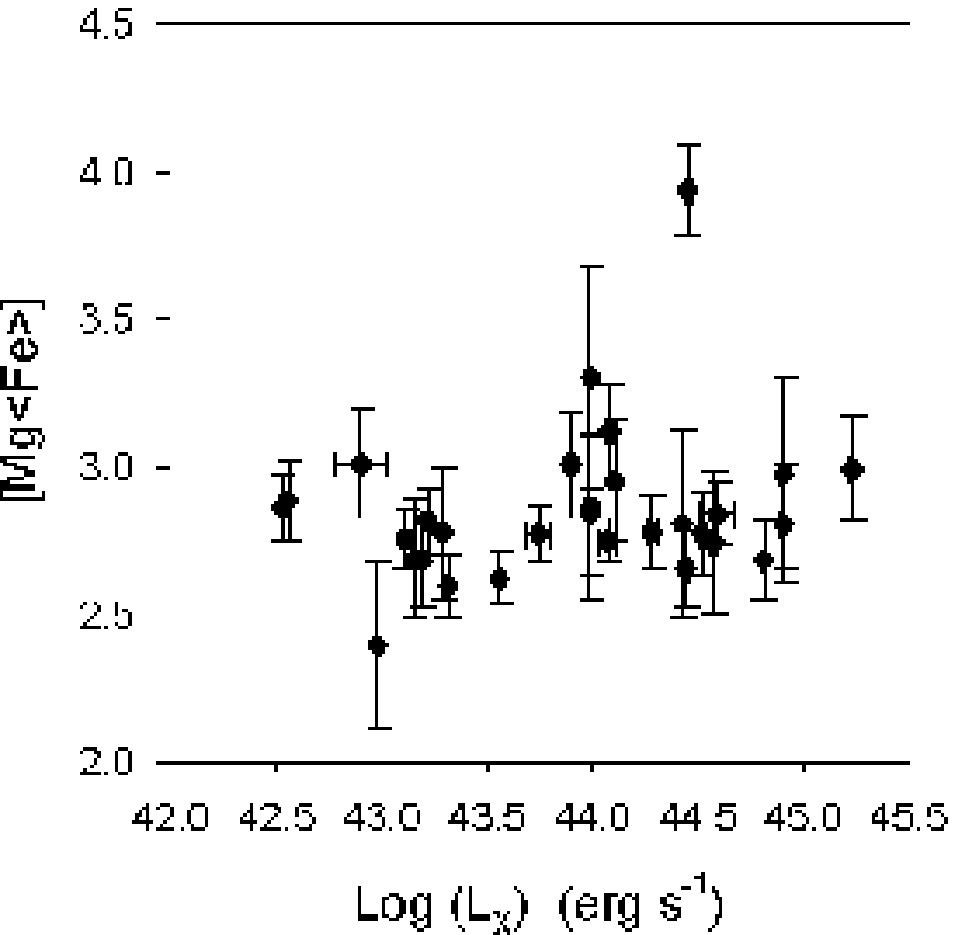}}}
   \caption[Indices vs $\log$ L$_{\rm X}$.]{Examples of indices against $\log$ L$_{\rm X}$ for the BCGs.}
   \label{Indices_loglx}
\end{figure}

Figure \ref{Indices_loglx} shows the relations of some of the central Lick/IDS indices measured in our sample of BCGs with the host cluster X-ray luminosity. We do not find any significant correlation for any of the indices. However, the plots of the Balmer lines against X-ray luminosity seem to suggest a break in the relationships at $\log$ L$_{\rm X}$ $\sim$ 44 erg s$^{-1}$. For low-L$_{\rm X}$ clusters, Balmer-line strengths seem to increase with X-ray luminosity. However, for clusters with X-ray luminosity greater than $\log \rm\ L_{\rm X} \sim$ 44 erg s$^{-1}$, this trend seems to reverse (with the exception of the most dense cluster). However, this is a weak trend, and not confirmed by the relationship between the derived age with cluster X-ray luminosity (not shown here). As discussed in the introduction, previous photometric studies reported different evolutionary histories in X-ray bright and dim clusters. Brough et al.\ (2002) places this break at L$_{\rm X}$ = 1.9 $\times 10^{44}$ erg s$^{-1}$, and conclude that BCGs in high-L$_{\rm X}$ clusters assemble their mass at $z > 1$ and have been passively evolving since, whereas BCGs in low-L$_{\rm X}$ clusters appear to be in the process of assembling their mass. Figure \ref{Indices_loglx} suggests two different regimes in the cluster X-ray luminosity and the BCG line strengths, but this is not conclusive due to the scatter and large errors on the measurements. 

Host cluster velocity dispersion data were also collected from the literature as shown in Table \ref{XRAY}, and no clear correlations between host cluster velocity dispersion (indicative of the mass of the host cluster) and any of the derived parameters were found.

\subsection{Cooling flow clusters}

\begin{figure}
   \centering
   \includegraphics[scale=0.4]{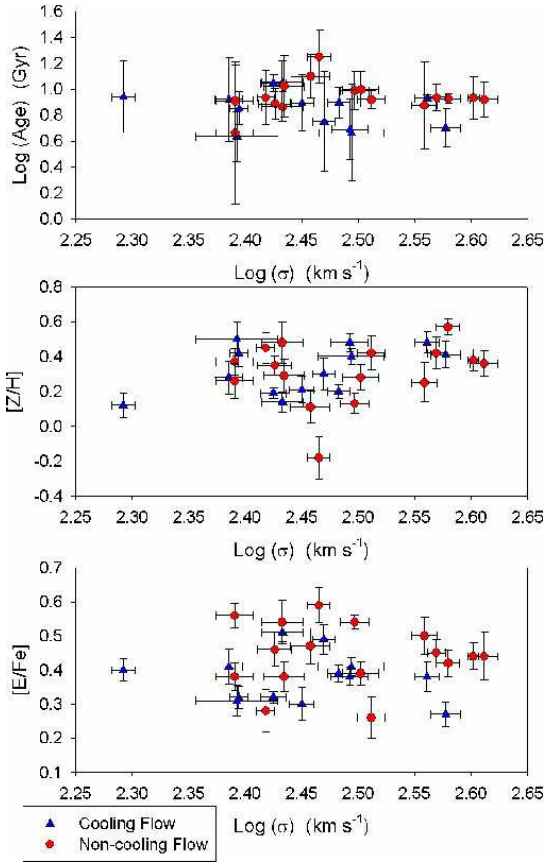}
   \caption[SSP Parameters vs $\log$ $\sigma$ for Cooling Flow and Non-Cooling Flow Clusters.]{The derived SSP-parameters against velocity dispersion for the BCGs. The blue symbols are BCGs in host clusters with cooling flows, and the red symbols those in clusters without cooling flows.}
   \label{sigma_cool}
\end{figure}

A very interesting aspect in the evolution of BCGs is the influence of cluster cooling flows. Several studies reported examples of recent or ongoing star formation in BCGs hosted by cooling flow clusters (Cardiel et al.\ 1998a; Crawford et al.\ 1999; McNamara et al.\ 2006; Edwards et al.\ 2007; O'Dea et al.\ 2008; Bildfell et al.\ 2008; Pipino et al.\ 2009). Cooling flow information was collected from the literature as shown in Table \ref{XRAY}. Figure \ref{sigma_cool} shows the derived SSP-parameters against velocity dispersion for clusters with cooling flow or non-cooling flow data available in the literature. For the six intermediate-aged galaxies (younger than $\sim$ 6 Gyr) for which cooling flow information is available, only one is hosted by a cluster without a cooling flow. Thus, there is a tendency that the BCGs with younger mean ages tend to be in clusters with cooling flows, in agreement with the previous photometric results by Bildfell et al.\ (2008).

\begin{figure}
   \centering
   \mbox{\subfigure{\includegraphics[scale=0.36]{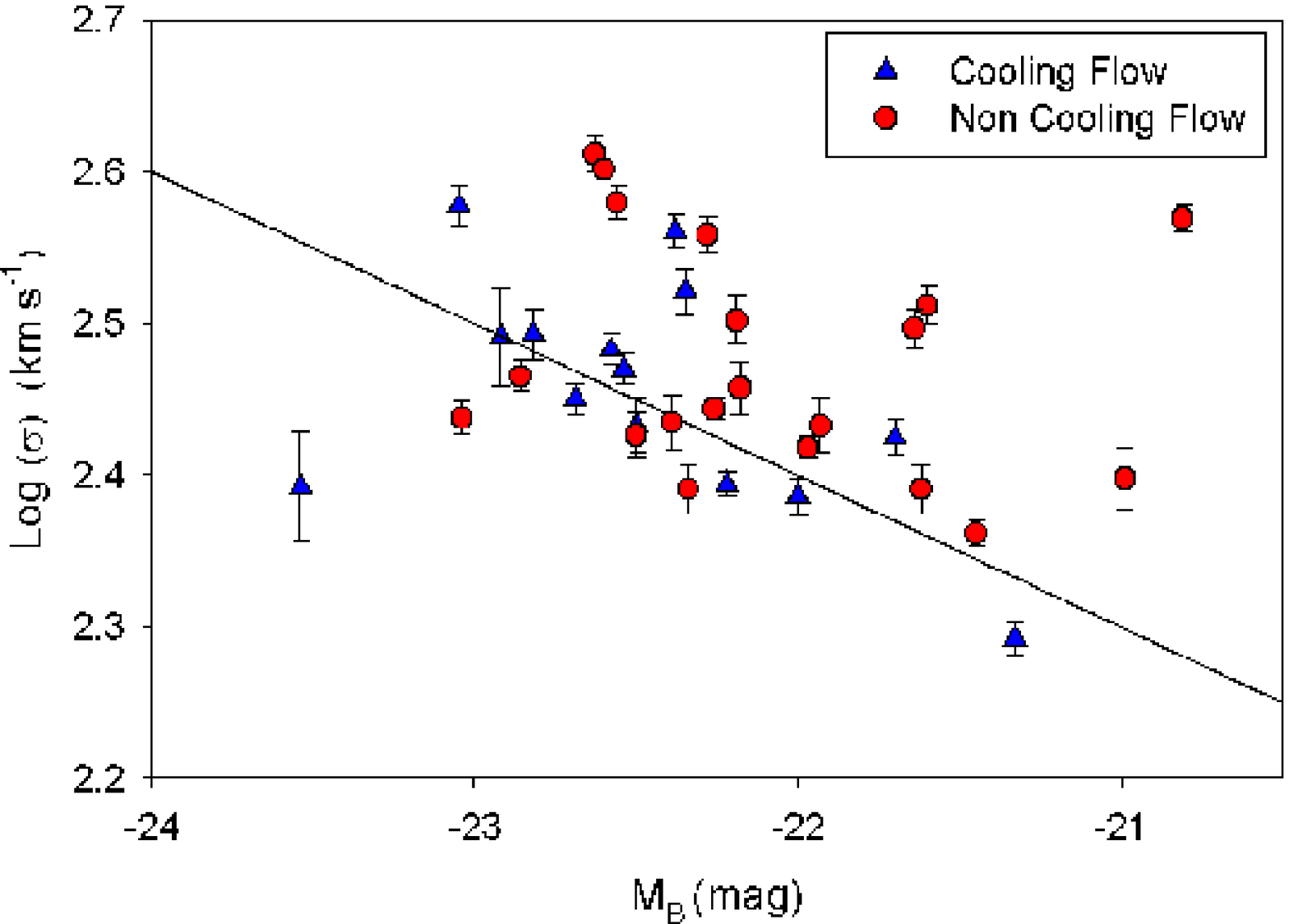}}}
   \mbox{\subfigure{\includegraphics[scale=0.37]{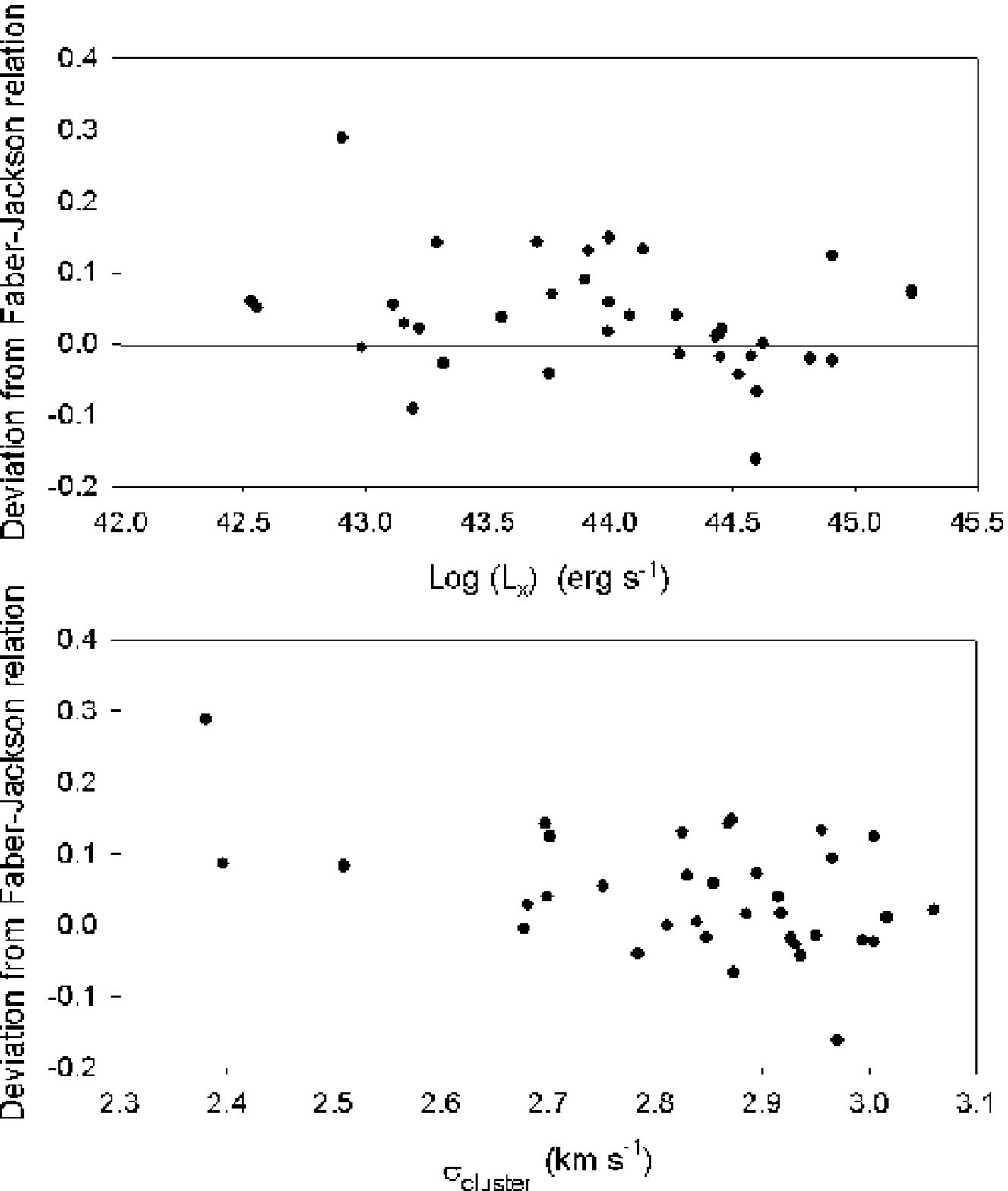}}}
   \caption[BCG $\log \sigma$ for Cooling Flow and Non-Cooling Flow Clusters on the Faber Jackson Relation.]{The $\log \sigma$ versus absolute $B$-band magnitude relation for BCGs in cooling-flow and non-cooling flow clusters. The straight line is the Faber--Jackson relation for normal ellipticals The bottom two plots show the deviation from the Faber--Jackson relation as a function of host cluster X-ray luminosity and velocity dispersion (proxies for host cluster density and mass, respectively).}
   \label{cool_Faber}
\end{figure} 

Several previous studies have shown that BCGs lie above the Faber-Jackson relation (Faber $\&$ Jackson 1976) defined by ordinary elliptical galaxies (Tonry 1984; Oegerle $\&$ Hoessel 1991; Bernardi et al.\ 2006; Desroches et al.\ 2007; Lauer et al.\ 2007; Von der Linder et al.\ 2007). Figure \ref{cool_Faber} shows the Faber--Jackson relation for normal ellipticals, corresponding to $L \propto \sigma^{4}$, and data points for BCGs in cooling-flow and non-cooling flow clusters on the same graph. The lower panels in Figure \ref{cool_Faber} show the deviation from the Faber--Jackson relation against cluster X-ray luminosity and cluster velocity dispersion, respectively. No real difference can be seen between the location of the cooling and non-cooling clusters on the relation (this is also the case if 2MASS $K$-magnitudes are used). Hence, the presence of cooling flows in clusters does not affect the position of the BCG in this scaling relationship. This is to be expected, as the deviation from the Faber--Jackson relation by BCGs is naturally explained by models of dissipationless mergers of elliptical galaxies, provided that the merger orbits become preferentially more radial for the most massive galaxies (Boylan-Kolchin et al.\ 2006). Hence, this deviation is not related to the presence of cooling flows in the cluster centre.

Of the nine emission-line galaxies in this sample for which cooling-flow information is available (see Table \ref{XRAY}), five are hosted by clusters with cooling flows and four are hosted by clusters without cooling flows. Edwards et al.\ (2007) found that the frequency of BCGs showing optical emission lines in their sample increased in cooling-flow clusters (70 per cent versus 10 per cent in non-cooling flow clusters), regardless of the mass density or velocity dispersion of the cluster. This is also true here, where the corresponding fractions are 33 per cent in cooling-flow clusters and 21 per cent in non-cooling clusters, although the difference in the fractions is not as pronounced. One possible cause of this difference could be that the fraction of BCGs with emission lines changes with the distance between the BCG and the cluster X-ray centre (Edwards et al.\ 2007; Best et al.\ 2007).

\subsection{Log L$_{\rm X}$ -- log T$_{\rm X}$}

It is believed that intrinsic scatter in the cluster X-ray luminosity--temperature (L$_{\rm X}$--T$_{\rm X}$) relation is physical in origin, caused by processes such as radiative cooling, and those associated with AGN (McCarthy et al.\ 2004; Bildfell et al.\ 2008).

\begin{figure}
   \centering
   \includegraphics[scale=0.38]{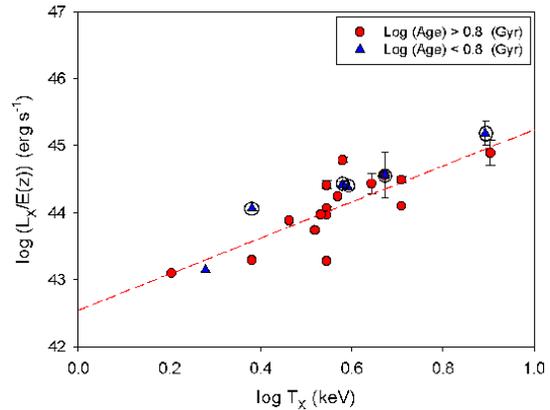}
   \caption[$\log$ L$_{\rm X}$ vs $\log$ T$_{\rm X}$.]{Log (L$_{\rm X}$/E($\textit{z})$) against $\log$ T$_{\rm X}$ for the BCGs. The clusters hosting BCGs with ages $\log$ (age) $<$ 0.8 Gyr are shown in blue, where the black circles denote young galaxies in cooling-flow clusters. The majority of the L$_{\rm X}$ values from the literature does not have errorbars.}
   \label{LxTx}
\end{figure}

Figure \ref{LxTx} shows the $\log$ L$_{\rm X}$ -- $\log$ T$_{\rm X}$ plot for the host clusters for which the measurements of T$_{\rm X}$ and L$_{\rm X}$ were available in the literature. We normalise L$_{\rm X}$ with E($z) = [\Omega_{\rm m}(1+z)^{3} + \Omega_{\Lambda}]^{\frac{1}{2}}$ to correct for the evolution of the mean background density, where $z$ is the redshift of the cluster. We follow Bildfell et al.\ (2008) and fit a power law of the form $\frac{\rm L_{\rm X}}{\rm E(\textit{z})}=\beta \rm T_{\rm X}^{\alpha}$ to the regular (i.e.\ old) BCGs (red dashed line in Figure \ref{LxTx}). We find a strong correlation with $t$ and $P$ values of 6.66 and $<$ 0.0001, respectively. Younger BCGs tend to be located above older BCGs in the diagram, i.e.\ they are predominantly in clusters with X-ray excess. Five of the young galaxies for which the host cluster information is available are hosted by cooling-flow clusters (denoted by a black circle). Only one of the younger galaxies, located at the bottom of the plot, NGC0541, is hosted by a cluster without cooling flows. The result that younger BCGs tend to be hosted by clusters with X-ray excess agrees with the photometric result from Bildfell et al.\ (2008) who showed that their star forming BCGs are exclusively located in clusters with a high-L$_{\rm X}$ deviation from the L$_{\rm X}$--T$_{\rm X}$ relation -- the region of the diagram usually populated by cool-core clusters. This implies that the origin of the cold gas fuelling the star formation may be linked to the processes that give rise to the L$_{\rm X}$ excess, and points to cooling flows as the source of the cold gas in galaxies with young stellar populations.

\subsection{BCG offset from X-ray peak}

If the star formation in all young BCGs is a result of cooling flows, then the young BCGs are expected to be located exclusively at the centres of relaxed clusters, where the cold gas is deposited. Furthermore, numerical simulations predict that the offset of the BCG from the peak of the cluster X-ray emission is an indication of how close the cluster is to the dynamical equilibrium state, and decreases as the cluster evolves (Katayama et al.\ 2003).

We collected the projected angular separations between BCGs and the peak of the X-ray emission from the literature as shown in Table \ref{XRAY}. For those clusters for which it was not available, we calculated it from the BCG and published X-ray peak coordinates. However, this was not possible for those clusters, e.g.\ Coma, where a BCG is not in the centre and where the coordinates of a corresponding local X-ray maximum were not available. 

\begin{figure}
   \centering
   \includegraphics[scale=0.38]{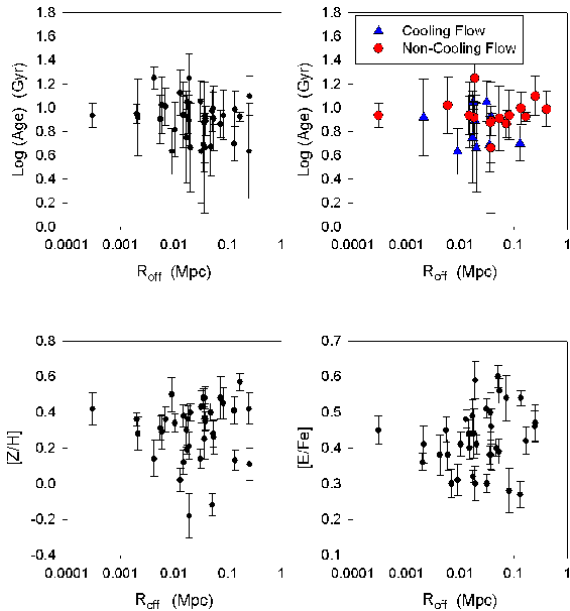}
   \caption[SSP Parameters vs X-ray Offset.]{Derived SSP-parameters plotted against BCG offset from the X-ray peak. The upper right plot shows the derived ages plotted against the X-ray offsets, with separate symbols for galaxies in cooling and non-cooling flow clusters. The other three plots show the derived SSP-parameters against the offsets for all BCGs for which the offsets were available.}
   \label{SSP_offset}
\end{figure}

Figure \ref{SSP_offset} shows the derived ages plotted against the X-ray offsets, with separate symbols for galaxies in cooling and non-cooling flow clusters, as well as plots of the derived SSP-parameters against the offsets for all BCGs for which the offsets were available. Contrary to what was found by Bildfell et al.\ (2008), there is no significant difference in mean X-ray offsets for young and old galaxies. Thus, the younger BCGs are preferentially found in cooling-flow clusters, but they are not necessarily closer to the centres of the clusters, where the cooling flows are deposited, than galaxies with old stellar populations. 

Rafferty, McNamara $\&$ Nulsen (2008) found that the central galaxy is likely to experience significant star formation when: 1) the X-ray and galaxy centroids are within $\sim$ 20 kpc of each other, and 2) the central cooling time of the hot atmosphere is much less than $\sim$ 8 $\times$ 10$^{8}$ yr (an entropy of less than $\sim$ 30 keV cm$^{2}$). Therefore, the younger BCGs need to be very close to the X-ray peaks, even though they are in cooling-flow clusters, for the recent star formation to be a result of the cooling flows. As only three of the five younger galaxies hosted by cooling-flow clusters are located within $\sim$ 20 kpc of the X-ray peak in the present sample, all the recent star formation found in this sample is not necessarily a consequence of the cooling flows.

\begin{figure}
   \centering
   \mbox{\subfigure{\includegraphics[scale=0.35]{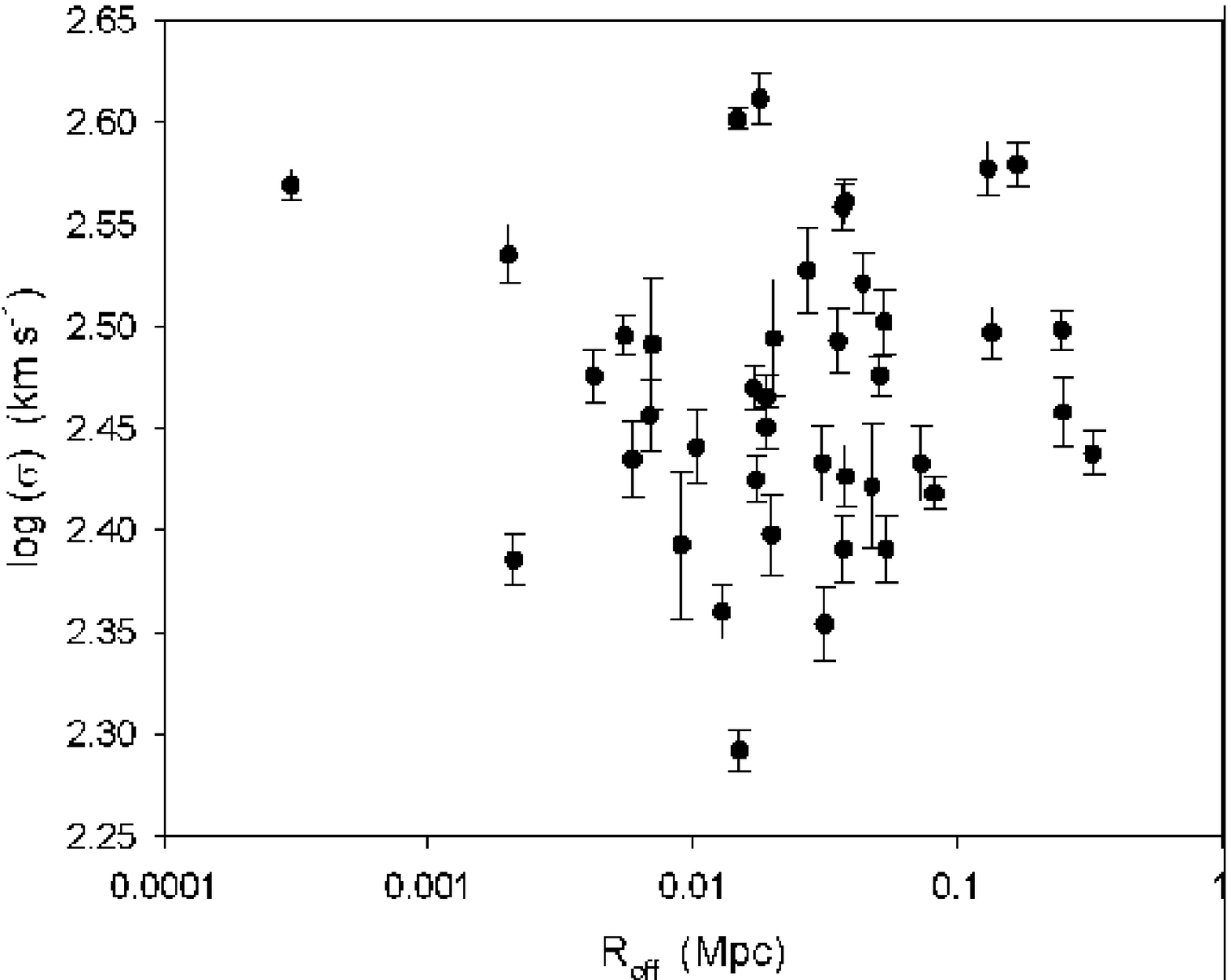}}}
   \mbox{\subfigure{\includegraphics[scale=0.35]{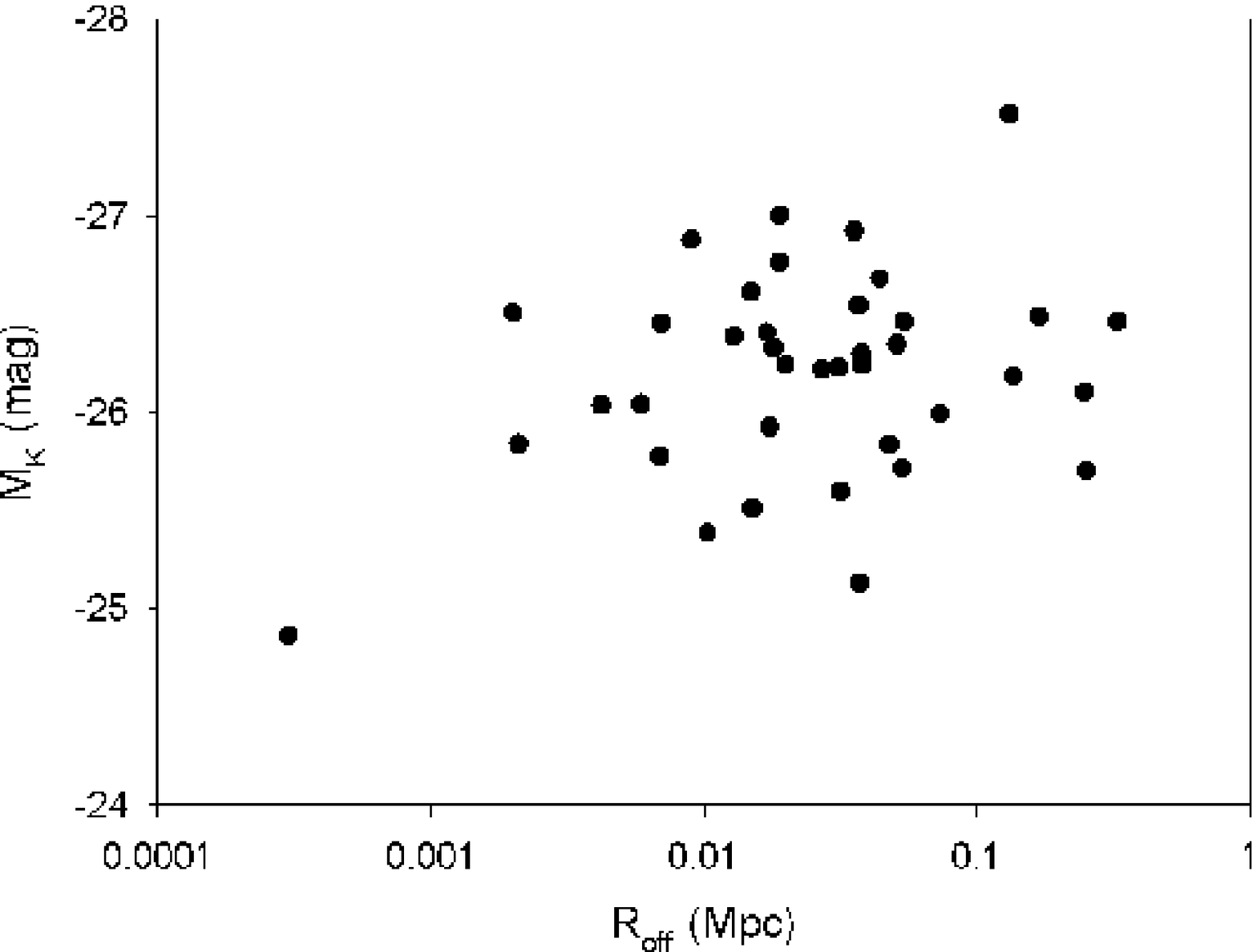}}}
   \caption[Galaxy Velocity Dispersion vs X-ray Offset.]{Galaxy velocity dispersion and absolute $K$-band magnitude plotted against X-ray offset.}
   \label{Sigma_offset}
\end{figure}

Figure \ref{Sigma_offset} shows the X-ray offsets plotted against the galaxy velocity dispersion and absolute $K$-band magnitude. Bildfell et al.\ (2008) found a weak tendency for brightest BCGs to lie closest to their host cluster's X-ray peak while the faintest BCGs are the furthest. This might be expected since massive BCGs will locate at the centre of the cluster potential well on a shorter time-scale than less massive BCGs. They will also have an enhanced probability to merge with small galaxies and will grow more rapidly than BCGs further away from the centre and they will be close to where the cool gas is deposited (Bildfell et al.\ 2008). However, as can be seen in Figure \ref{Sigma_offset}, no such trend can be seen for the present sample, within the uncertainties. 

\section{Conclusions}

We have obtained high signal-to-noise ratio, long-slit spectra of 49 BCGs in the nearby Universe, and derived SSP-equivalent ages, metallicities and $\alpha$-enhancements in the centres of 43 galaxies using the Lick/IDS system of absorption line indices. We compared the indices and parameters derived for these BCGs with those of ordinary ellipticals in the same mass range. We tested the derived properties for possible correlations with the kinematic properties (velocity dispersion, rotation) and luminosity of the galaxies and the X-ray properties of the host clusters (density, mass, distance to X-ray peak, presence of cooling flows).

\begin{itemize}
\item With the exception of G4300, we find no significant discrepancies between the index relations with velocity dispersion of the BCG data and those of normal ellipticals in the same mass range. Only six (Ca4227, H$\gamma_{\rm F}$, Ca4455, Fe4531, C$_{2}$4668 and Mg$_{\rm b}$) out of 18 indices possess a statistically significant slope different from zero. The BCG data are much more scattered around the derived index--velocity dispersion slopes than the elliptical galaxies in that mass range.

\item In general, most BCGs are very old. However, 11 galaxies (26 per cent of the sample for which ages were derived) were found with SSP-equivalent ages $\log$ (age) $<$ 0.8 (in Gyr). Young BCGs are not unusual. Other examples of observational signs of recent star formation and accretion activity in massive central galaxies in clusters are young globular clusters in NGC1275 and multiple nuclei in NGC6166 (Trager et al.\ 2008, and references therein). Bildfell et al.\ (2008) also found 25 per cent of the 48 BCGs in their photometric sample to have blue cores. Active star formation in BCGs pose a challenge for simulations, which predict virtually no recent star formation due to the assumption of extremely efficient AGN feedback (for example De Lucia $\&$ Blaizot 2007).

\item No significant age--velocity dispersion relation is found for the BCG sample in this mass range. The peaks of the age distributions, and the slope of the age--velocity dispersion relation of the BCG and elliptical samples are similar. However, the BCG age distribution also shows a smaller second peak at $\log$ (age) $\sim$ 0.65 which, although not statistically significant according to a Kolmogorov--Smirnov test, is absent in the elliptical galaxy distributions. Thus, it can be argued that the mean, generally old ages of BCGs are similar to normal giant elliptical galaxies, but there is a weak indication that for a small fraction of BCGs, recent star formation occurred as a result of their privileged location in the centre of the cluster. 

\item The peak of the metallicity distribution occurs at a higher value for the BCGs than for the samples of massive elliptical galaxies. The BCG metallicity--velocity dispersion relation is similar in slope than that of normal ellipticals (Figures \ref{Histogram_age} and  \ref{Correlation}). 

\item No significant [E/Fe]--velocity dispersion relation is found for the BCG sample in this mass range. The BCGs have higher [E/Fe] values than the comparison elliptical samples (Figures \ref{Histogram_age} and  \ref{Correlation}). This can naively be interpreted as a consequence of shorter formation time-scales in BCGs. However, other differences such as the IMF, differences in the binary fractions, or selective winds that drive most of the Fe-group elements to the intracluster medium cannot be conclusively eliminated.

\item No real difference is visible in the ages of rotating and non-rotating BCGs, and a wide range of ages were derived for galaxies that show a lack of rotation. No relations between the derived parameters and absolute magnitude were found for the BCG sample. The strong age--metallicity relation found for BCGs is almost entirely due to the correlation between errors on the parameters.

\item No correlation is found between the X-ray luminosity of the clusters and the velocity dispersion of the BCGs (Figure \ref{logsigma_loglx}). Thus, galaxies in higher density clusters are not necessarily more massive. BCGs with younger SSP-equivalent ages are found in both dense and less dense host clusters, however, the relationships between the Balmer line indices and cluster X-ray luminosity possibly suggest two different regimes, and therefore an evolutionary history which is dependent on environment (Figure \ref{Indices_loglx}). 

\item No clear correlation between host cluster velocity dispersion (cluster mass) and any of the derived parameters (age, metallicity and $\alpha$-abundance) were found.

\item Several sources of gas for fuelling the recent star formation in BCGs have been postulated (discussed by Pipino et al.\ 2009). These are: 1) cooling flows; 2) recycling of stellar ejecta; and 3) accretion of satellites. For the six young galaxies (younger than $\sim$ 6 Gyr) for which cooling flow information is available, only one is hosted by a cluster without a cooling flow. This suggest that the recent star formation in this galaxy (NGC0541) might have been fuelled by gas deposited in a merger event, or perhaps triggered by the radio jet originating from the galaxy's supermassive black hole. NGC0541 is known to be very peculiar and is associated with Minkowski's Object, an irregular dwarf located 16 kpc from NGC0541 and in the path of the galaxy's radio jet. The radio jet is thought to have triggered the starburst in Minkowski's Object (Schaerer, Contini $\&$ Pindao 1999; Verdoes Kleijn et al.\ 1999; Croft et al.\ 2006). 

\item As mentioned above, there is a tendency that the younger galaxies in this BCG sample are hosted by clusters with cooling flows. In addition, the L$_{\rm X}$--T$_{\rm X}$ cluster relation shows that the younger BCGs are located in clusters with large values of X-ray excess. However, some of the younger galaxies were found to be slightly further away from the X-ray peak of the cluster. This will limit the influence of the cluster cooling flows, which in turn suggests that the possibility that the gas fuelling the star formation had its origin in mergers cannot be discarded. On the other hand, potential merging and capturing of less massive galaxies would be enhanced by the galaxy being close to the cluster potential well. 

\end{itemize}

In summary: There are differences -- albeit small -- between the stellar populations in BCGs and ordinary elliptical galaxies over the same mass range. The BCGs show higher metallicity and $\alpha$-enhancement values. The former possibly indicates more efficient star formation, and the latter is most commonly interpreted as a consequence of shorter formation time-scales in BCGs. The SSP-equivalent parameters show very little dependence on the mass or brightness of the galaxies, or the mass or density of the host clusters. No real differences are found between the ages of rotating and non-rotating BCGs. Most of BCGs are very old, as expected. However, 11 galaxies (26 per cent of the sample for which ages were derived) were found to have intermediate ages (SSP-equivalent ages of $<$ 6 Gyr). We have shown that the younger BCGs tend to be found in cooling flow clusters which lies above the L$_{\rm X}$--T$_{\rm X}$ relation. However, this is not exclusively the case. In at least one of the young BCGs (NGC0541 -- hosted by a cluster without cooling flows), the gas must have a different origin. This means that there is likely to be more than one process responsible for the recent star formation in BCGs. In addition, the younger BCGs seem to be slightly further away from the cluster X-ray peak, limiting the influence of the cooling flows. As a result, the possibility that more of the young BCGs experienced recent mergers or accretion events involving gas, and leading to star formation, should not be discounted. Overall, the recent star formation in BCGs, and the connection with the processes in the cluster centres, is very complex.

All the indices and SSP results discussed here are only for the central regions of the galaxies. The radial SSP gradients in BCGs will be investigated in a subsequent paper.

\section*{Acknowledgments}
SIL thanks the South African National Research Foundation and the University of Central Lancashire for a Stobie-SALT Scholarship, and gratefully acknowledges helpful discussions with Rob Proctor and Scott Trager. We also thanks the anonymous referee for constructive comments which contributed to the improvement of this paper. PSB is supported by a Marie Curie Intra-European Fellowship within the 6th European Community Framework Programme.

Based on observations obtained on the WHT and Gemini North and South telescopes (Gemini program numbers GS-2006B-Q-71, GN-2006B-Q-88, GS-2007A-Q-73, GN-2007A-Q-123, GS-2007B-Q-43 and GN-2007B-Q-101). The WHT is operated on the island of La Palma by the Royal Greenwich Observatory at the Observatorio del Roque de los Muchachos of the Instituto de Astrof\'{\i}sica de Canarias. The Gemini Observatory is operated by the Association of Universities for Research in Astronomy, Inc., under cooperative agreement with the NSF on behalf of the Gemini Partnership: the National Science Foundation (USA), the Science and Technology Facilities Council (UK), the National Research Council (Canada), CONICYT (Chile), the Australian Research Council (Australia), CNPq (Brazil) and CONICET (Argentina). This research has made use of the NASA/IPAC Extragalactic Database (NED) which is operated by the Jet Propulsion Laboratory, California Institute of Technology.

\appendix

\section{Converting to the Lick/IDS system}
\label{Appendix2}

Index measurements depend on the broadening of the absorption lines caused by both the instrumental spectral resolution and the line-of-sight velocities of the stars. In order to use model predictions based on the Lick system, the observed spectra need to be degraded to the wavelength-dependent resolution of the Lick/IDS spectrograph, and the indices need to be corrected for the broadening caused by the velocity dispersion of the galaxies. 

This section entails the calibration to the Lick/IDS system which consists of three effects that should be accounted for to compare the observed values with those predicted by models:
\begin{itemize}
 \item The difference in spectral resolution between the observational data, and that of the Lick stars observed with the Lick set-up (Faber et al.\ 1985).
 \item The internal velocity dispersion of the galaxies.
 \item Small systematic differences caused by the fact that the Lick/IDS spectra were not flux-calibrated. 
\end{itemize} 

\subsection{Correction to Lick spectral resolution and velocity dispersion corrections}

A total of 22 Lick calibration stars were observed with the 5300 \AA{} dichroic and 10 with the 6100 \AA{} dichroic at the WHT, as described in Paper 1. The Lick stars used for the Gemini data are from the Gemini Science Archive and were observed with an observational set-up corresponding to that of the galaxy data presented here (PI: Bryan Miller, private communication).

As the first step, the Lick star spectra observed with both sets of observations in this work (the WHT and the Gemini data) were shifted to zero radial velocity (Doppler correction), and broadened to the wavelength-dependent resolution of the original Lick/IDS spectra (Worthey $\&$ Ottaviani 1997).

The instrumental broadening ($\sigma_{\rm I}$) of the WHT telescope and ISIS CCD as well as that of the Gemini telescopes and GMOS-N and GMOS-S were measured from the emission lines of the arc spectra. The lines were fitted by Gaussian functions from which the FWHMs were measured.

The stellar velocity dispersion weakens the strength of most absorption features. To calculate the effects of velocity dispersion on the index measurements, the Lick star spectra were smoothed to varying widths in the range $\sigma_{c}$ = 0 to 300 km s$^{-1}$, in intervals of 20 km s$^{-1}$, where $\sigma_{c}$ is the width of the broadening Gaussian. Following Proctor $\&$ Sansom (2002), correction factors $C_{i}(\sigma_{c})$ were determined.

For molecular line indices (CN$_{1}$, CN$_{2}$, Mg$_{1}$, Mg$_{2}$, TiO$_{1}$ and TiO$_{2}$) and for indices with ranges which include negative values (i.e.\ H$\gamma$ and H$\delta$):
\begin{equation}
 C_{i}(\sigma_{c})=I_{\rm L}-I_{\rm Measured},
\end{equation} 
where $I_{\rm L}$ is the index value at the calibration resolution and $I_{\rm Measured}$ the index value measured in the broadened star spectrum.

For all the other atomic line indices:
 \begin{equation}
 C_{i}(\sigma_{c})=I_{\rm L}/I_{\rm Measured}.
\end{equation} 

For each of the indices, $C_{i}$ was plotted against $\sigma_{c}$, and a third order polynomial ($C_{i}=x_{0}+x_{1}\sigma_{c}+x_{2}\sigma_{c} ^{2} +x_{3}\sigma_{c} ^{3}$) was fitted. The value of $x_{0}$ was fixed at zero for the molecular, H$\gamma$ and H$\delta$ indices and at one for the other atomic indices.

\begin{figure*}
   \centering
   \mbox{\subfigure{\includegraphics[width=3.2cm,height=3.2cm]{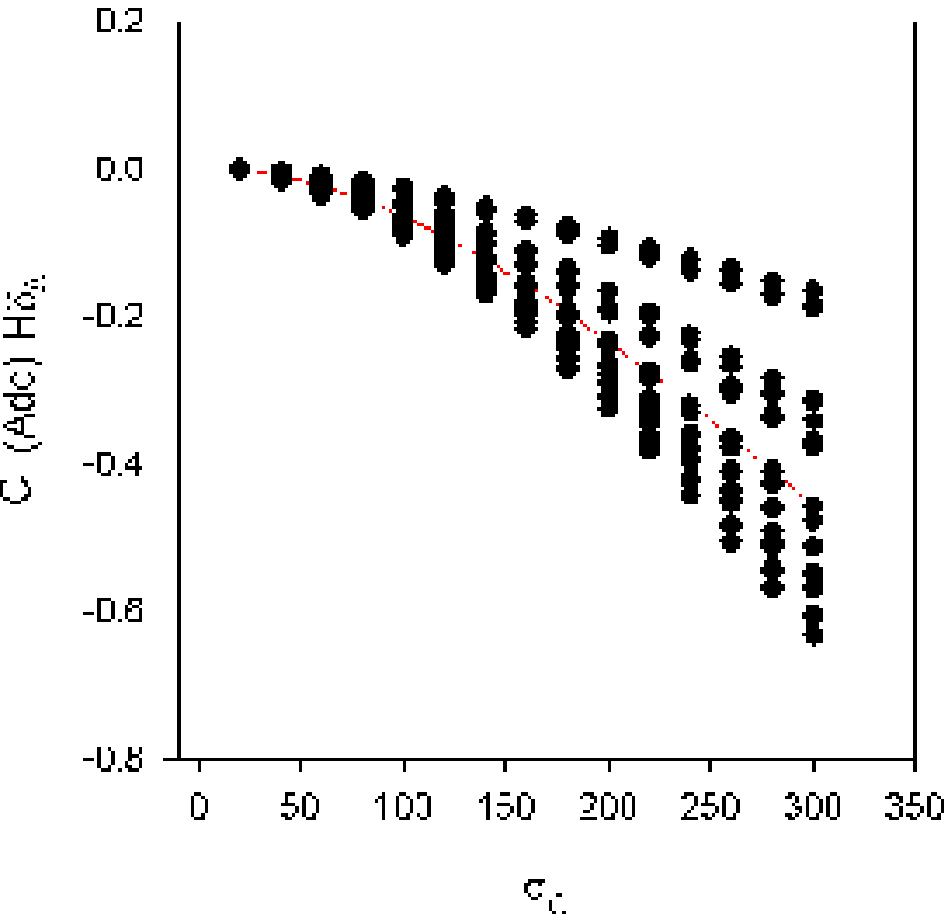}}\quad
\subfigure{\includegraphics[width=3.2cm,height=3.2cm]{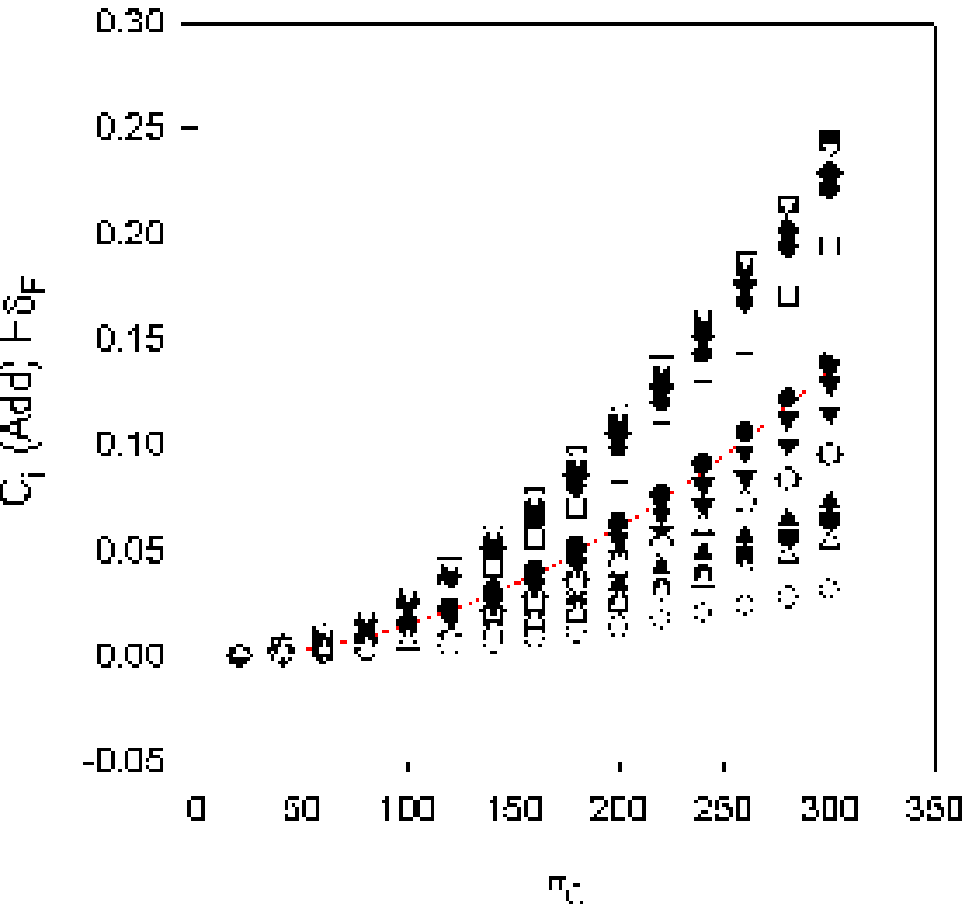}}\quad
\subfigure{\includegraphics[width=3.2cm,height=3.2cm]{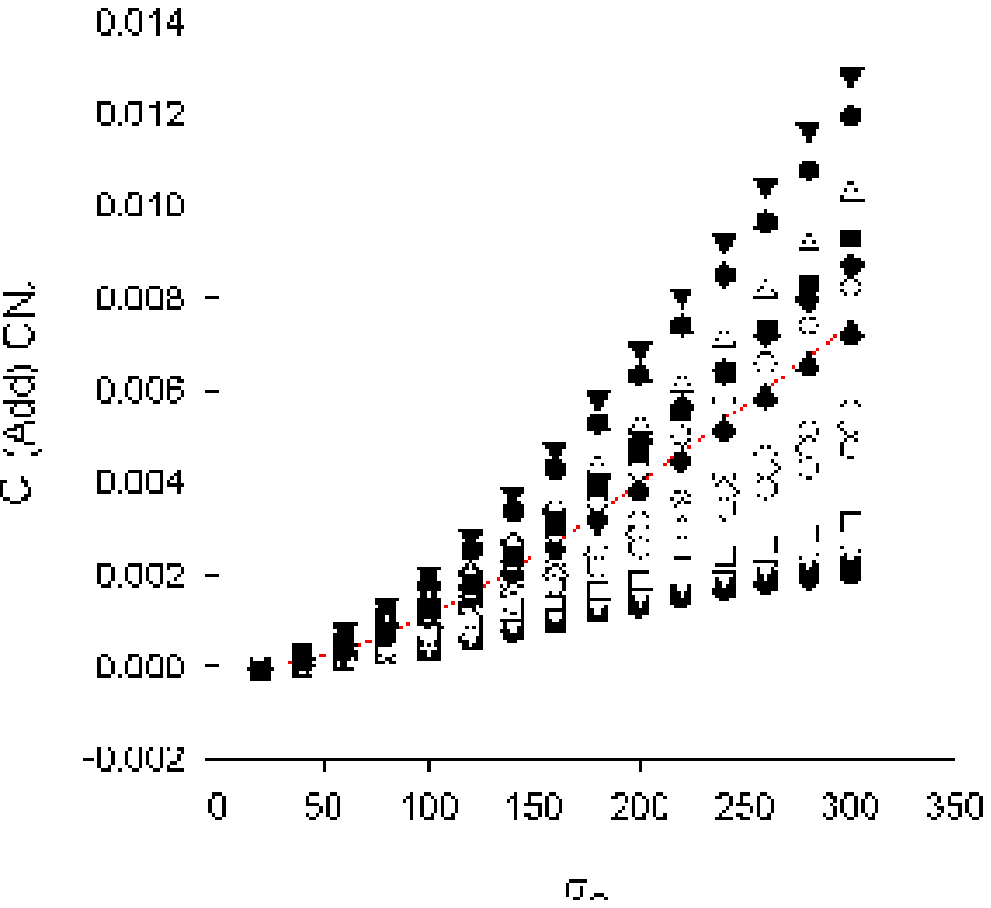}}\quad
\subfigure{\includegraphics[width=3.2cm,height=3.2cm]{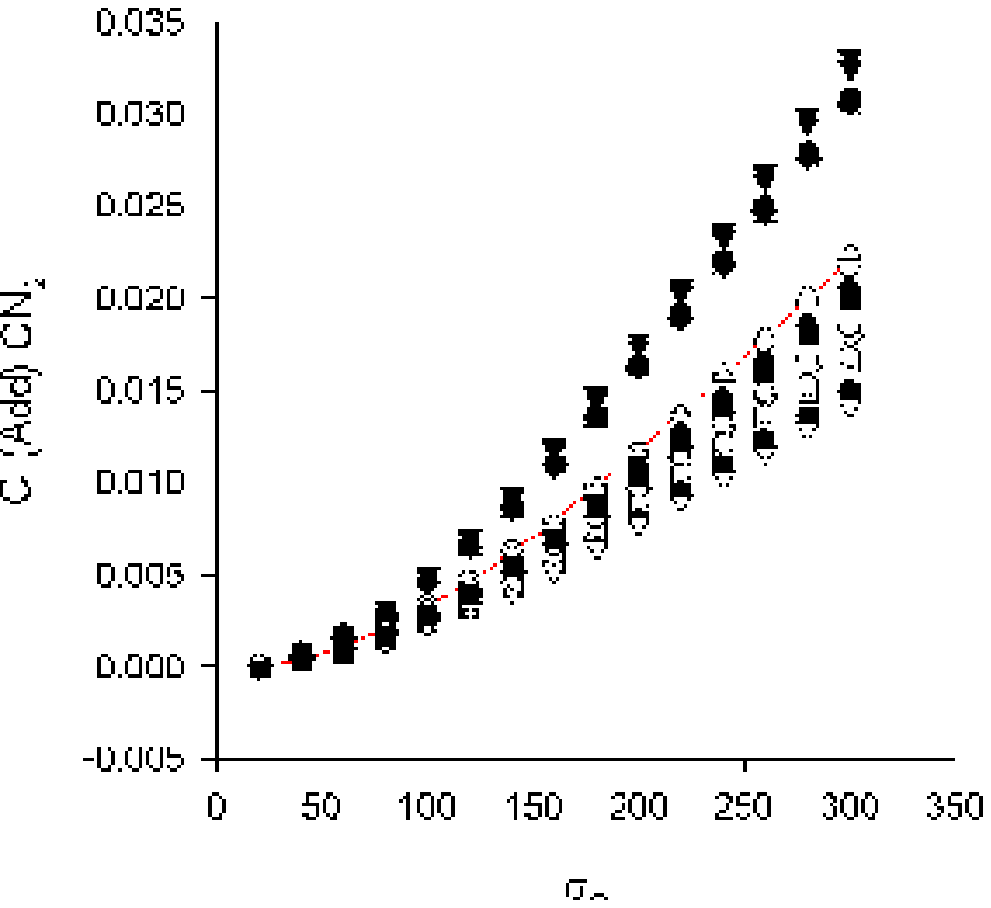}}}
         \mbox{\subfigure{\includegraphics[width=3.2cm,height=3.2cm]{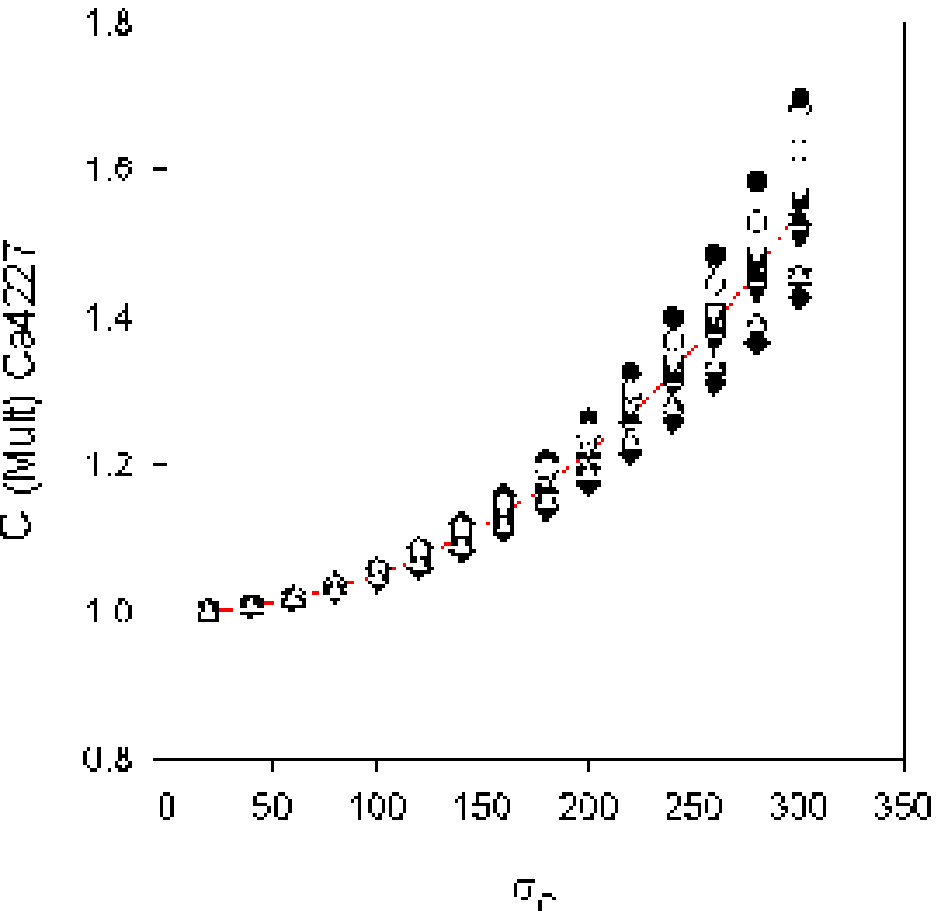}}\quad
\subfigure{\includegraphics[width=3.2cm,height=3.2cm]{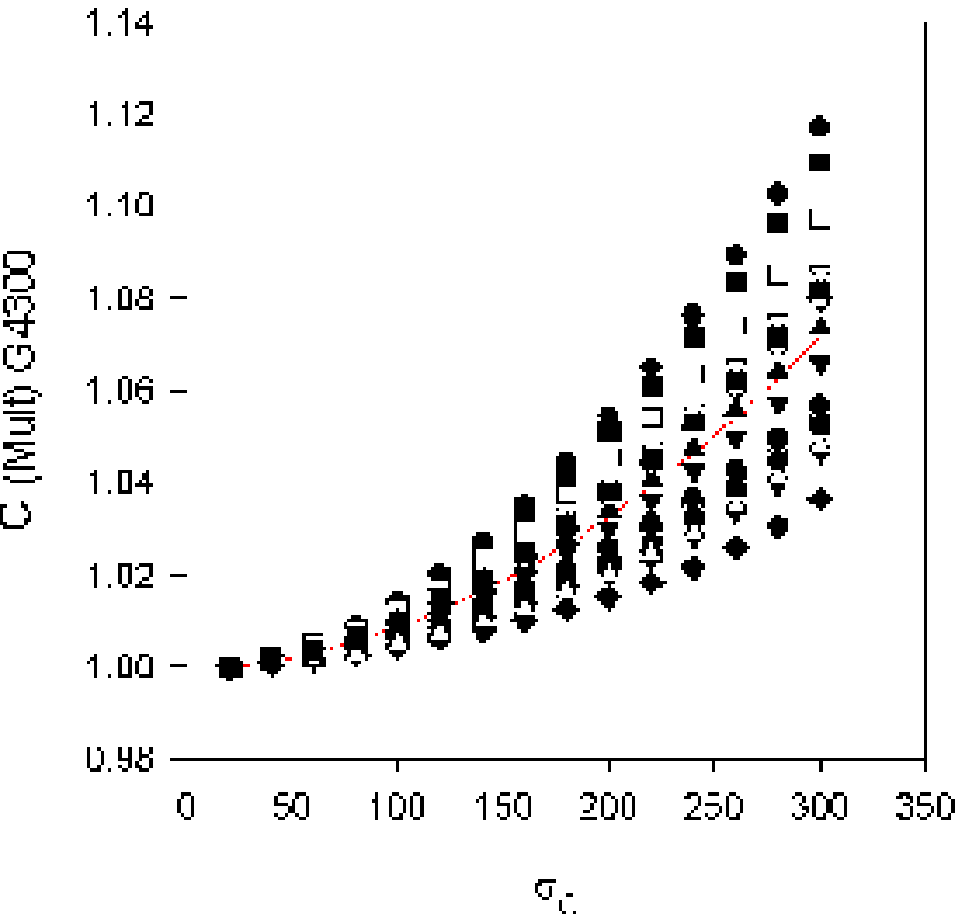}}\quad
\subfigure{\includegraphics[width=3.2cm,height=3.2cm]{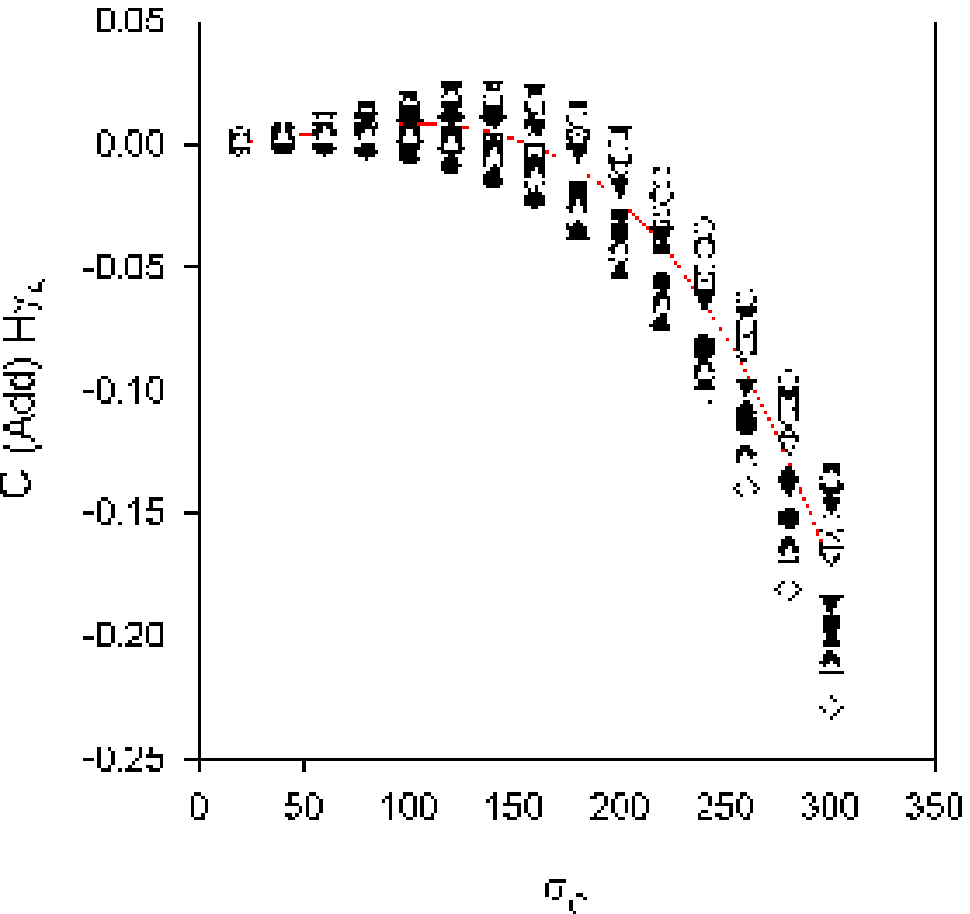}}\quad
\subfigure{\includegraphics[width=3.2cm,height=3.2cm]{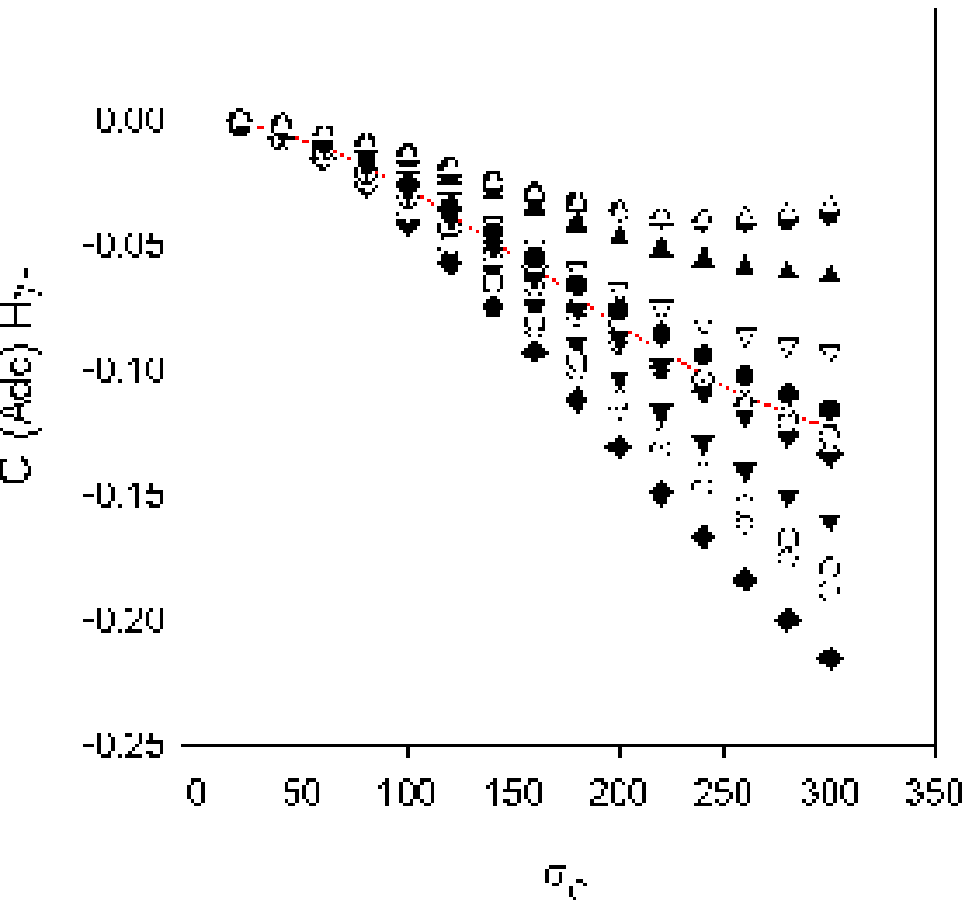}}}
   \mbox{\subfigure{\includegraphics[width=3.2cm,height=3.2cm]{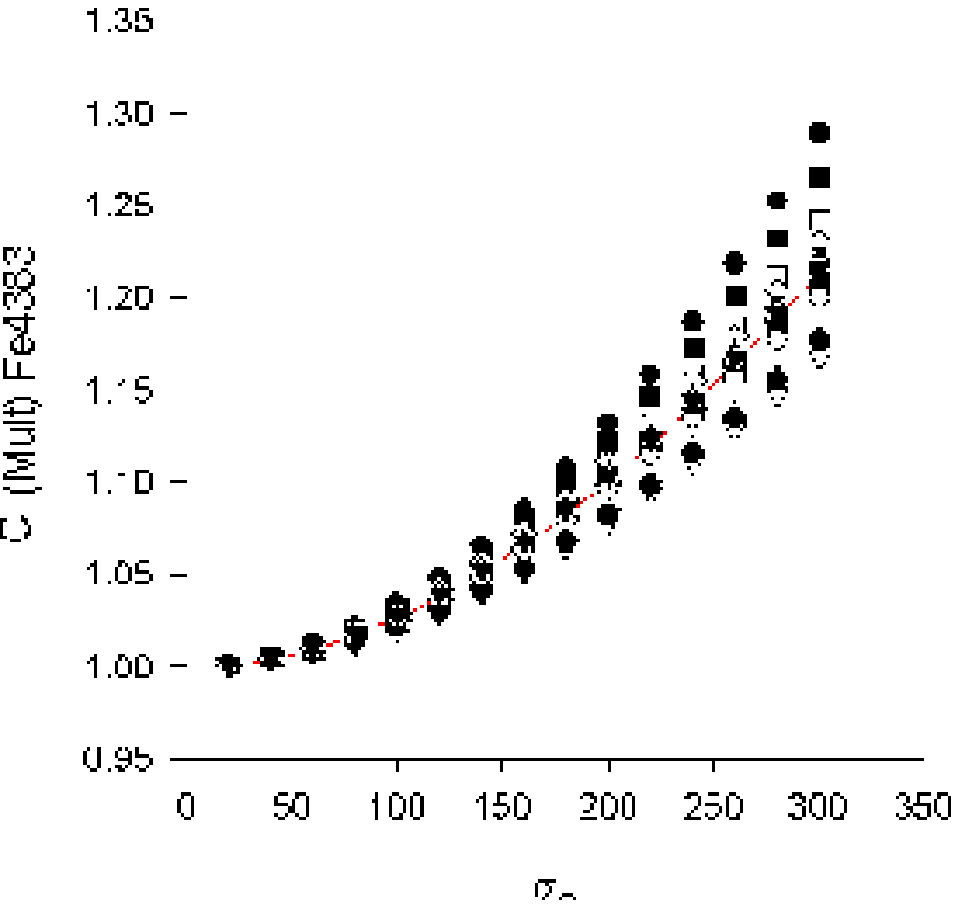}}\quad
\subfigure{\includegraphics[width=3.2cm,height=3.2cm]{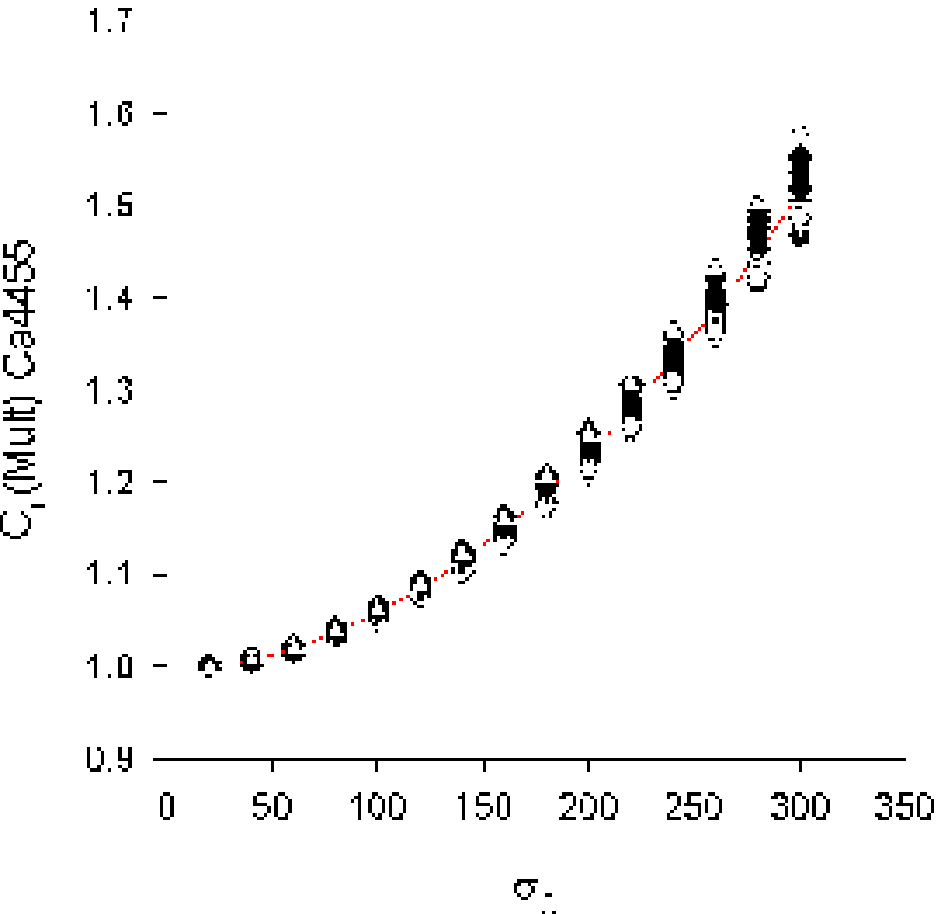}}\quad
\subfigure{\includegraphics[width=3.2cm,height=3.2cm]{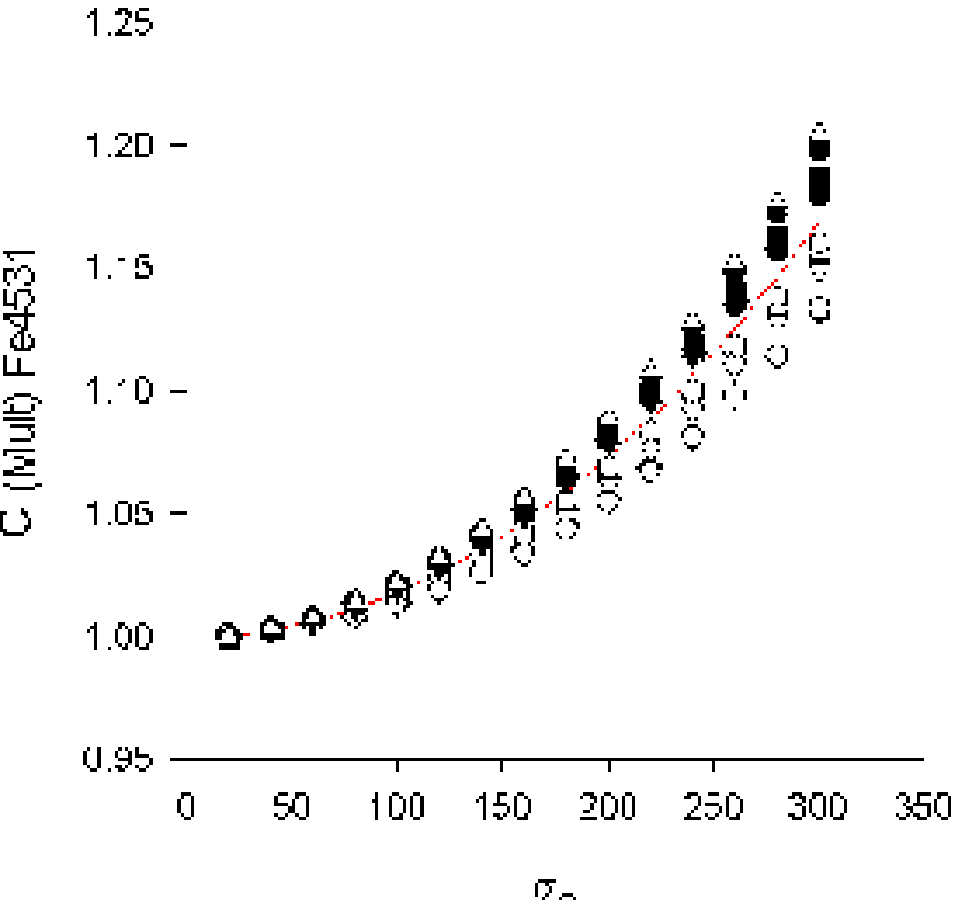}}\quad
\subfigure{\includegraphics[width=3.2cm,height=3.2cm]{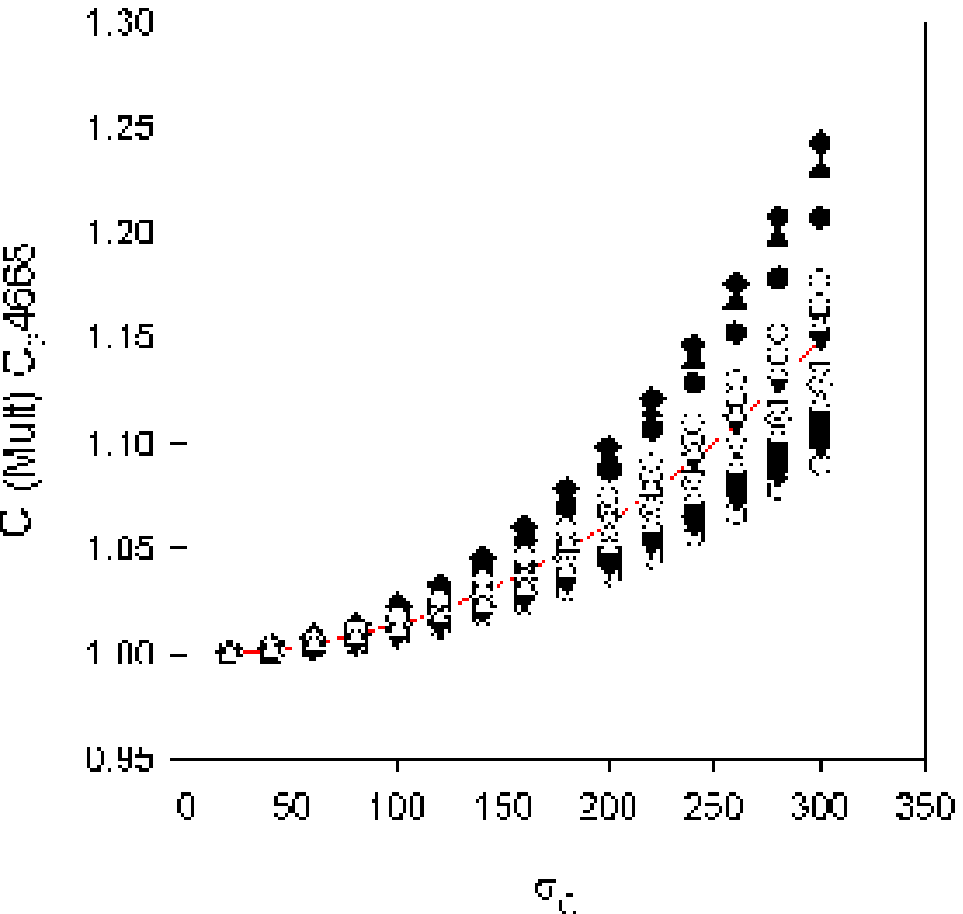}}}
         \mbox{\subfigure{\includegraphics[width=3.2cm,height=3.2cm]{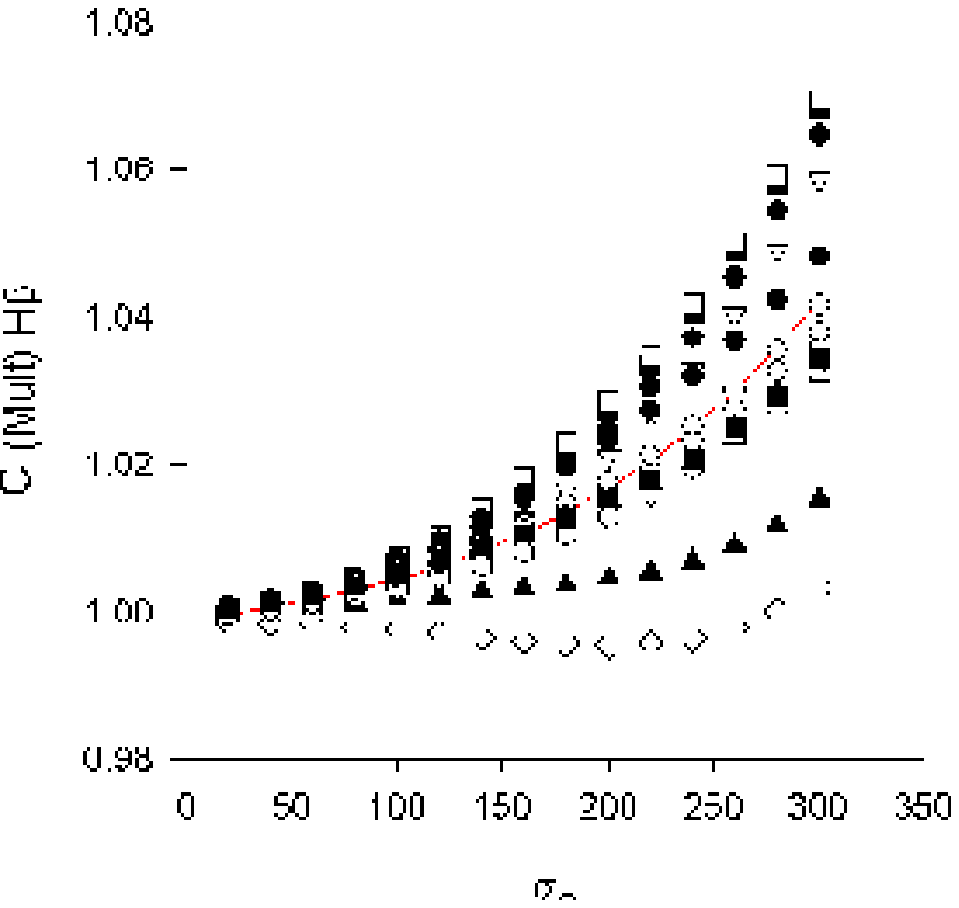}}\quad
\subfigure{\includegraphics[width=3.2cm,height=3.2cm]{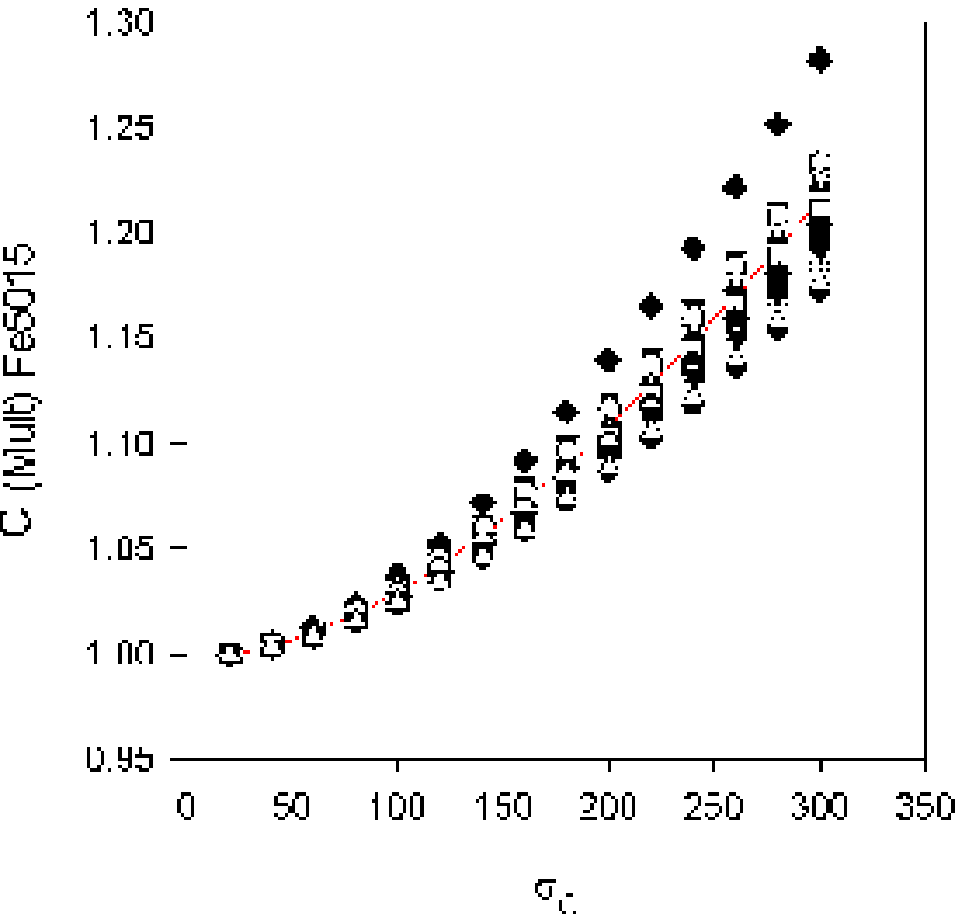}}\quad
\subfigure{\includegraphics[width=3.2cm,height=3.2cm]{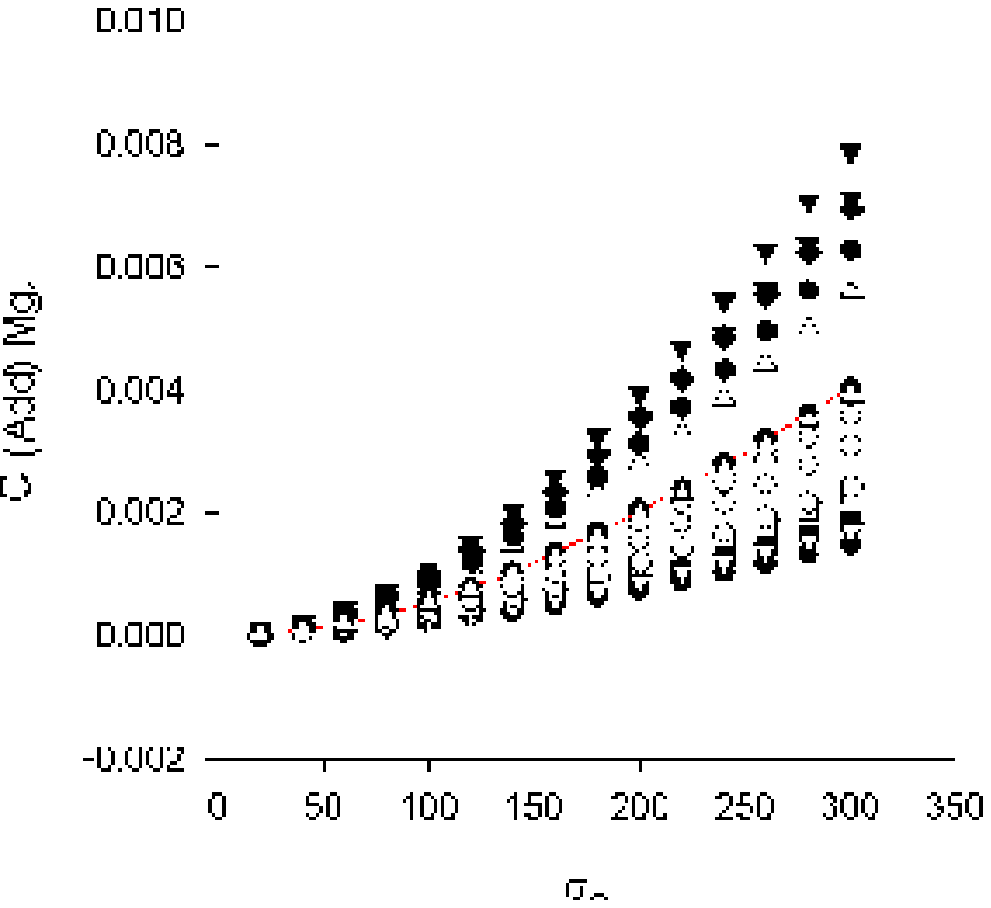}}\quad
\subfigure{\includegraphics[width=3.2cm,height=3.2cm]{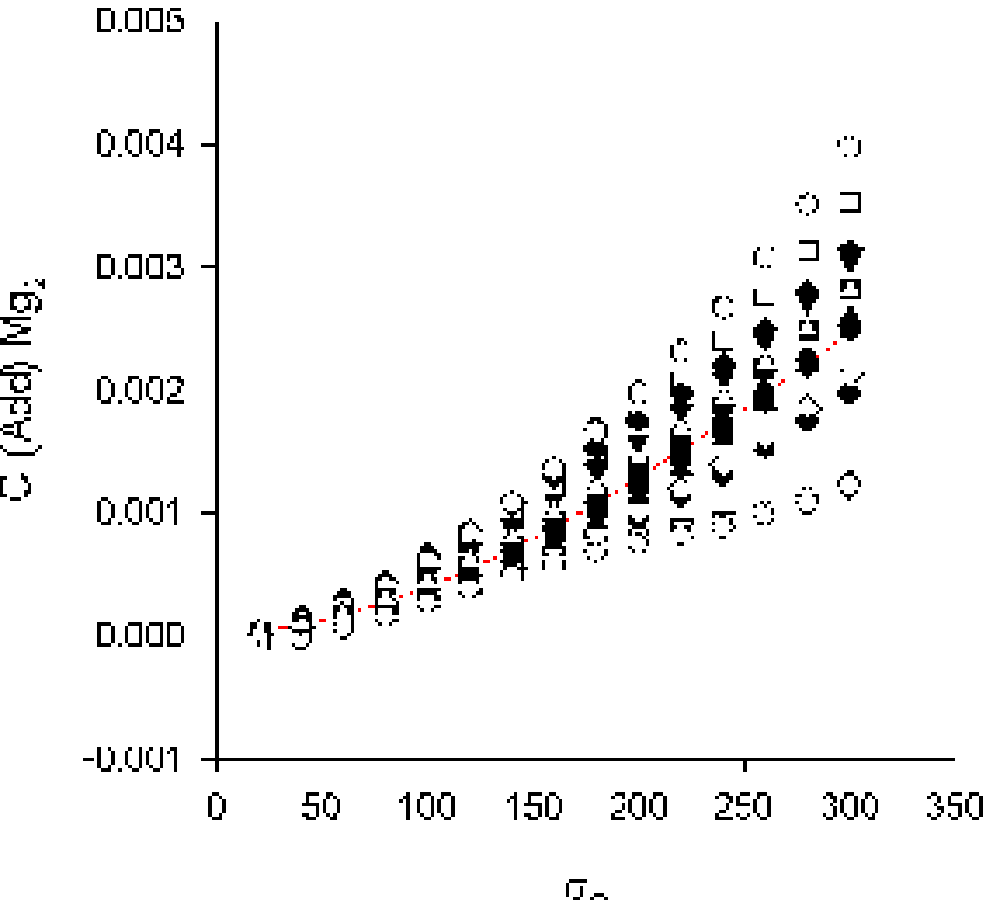}}}
  \mbox{\subfigure{\includegraphics[width=3.2cm,height=3.2cm]{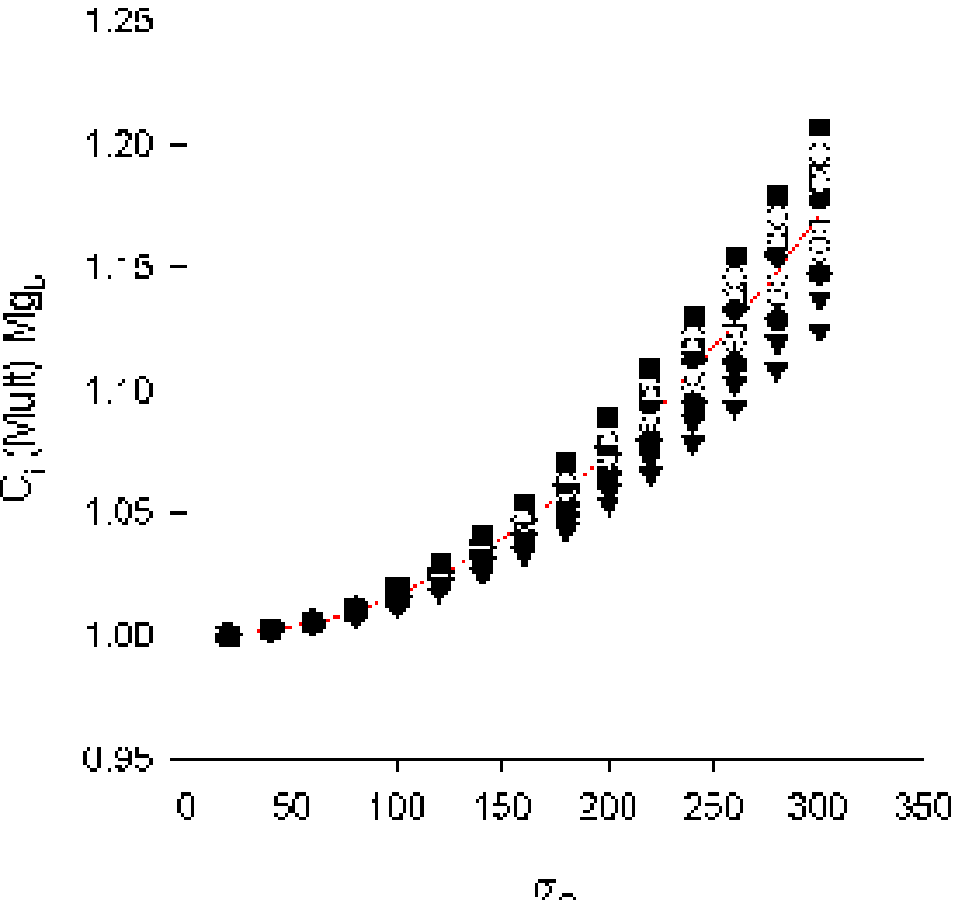}}\quad
\subfigure{\includegraphics[width=3.2cm,height=3.2cm]{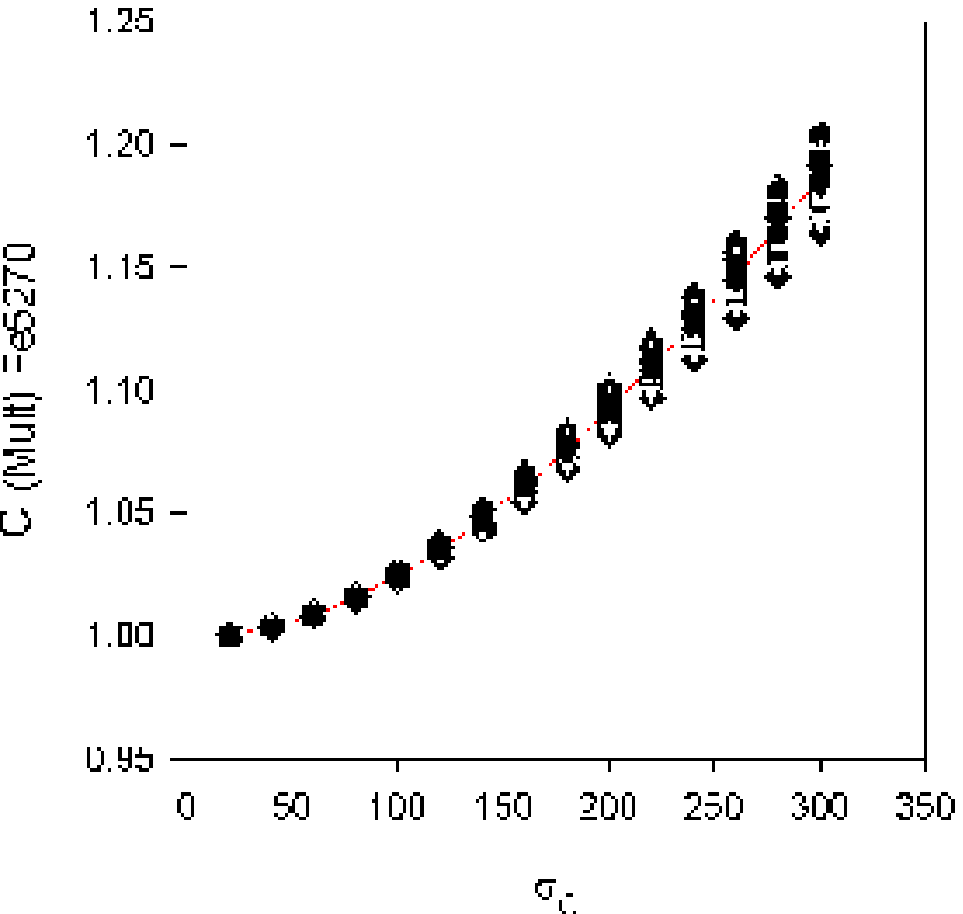}}\quad
\subfigure{\includegraphics[width=3.2cm,height=3.2cm]{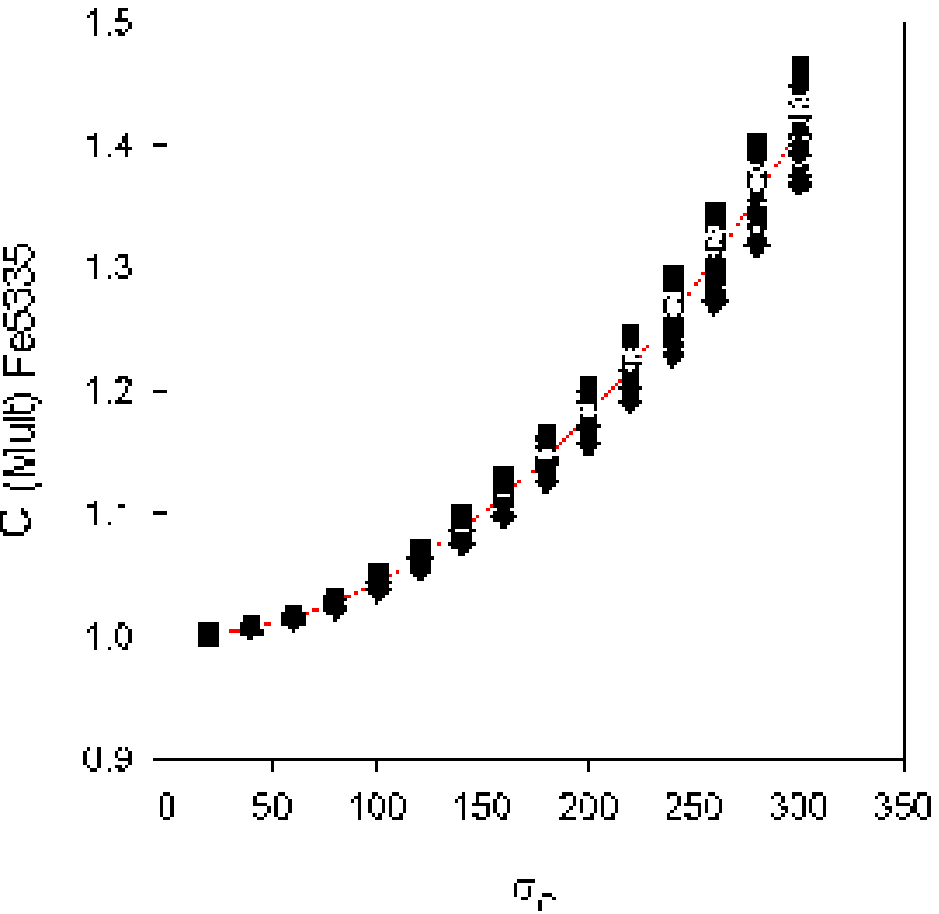}}\quad
\subfigure{\includegraphics[width=3.2cm,height=3.2cm]{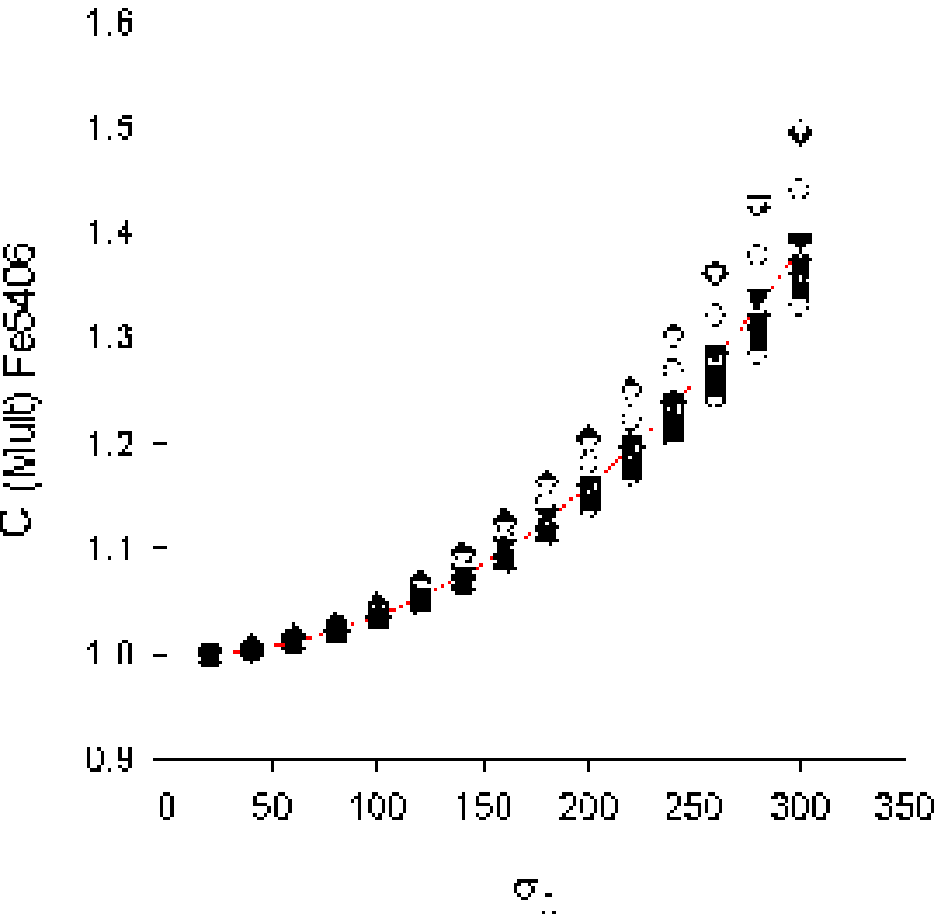}}}
         \mbox{\subfigure{\includegraphics[width=3.2cm,height=3.2cm]{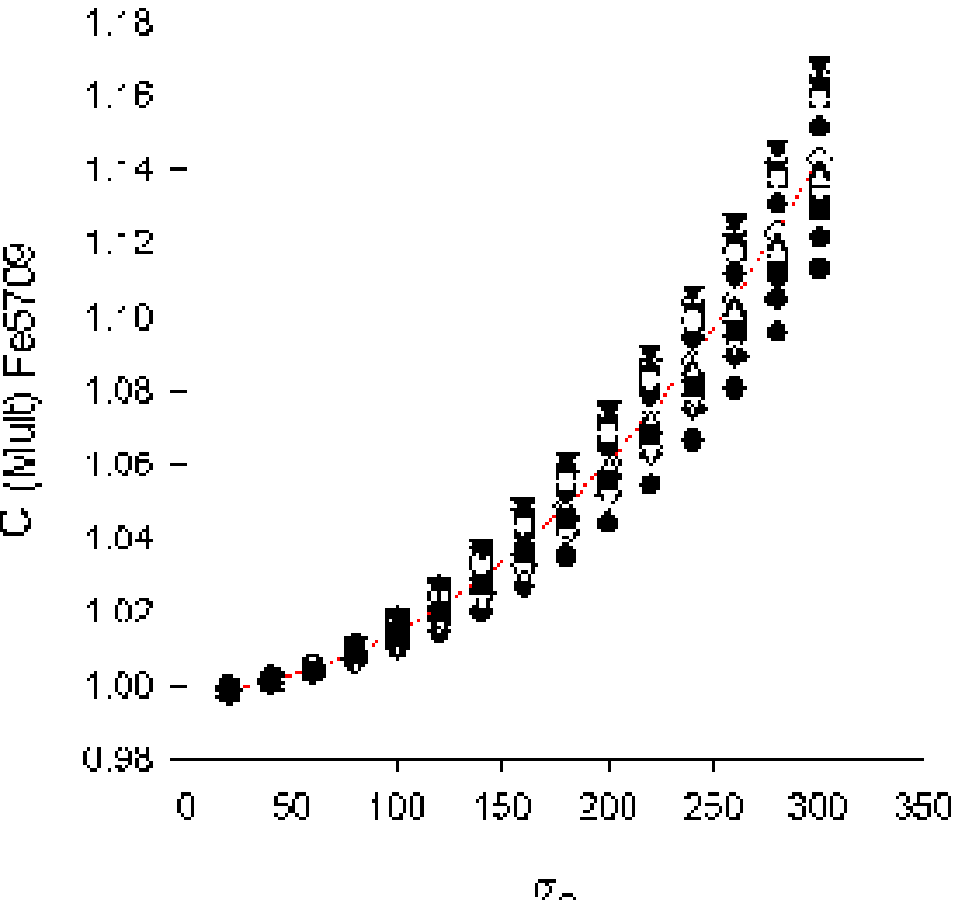}}\quad
\subfigure{\includegraphics[width=3.2cm,height=3.2cm]{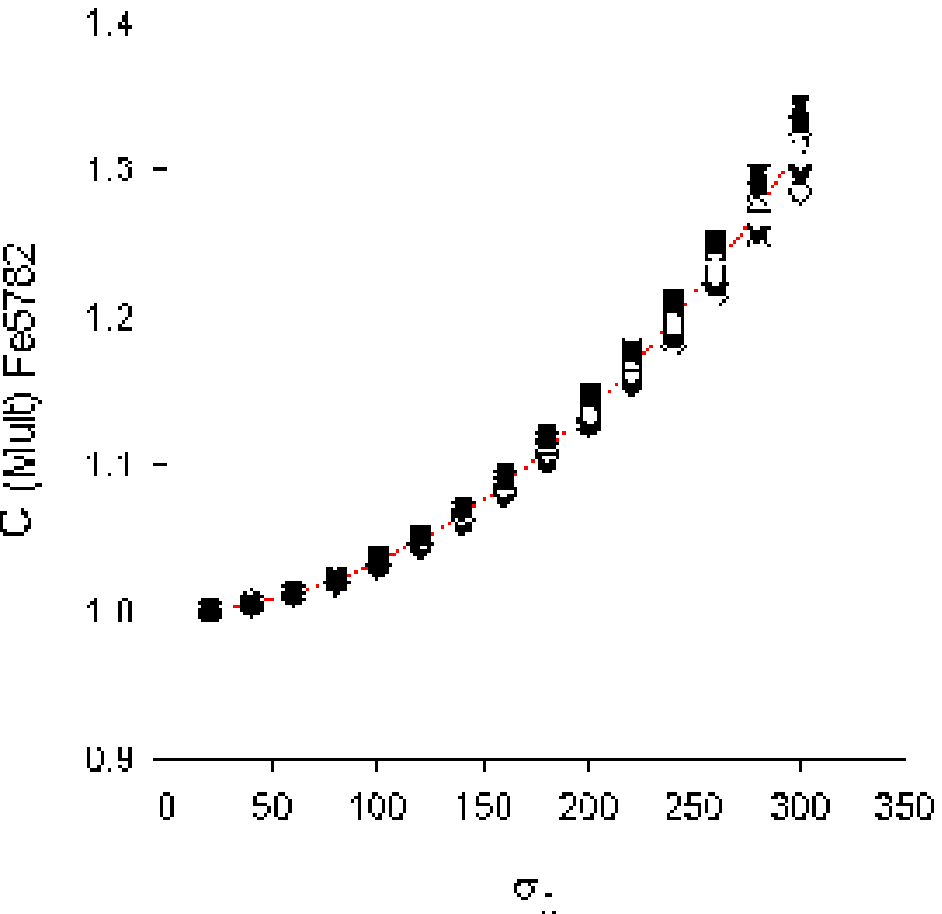}}\quad
\subfigure{\includegraphics[width=3.2cm,height=3.2cm]{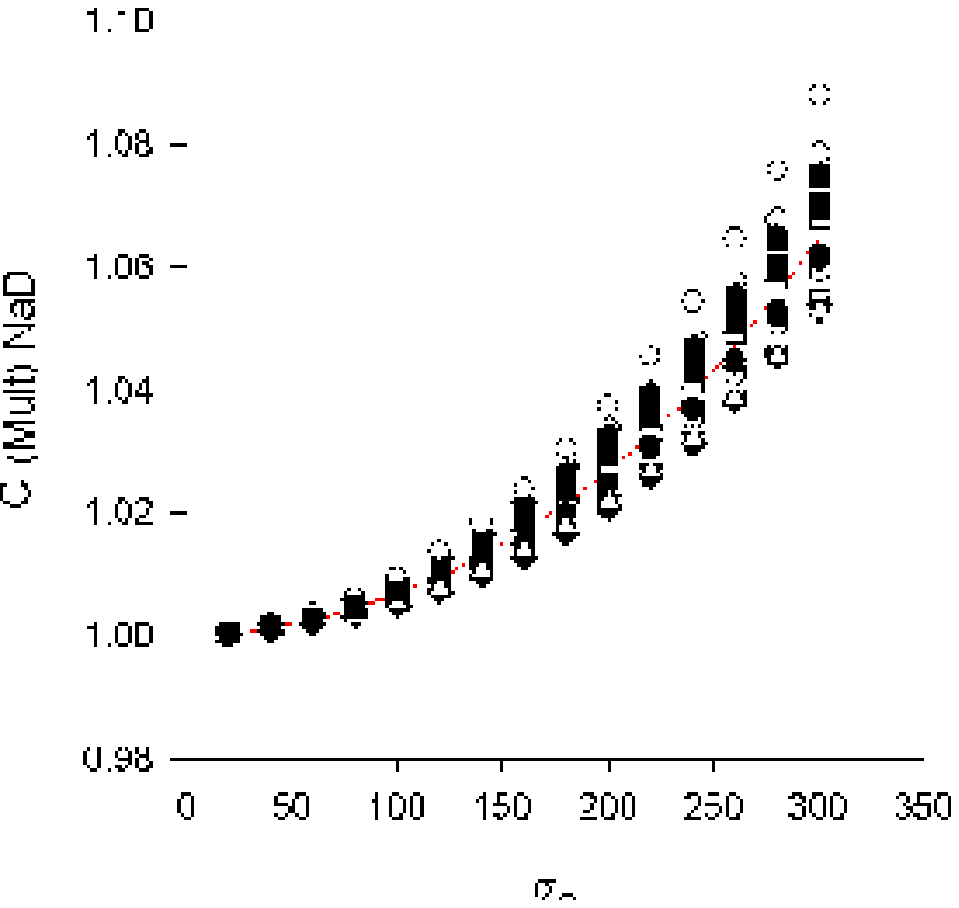}}\quad
\subfigure{\includegraphics[width=3.2cm,height=3.2cm]{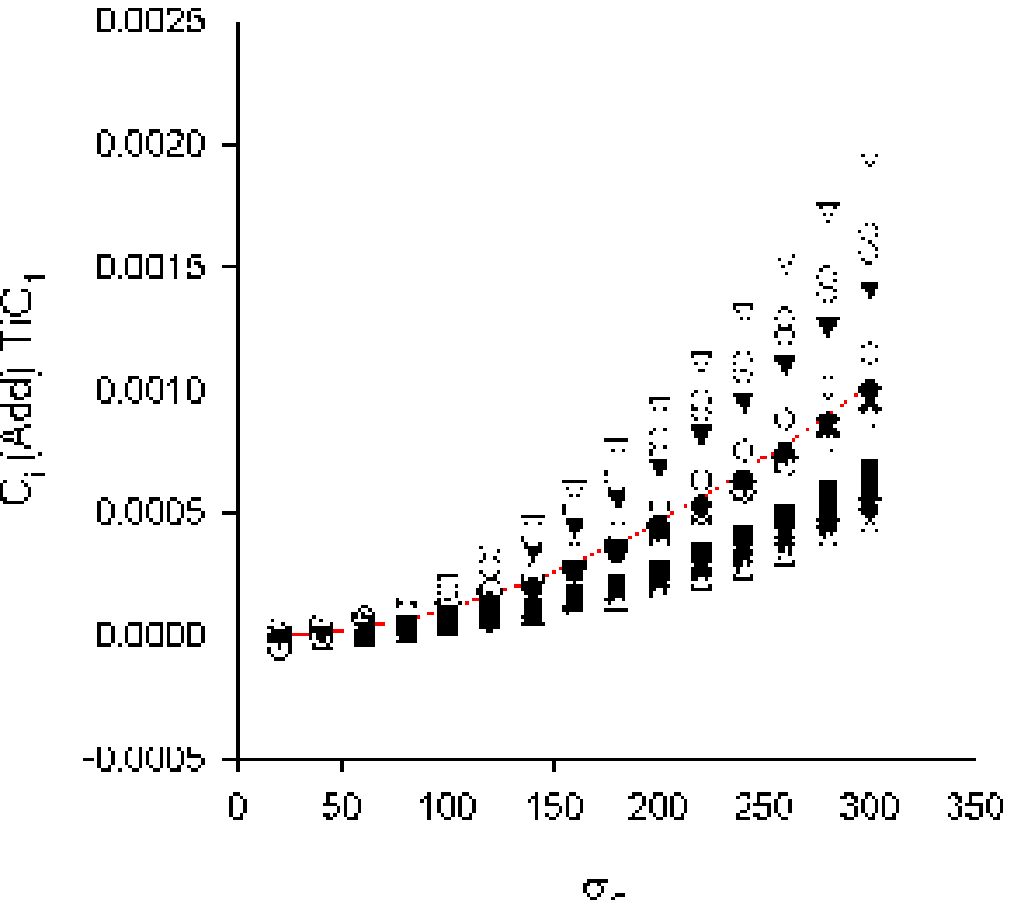}}}
   \caption[$C_{i}$ Plotted as a Function of $\sigma_{c}$ (Gemini Data).]{Broadening functions: $C_{i}$ plotted as a function of $\sigma_{c}$ (Gemini data) for all the indices. The best-fitting third order polynomial is indicated in each figure by the continuous curve. The different symbols represent the different stars used to obtain the best-fitting curve. The abbreviations ``Add'' and ``Mult'' indicates whether the index values were added (molecular and higher order Balmer indices) or multiplied (atomic indices) to calculate $C_{i}$, and $\sigma_{c}$ is in km s$^{-1}$.}
   \label{fig:Polinome1}
\end{figure*}

Figure \ref{fig:Polinome1} shows the broadening functions of the Gemini data as an example, and the plots illustrate how the index measurements are affected by velocity dispersion broadening. The derived WHT broadening functions agreed very well when compared with those published by Proctor $\&$ Sansom (2002) and Kuntschner (2004). 

Depending on whether galaxies have a total broadening ($\sqrt{\sigma_{\rm I}^{2}+\sigma_{\rm V}^{2}}$) greater or smaller than $\sigma_{\rm L}$, where $\sigma_{\rm I}$ is the instrumental broadening, $\sigma_{\rm V}$ is the galaxy velocity dispersion and $\sigma_{\rm L}$ is the Lick resolution, their spectra were effectively debroadened or broadened to the Lick resolution, as well as corrected for the broadening caused by the velocity dispersion of the galaxies. 

\begin{table*}
\centering
\begin{footnotesize}
\begin{tabular}{l c c r r r r c}
\hline Index & FWHM  & $x_{0}$ & $x_{1}$ & $x_{2}$ & $x_{3}$ & $\Delta$I  & RMS \\
             & ($\sigma_{\rm L}$) & &        &         &         &  &  \\
\hline H$\delta_{\rm A}$ & 10.9 & 0 & 3.5695e-5 & --7.7694e-6 & 8.5777e-9 &--0.139 &0.154 \\
H$\delta_{\rm F}$ & 10.9 & 0 & --3.0135e-5 & 1.7720e-6 & --5.6082e-10 &--0.065 & 0.073\\
CN$_{1}$ & 10.6 & 0 & --1.3830e-6 & 1.4721e-7 & --2.0204e-10 & $*$ & 0.013 \\
CN$_{2}$ & 10.6 & 0 & --3.6741e-6 & 4.2346e-7 & --5.5390e-10 & $*$ & 0.015 \\
Ca4227 & 10.2 & 1 & 1.7865e-4 & 2.9655e-6 & 9.0006e-9 & 0.021 & 0.050 \\
G4300 & 9.8 & 1 & 3.2797e-5 & 6.4488e-7 & 1.7805e-10 & --0.058 & 0.061\\
H$\gamma_{\rm A}$ & 9.5 & 0 & 2.8251e-6 & 2.0551e-6 & --1.3255e-8 & 0.344 & 0.087 \\
H$\gamma_{\rm F}$ & 9.5 & 0 & --1.4965e-5 & --3.3005e-6 & 6.6191e-9 & 0.071 & 0.061 \\
Fe4383 & 9.2 & 1 & 4.7860e-6 &2.7446e-6 &--1.2824e-9 & --0.016 & 0.094\\
Ca4455 & 9.1 & 1 &5.0919e-5 &5.8155e-6 &--8.1350e-10 &0.452 &0.041 \\
Fe4531 & 9.0 & 1 &3.1621e-5 &1.5188e-6 &8.8514e-10 &--0.045 &0.072 \\
C$_{2}$4668 & 8.8 & 1 & --1.9459e-6 &1.3580e-6 &1.0041e-9 &0.018 &0.106 \\
H$\beta$ & 8.5 & 1 & 4.5530e-5&2.4319e-9 &1.0960e-9 & --0.051&0.041 \\
Fe5015 & 8.4 & 1 &1.5787e-5 &3.2697e-6 &--3.0301e-9 &--0.066 & 0.090\\
Mg$_{1}$ & 8.4 & 0 &--2.9836e-7 &6.4496e-8 &--6.1692e-11 &0.005 &0.003 \\
Mg$_{2}$ & 8.4 & 0 &1.0236e-6 &3.3235e-8 &--3.0902e-11 &0.005 &0.003 \\
Mg$_{\rm b}$ & 8.4 & 1 &--4.1861e-5 &2.0200e-6 &4.1426e-11 & --0.001 &0.042 \\
Fe5270 & 8.4 & 1 &--8.9369e-6 &2.8442e-6 &--2.4675e-9 &--0.073 &0.053 \\
Fe5335 & 8.4 & 1 &2.0324e-5 &4.1990e-6 &1.1364e-9 &--0.017 &0.045 \\
Fe5406 & 8.4 & 1 &3.1157e-5 &3.2510e-6 &3.1952e-9 &0.001 &0.043 \\
Fe5709 & 9.1 & 1 &5.4235e-5 &1.0542e-6 &1.2787e-9 & 0.005 &0.030 \\
Fe5782 & 9.3 & 1 &3.3839e-5 &2.9915e-6 &1.3392e-9 & --0.003 &0.041 \\
NaD & 9.5 & 1 &3.0030e-5 &3.4807e-7 &9.2784e-10 &0.097 &0.055 \\
TiO$_{1}$ & 9.7 & 0 &--2.8748e-7 &1.4645e-8 &--7.8336e-12 &0.006 &0.002 \\
TiO$_{2}$ & 10.3 & 0 &2.7041e-7 &--1.8201e-9 &1.1353e-11 &--0.007 &0.003 \\
\hline
\end{tabular} 
\end{footnotesize}
\caption[Values for the Calibration of the Gemini Data to Lick Resolution.]{Values for the calibration of the Gemini data to Lick resolution and offsets. The values of the Lick spectral resolution ($\sigma_{\rm L}$) were taken from Worthey $\&$ Ottaviani (1997) and are given as FWHM values, where $\mathrm{FWHM}=2.35 \sigma_{\rm L}$ (in \AA{}). The values of $\Delta$I are Lick -- this work, and the RMS are the error on the mean. $^{*}$ Relations were derived for the two CN offsets (as described in Section \ref{Eq_section}), whereas all the other offsets were found to be independent of the strength of the index.}
\label{table:GeminitoLick}
\end{table*}

\subsection{Corrections to the Lick flux scale}
\label{Eq_section}

The original Lick/IDS spectra were not flux-calibrated by means of spectrophotometric standard stars but instead were normalised to a calibration lamp. This causes small differences in the indices compared with those measured from flux-calibrated data. By comparing the measured indices in the observed, flux-calibrated stars with those in the Lick/IDS database, the mean differences in index measurements caused by this flux scale difference can be derived for all the indices. 

Index values were measured from the Lick star spectra observed in this work (WHT and Gemini data) broadened up to the Lick/IDS resolution. Comparisons to the Lick star data presented in Worthey (1994) and Worthey and Ottaviani (1997) for the same stars enabled the calculation of the mean Lick index differences (hereafter called offsets, $\Delta$I). For all the Lick index measurements presented here, the flux calibration correction was performed by adding the appropriate $\Delta$I to the measured index value.

\begin{figure*}
   \centering
 \mbox{\subfigure{\includegraphics[width=4cm,height=2.4cm]{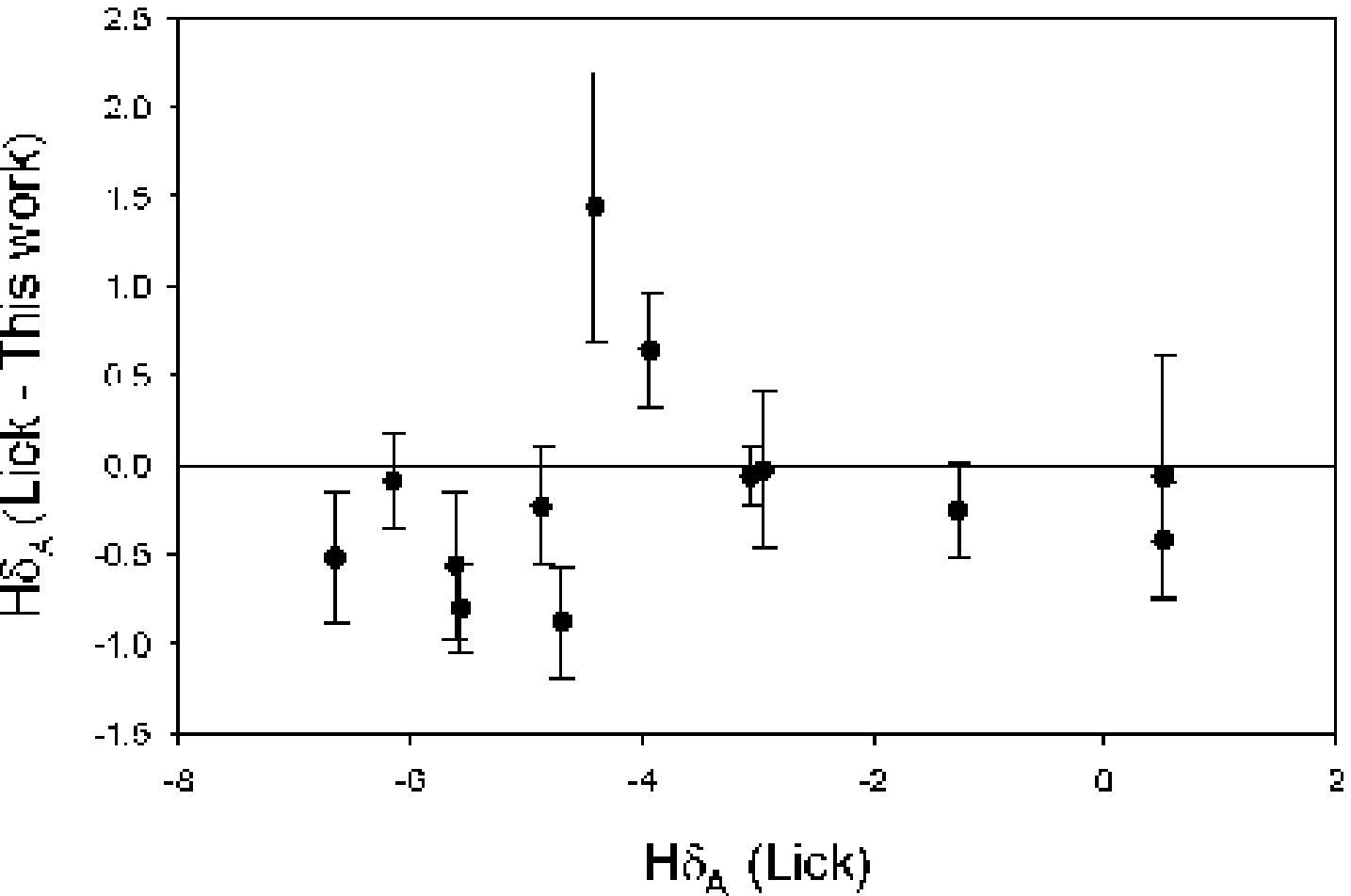}}\quad
   \subfigure{\includegraphics[width=4cm,height=2.4cm]{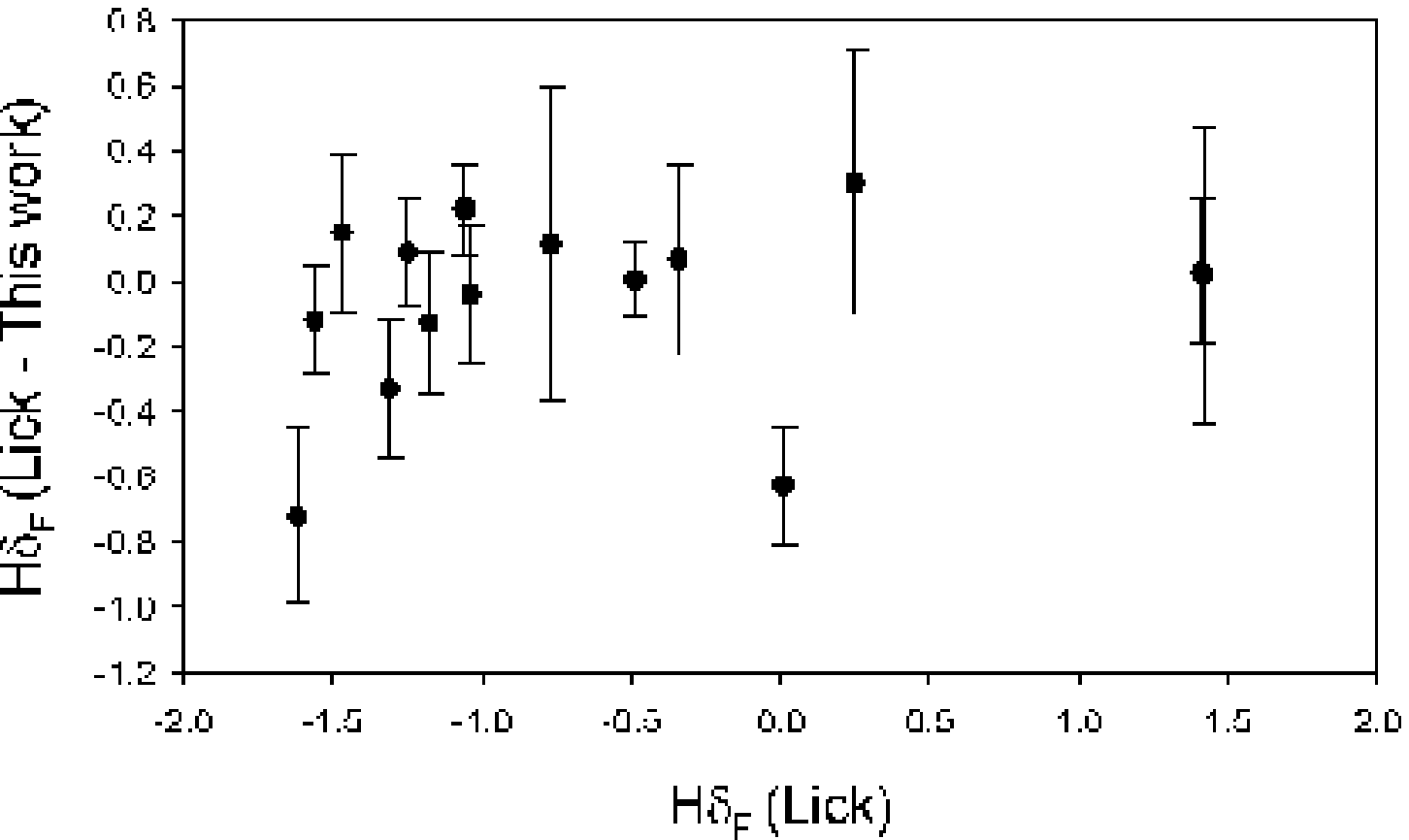}}\quad
\subfigure{\includegraphics[width=4cm,height=2.4cm]{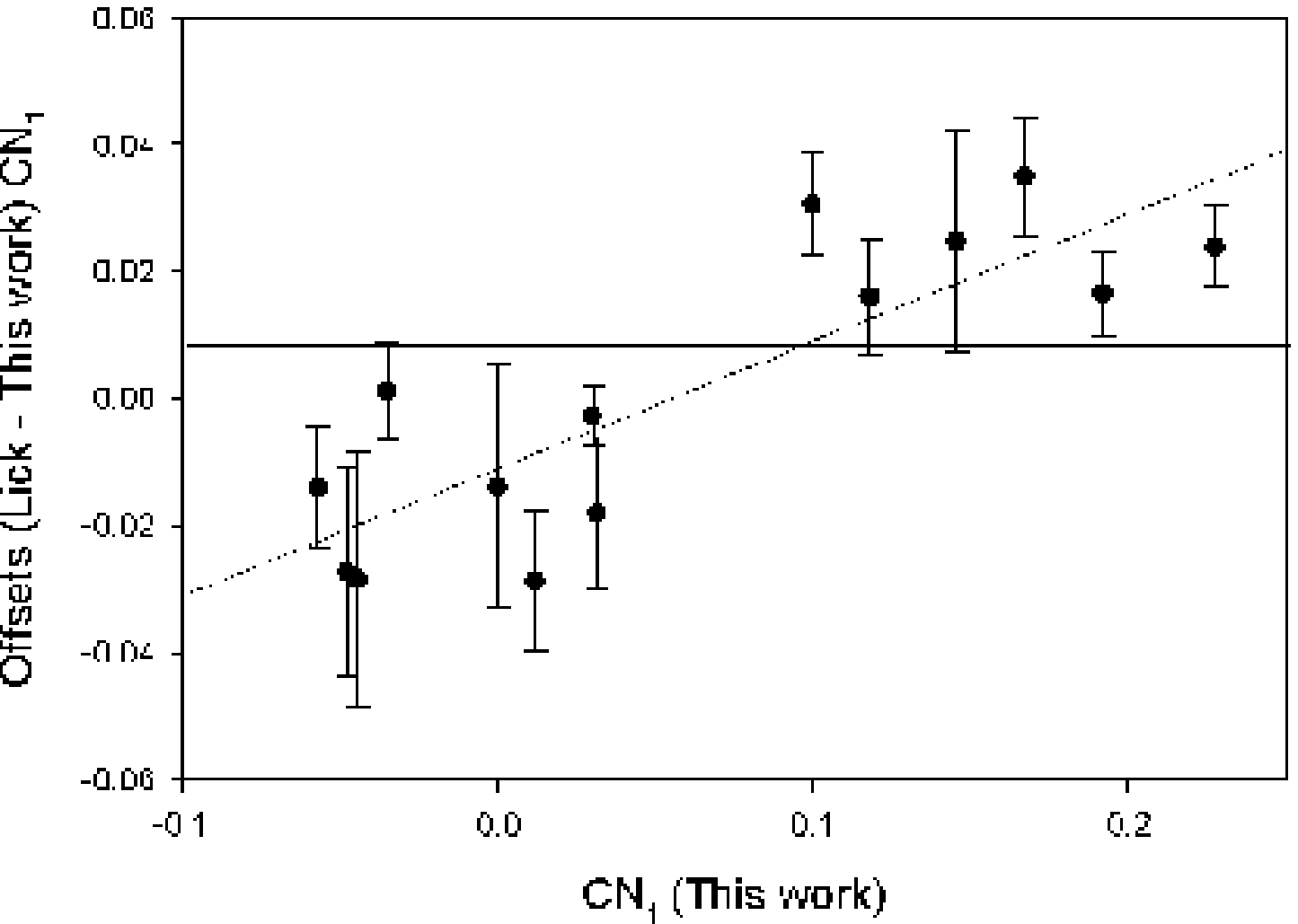}}}
\mbox{\subfigure{\includegraphics[width=4cm,height=2.4cm]{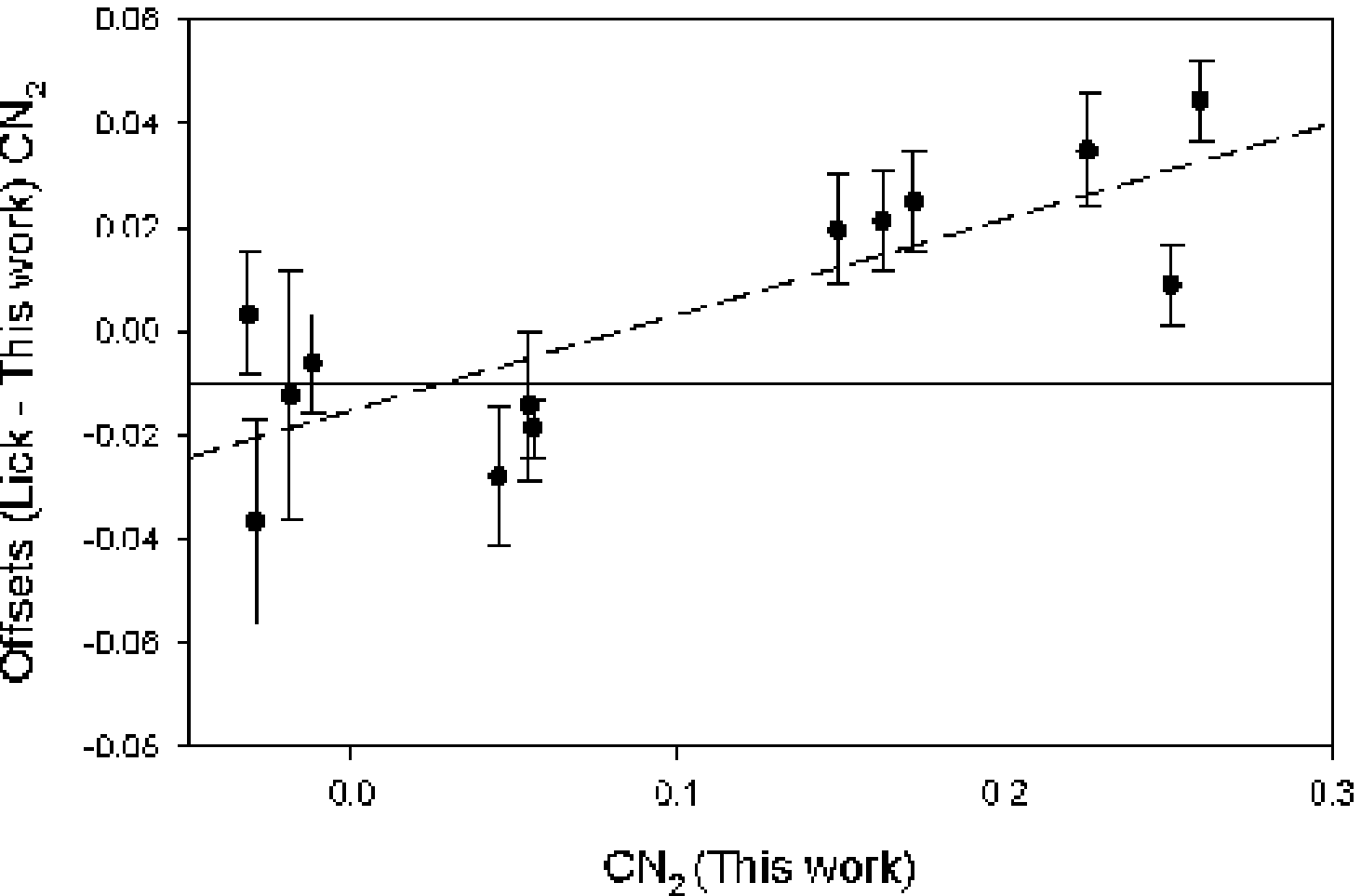}}\quad
   \subfigure{\includegraphics[width=4cm,height=2.4cm]{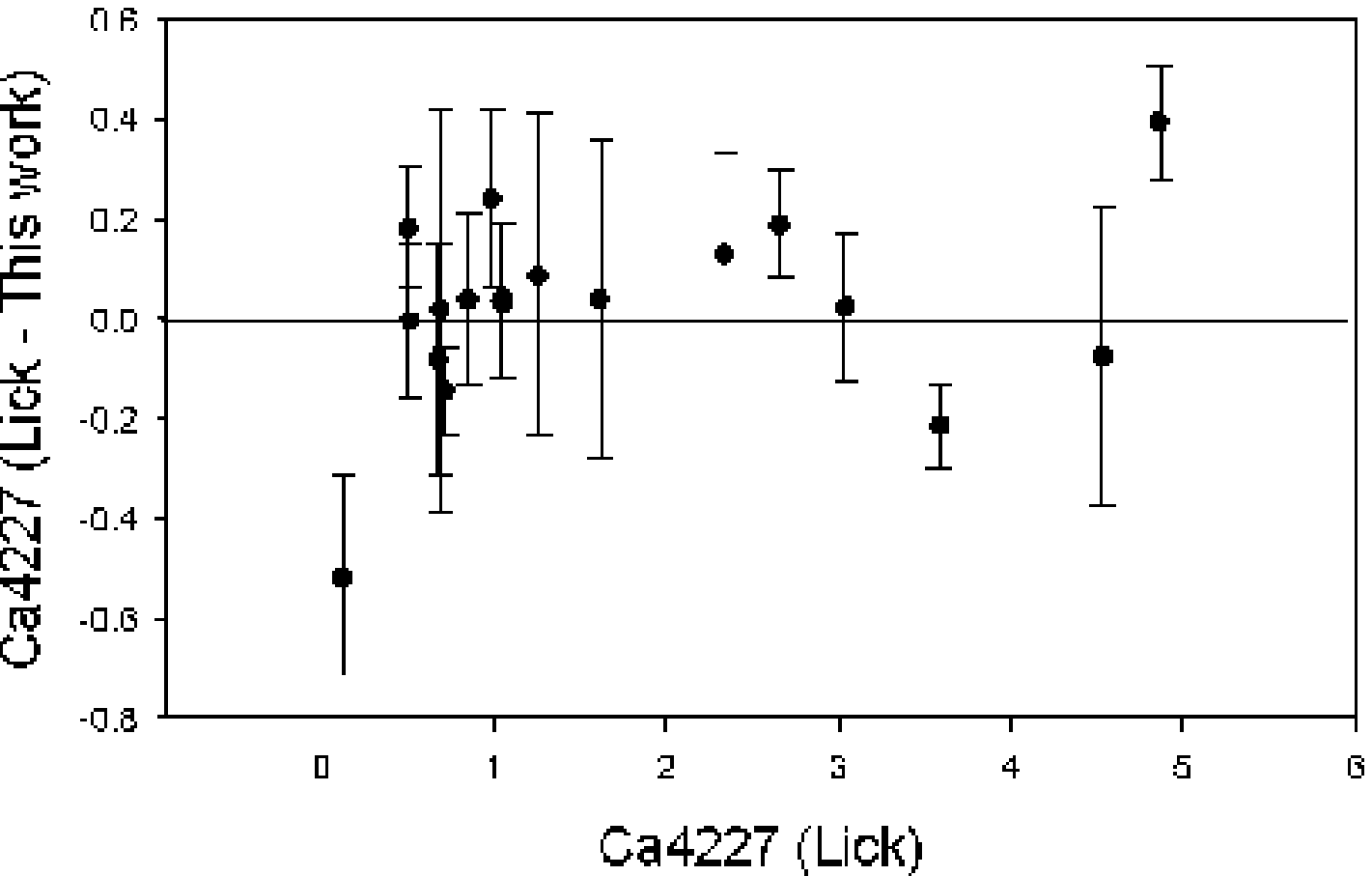}}\quad
\subfigure{\includegraphics[width=4cm,height=2.4cm]{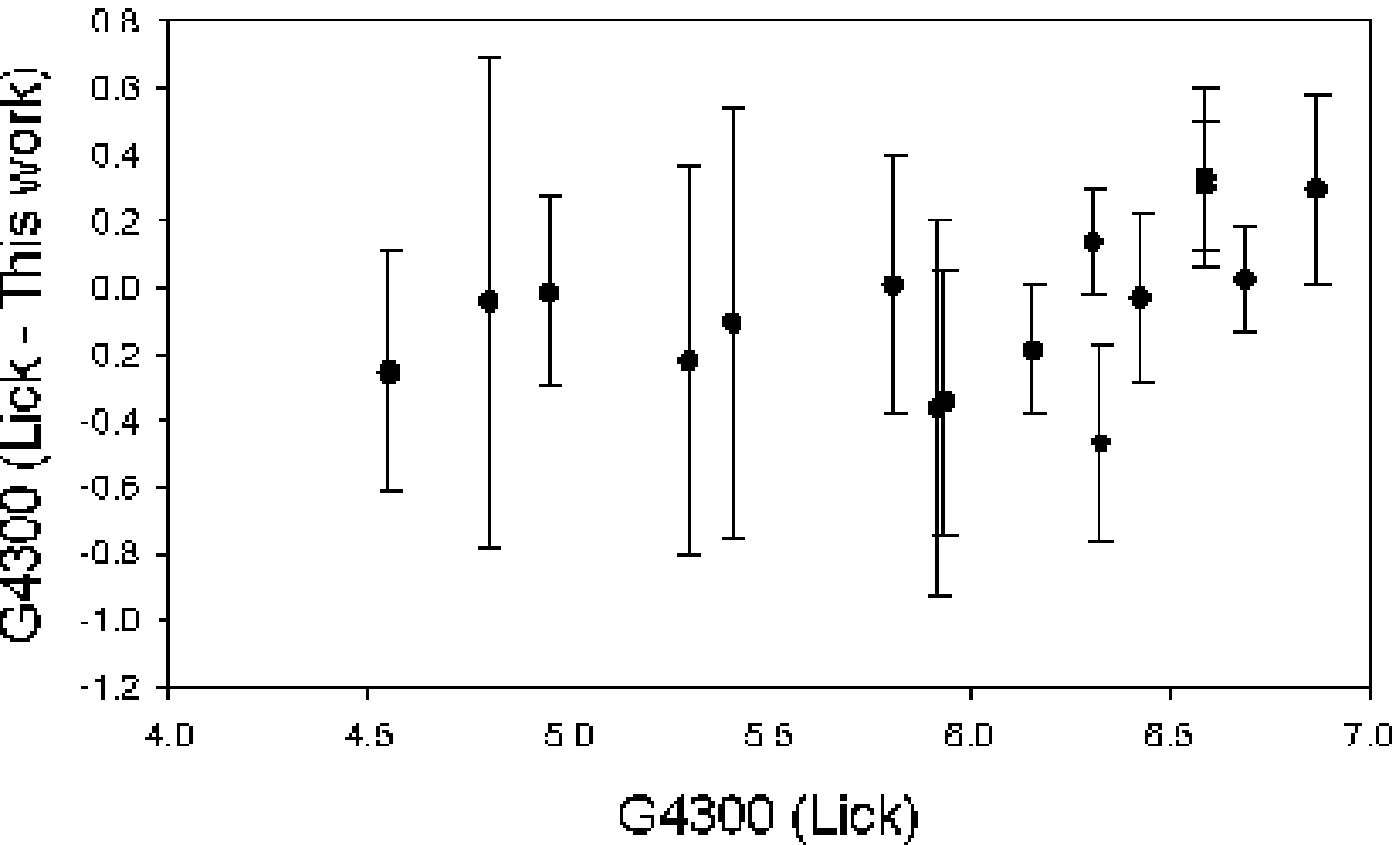}}}
\mbox{\subfigure{\includegraphics[width=4cm,height=2.4cm]{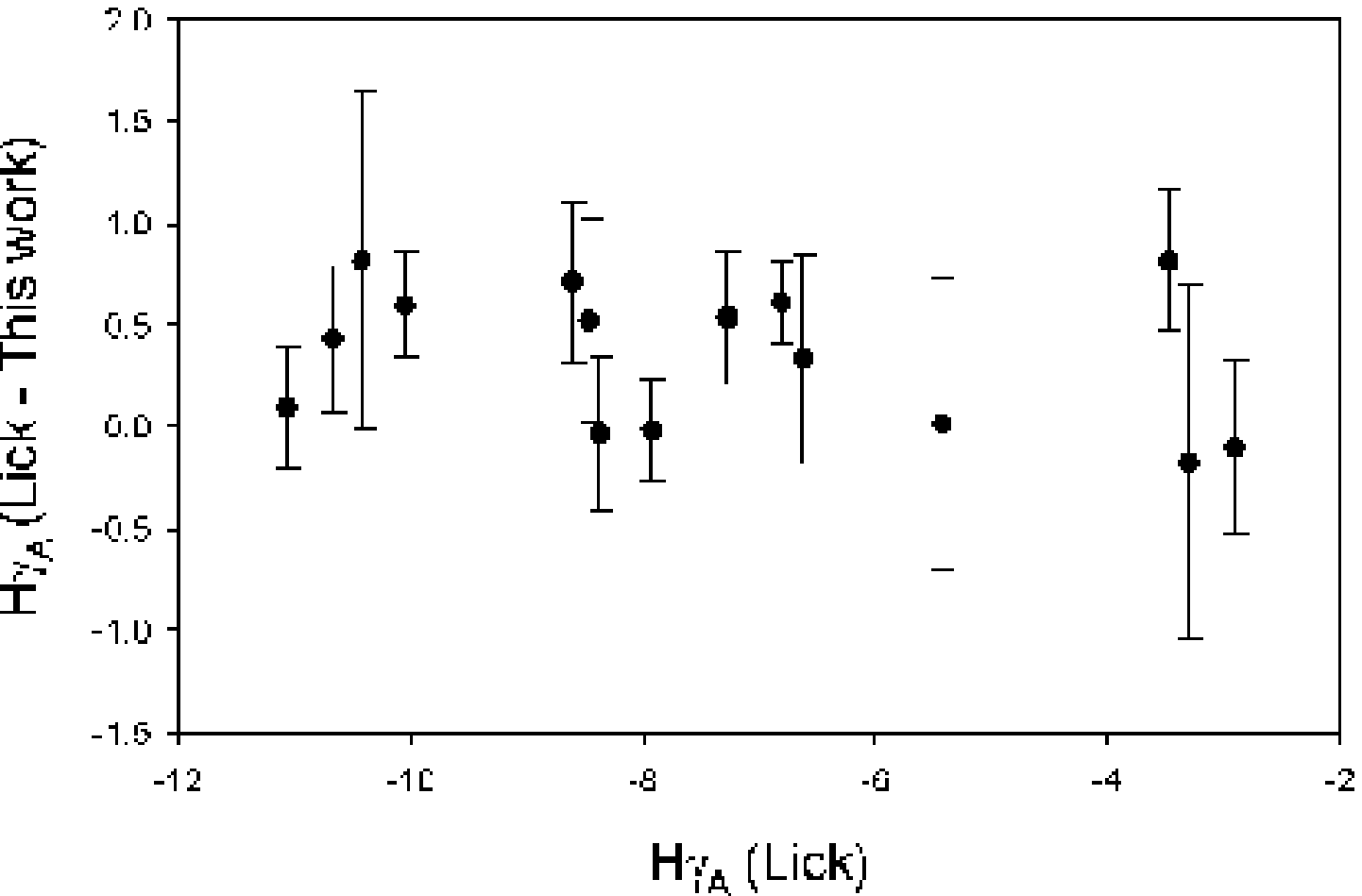}}\quad
   \subfigure{\includegraphics[width=4cm,height=2.4cm]{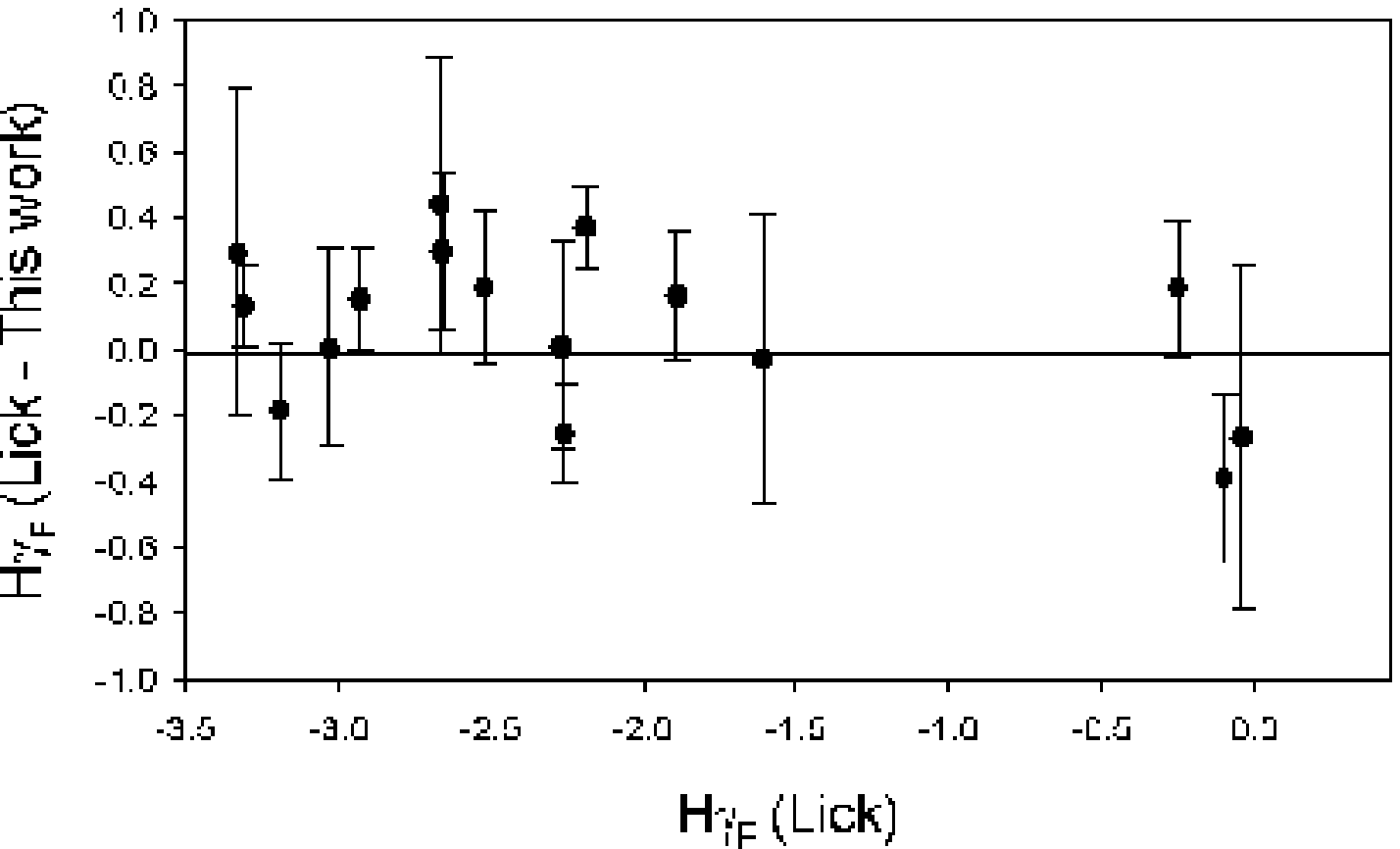}}\quad
\subfigure{\includegraphics[width=4cm,height=2.4cm]{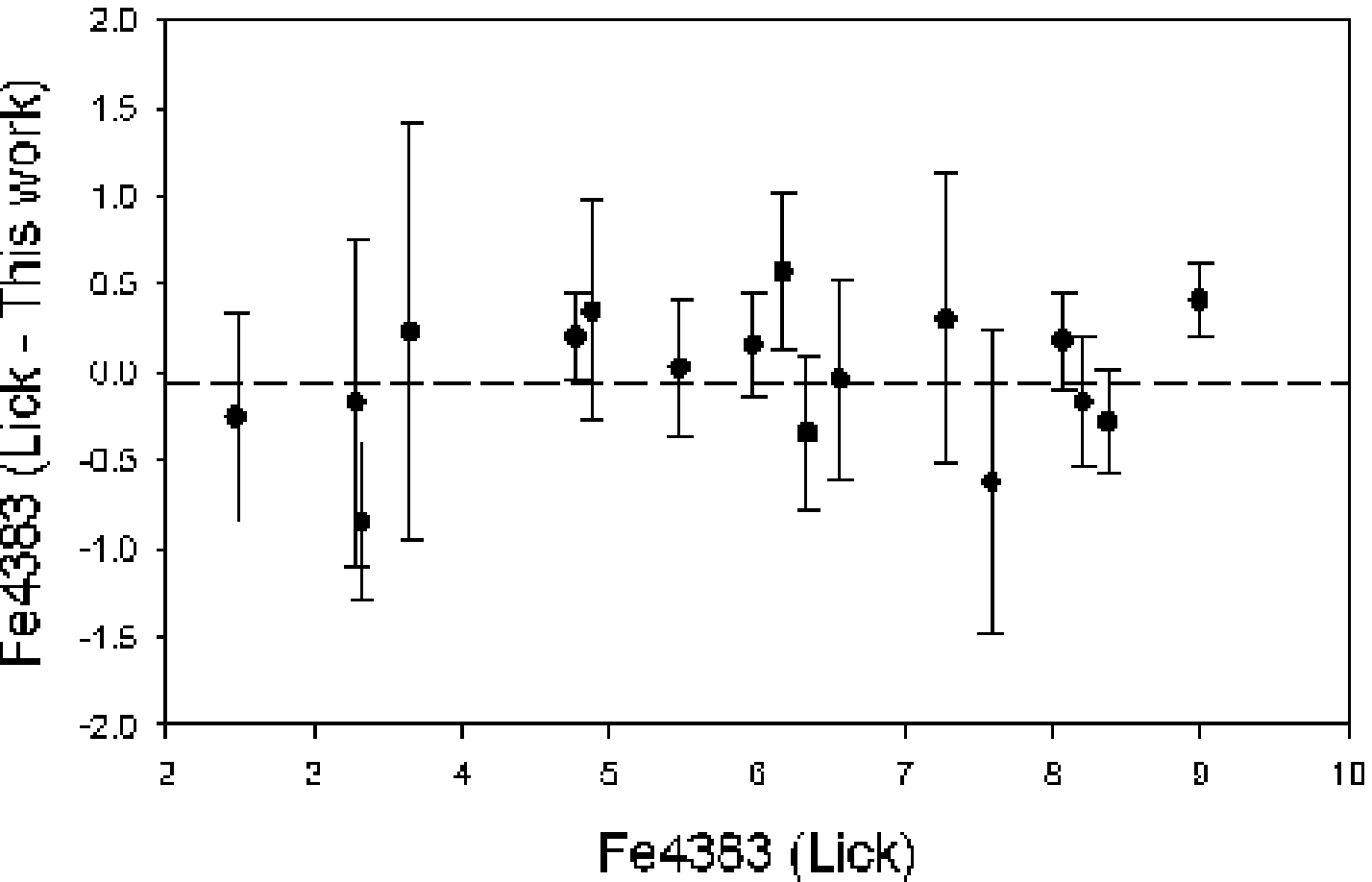}}}
\mbox{\subfigure{\includegraphics[width=4cm,height=2.4cm]{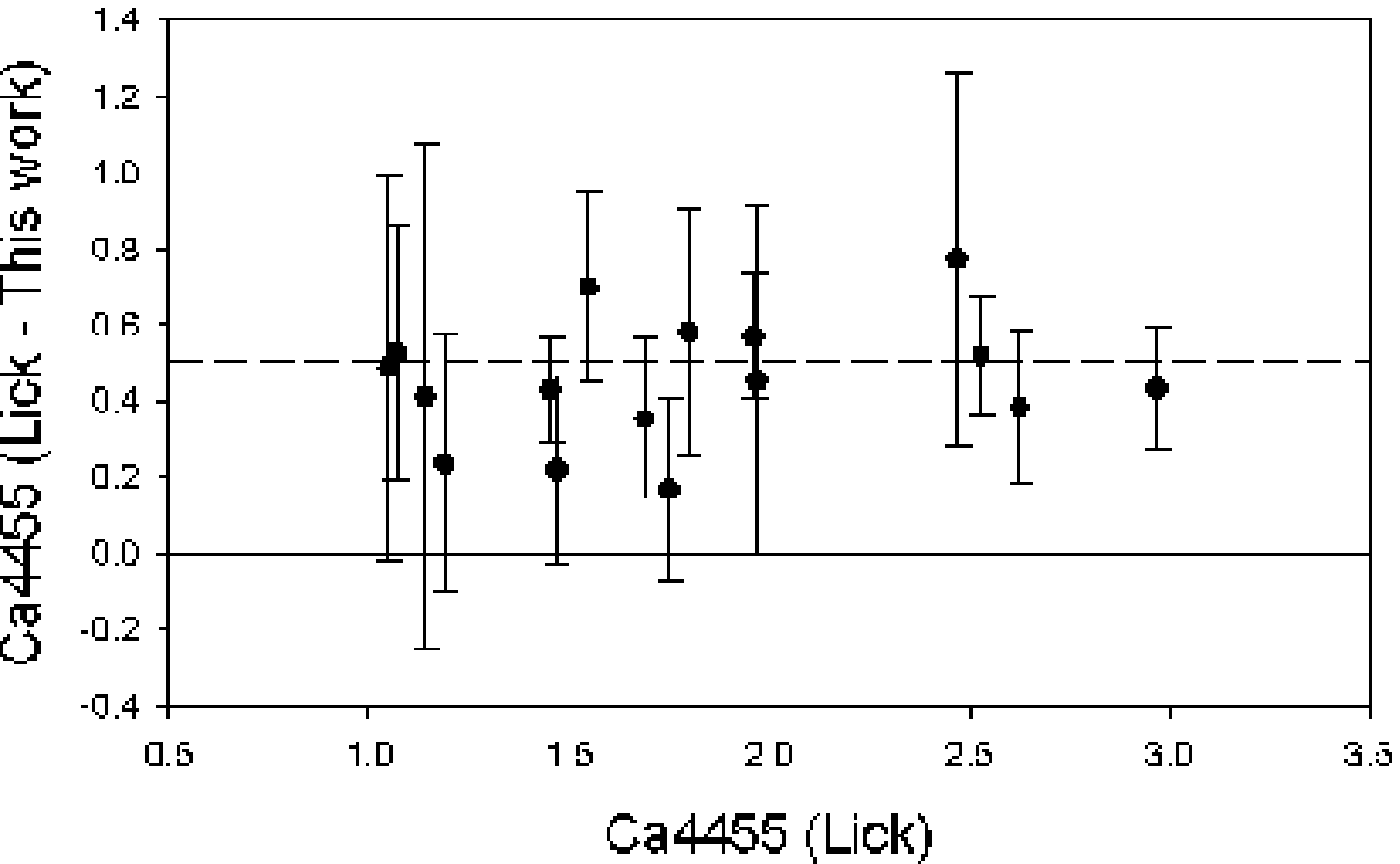}}\quad
   \subfigure{\includegraphics[width=4cm,height=2.4cm]{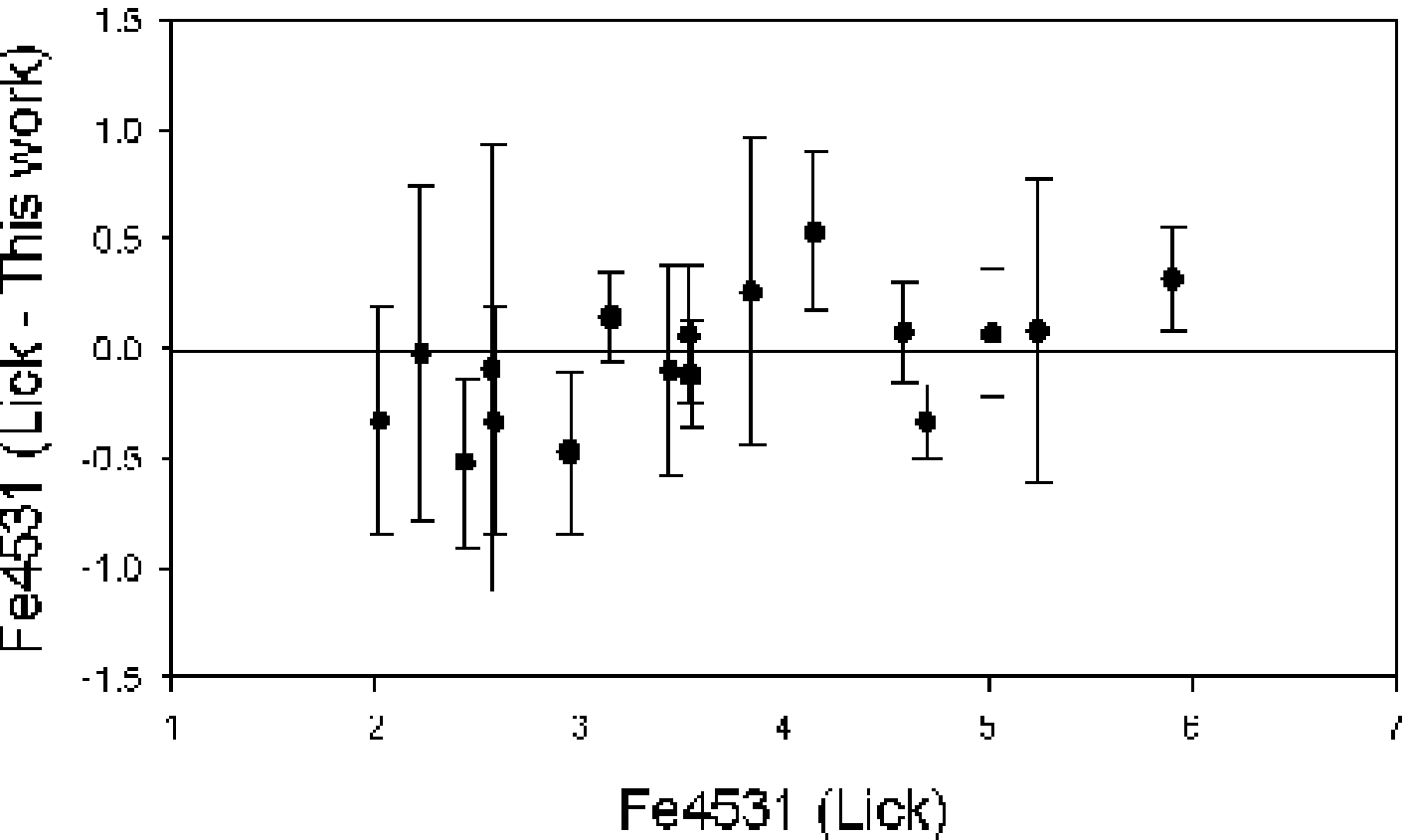}}\quad
\subfigure{\includegraphics[width=4cm,height=2.4cm]{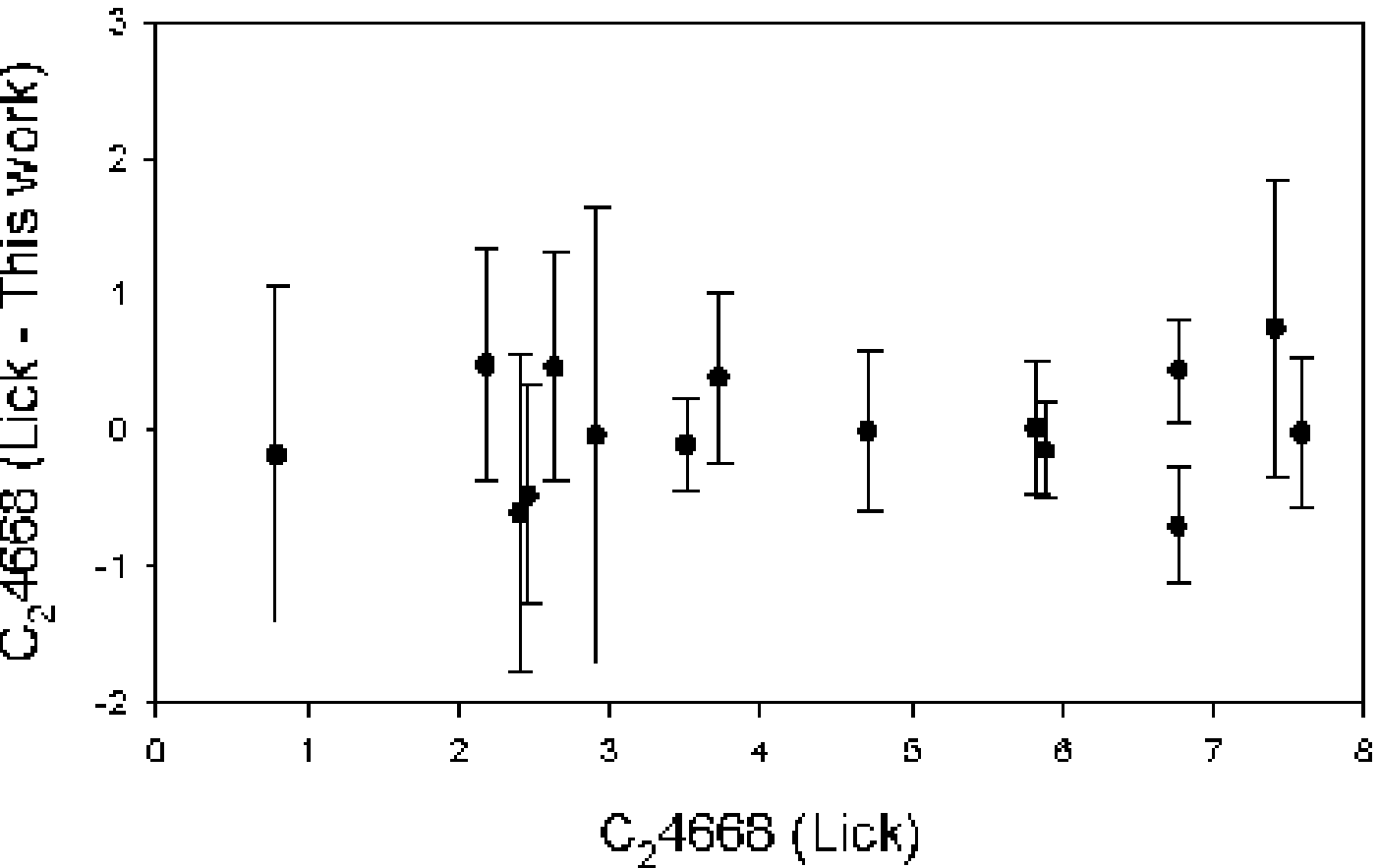}}}
\mbox{\subfigure{\includegraphics[width=4cm,height=2.4cm]{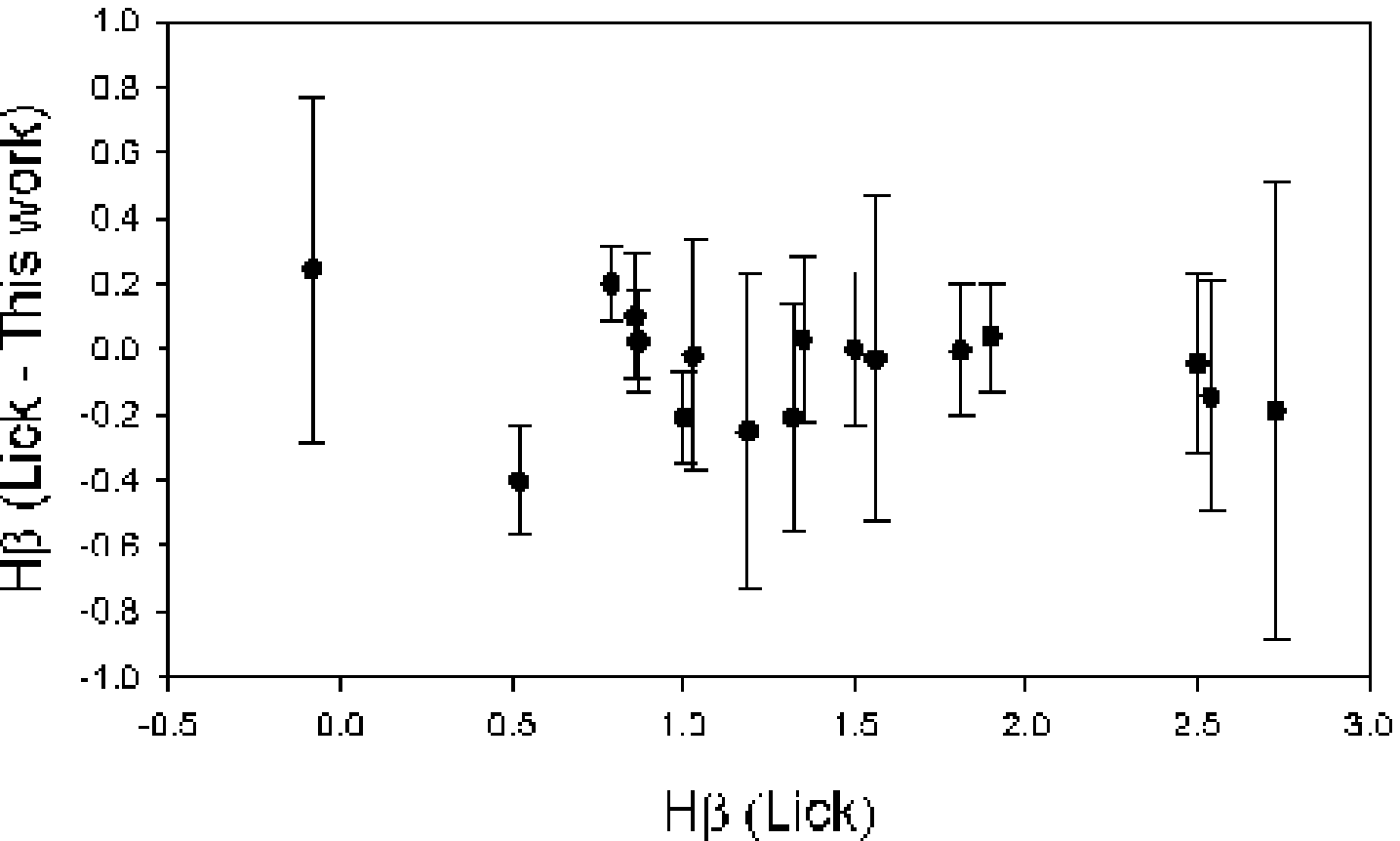}}\quad
   \subfigure{\includegraphics[width=4cm,height=2.4cm]{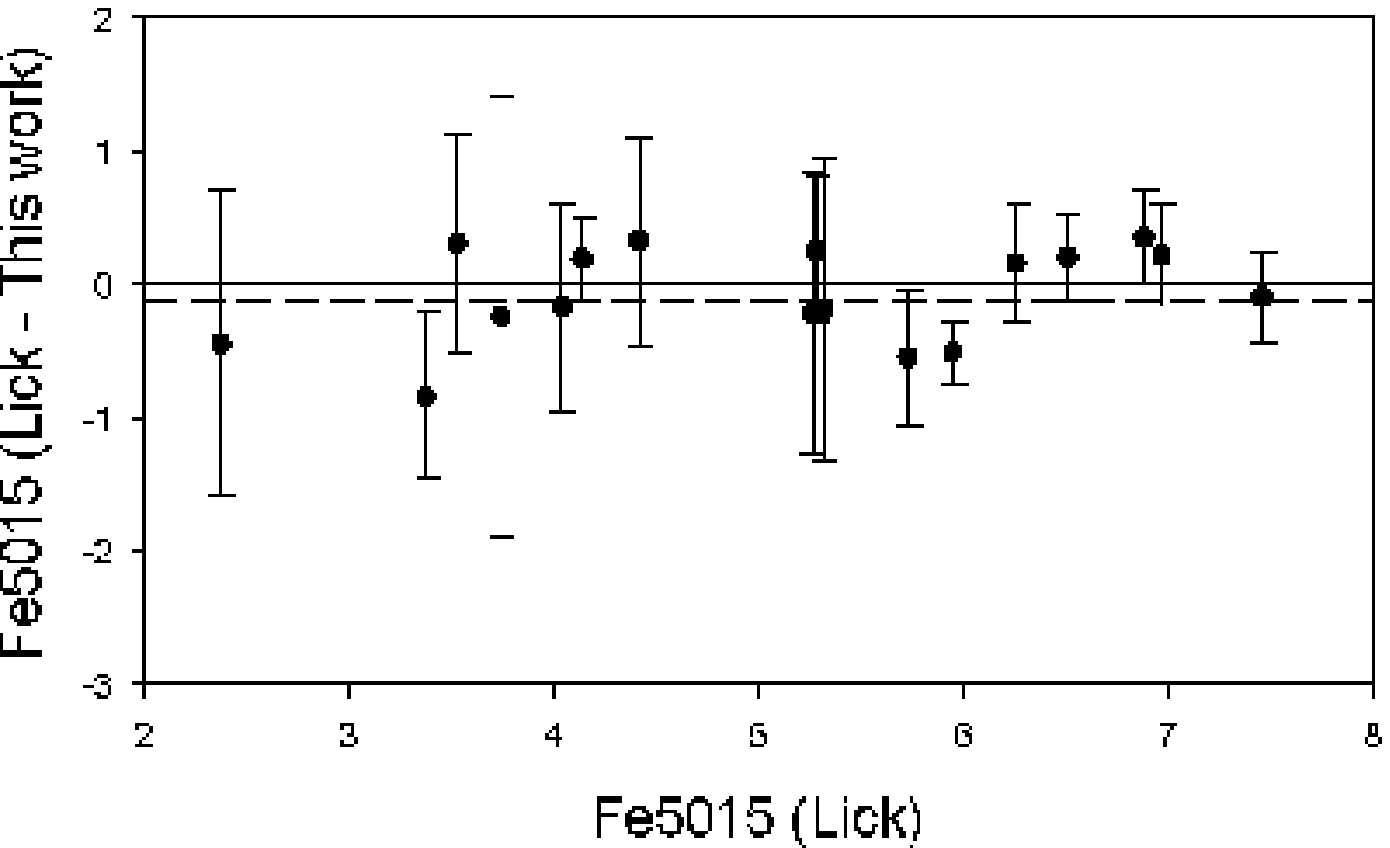}}\quad
\subfigure{\includegraphics[width=4cm,height=2.4cm]{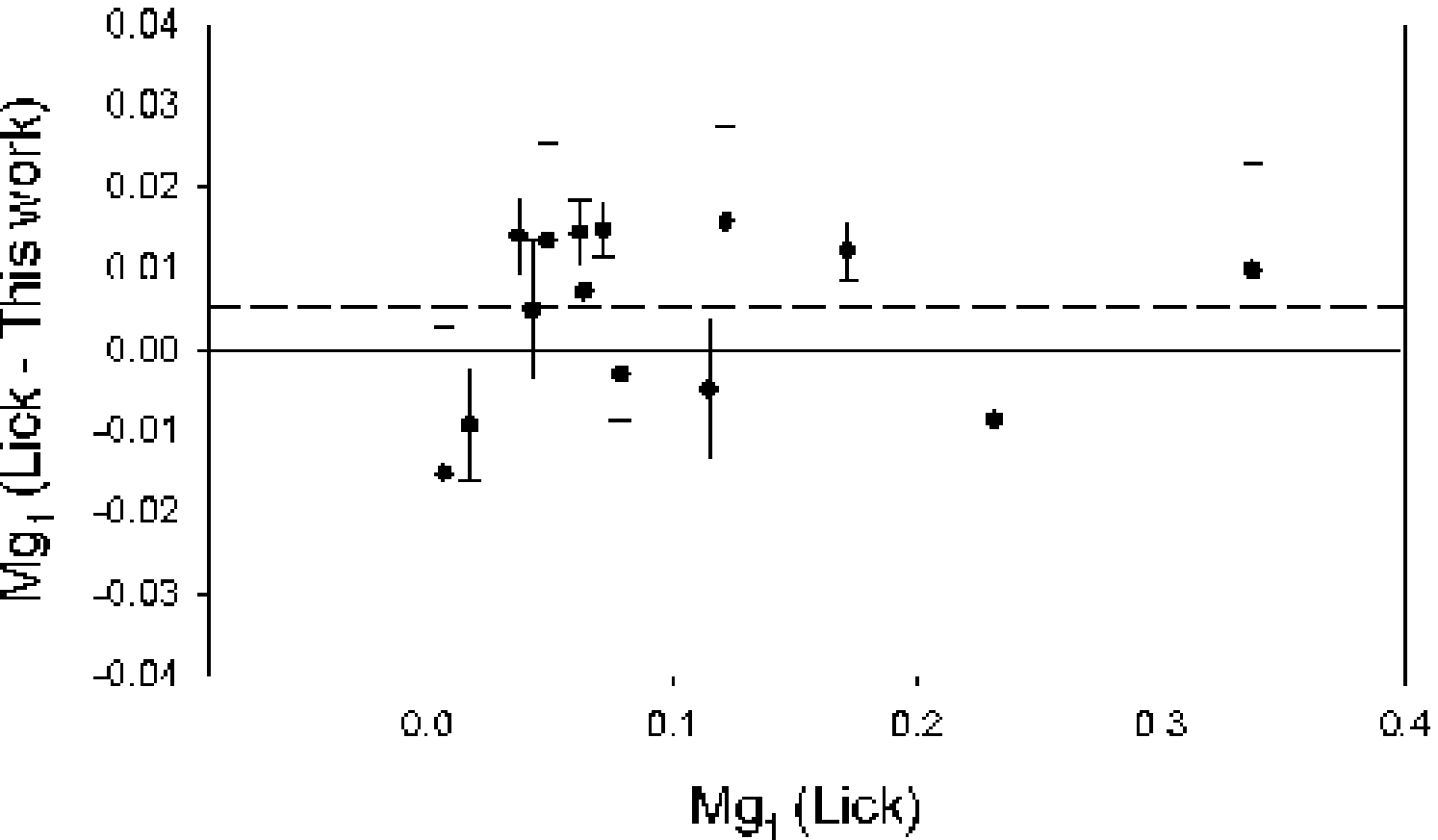}}}
\mbox{\subfigure{\includegraphics[width=4cm,height=2.4cm]{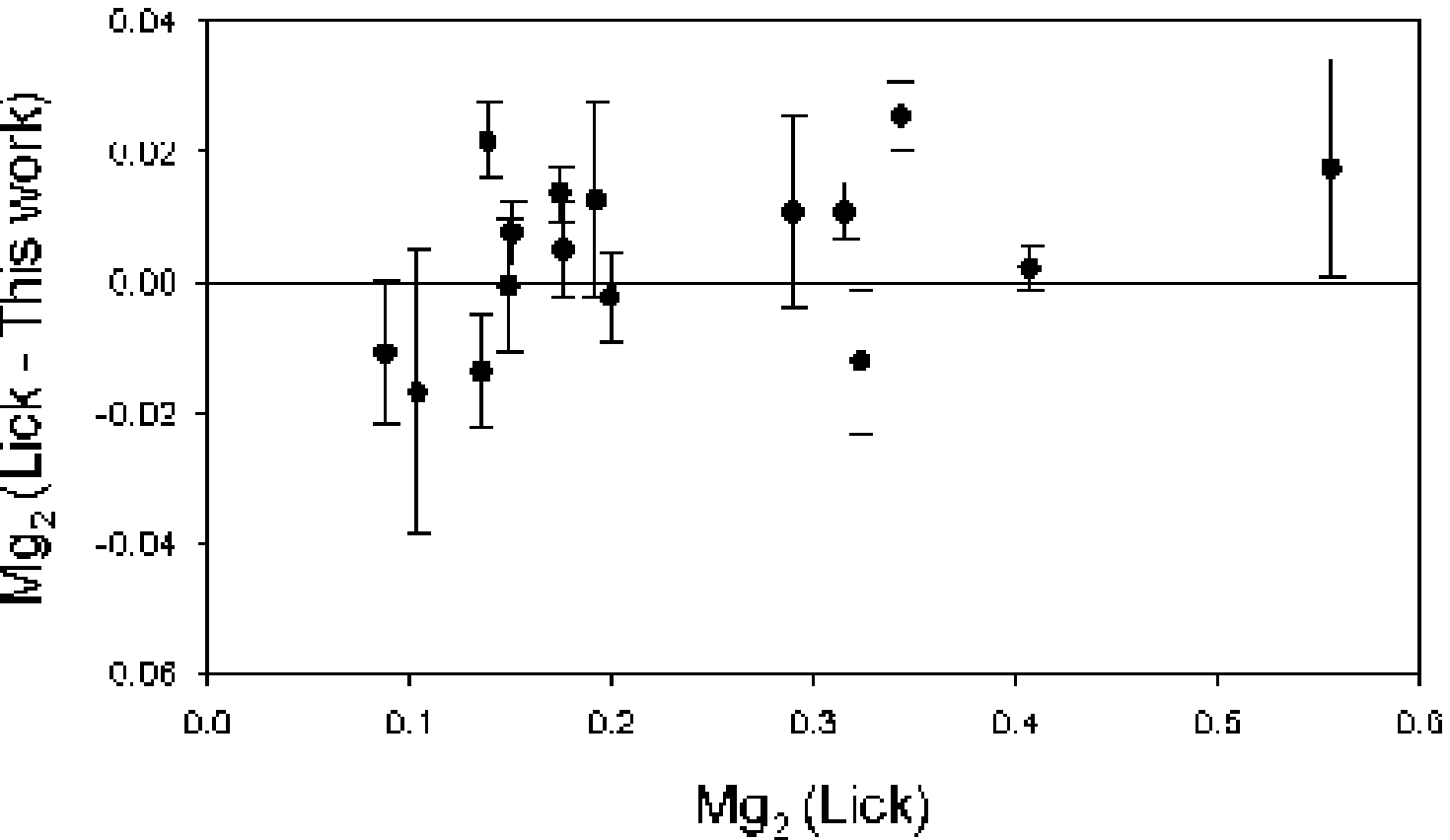}}\quad
   \subfigure{\includegraphics[width=4cm,height=2.4cm]{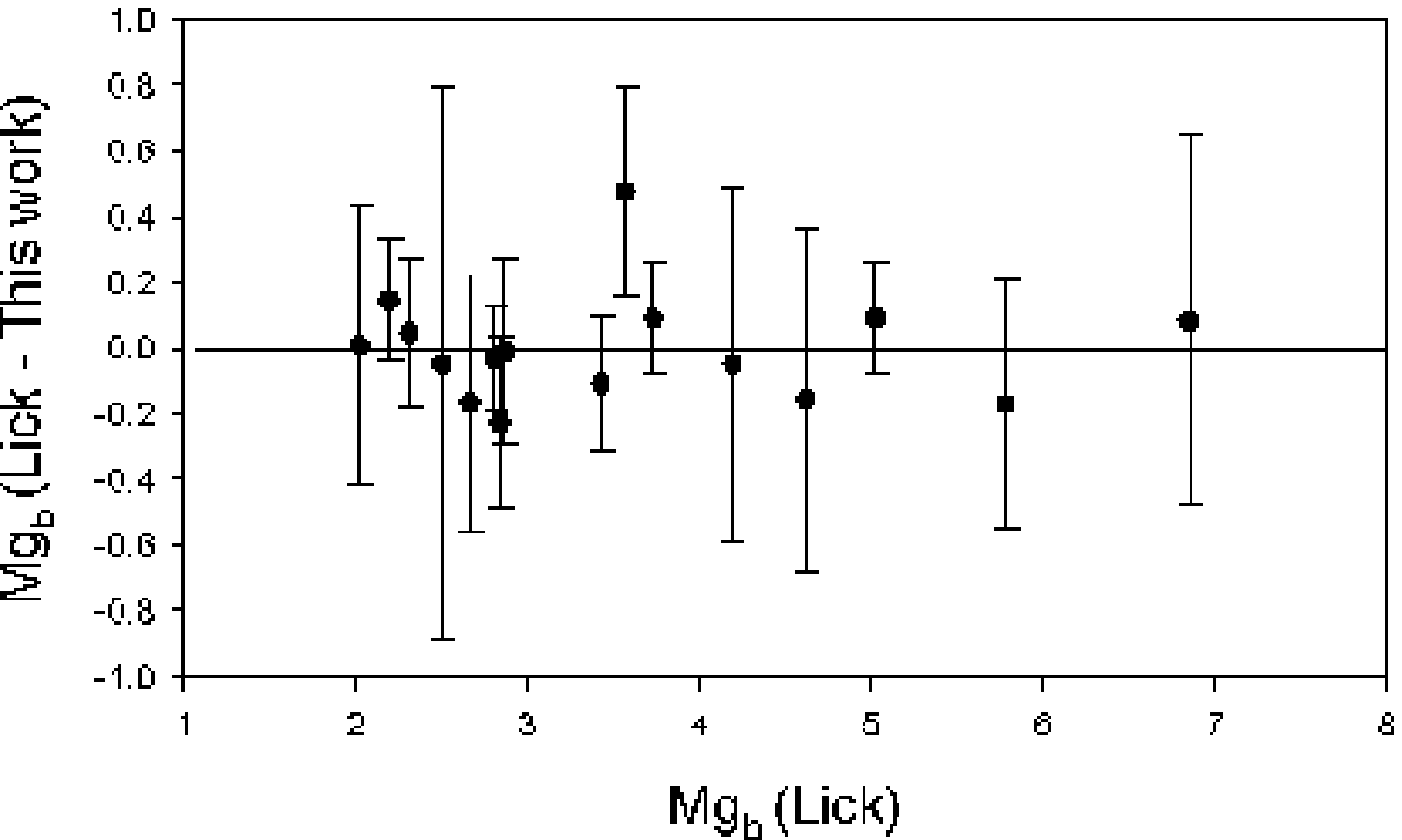}}\quad
\subfigure{\includegraphics[width=4cm,height=2.4cm]{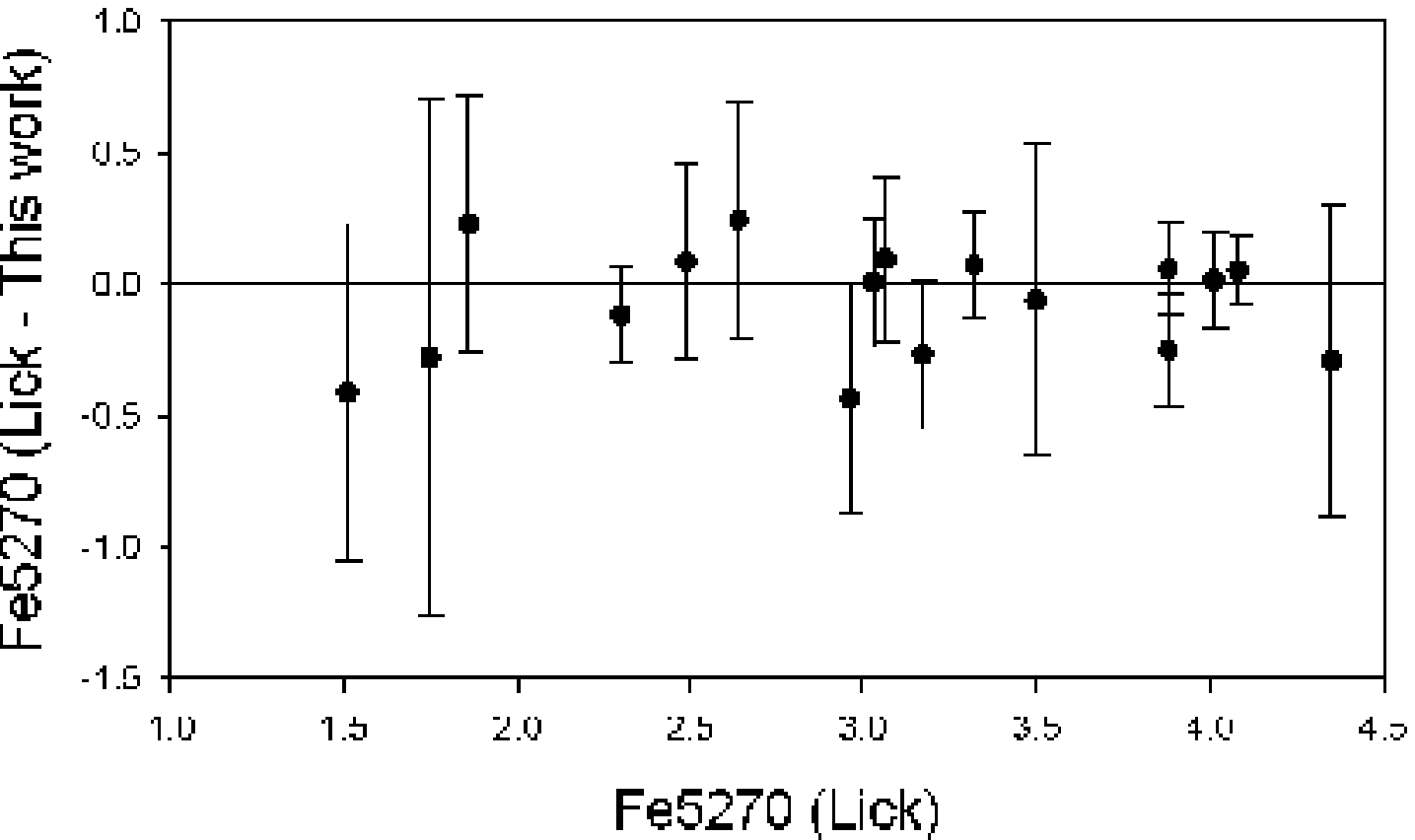}}}
\mbox{\subfigure{\includegraphics[width=4cm,height=2.4cm]{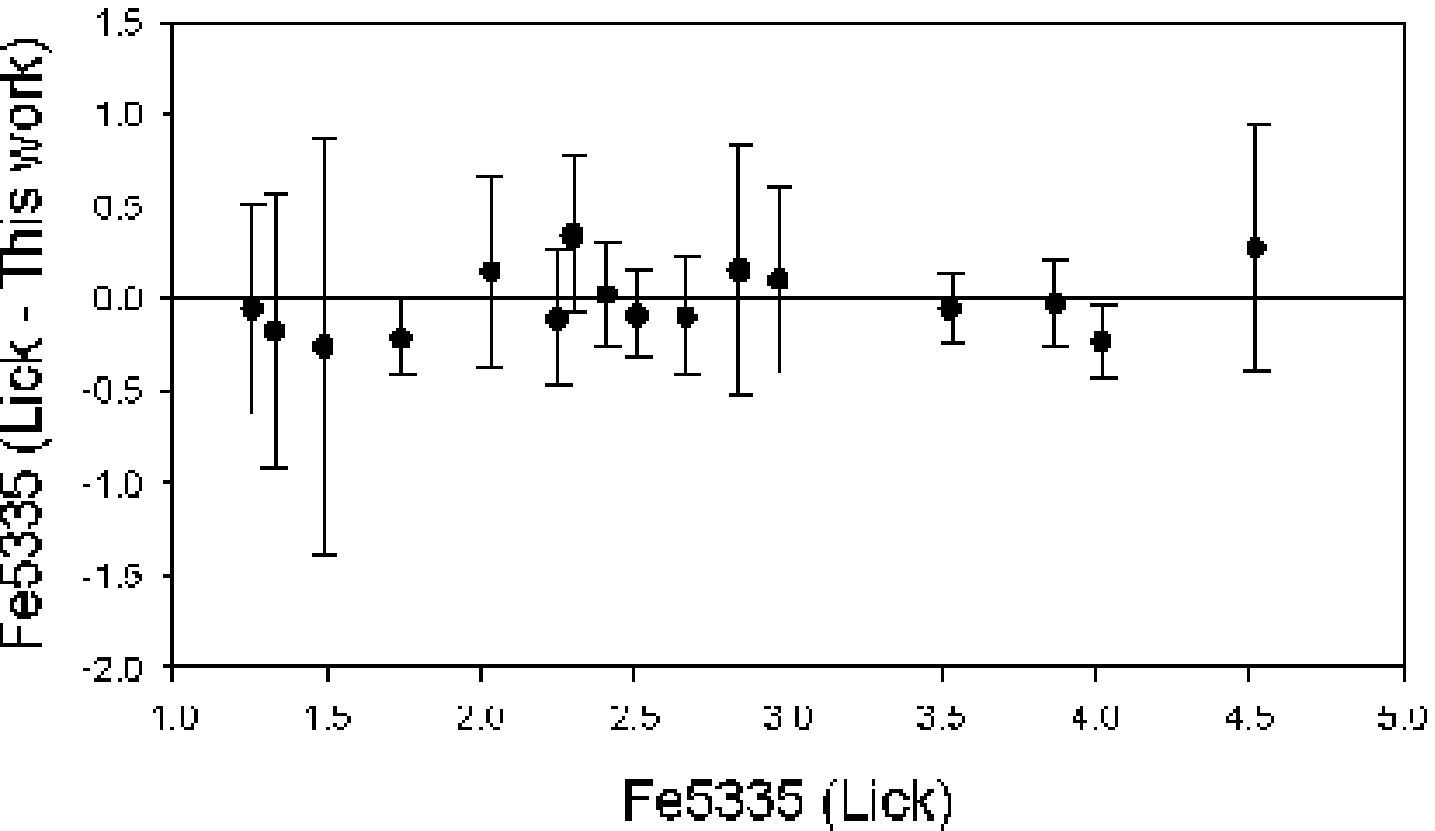}}\quad
   \subfigure{\includegraphics[width=4cm,height=2.4cm]{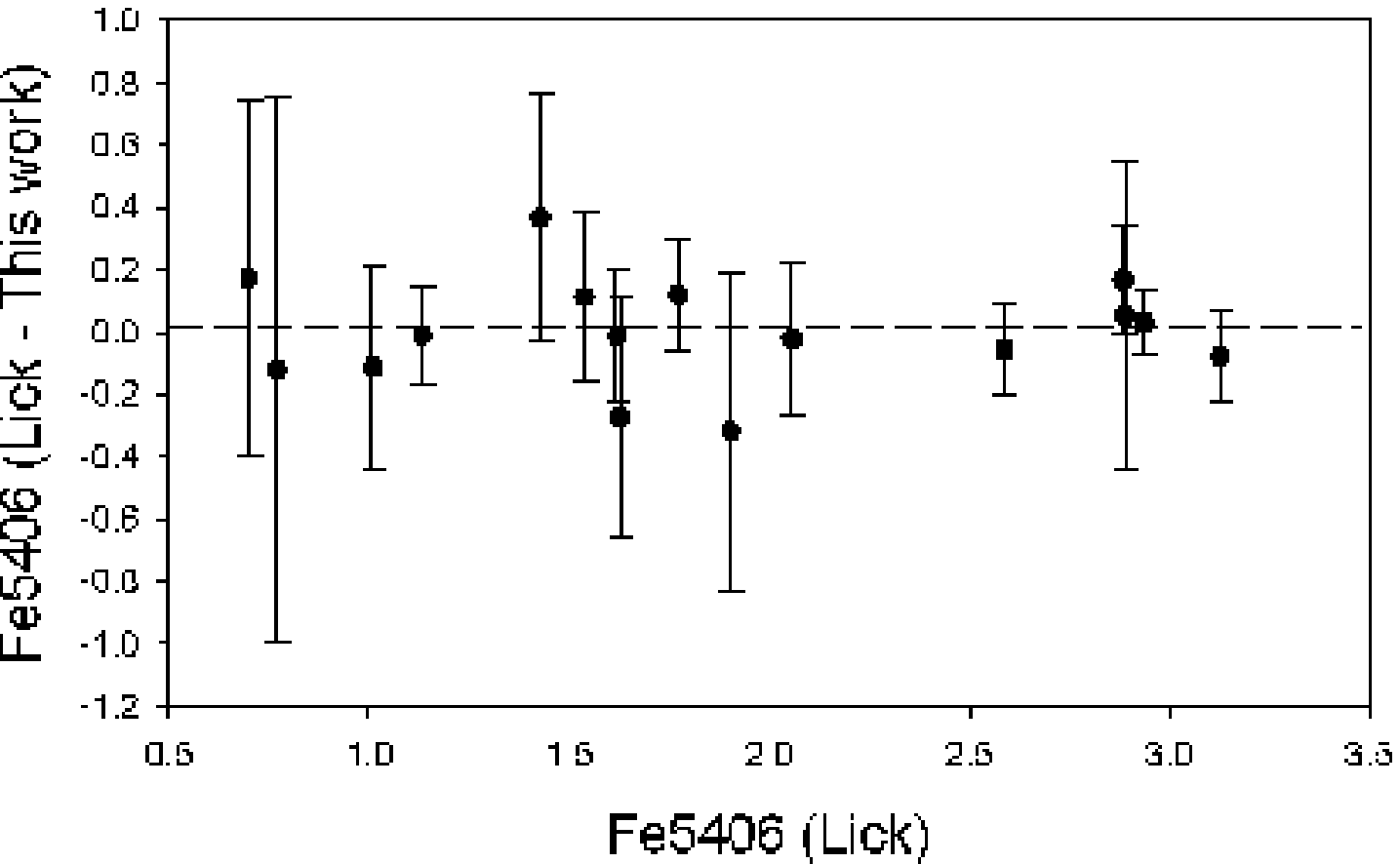}}\quad
\subfigure{\includegraphics[width=4cm,height=2.4cm]{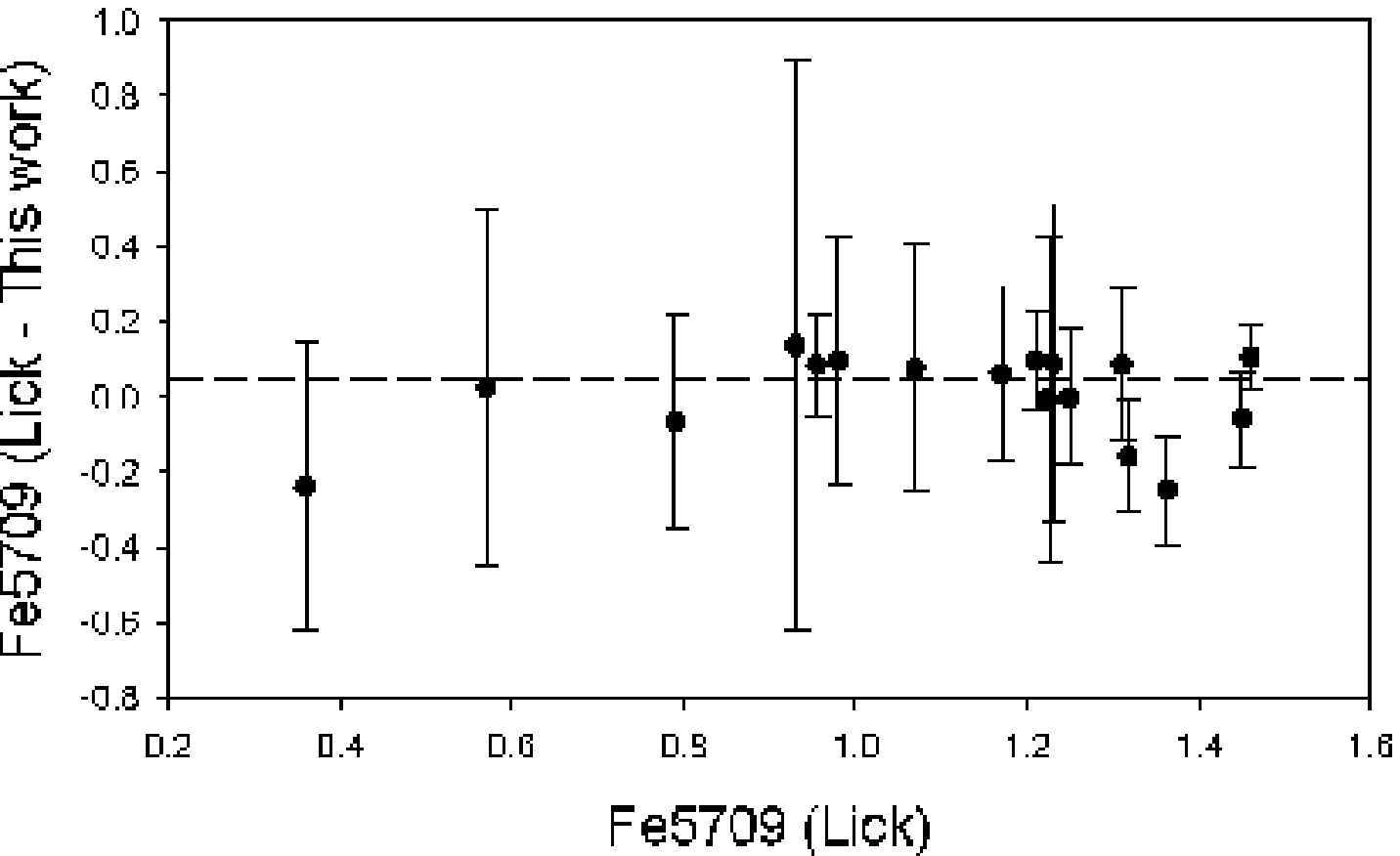}}}
\mbox{\subfigure{\includegraphics[width=4cm,height=2.4cm]{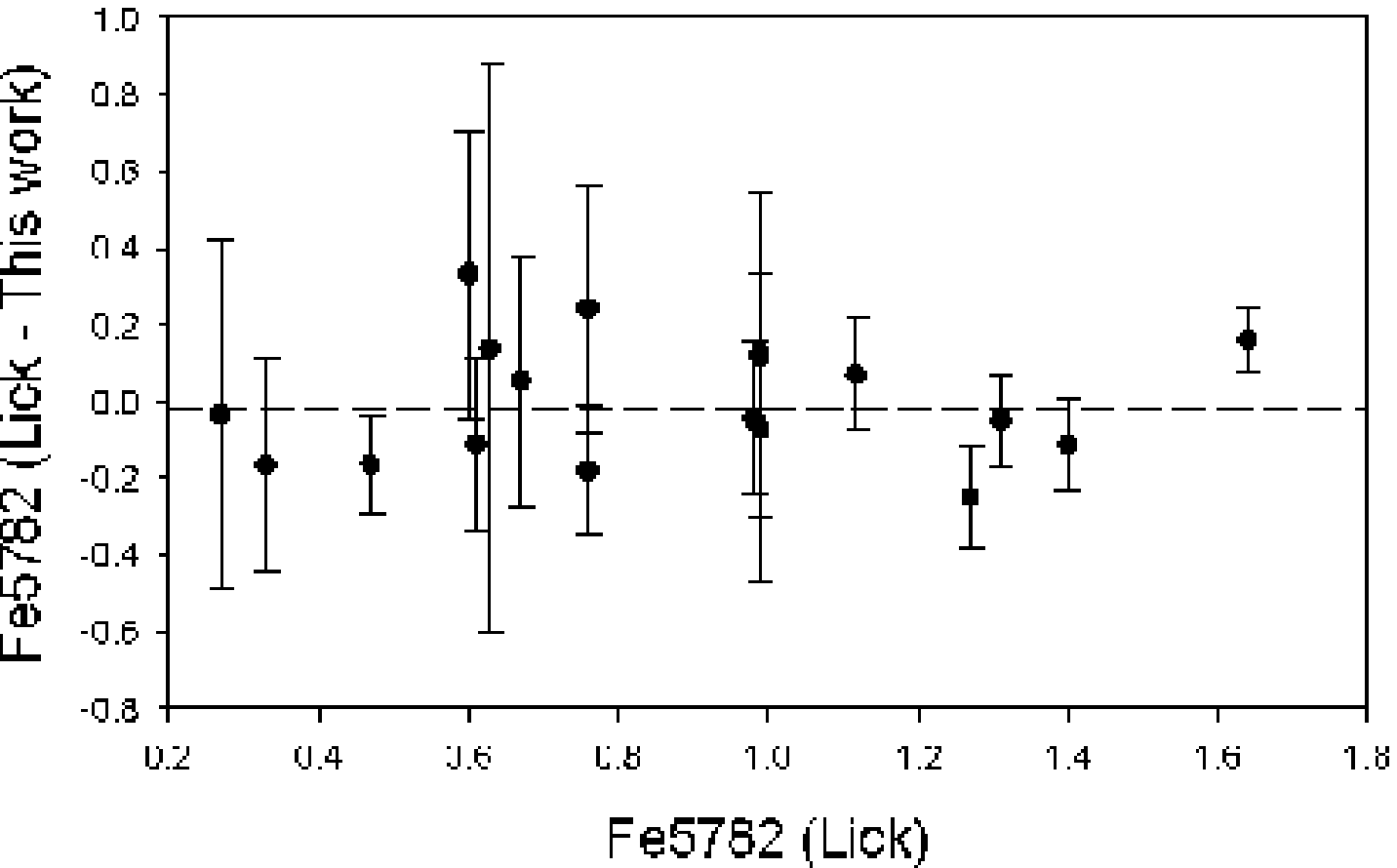}}\quad
   \subfigure{\includegraphics[width=4cm,height=2.4cm]{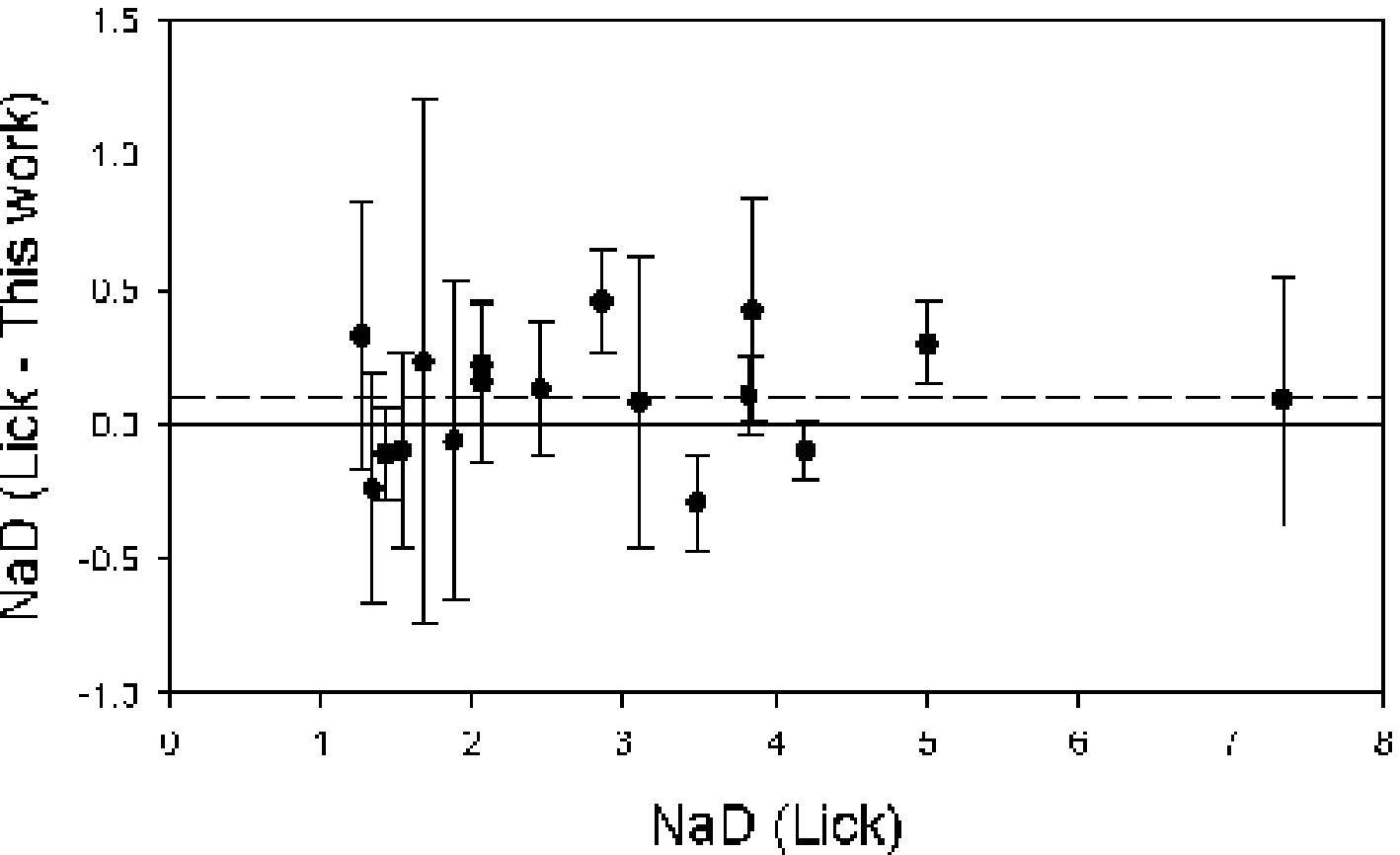}}\quad
\subfigure{\includegraphics[width=4cm,height=2.4cm]{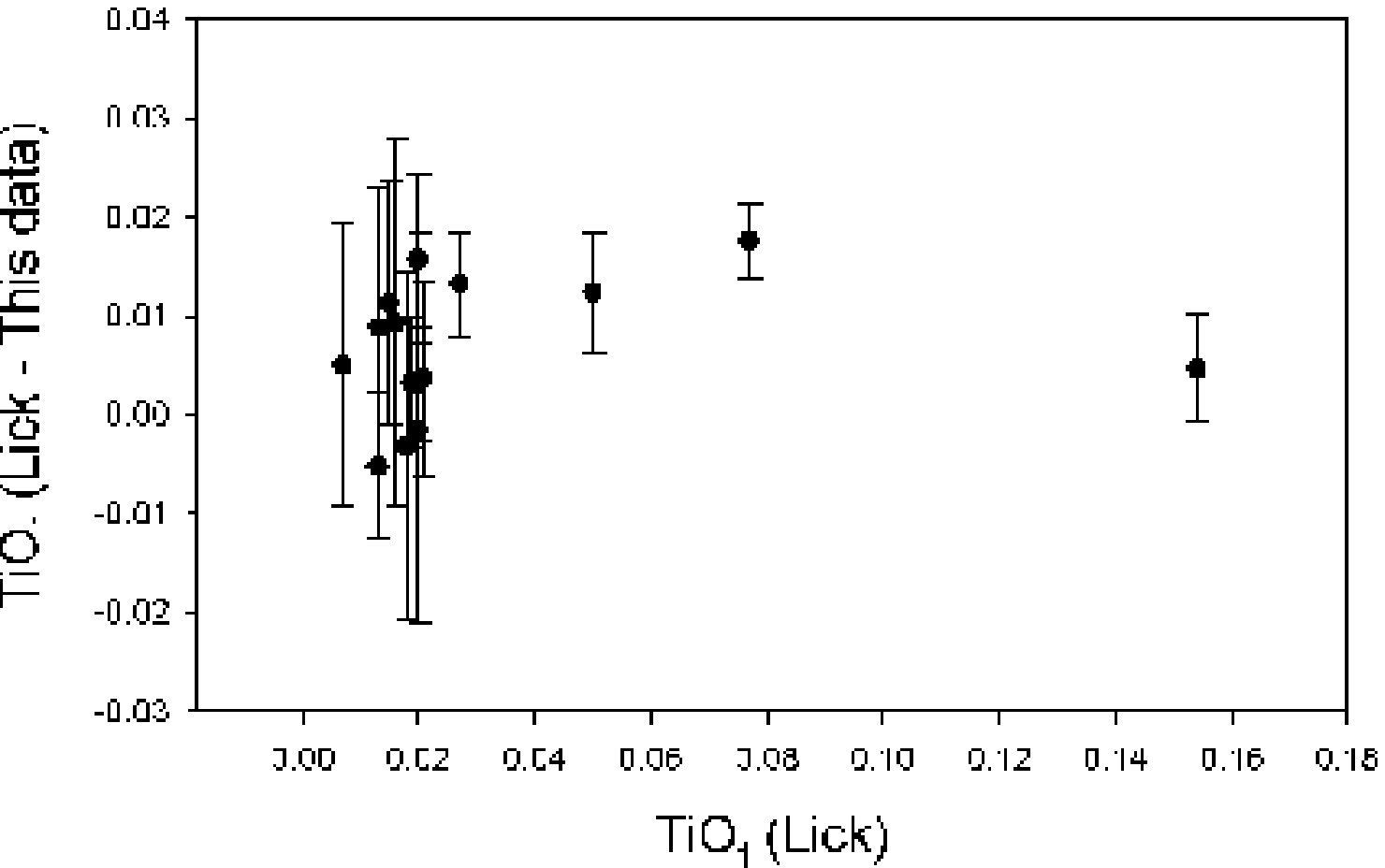}}}
\caption[Gemini Offsets]{Gemini offsets derived from index measurements (original Lick data -- this work) plotted against the original Lick measurements. The dashed horizontal line represents the Lick offset.}
\label{fig:Offsets1}
\end{figure*}

All the offsets derived for the WHT and Gemini data were independent of the strength of the index itself, with the exception of the two Gemini CN indices. For all other indices the average differences were computed and used as final corrections. For the CN indices, correlations between the offsets and the measured index values (from this work) were found: 
\begin{equation}
 \Delta \rm I = -0.0109 + 0.201 \times \rm CN_{1}
\end{equation}  
\begin{equation}
 \Delta \rm I = -0.0150 + 0.184 \times \rm CN_{2}
\end{equation}  
For all indices, the error in calibration to the Lick system was calculated as RMS/$\sqrt{N-1}$, where $N$ is the number of calibration stars ($N = 17$ for the Gemini data, $N = 20$ for the WHT data 5300 dichroic and $N = 10$ for the WHT data 6100 dichroic). The Gemini offsets are shown in Figure \ref{fig:Offsets1} as an example. As can be seen in the plots, the offsets are typically smaller than the error on the index measurement. A comparison was made between the Gemini offsets derived in this work with those derived by M. Norris (private communication), using the same archive data of the Lick stars. No systematic differences between the two sets of offsets were visible. Most of the offsets used here are in agreement, with the exception of C$_{2}$4668 and Ca4455 which are slightly higher than those derived by M. Norris.

Table \ref{table:GeminitoLick} shows all the parameters of the calibration to the Lick system for the Gemini data as an example, and similar parameters were derived for the WHT data.

\section{Indices: comparison with previous measurements}
\label{appendix3}

The central index measurements were compared to those of Trager (1998, hereafter T98) for 14 galaxies which the samples had in common. The measurements of the original 21 Lick indices were presented in T98, and the four higher-order Balmer indices (of the same galaxy sample) in Lee $\&$ Worthey 2005 (hereafter LW05). The central values of the T98 sample were measured in an aperture of 1.4 $\times$ 4 arcsec$^{2}$. Only one of the 14 galaxies had TiO$_{2}$ measurements in both T98 and this sample, so no comparison of TiO$_{2}$ measurements were made. Galaxies with large differences in certain indices (for example a few Fe5270 measurements) were compared on an individual basis (S. Trager, private communication), and can be attributed to lower S/N data.

\begin{figure*}
   \centering
   \mbox{\subfigure{\includegraphics[width=3.2cm,height=3.2cm]{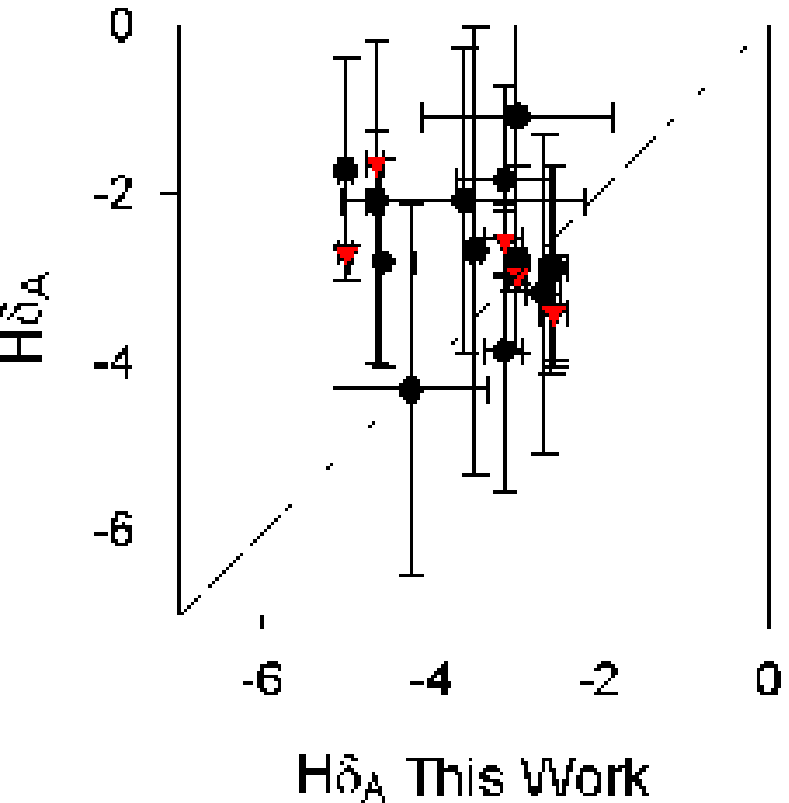}}\quad
\subfigure{\includegraphics[width=3.2cm,height=3.2cm]{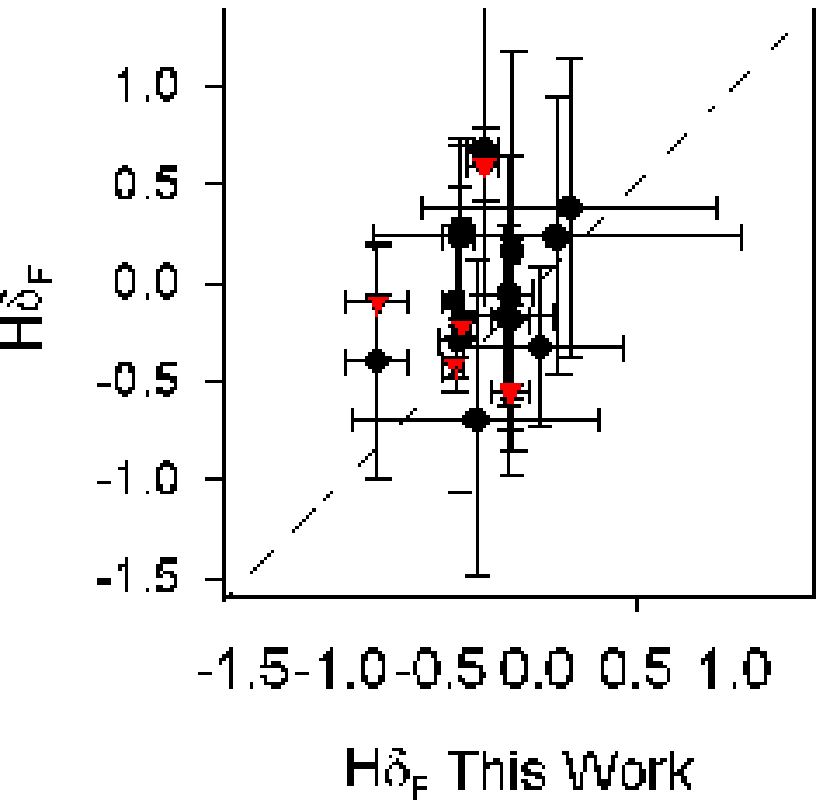}}\quad
\subfigure{\includegraphics[width=3.2cm,height=3.2cm]{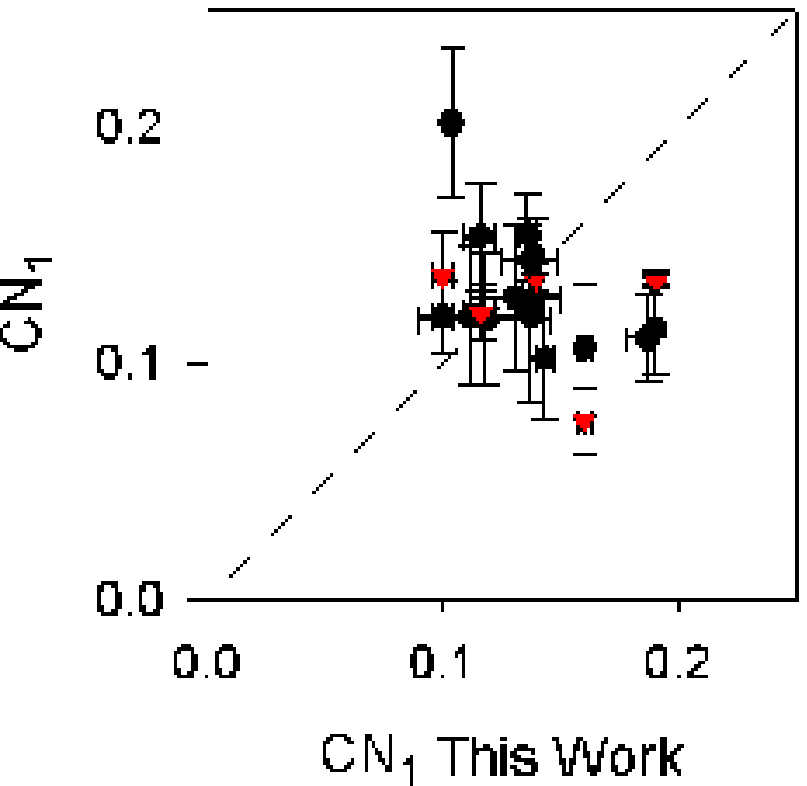}}\quad
\subfigure{\includegraphics[width=3.2cm,height=3.2cm]{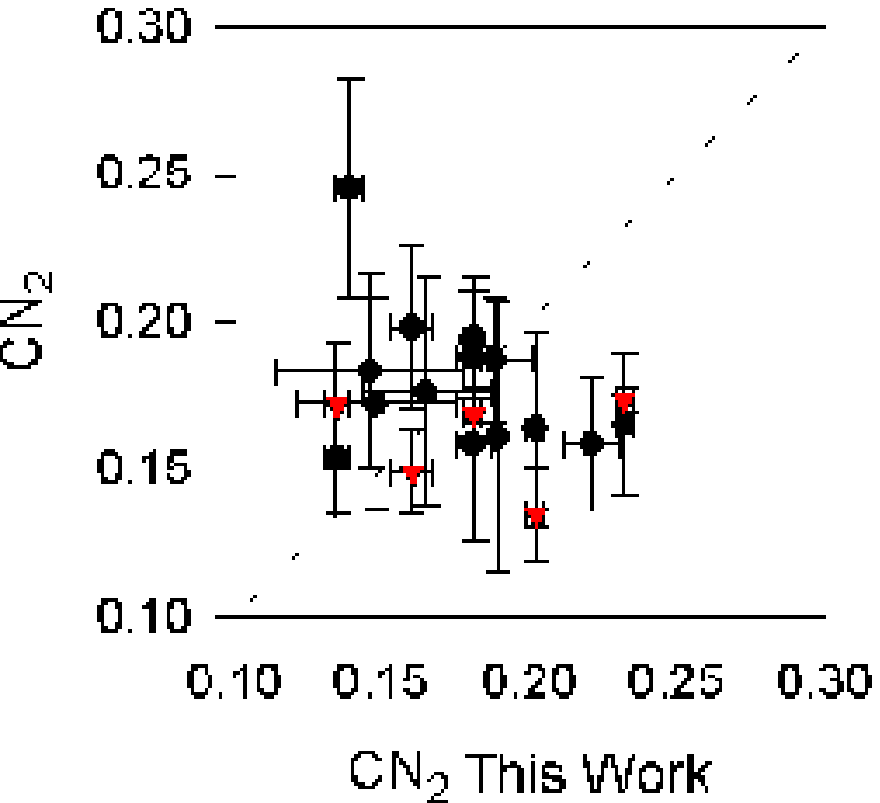}}}
\mbox{\subfigure{\includegraphics[width=3.2cm,height=3.2cm]{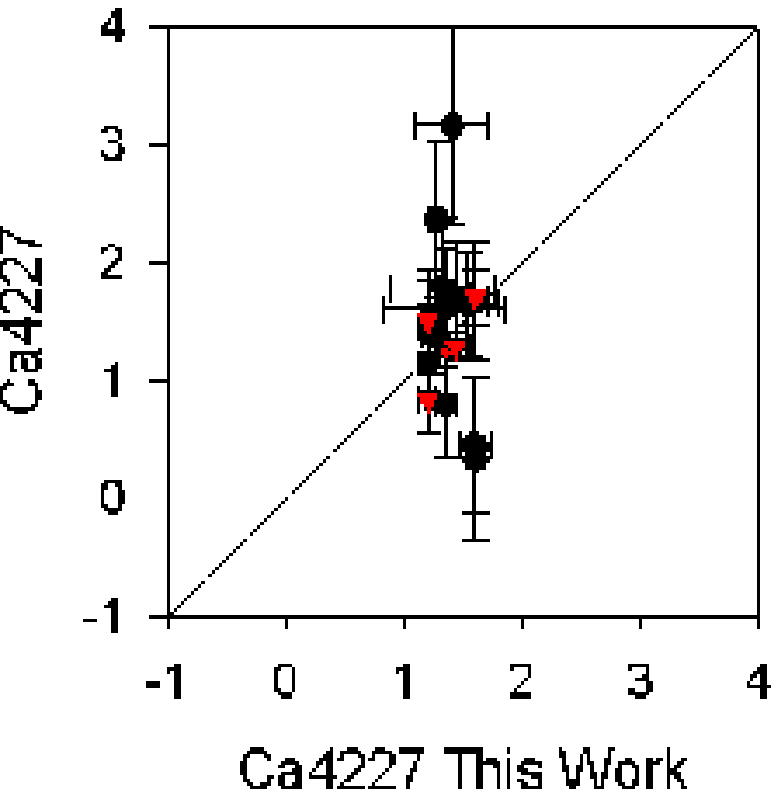}}\quad
\subfigure{\includegraphics[width=3.2cm,height=3.2cm]{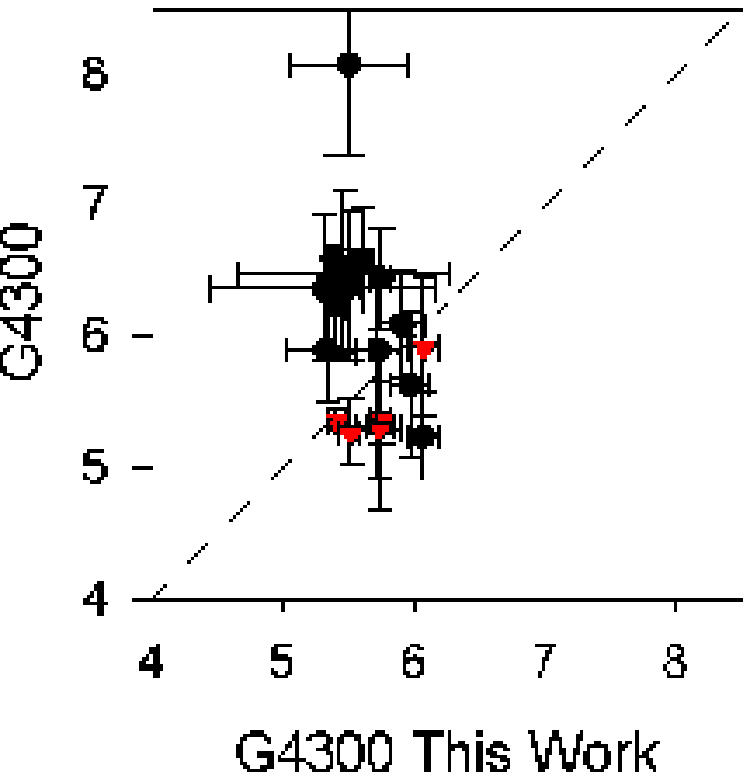}}\quad
\subfigure{\includegraphics[width=3.2cm,height=3.2cm]{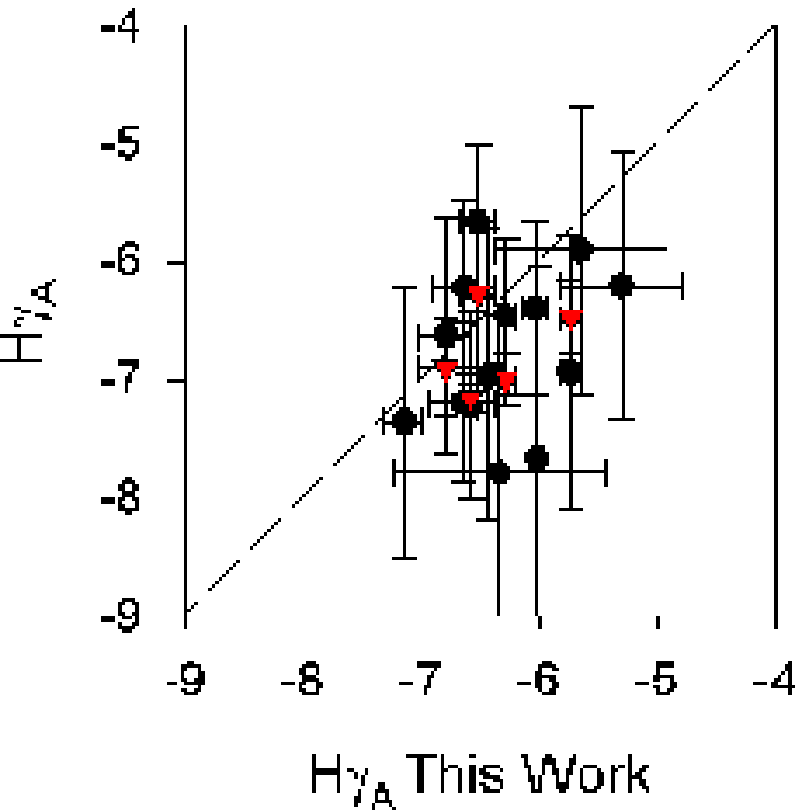}}\quad
\subfigure{\includegraphics[width=3.2cm,height=3.2cm]{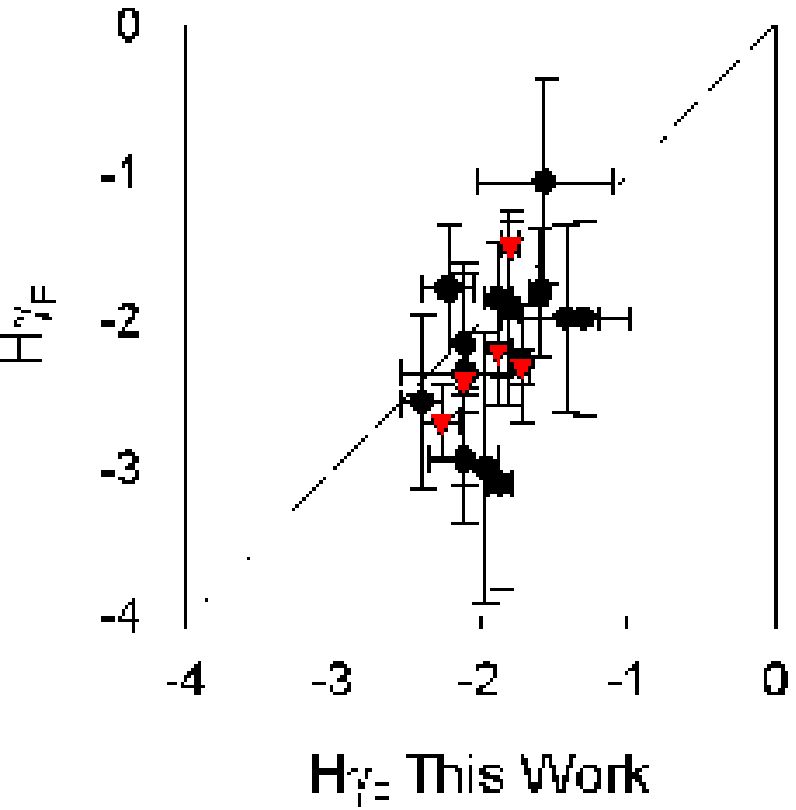}}}
\mbox{\subfigure{\includegraphics[width=3.2cm,height=3.2cm]{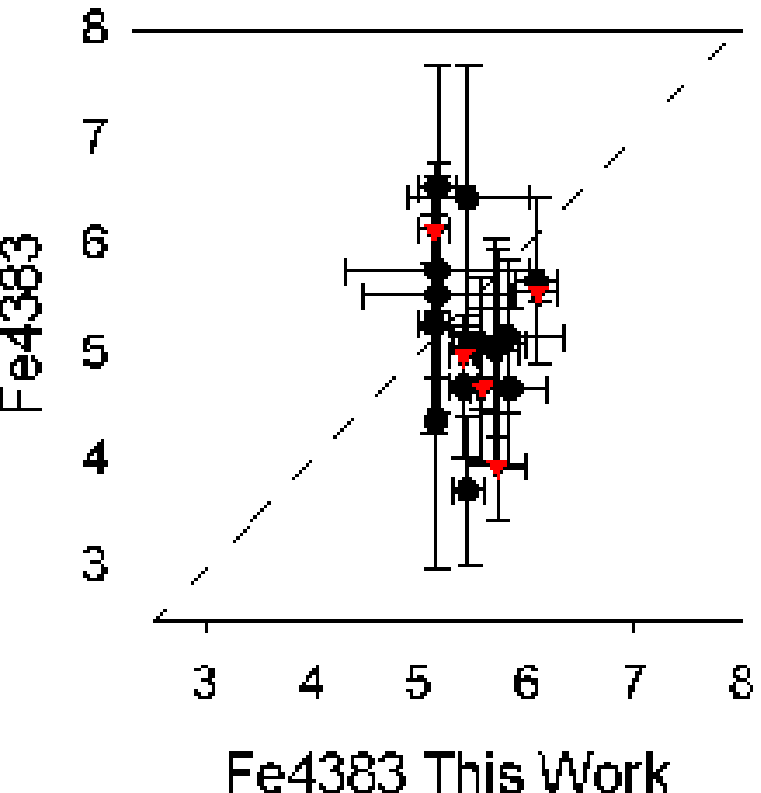}}\quad
\subfigure{\includegraphics[width=3.2cm,height=3.2cm]{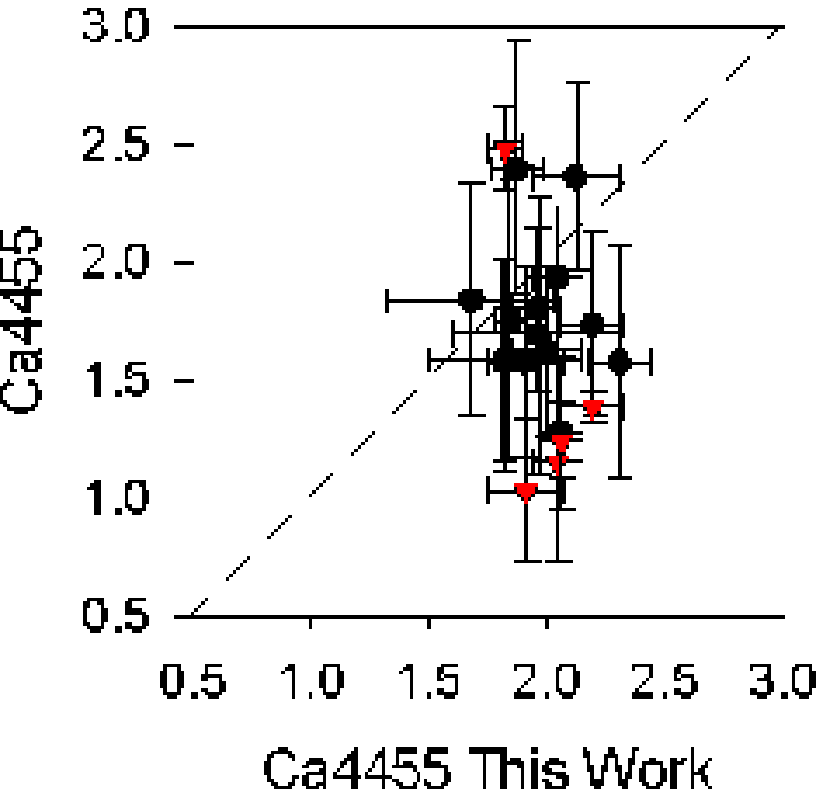}}\quad
\subfigure{\includegraphics[width=3.2cm,height=3.2cm]{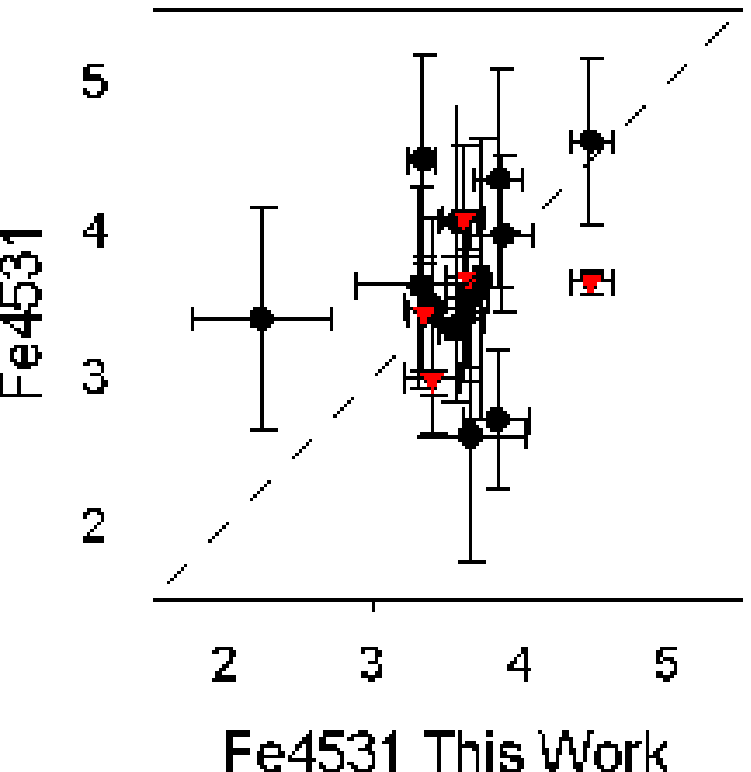}}\quad
\subfigure{\includegraphics[width=3.2cm,height=3.2cm]{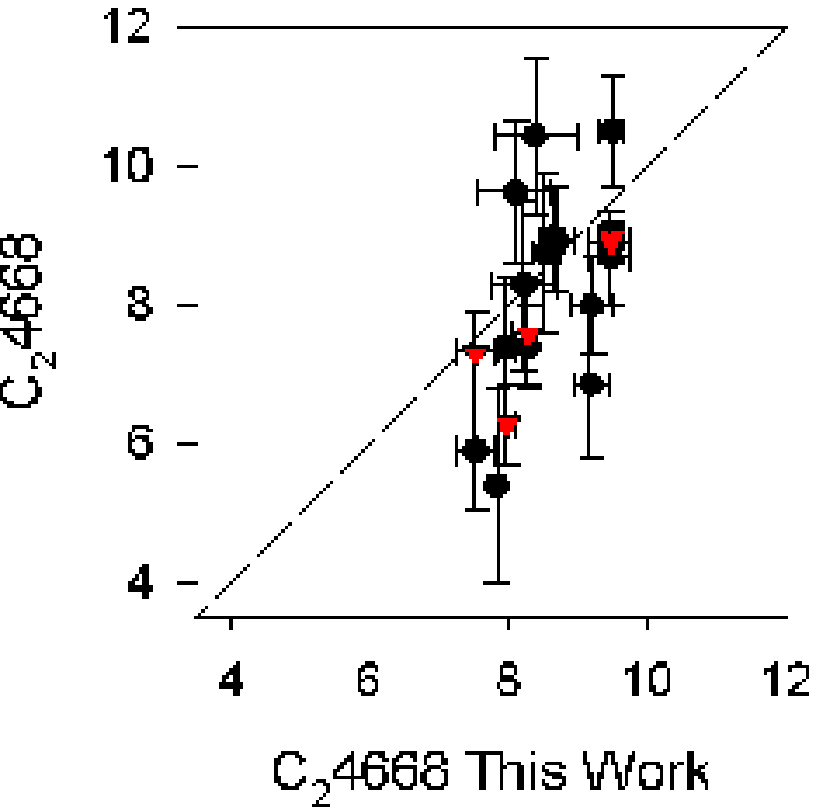}}}
\mbox{\subfigure{\includegraphics[width=3.2cm,height=3.2cm]{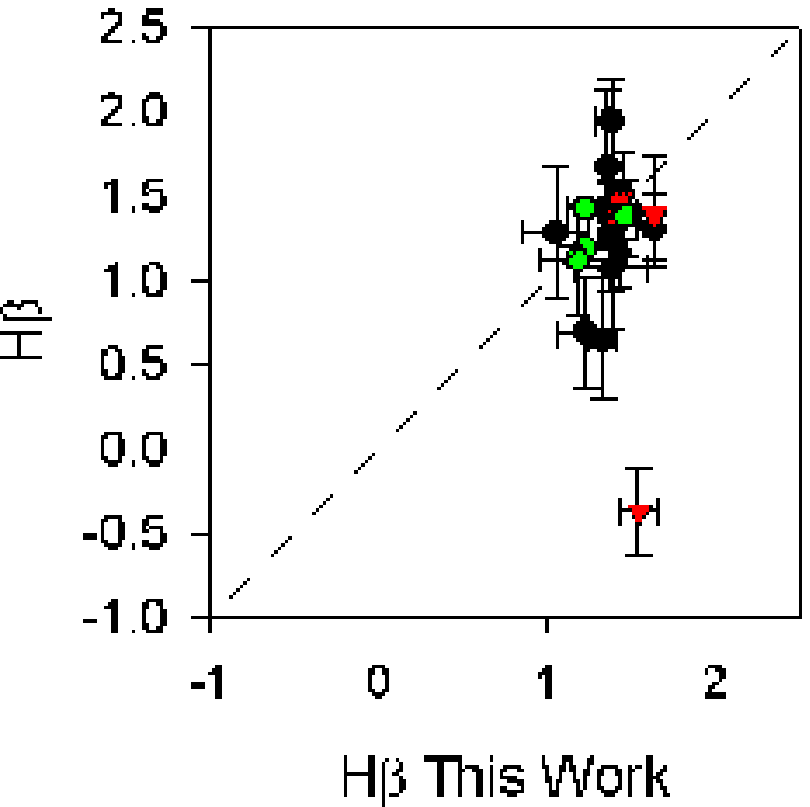}}\quad
\subfigure{\includegraphics[width=3.2cm,height=3.2cm]{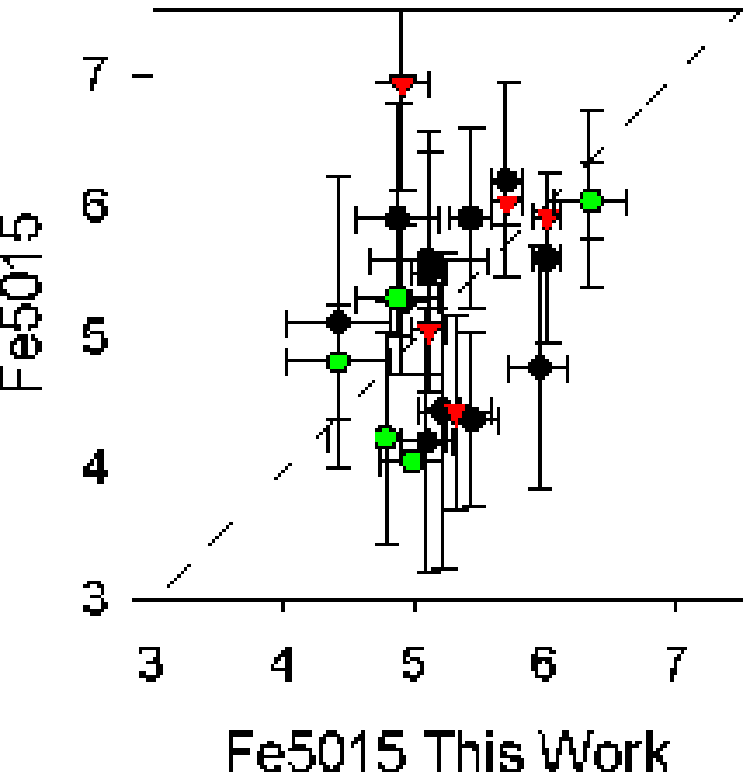}}\quad
\subfigure{\includegraphics[width=3.2cm,height=3.2cm]{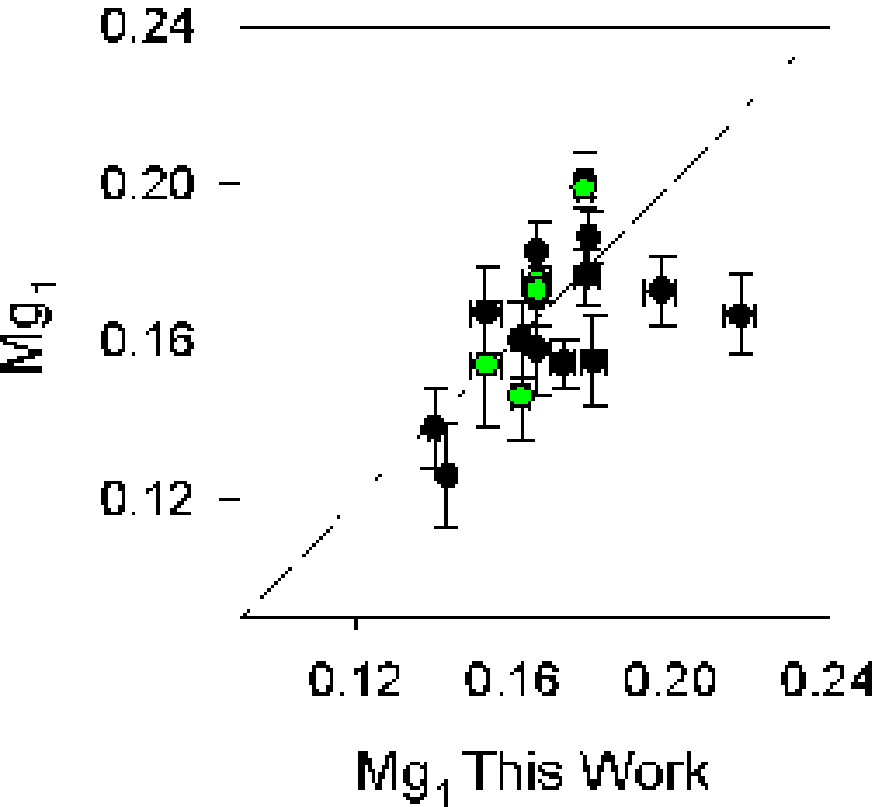}}\quad
\subfigure{\includegraphics[width=3.2cm,height=3.2cm]{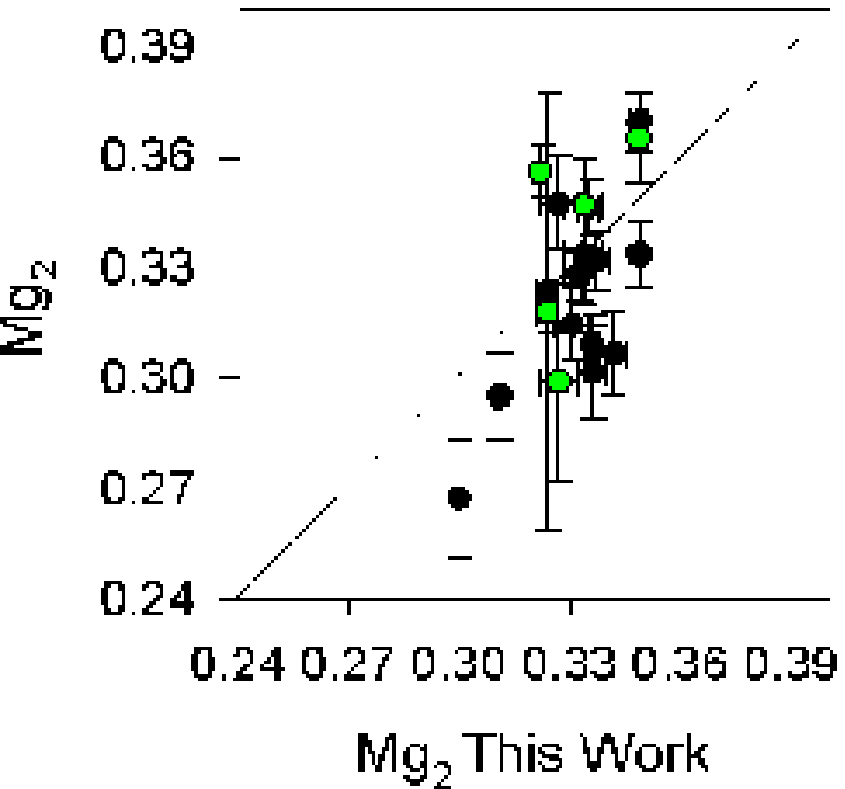}}}
\mbox{\subfigure{\includegraphics[width=3.2cm,height=3.2cm]{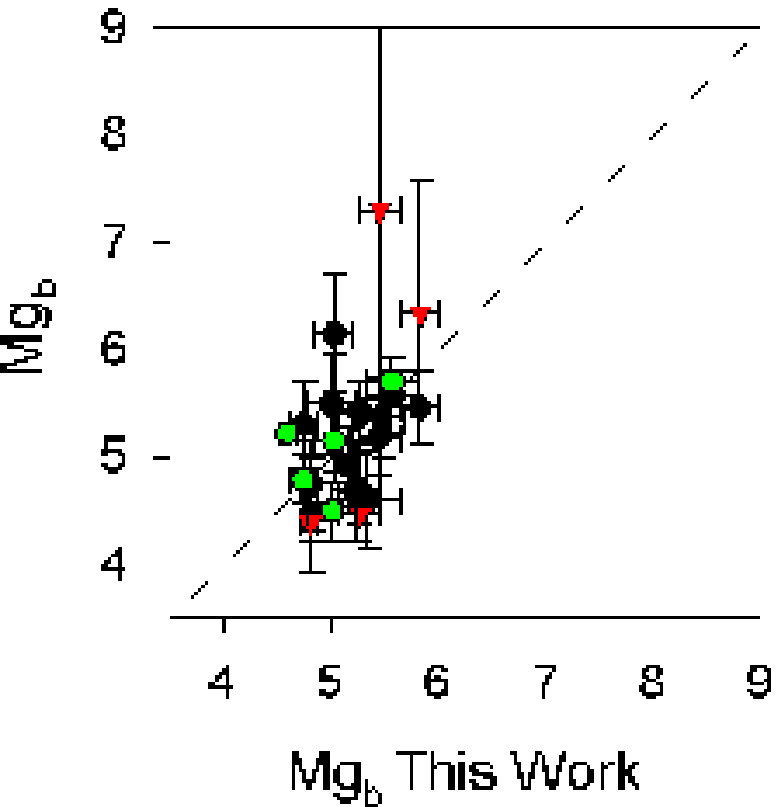}}\quad
\subfigure{\includegraphics[width=3.2cm,height=3.2cm]{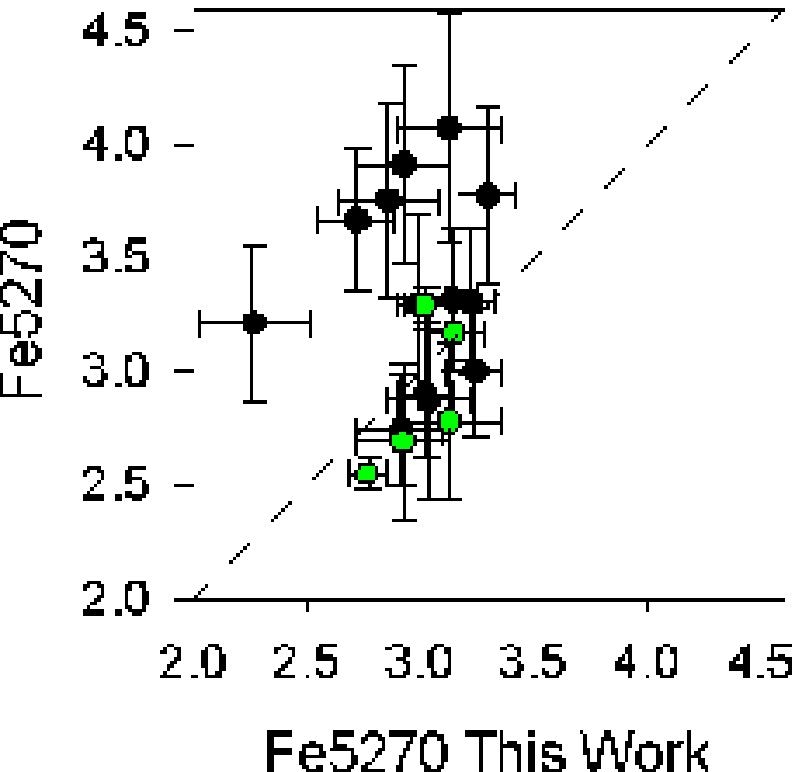}}\quad
\subfigure{\includegraphics[width=3.2cm,height=3.2cm]{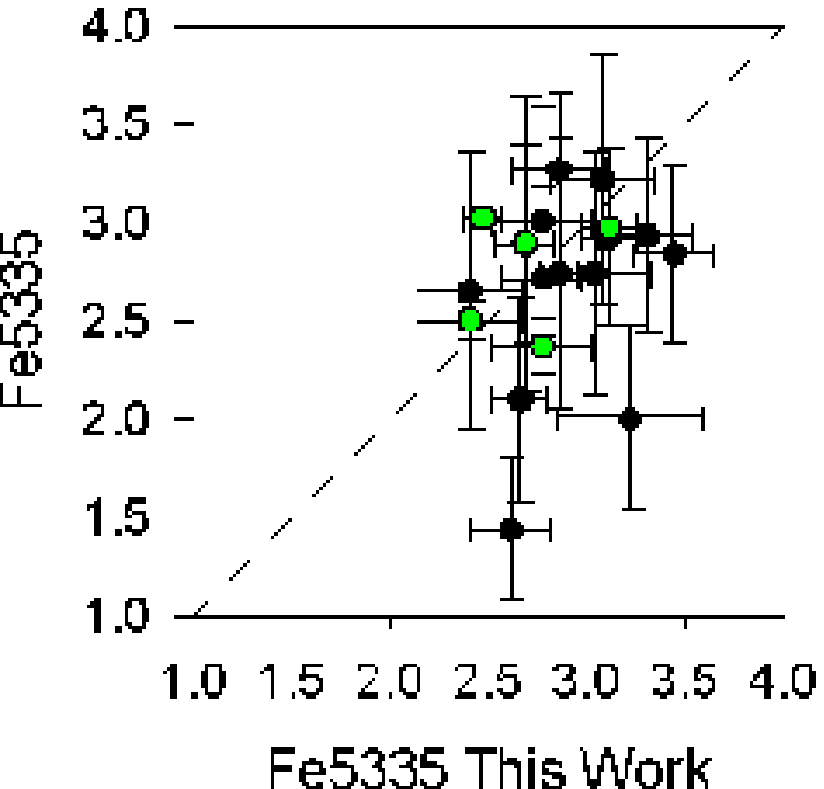}}\quad
\subfigure{\includegraphics[width=3.2cm,height=3.2cm]{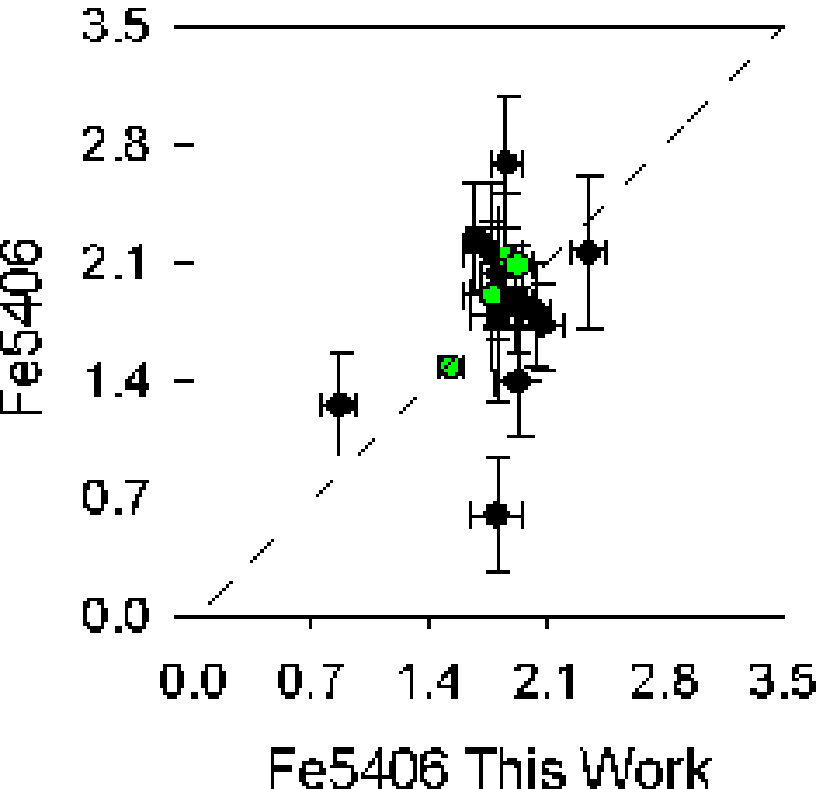}}}
\mbox{\subfigure{\includegraphics[width=3.2cm,height=3.2cm]{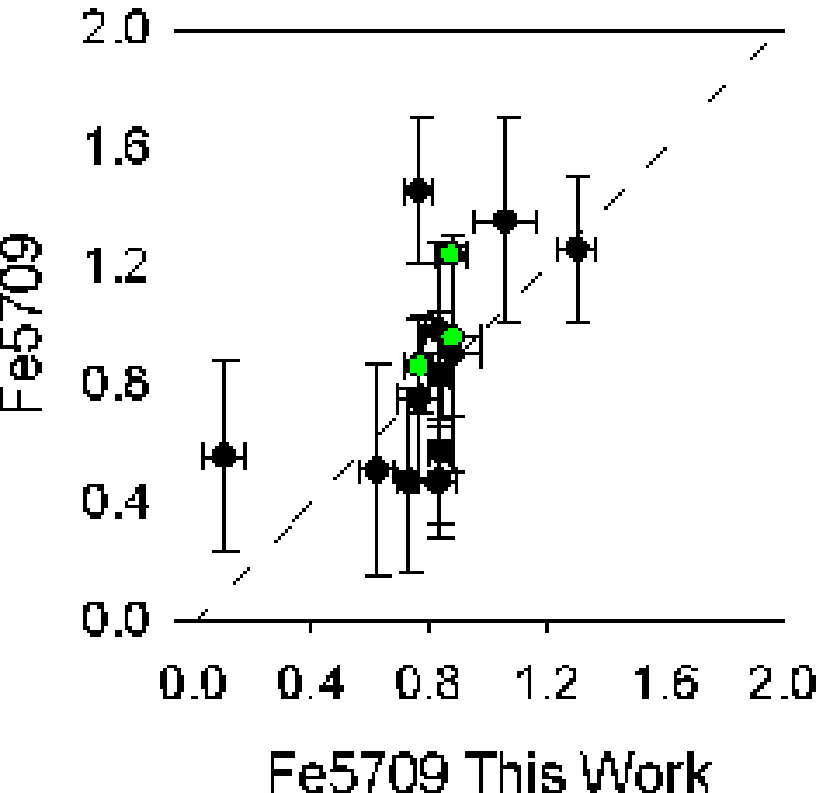}}\quad
\subfigure{\includegraphics[width=3.2cm,height=3.2cm]{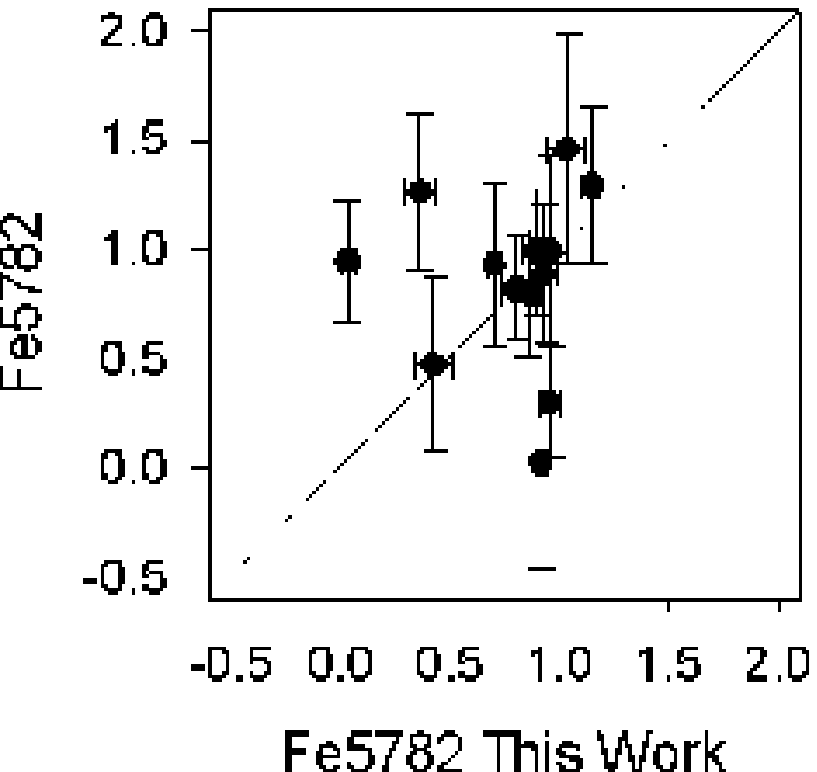}}\quad
\subfigure{\includegraphics[width=3.2cm,height=3.2cm]{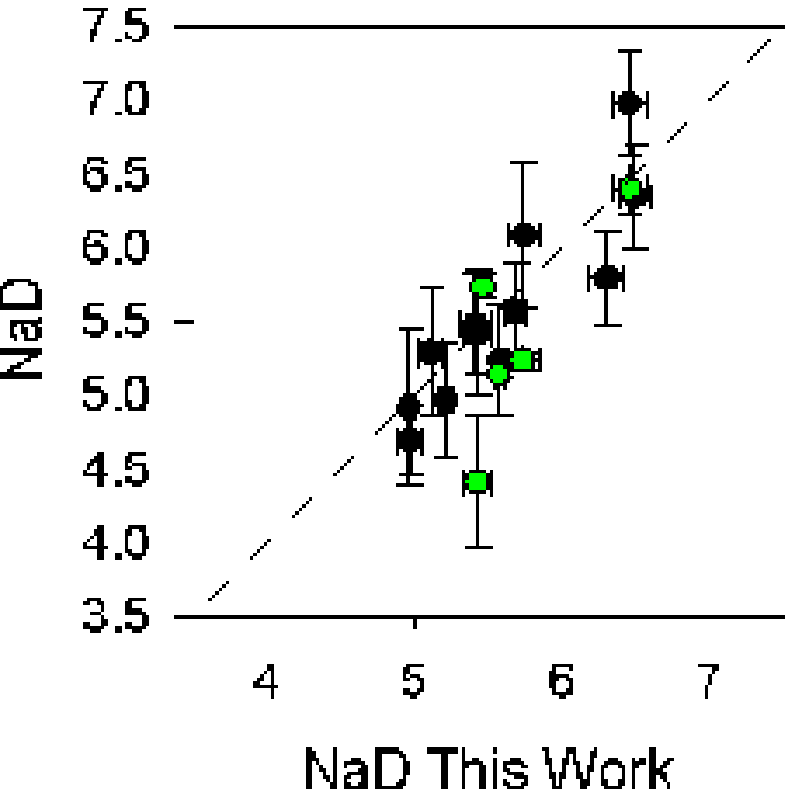}}\quad
\subfigure{\includegraphics[width=3.2cm,height=3.2cm]{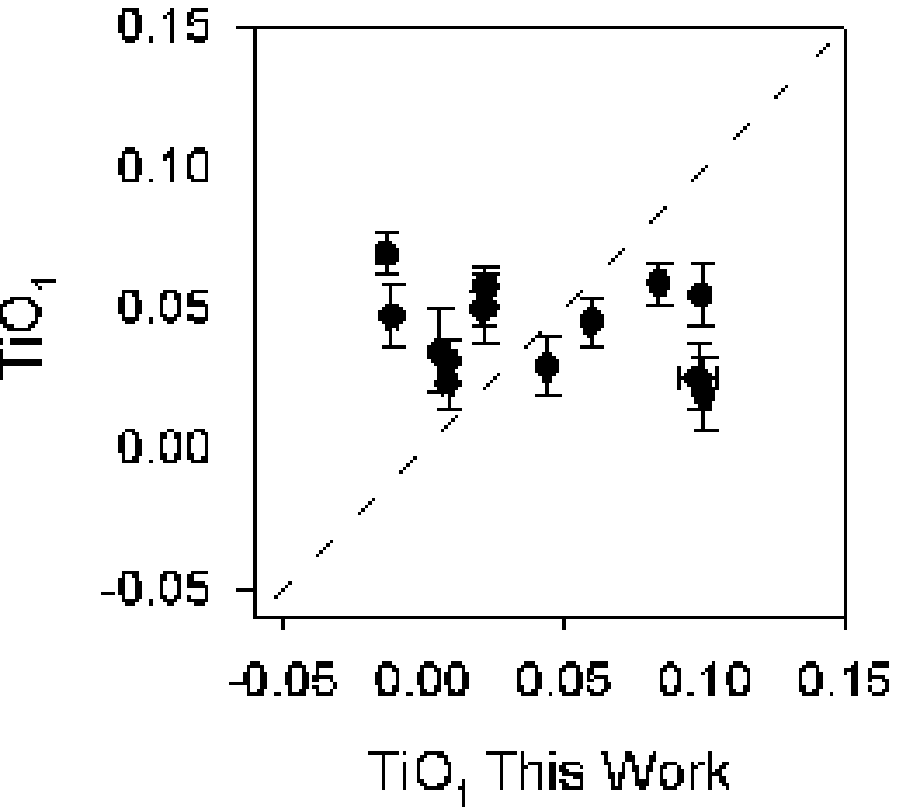}}}
\caption[Central Index Measurement Comparison]{Central index measurements compared to T98 and LW05 (black circles) for 14 galaxies in common, SB06 (red triangles) for five galaxies in common and OG08 (green circles) for five galaxies in common. All index measurements are in \AA{}, except CN, Mg and TiO which are in magnitudes.}
\label{Ind_Comp}
\end{figure*}

The central index measurements were also compared to those of SB06 for five galaxies in common (also indicated in Figure \ref{Ind_Comp}). For this comparison, the same central aperture of the galaxies was extracted, and 15 indices measured from the SB06 data. Nine index measurements for five galaxies were also compared with those measured by Ogando et al.\ (2008, hereafter OG08). The indices from OG08 were measured within a metric circular aperture equivalent to 1.19 kpc at the distance of the Coma cluster. No flux calibration was applied to these galaxies. No comparison between the six different observing runs (WHT, Gemini North 06B and 07A, Gemini South 06B, 07A and 07B) and T98, SB06 or OG08 could be made since there were too few galaxies in common. In total, index measurements of 15 galaxies (some of the galaxies being in common with more than one of the above-mentioned samples) could be compared, and are all shown in Figure \ref{Ind_Comp}. This comparison shows that the H$\delta_{\rm A}$ index measurements in the Gemini data are systematically higher (by 1 -- 2 \AA{}) compared with previous measurements, and shows a large amount of scatter. No other systematic differences were found. With the exception of the TiO$_{1}$ index, which was often very close to the edge of the spectrum in the BCG data, all the index comparison plots show scatter that is smaller than that expected from the errors on the index measurements.

\section{Notes on individual objects}
This section contains notes of some of the individual objects that could be compared with existing SSP-equivalent measurements from the literature.

\subsubsection*{IC1633}
The surface brightness profile of this galaxy was published by Schombert (1986), and HST imaging by Laine et al.\ (2003). The radial velocity profile reveals some evidence for substructure at the centre of this galaxy (Paper 1). The SSP equivalent parameters derived for the central $\frac{1}{8}$R$_{\rm e}$ in this study ($\log$ (age) = 0.935 $\pm$ 0.187; [E/Fe] = 0.440 $\pm$ 0.039; [Z/H] = 0.380 $\pm$ 0.061) could be compared with those presented in T05 for the central $\frac{1}{10}$R$_{\rm e}$ ($\log$ (age) = 0.716 $\pm$ 0.308; [E/Fe] = 0.366 $\pm$ 0.024; [Z/H] = 0.563 $\pm$ 0.057).

\subsubsection*{NGC1399}
The surface brightness profile was published by Schombert (1986). Paper 1 showed a steep decreasing velocity dispersion profile and a flat radial velocity profile with a small dip in the centre. Lyubenova, Kuntschner $\&$ Silva (2008) were able to investigate the centres of the kinematic profiles of NGC1399 in much more detail, and interpreted this dip ($r \leqslant 0.2$ arcsec) from high resolution $K$-band maps of the central kinematics as a dynamically cold subsystem in the centre. This cold subsystem could be a central stellar disc or globular cluster having fallen into the centre on a purely radial orbit (Lyubenova et al.\ 2008). An age of 8.6 $\pm$ 0.9 Gyr was found for this galaxy in this study, which is consistent with the age of 10 $\pm$ 2 Gyr found by Forbes et al.\ (2001). 

\subsubsection*{NGC2832}
The surface brightness profiles were published by Schombert (1986) and Jord\'an et al.\ (2004), and HST imaging by Laine et al.\ (2003). The radial velocity profile shows the presence of a KDC in the centre of this galaxy (Paper 1). The SSP equivalent parameters derived for the central $\frac{1}{8}$R$_{\rm e}$ in this study ( $\log$ (age) = 0.929 $\pm$ 0.024; [E/Fe] = 0.380 $\pm$ 0.043; [Z/H] = 0.480 $\pm$ 0.062) are in very good agreement with that derived by other authors. Proctor $\&$ Sansom (2002) derived $\log$ (age) = 0.875 $\pm$ 0.069; [E/Fe] = 0.300 $\pm$ 0.025; [Z/H] = 0.457 $\pm$ 0.074 for the central 3.6 $\times$ 1.25 arcsec$^{2}$. S\'{a}nchez-Bl\'{a}zquez et al.\ (2006b) derived $\log$ (age) = 0.995 $\pm$ 0.007 in the centre (equivalent aperture of 4'' at a redshift of $z=0.016$).

\subsubsection*{NGC4839}
Surface brightness profiles were published by Schombert (1986), Oemler (1976) as well as Jord\'an et al.\ (2004), who all confirm the presence of a very prominent cD envelope. Rotation of the order 44 km s$^{-1}$ and a KDC in the centre of the galaxy are detected (Paper 1). Table \ref{Toorberg} shows the SSP-equivalent ages and metallicities for the three BCGs in the Coma cluster from the literature (as summarised by Trager et al.\ 2008). The data from this study is for $\frac{1}{8}$R$_{\rm e}$. For Mehlert et al.\ (2000), S\'{a}nchez-Bl\'{a}zquez et al.\ (2006b) and Trager et al.\ (2008) the data is for 2.7 arcsec diameter equivalent circular apertures (by Trager et al.\ 2008), and for the other references as originally published. It can be seen that the literature SSP-equivalent age falls in a very large range, and the value derived in this study is within this range.

\subsubsection*{NGC4874} 
This galaxy has a large, extended envelope, and is the second brightest galaxy of the famous pair of BCGs at the centre of the Coma cluster. The surface brightness profile was published by Peletier et al.\ (1990). No significant rotation or velocity substructure is detected for this galaxy (Paper 1). Trager et al.\ (2008) compiled a list of all the central H$\beta$ measurements for this galaxy, and found an error weighted mean value of 1.57 $\pm$ 0.05 \AA{} (including data from Fisher et al.\ 1995; Trager et al.\ 1998; J\o{}rgensen 1999; Kuntschner et al.\ 2001; Moore et al.\ 2002; Nelan et al.\ 2005; S\'{a}nchez-Bl\'{a}zquez et al.\ 2006b). The H$\beta$ index measured here (1.630 $\pm$ 0.082 \AA{}) compares very well with the literature data. Table \ref{Toorberg} shows the SSP-equivalent parameters compared with those from the literature.

\begin{table*}
\centering
\begin{tabular}{l c c c}
\hline \multicolumn{4}{c}{NGC4889} \\
 Reference & $\log$ (age) & [Z/H] & [E/Fe] \\
\hline This work & 0.92$^{+0.04}_{-0.04}$ & 0.57$^{+0.05}_{-0.05}$ & 0.42$^{+0.04}_{-0.04}$ \\
S\'{a}nchez-Bl\'{a}zquez et al.\ 2006b & 1.08$^{+0.27}_{-0.08}$ & 0.30$^{+0.13}_{-0.14}$ & 0.18$^{+0.04}_{-0.04}$ \\
J\o{}rgensen 1999 & 0.23$^{+0.05}_{-0.01}$ & 1.07$^{+0.08}_{-0.04}$ & 0.29$^{+0.01}_{-0.01}$\\
Mehlert et al.\ 2000 & 0.12$^{+0.16}_{-0.01}$ & 1.16$^{+0.02}_{-0.19}$ & 0.39$^{+0.01}_{-0.13}$\\
Moore et al.\ 2002 & 0.14$^{+0.02}_{-0.01}$ & 1.13$^{+0.02}_{-0.01}$ & 0.38$^{+0.04}_{-0.04}$\\
Nelan et al.\ 2005 & 1.35$^{+0.02}_{-0.01}$ & --0.18$^{+0.05}_{-0.07}$ & 0.27$^{+0.01}_{-0.01}$\\
\hline Mean $\pm$ std. err. & 0.64 $\pm$ 0.54 & 0.68 $\pm$ 0.54 & 0.32 $\pm$ 0.09 \\ 
\hline \multicolumn{4}{c}{NGC4874} \\
 Reference & $\log$ (age) & [Z/H] & [E/Fe] \\
\hline This work & 0.89$^{+0.12}_{-0.12}$ & 0.35$^{+0.05}_{-0.05}$ & 0.46$^{+0.05}_{-0.05}$ \\
S\'{a}nchez-Bl\'{a}zquez et al.\ 2006b & 1.02$^{+0.20}_{-0.14}$ & 0.21$^{+0.11}_{-0.13}$ & 0.10$^{+0.04}_{-0.04}$\\
J\o{}rgensen 1999 & 0.31$^{+0.23}_{-0.10}$ & 0.86$^{+0.14}_{-0.16}$ & 0.25$^{+0.04}_{-0.03}$ \\
Mehlert et al.\ 2000 & 0.12$^{+0.02}_{-0.01}$ & 1.09$^{+0.02}_{-0.01}$ & 0.35$^{+0.04}_{-0.04}$\\
Moore et al.\ 2002 & 0.62$^{+0.29}_{-0.33}$ & 0.54$^{+0.17}_{-0.16}$ & 0.18$^{+0.04}_{-0.04}$\\
Nelan et al.\ 2005 & 1.11$^{+0.25}_{-0.22}$ & 0.08$^{+0.25}_{-0.22}$ & 0.18$^{+0.11}_{-0.08}$ \\
Trager et al.\ 2008 & 0.90$^{+0.05}_{-0.04}$ & 0.38$^{+0.04}_{-0.01}$ & 0.17$^{+0.01}_{-0.01}$\\
\hline Mean $\pm$ std. err. & 0.71 $\pm$ 0.37 & 0.50 $\pm$ 0.36 & 0.24 $\pm$ 0.13 \\
\hline \multicolumn{4}{c}{NGC4839} \\
 Reference & $\log$ (age) & [Z/H] & [E/Fe] \\
\hline This work &  1.07$^{+0.12}_{-0.12}$ & 0.13$^{+0.05}_{-0.05}$ & 0.35$^{+0.03}_{-0.03}$\\
S\'{a}nchez-Bl\'{a}zquez et al.\ 2006b & 0.66$^{+0.42}_{-0.39}$ & 0.63$^{+0.27}_{-0.23}$ & 0.14$^{+0.07}_{-0.06}$\\
J\o{}rgensen 1999 & 1.24$^{+0.19}_{-0.75}$ & 0.05$^{+0.22}_{-0.17}$ & 0.20$^{+0.07}_{-0.07}$\\ 
Mehlert et al.\ 2000 &  1.07$^{+0.34}_{-0.75}$ & 0.21$^{+0.22}_{-0.25}$ & 0.19$^{+0.08}_{-0.06}$\\
\hline Mean $\pm$ std. err. & 1.01 $\pm$ 0.25 & 0.26 $\pm$ 0.26 & 0.22 $\pm$ 0.09 \\
\hline
\end{tabular} 
\caption[SSP Parameters of the Coma BCG galaxies.]{SSP parameters of the Coma BCG galaxies compared to that in the literature (as summarised by Trager et al.\ 2008).}
\label{Toorberg}
\end{table*}

\subsubsection*{NGC4889}
A large BCG with a very extended envelope, and the brightest galaxy of the Coma cluster. The surface brightness profile was published by Peletier et al.\ (1990), and HST imaging by Laine et al.\ (2003). The radial velocity profile clearly shows a KDC (Paper 1). Table \ref{Toorberg} shows the SSP-equivalent parameters compared to that of the literature. Similarly to the other Coma BCGs, the ages fall in a very large range, and the value derived in this study is within this range.

\section{Central index measurements}

Table \ref{table:Indices} contains the index measurements from H$\delta_{\rm A}$ to C$_{2}$4668, and Table \ref{table:Indices_cont} from H$\beta$ to TiO$_{\rm 2}$ for all the galaxies. The complete tables are available as supplemetary material.

\begin{table*}
\begin{scriptsize}
\begin{tabular}{l c c c c c c c c c c c c}
\hline Object & H$\delta_{\rm A}$ & H$\delta_{\rm F}$ & CN$_{1}$ & CN$_{2}$ & Ca4227 & G4300 & H$\gamma_{\rm A}$ & H$\gamma_{\rm F}$ & Fe4383 & Ca4455 & Fe4531 & C$_{2}$4668 \\
Name & \AA{} & \AA{} & mag & mag & \AA{} & \AA{} & \AA{} & \AA{} & \AA{} & \AA{} & \AA{} & \AA{} \\
\hline
ESO146-028 & --2.582&	0.242&	0.102&	0.143&	1.375&	5.808&	--5.969&	--1.783&	4.193&	2.228&	3.562&	7.903\\
       & 0.407&	0.285&	0.009&	0.011&	0.145&	0.204&	0.314&	0.166&	0.314&	0.142&	0.159&	0.220\\	
ESO202-043 & --3.117&	0.394&	0.140&	0.189&	1.331&	5.861&	--6.102&	--1.999&	5.888&	2.293&	3.517&	7.892\\
       & 0.559&	0.396&	0.012&	0.013&	0.219&	0.439&	0.345&	0.198&	0.329&	0.165&	0.218&	0.287\\	
ESO303-005 & --3.642&	0.070&	0.130&	0.167&	1.385&	5.868&	--6.853&	--2.122&	5.256&	1.563&	3.796&	8.925\\	
       & 0.376&	0.250&	0.008&	0.010&	0.142&	0.216&	0.403&	0.214&	0.180&	0.135&	0.172&	0.259\\
ESO346-003 & --3.039&	--0.185&	0.067&	0.107&	1.359&	5.703&	--6.384&	--2.015&	5.672&	1.435&	3.151&	7.993\\	
       & 0.613&	0.401&	0.013&	0.015&	0.192&	0.291&	0.411&	0.255&	0.352&	0.176&	0.261&	1.109\\	
ESO349-010 & --3.574&	--0.478&	0.127&	0.154&	1.457&	5.595&	--5.977&	--1.449&	5.046&	1.697&	3.796&	8.665\\	
       & 0.241&	0.163&	0.005&	0.006&	0.100&	0.158&	0.282&	0.140&	0.170&	0.129&	0.160&	0.249\\	
\hline
\end{tabular}
\end{scriptsize}
\caption[Index measurements: H$\delta_{\rm A}$ to C$_{2}$4668.]{The index measurements from H$\delta_{\rm A}$ to C$_{2}$4668 for all 51 galaxies (49 BCG and 2 ellipticals). The first line for each galaxy is the index measurement, and the second is the error on the index measurement. The complete table is available as an on-line table.}
\label{table:Indices}
\end{table*}

\begin{table*}
\begin{scriptsize}
\begin{tabular}{l c c c c c c c c c c c c c}
\hline Object & H$\beta$ & Fe5015 & Mg$_{\rm 1}$ & Mg$_{\rm 2}$ & Mg$_{\rm b}$ & Fe5270 & Fe5335 & Fe5406 & Fe5709 & Fe5782 & NaD & TiO$_{\rm 1}$ & TiO$_{\rm 2}$\\
Name & \AA{} & \AA{} & mag & mag & \AA{} & \AA{} & \AA{} & \AA{} & \AA{} & \AA{} & \AA{} & mag & mag\\
\hline
ESO146-028&1.707&       5.570&	0.178&	0.336&	5.004&	2.923&	1.035&	1.832&	0.740&	0.944&	5.130&	0.009&	---\\ 
	 &0.075&	0.220&	0.001&	0.002&	0.099&	0.075&	0.083&	0.077&	0.034&	0.038&	0.039&	0.001&	---\\ 
ESO202-043&1.849&       5.325&	0.177&	0.346&	5.285&	3.061&	2.983&	2.684&	0.745&	1.011&	5.062&	--0.011&	---\\
	 &0.109&	0.227&	0.002&	0.002&	0.116&	0.106&	0.155&	0.099&	0.046&	0.047&	0.047&	0.002&	---\\
ESO303-005 &1.655&      5.942&	0.178&	0.309&	5.026&	2.890&	2.914&	2.034&	0.942&	1.030&	5.857&	0.057&	---\\
	 &0.097&	0.234&	0.002&	0.003&	0.122&	0.158&	0.162&	0.134&	0.073&	0.074&	0.088&	0.002&	---\\
ESO346-003 &1.691&      5.514&	0.160&	0.327&	4.626&	3.345&	2.899&	0.293&	0.711&	0.973&	5.254&	0.022&	---\\
	 &0.109&	0.229&	0.002&	0.002&	0.104&	0.101&	0.124&	0.085&	0.055&	0.047&	0.050&	0.001&	---\\
ESO349-010 &1.625&     5.890&  0.163&	0.290&	4.379&	2.908&	2.799&	1.948&	0.927&	0.842&	6.010&	0.054&	---\\
	 &0.092&	0.196&	0.002&	0.002&	0.171&	0.156&	0.193&	0.128&	0.066&	0.071&	0.077&	0.002&	---\\
\hline
\end{tabular}
\end{scriptsize}
\caption[Index measurements: H$\beta$ to TiO$_{\rm 2}$]{Index measurements: H$\beta$ to TiO$_{\rm 2}$ for all 51 galaxies (49 BCG and 2 ellipticals). The first line for each galaxy is the index measurement, and the second is the error on the index measurement. The complete table is available as an on-line table.}
\label{table:Indices_cont}
\end{table*}

\bsp

\label{lastpage}

\end{document}